\documentclass[numberedappendix]{emulateapj}

\def\bmath#1{{\bf #1}} 

\def\coltab#1{\colhead{\parbox{1cm}{\begin{center}#1\end{center}}}}

\def\vrho{\bmath{\rho}}
\def\vdel{\bmath{\nabla}}
\def\vPi{\bmath{\hat\Pi}}
\def\vGamma{\bmath{\Gamma}}
\def\gktheta{\vartheta}
\def\ephi{\varphi}

\def\phantfrac{\vphantom{1\over2}}
\def\order#1#2{\underbrace{\vphantom{\Biggl(}#2}_{\lefteqn{\bigcirc}{\hspace{0.325em}\bf #1}}}

\def\ADD#1{}
\def\DEL#1{}
\def\NB#1{}
\def\REM#1{}

\def\pseudofigureone#1#2#3#4{
\begin{figure}
\plotone{#3}
\caption{#4
\label{#1}}
\end{figure}}

\def\pseudofigurewide#1#2#3#4{
\begin{figure}
\plotone{#3}
\caption{#4
\label{#1}}
\end{figure}}

\usepackage{apjfonts}
\usepackage{color}

\def\pseudofigureone#1#2#3#4{
\begin{figure}
\plotone{#2}
\caption{#4
\label{#1}\\\\}
\end{figure}}

\def\pseudofigurewide#1#2#3#4{
\begin{figure*}
\plotone{#2}
\caption{#4
\label{#1}\\}
\end{figure*}}

\journalinfo{The Astrophysical Journal Supplement Series, 182:310 (2009) 
{\rm [e-print~{\tt arXiv:0704.0044}]}
}

\def\secref#1{\hbox{\S\,\ref{#1}}}
\def\secsref#1#2{\secref{#1}, \secref{#2}}
\def\secsand#1#2{\secref{#1} and \secref{#2}}
\def\secsdash#1#2{\hbox{\S\S\,\ref{#1}-\ref{#2}}}
\def\apref#1{Appendix~\ref{#1}}

\def\apsand#1#2{Appendices \ref{#1} and \ref{#2}}
\def\exref#1{(\ref{#1})}
\def\exsand#1#2{(\ref{#1}) and~(\ref{#2})}
\def\exsdash#1#2{(\ref{#1}-\ref{#2})}
\def\eqref#1{Eq.~(\ref{#1})}
\def\eqsref#1{Eqs.~(\ref{#1})}
\def\eqsand#1#2{Eqs.~(\ref{#1}) and~(\ref{#2})}
\def\eqsdash#1#2{Eqs.~(\ref{#1}-\ref{#2})}
\def\Eqref#1{Equation~(\ref{#1})}
\def\Eqsref#1{Equations~(\ref{#1})}
\def\Eqsdash#1#2{Equations~(\ref{#1}-\ref{#2})}
\def\Eqsand#1#2{Equations~(\ref{#1}) and~(\ref{#2})}
\def\figref#1{Fig.~\ref{#1}}

\def\Figref#1{Figure~\ref{#1}}

\def\tabref#1{Table~\ref{#1}}

\def\const{{\rm const}}
\def\erf{{\rm erf\,}}

\def\beq{\begin{equation}}
\def\eeq{\end{equation}}
\def\bea{\begin{eqnarray}}
\def\eea{\end{eqnarray}}

\def\({\left(}
\def\){\right)}
\def\<{\langle}
\def\>{\rangle}
\def\lt{\left}
\def\rt{\right}
\def\bl{\bigl}
\def\br{\bigr}

\def\unity{\bmath{\hat I}}

\def\eps{\varepsilon}

\def\vr{\bmath{r}}
\def\vR{\bmath{R}}
\def\vx{\bmath{\hat x}}
\def\vy{\bmath{\hat y}}
\def\vz{\bmath{\hat z}}

\def\dd{\partial}
\def\vdperp{\vdel_{\perp}}
\def\dperp{\nabla_{\perp}}

\def\vb{\bmath{\hat b}} 
\def\zz{\parallel}
\def\dpar{{\dd\over\dd z}}
\def\Dpar{\vb\cdot\vdel}

\def\vk{{\bmath{k}}}
\def\kpar{k_{\zz}}
\def\kparo{k_{\parallel 0}}
\def\kparA{k_{\parallel A}}
\def\vkperp{\vk_{\perp}}
\def\kperp{k_{\perp}}
\def\kr{a}
\def\krsq{\alpha}

\def\Vsw{V_{\rm sw}}
\def\vVsw{\bmath{V}_{\rm sw}}

\def\vv{\bmath{v}}
\def\vw{\bmath{w}}
\def\vvperp{\vv_\perp}
\def\vperp{v_\perp}
\def\vpar{v_\zz}

\def\vthi{v_{{\rm th}i}}
\def\vthe{v_{{\rm th}e}}
\def\vths{v_{{\rm th}s}}
\def\vthsp{v_{{\rm th}s'}}

\def\vu{\bmath{u}}
\def\vueff{\vu_{\rm eff}}
\def\vuperp{\vu_\perp}
\def\vuperpe{\vu_{\perp e}}
\def\uperp{u_\perp}
\def\upar{u_\zz} 
\def\uzero{\upar^{(0)}}
\def\uone{\upar^{(1)}}
 
\def\upari{u_{\zz i}} 
\def\vuperpi{\vu_{\perp i}} 
\def\uparik{u_{\zz\vk i}} 
\def\upare{u_{\zz e}} 
\def\uparek{u_{\zz\vk e}} 
\def\zpar{z_\zz}

\def\vF{\bmath{F}}

\def\vE{\bmath{E}}

\def\Eperp{E_\perp}
\def\vEperp{\vE_\perp}
\def\vEperpk{\vE_{\perp\vk}}
\def\Epar{E_\parallel}

\def\vA{\bmath{A}} 
\def\vAperp{\vA_\perp} 
\def\vAperpk{\vA_{\perp\vk}} 
\def\Apar{A_\zz} 
\def\Apark{A_{\zz\vk}} 

\def\vB{\bmath{B}}
\def\dvB{\delta\vB}
\def\dB{\delta B}
\def\dvBperp{\delta\vB_\perp}
\def\dvBperpk{\delta\vB_{\perp\vk}}
\def\dBperp{\delta B_\perp}
\def\dBpar{\delta B_\zz}
\def\dBpark{\delta B_{\zz\vk}}

\def\dBzero{\dBpar^{(0)}}
\def\dBone{\dBpar^{(1)}}

\def\vj{\bmath{j}}
\def\vjperp{\vj_\perp}
\def\jpar{j_\zz}

\def\avchi{\<\chi\>_{\vR_s}}
\def\avchie{\<\chi\>_{\vR_e}}
\def\avchii{\<\chi\>_{\vR_i}}
\def\avchik{\<\chi\>_{\vR_s,\vk}}
\def\avchiik{\<\chi\>_{\vR_i,\vk}}
\def\avApari{\<\Apar\>_{\vR_i}}

\def\mfp{\lambda_{{\rm mfp}i}}
\def\mfpe{\lambda_{{\rm mfp}e}}

\def\nui{\nu_{ii}}
\def\nue{\nu_{ei}}
\def\nuie{\nu_{ie}}
\def\nuee{\nu_{ee}}
\def\nuss{\nu_{ss}}
\def\nuDi{\nu_D^{ii}}
\def\nuDei{\nu_D^{ei}}
\def\nuDs{\nu_D^{ss}}
\def\nuS{\nu_S}

\def\vU{\bmath{U}}
\def\Uperp{U_\perp}
\def\Upar{U_\zz}

\def\nupar{\nu_{\parallel i}}
\def\kappar{\kappa_{\parallel i}}
\def\kappare{\kappa_{\parallel e}}

\def\qi{q_i}
\def\qe{q_e}
\def\qs{q_s}
\def\qsp{q_{s'}}
\def\fMi{F_{0i}}
\def\fMe{F_{0e}}
\def\fMs{F_{0s}}
\def\fMsp{F_{0s'}}
\def\tdfi{\delta\!\tilde f_i}
\def\dfi{\delta f_i}
\def\dfe{\delta f_e}
\def\dfs{\delta f_s}
\def\dfzero{\tdfi^{(0)}}
\def\dfone{\tdfi^{(1)}}
\def\Ti{T_{0i}}
\def\Te{T_{0e}}
\def\Ts{T_{0s}}
\def\Tperp{T_{0\perp}}
\def\Tpar{T_{0\parallel}}
\def\ni{n_{0i}}
\def\ne{n_{0e}}
\def\ns{n_{0s}}
\def\nsp{n_{0s'}}
\def\pperp{p_\perp}
\def\ppar{p_\parallel}

\def\dn{\delta n} 
\def\dni{\delta n_i} 
\def\dTi{\delta T_i} 
\def\dTzero{\dTi^{(0)}}
\def\dTone{\dTi^{(1)}}
\def\dpi{\delta p_i} 
\def\dne{\delta n_e} 
\def\dnek{\delta n_{\vk e}} 
 
\def\dTe{\delta T_e} 
\def\dpe{\delta p_e} 
\def\dnzero{\dne^{(0)}}
\def\dnone{\dne^{(1)}}

\def\drho{\delta\rho}
\def\dpr{\delta p}
\def\ds{\delta s}
\def\dszero{\ds^{(0)}}

\def\Ws{W_s}
\def\Whi{W_{\hi}}
\def\Whe{W_{\he}}
\def\Wperp{W_{\rm AW}}
\def\Wpar{W_{\rm sw}}
\def\Wcompr{W_{\rm compr}}
\def\Is{I_s}
\def\Ii{I_i}
\def\Ie{I_e}

\def\hh{h}
\def\hs{h_s}
\def\hi{h_i}
\def\he{h_e}
\def\hk{h_\vk}
\def\hks{h_{s\vk}}
\def\hksp{h_{s'\vk}}
\def\hki{h_{i\vk}}
\def\hke{h_{e\vk}}
\def\hezero{h_e^{(0)}}
\def\heone{h_e^{(1)}}
\def\tH{h_{e,{\rm hom}}^{(0)}}

\def\dtcolls{\({\dd h_s\over\dd t}\)_{\rm c}}
\def\dtcolli{\({\dd h_i\over\dd t}\)_{\rm c}}
\def\dtcolle{\({\dd h_e\over\dd t}\)_{\rm c}}
\def\dC{C}
\def\Cpa{C_M}

\def\gi{g}
\def\gki{g_\vk}

\def\intvi{{1\over\ni}\int d^3\vv\,}
\def\intve{{1\over\ne}\int d^3\vv\,}
\def\intRs{\int d^3\vR_s\,}
\def\intRi{\int d^3\vR_i\,}
\def\intRe{\int d^3\vR_e\,}
\def\intr{\int d^3\vr\,}
\def\intv{\int d^3\vv}
\def\intvpar{\int d\vpar}
\def\Gn{G_n}
\def\GB{G_B}
\def\tGn{G^-}
\def\tGB{G^+}
\def\FM{F_M}

\def\ul{u_{\lambda}}
\def\upl{u_{\perp\lambda}}
\def\dBl{\delta B_{\lambda}}
\def\dBpl{\delta B_{\perp\lambda}}
\def\dBparl{\delta B_{\parallel\lambda}}
\def\dsl{\delta s_{\lambda}}
\def\epss{\eps_{s}}
\def\epsB{\eps_{\rm KAW}}
\def\epshel{\eps_H}
\def\epsh{\eps_{h}}

\def\taul{\tau_\lambda}

\def\lparl{l_{\parallel\lambda}}
\def\hl{h_{i\lambda}}
\def\hle{h_{e\lambda}}
\def\Psil{\Psi_{\lambda}}
\def\Phil{\Phi_{\lambda}}

\def\taur{\tau_{\rho_i}}
\def\taure{\tau_{\rho_e}}
\def\tKAW{\tau_{{\rm KAW}\lambda}}
\def\th{\tau_{h\lambda}}

\def\Do{{\rm Do}}

\def\uo{U}
\def\lf{L}
\def\lo{l_0}

\def\Pa{P_{\rm ext}}
\def\vja{\vj_{\rm ext}}
\def\vjak{\vj_{{\rm ext},\vk}}

\slugcomment{Submitted April 1, 2007; accepted February 21, 2009; published May 6, 2009}
\shorttitle{KINETIC TURBULENCE IN MAGNETIZED PLASMAS} 
\shortauthors{SCHEKOCHIHIN ET AL.}

\begin{document}

\title{ASTROPHYSICAL GYROKINETICS: KINETIC AND FLUID TURBULENT CASCADES 
IN MAGNETIZED WEAKLY COLLISIONAL PLASMAS} 
\author{A.~A.~Schekochihin,\altaffilmark{1,2}
S.~C.~Cowley,\altaffilmark{2,3} 
W.~Dorland,\altaffilmark{4}
G.~W.~Hammett,\altaffilmark{5}
G.~G.~Howes,\altaffilmark{6}
E.~Quataert,\altaffilmark{7} 
and T.~Tatsuno\altaffilmark{4}}
\email{a.schekochihin1@physics.ox.ac.uk}
\altaffiltext{1}{Rudolf Peierls Centre for Theoretical Physics, 
University of Oxford, Oxford~OX1~3NP, UK.}
\altaffiltext{2}{Plasma Physics, 
Blackett Laboratory, Imperial College, London~SW7~2AZ, UK.}
\altaffiltext{3}{Euratom/UKAEA Fusion Association, Culham Science Centre, Abington OX14 3DB, UK.}
\altaffiltext{4}{Department of Physics, 
University of Maryland, College Park, MD~20742-3511.}
\altaffiltext{5}{Princeton Plasma Physics Laboratory, 
Princeton, NJ~08543-0451.}
\altaffiltext{6}{Department of Physics and Astronomy, 
University of Iowa, Iowa City, IA~52242-1479.} 
\altaffiltext{7}{Department of Astronomy, 
University of California, Berkeley, CA 94720-3411.} 

\begin{abstract}
This paper presents a theoretical framework for understanding plasma 
turbulence in astrophysical plasmas. It is motivated by 
observations of electromagnetic and density fluctuations in the solar wind, 
interstellar medium 
and galaxy clusters, as well as by models of particle heating in accretion disks. 
All of these plasmas and many others have turbulent motions at weakly 
collisional and collisionless scales. 
The paper focuses on turbulence in a strong mean magnetic field. 
The key assumptions are that the turbulent fluctuations are small compared 
to the mean field, spatially anisotropic with respect 
to it and that their frequency is low compared to the ion cyclotron frequency. 
The turbulence is assumed to be forced at some system-specific outer scale. 
The energy injected at this scale has to be dissipated into heat, 
which ultimately cannot be accomplished without collisions. 
A {\em kinetic cascade} develops 
that brings the energy to collisional scales both in space 
and velocity. The nature of the kinetic cascade in various 
scale ranges depends on the physics of plasma fluctuations 
that exist there. There are four special scales that separate 
physically distinct regimes: the electron and ion gyroscales, 
the mean free path and the electron diffusion scale. 
In each of the scale ranges separated by these scales, the fully kinetic problem 
is systematically reduced to a more physically transparent and 
computationally tractable system of equations, which 
are derived in a rigorous way. In the {\em ``inertial range''} 
above the ion gyroscale, the kinetic cascade separates into two parts: 
a cascade of Alfv\'enic fluctuations and a passive cascade of density and 
magnetic-field-strength fluctuations. The former are 
governed by the Reduced Magnetohydrodynamic 
(RMHD) equations at both the collisional and collisionless scales; 
the latter obey a linear kinetic equation along the (moving) 
field lines associated with the Alfv\'enic component (in the 
collisional limit, these compressive fluctuations become 
the slow and entropy modes of the conventional MHD). 
In the {\em ``dissipation range''} 
below ion gyroscale, there are again two cascades: 
the kinetic-Alfv\'en-wave (KAW) cascade governed by two fluid-like 
Electron Reduced Magnetohydrodynamic (ERMHD) equations and a passive 
cascade of ion entropy fluctuations both in space and velocity. 
The latter cascade brings the energy of the inertial-range 
fluctuations that was Landau-damped 
at the ion gyroscale to collisional scales in the phase space 
and leads to ion heating. 
The KAW energy is similarly damped at the electron gyroscale 
and converted into electron heat. 
Kolmogorov-style scaling relations are derived for all of these 
cascades. The relationship between the theoretical models 
proposed in this paper and astrophysical applications 
and observations is discussed in detail. 
\end{abstract}


\section{Introduction}
\label{sec_intro}

As observations of velocity, density and magnetic fields in astrophysical 
plasmas probe ever smaller scales, turbulence---i.e., 
broadband disordered fluctuations usually characterized 
by power-law energy spectra---emerges as a fundamental and ubiquitous feature. 

One of the earliest examples of observed turbulence in space was 
the detection of a Kolmogorov $k^{-5/3}$ spectrum of magnetic 
fluctuations in the solar wind over a frequency range of about three decades 
(first reported by \citealt{Matthaeus_Goldstein,Bavassano_etal} and confirmed 
to a high degree of accuracy by a multitude of subsequent observations,  
e.g., \citealt{Marsch_Tu_z,Horbury_etal96,Leamon_etal98,Bale_etal}; 
see \figref{fig_bale}). Another famous example in which 
the Kolmogorov power law appears to hold is the electron density spectrum in 
the interstellar medium (ISM)---in this case it emerges from observations 
by various methods in several scale intervals and, when these are pieced 
together, the power law famously extends over as many as 12 decades 
of scales \citep{Armstrong_Cordes_Rickett,Armstrong_Rickett_Spangler,Lazio_etal_review}, 
a record that has earned it the name of ``the Great Power Law in the Sky.'' 
Numerous other measurements in space and astrophysical plasmas, from the 
magnetosphere to galaxy clusters, result in Kolmogorov (or consistent with Kolmogorov) 
spectra but also show steeper power laws at very small (microphysical) scales (these 
observations are discussed in more detail in \secref{sec_astro}). 

Power-law spectra spanning broad bands of scales are symptomatic of the fundamental role 
of turbulence as a mechanism of transferring energy from the {\em outer scale(s)} 
(henceforth denoted $\lf$), where the energy is injected to the {\em inner scale(s)}, 
where it is dissipated. As these scales tend to be widely separated in astrophysical systems, 
one way for the system to bridge this scale gap is to fill it with fluctuations; 
the power-law spectra then arise due to scale invariance at the intermediate scales. 
Besides being one of the more easily measurable characteristics of the multi-scale 
nature of turbulence, power-law (and, particularly, Kolmogorov) spectra evoke 
a number of fundamental physical ideas that lie at the 
heart of the turbulence theory: universality of 
small-scale physics, energy cascade, locality of interactions, etc. 
In this paper, we shall revisit and generalize these ideas for the problem of 
{\em kinetic} plasma turbulence,\footnote{An outline of a Kolmogorov-style approach 
to kinetic turbulence was given in a recent paper by \citet{SCDHHPQT_crete}. It can be read as a conceptual 
introduction to the present paper, which is much more detailed and covers a much broader set of topics.} 
so it is perhaps useful to remind the reader how they 
emerge in a standard argument that leads to the $k^{-5/3}$ spectrum 
\citep{K41,Obukhov_K41}. 

\subsection{Kolmogorov Turbulence}
\label{sec_K41}

Suppose the average energy per unit 
time per unit volume that the system dissipates is $\varepsilon$. 
This energy has to be transferred from some (large) outer scale 
$\lf$ at which it is injected to some (small) inner scale(s) 
at which the dissipation occurs (see \secref{sec_scales}). 
It is assumed that in the range of scales intermediate between 
the outer and the inner (the {\em inertial range}), the statistical 
properties of the turbulence are universal (independent of the macrophysics of 
injection or of the microphysics of dissipation), spatially homogeneous 
and isotropic and the energy transfer is local in scale space. 
The flux of kinetic energy through any inertial-range scale $\lambda$ is 
independent of~$\lambda$:
\bea
{\ul^2\over\taul}\sim\varepsilon = \const,
\label{const_flux}
\eea
where the (constant) density of the medium is absorbed into $\varepsilon$, 
$\ul$ is the typical velocity fluctuation associated with 
the scale $\lambda$, and $\taul$ is the cascade time.\footnote{This is the version 
of Kolmogorov's theory due to \citealt{Obukhov_K41}.} 
Since interactions are assumed local, $\taul$ must be 
expressed in terms of quantities associated with scale $\lambda$. 
It is then dimensionally inevitable that $\taul\sim\lambda/\ul$ 
(the nonlinear interaction time, or turnover time), 
so we get
\bea
\ul\sim (\varepsilon\lambda)^{1/3}. 
\eea
This corresponds to a $k^{-5/3}$ spectrum of kinetic energy. 

\subsection{MHD Turbulence and Critical Balance}
\label{sec_GS}

That astronomical data appear to point to a 
ubiquitous nature of what, in its origin, is a dimensional 
result for the turbulence in a neutral fluid, 
might appear surprising. Indeed, the astrophysical plasmas 
in question are highly conducting and support magnetic fields 
whose energy is at least comparable to the kinetic energy of the 
motions. Let us consider a situation where the plasma is threaded 
by a uniform dynamically strong magnetic field $B_0$ 
(the {\em mean}, or {\em guide, field}; see \secref{sec_two_regimes} 
for a brief discussion of the validity of this assumption). 
In the presence of such a field, there is no dimensionally 
unique way of determining the cascade time $\taul$ because 
besides the nonlinear interaction time $\lambda/\ul$, 
there is a second characteristic time associated with 
the fluctuation of size $\lambda$, namely the 
Alfv\'en time $\lparl/v_A$, where $v_A$ is the 
Alfv\'en speed and $\lparl$ is the typical scale 
of the fluctuation along the magnetic field. 

The first theories of magnetohydrodynamic (MHD) turbulence 
\citep{Iroshnikov,Kraichnan,Dobrowolny_Mangeney_Veltri} 
calculated $\taul$ by assuming 
an isotropic cascade ($\lparl\sim\lambda$) of 
weakly interacting Alfv\'en-wave packets ($\taul\gg\lparl/v_A$) 
and obtained a $k^{-3/2}$ spectrum. 
The failure of the observed spectra to conform to this law 
(see references above) and especially the observational (see references 
at the end of this subsection) and 
experimental \citep{Robinson_Rusbridge,Zweben_Menyuk_Taylor}
evidence of anisotropy of MHD fluctuations 
led to the isotropy assumption being discarded \citep{Montgomery_Turner}.

\pseudofigureone{fig_bale}{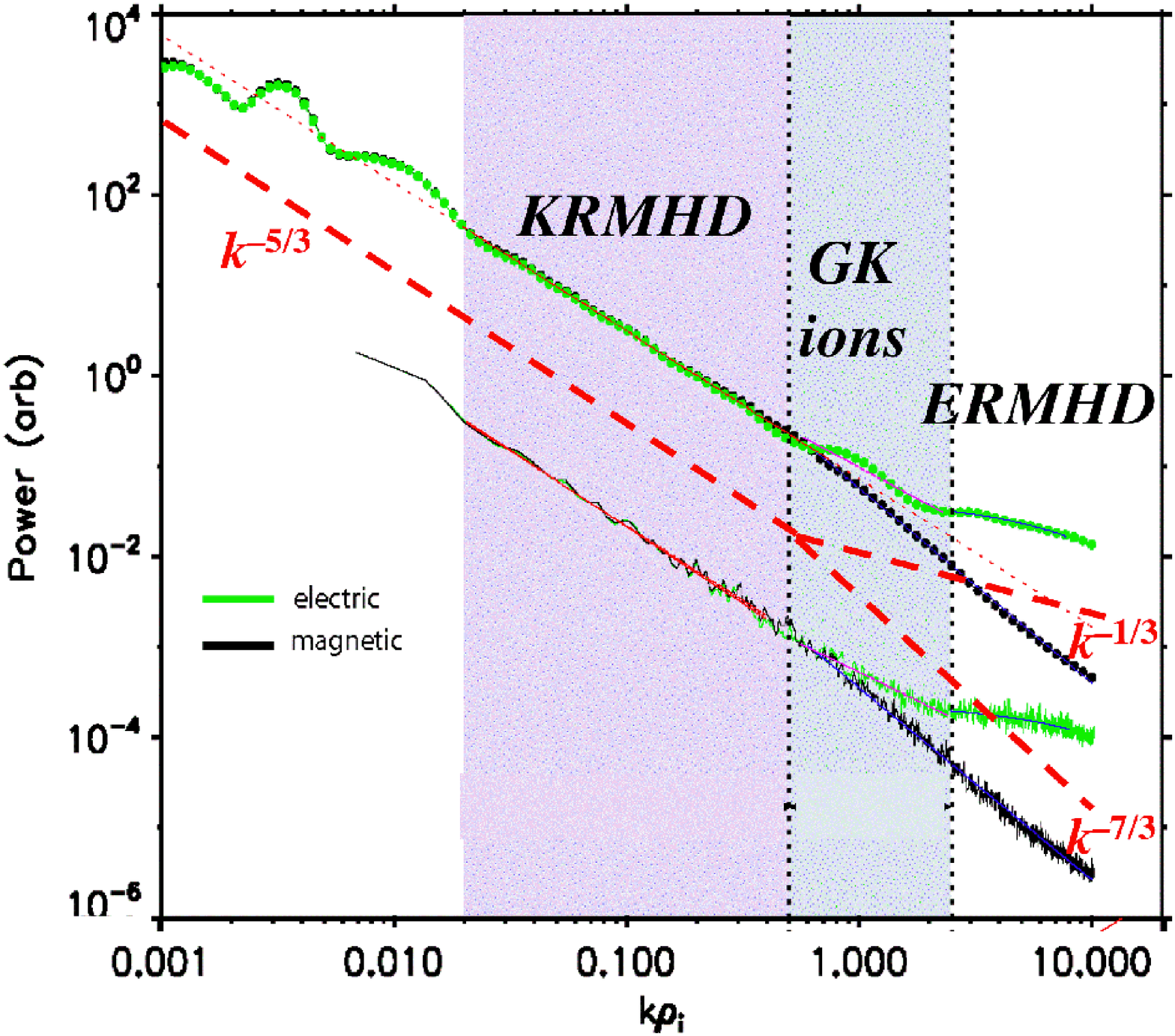}{f1.ps}{Spectra of electric and magnetic fluctuations 
in the solar wind at 1~AU (see \tabref{tab_scales} for the solar-wind parameters 
corresponding to this plot). This figure is adapted with permission from 
Fig.~3 of \citet{Bale_etal} (copyright 2005 by the American Physical Society). 
We have added the reference slopes for Alfv\'en-wave 
and kinetic-Alfv\'en-wave turbulence in bold dashed (red) lines and labeled ``KRMHD,''
``GK ions,'' and ``ERMHD'' the wavenumber intervals where these analytical descriptions 
are valid (see \secref{sec_GK}, \secref{sec_KRMHD} and \secref{sec_ERMHD}).}

The modern form of MHD turbulence theory is commonly associated 
with the names of \citet[][henceforth, GS]{GS95,GS97}. 
It can be summarized as follows. Assume that 
(a) all electromagnetic perturbations are strongly anisotropic, 
so that their characteristic scales along the mean field 
are much larger than those across it, $\lparl\gg\lambda$, 
or, in terms of wavenumbers, $\kpar\ll\kperp$; 
(b) the interactions between the Alfv\'en-wave packets are 
strong and the turbulence at sufficiently small scales always arranges itself in such a way 
that the Alfv\'en timescale and the perpendicular nonlinear interaction timescale 
are comparable to each other, i.e., 
\bea
\label{crit_bal}
\omega\sim\kpar v_A\sim\kperp\uperp,
\eea
where $\omega$ is the typical frequency of the fluctuations 
and $\uperp$ is the velocity fluctuation perpendicular to the mean field. 
Taken scale by scale, this assumption, known as the {\em critical balance}, 
removes the dimensional ambiguity of the MHD turbulence theory. 
Thus, the cascade time is $\taul\sim {\lparl/v_A} \sim \lambda/\ul$, whence
\bea
\label{GS_scaling}
\ul&\sim&(\eps\lparl/v_A)^{1/2}\sim \(\eps\lambda\)^{1/3},\\
\lparl&\sim&\lo^{1/3}\lambda^{2/3},
\label{GS_aniso} 
\eea
where $\lo=v_A^3/\eps$.
The scaling relation \exref{GS_scaling} is equivalent to 
a $\kperp^{-5/3}$ spectrum of kinetic energy,
while \eqref{GS_aniso} quantifies the anisotropy by 
establishing the relationship between the perpendicular and parallel scales. 
Note that \eqref{GS_scaling} implies 
that in terms of the parallel wavenumbers, 
the kinetic-energy spectrum is $\sim\kpar^{-2}$. 

The above considerations apply to Alfv\'enic fluctuations, 
i.e., {\em perpendicular} velocities and magnetic-field perturbations 
from the mean given (at each scale) 
by $\dBperp\sim\uperp\sqrt{4\pi\rho_0}$, where $\rho_0$ is the mean mass density
of the plasma (see \figref{fig_bale} and discussion in \secref{sec_SW_Alfvenic}). 
Other low-frequency MHD modes---slow waves and the entropy mode---turn out 
to be passively advected by the Alfv\'enic component of the turbulence 
\citep[this follows from the anisotropy; see][and \secsdash{sec_sw_fluid}{sec_scalings_passive}, 
\secref{sec_sw}, and \secref{sec_par_cascade} for 
further discussion of the compressive fluctuations]{Lithwick_Goldreich}. 

As we have mentioned above, the anisotropy was, in fact, 
incorporated into MHD turbulence theory already by \citet{Montgomery_Turner}. 
However, these authors' view differed from the GS theory in that they 
thought of MHD turbulence as essentially two dimensional, described 
by a Kolmogorov-like cascade \citep{Fyfe_Joyce_Montgomery}, with an admixture 
of Alfv\'en waves having some spectrum in $\kpar$ unrelated to the perpendicular 
structure of the turbulence (note that \citealt{Higdon}, while adopting a similar 
view, anticipated the scaling relation \exref{GS_aniso}, but did not seem to 
consider it to be anything more than the confirmation of an essentially 2D 
nature of the turbulence). In what we are referring to here as GS turbulence, 
the 2D and Alfv\'enic fluctuations are not separate components of the turbulence. 
The turbulence is three dimensional, with correlations parallel and perpendicular 
to the (local) mean field related at each scale by the critical balance assumption. 

Indeed, intuitively, we cannot have $\kpar v_A\ll\kperp\uperp$:  
the turbulence cannot be any more 2D than 
allowed by the critical balance because fluctuations in any two planes perpendicular 
to the mean field can only remain correlated if an Alfv\'en wave can 
propagate between them in less than their perpendicular decorrelation time. 
In the opposite limit, weakly interacting Alf\'en waves with fixed $\kpar$ 
and $\omega=\kpar v_A\gg\kperp\uperp$ can be shown to give rise to an 
energy cascade towards smaller perpendicular scales where the turbulence 
becomes strong and \eqref{crit_bal} is satisfied \citep{GS97,Galtier_etal,YSN_aw}.
Thus, there is a natural tendency towards critical balance in a system containing 
nonlinearly interacting Alfv\'en waves. We will see in what follows that critical 
balance may, in fact, be taken as a general physical principle relating 
parallel scales (associated with linear propagation) and perpendicular scales 
(associated with nonlinear interaction) in anisotropic plasma turbulence 
(see \secref{sec_KAW_turb}, \secref{sec_par_with_KAW}, \secref{sec_par_no_KAW}). 

We emphasize that, the anisotropy of astrophysical 
plasma turbulence is an observed phenomenon. It is seen most clearly 
in the spacecraft measurements of the turbulent fluctuations 
in the solar wind \citep{Belcher_Davis,Matthaeus_Goldstein_Roberts,Bieber_etal,Dasso_etal,Bigazzi_etal,SorrisoValvo_etal,Horbury_etal_review,Horbury_etal_aniso,Osman_Horbury,Hamilton_etal} 
and in the magnetosheath \cite{Sahraoui_etal,Alexandrova_msheath}. 
In a recent key development, solar-wind data analysis by \citet{Horbury_etal_aniso} 
approaches quantitative corroboration of the critical balance conjecture by 
confirming the scaling of the spectrum with the parallel 
wavenumber $\sim\kpar^{-2}$ that follows from the first 
scaling relation in \eqref{GS_scaling}. Anisotropy is also observed indirectly 
in the ISM \citep{Wilkinson_Narayan_Spencer,Trotter_Moran_Rodriguez,Rickett_etal_aniso,DennettThorpe_deBruyn}, including recently in molecular clouds \citep{Heyer_etal}, 
and, with unambiguous consistency, in numerical simulations of MHD turbulence 
\citep{Shebalin_Matthaeus_Montgomery,Oughton_Priest_Matthaeus,CV_aniso,Maron_Goldreich,CLV_aniso,Mueller_Biskamp_Grappin}.\footnote{The numerical evidence is much less clear on the scaling of the spectrum. 
The fact that the spectrum is closer to $\kperp^{-3/2}$ than to $\kperp^{-5/3}$ 
in numerical simulations 
\citep{Maron_Goldreich,Mueller_Biskamp_Grappin,Mason_Cattaneo_Boldyrev2,Perez_Boldyrev,Perez_Boldyrev_imb,Beresnyak_Lazarian2} 
prompted \citet{Boldyrev_spectrum2} to propose a scaling argument 
that allows an anisotropic Alfv\'enic turbulence with a $\kperp^{-3/2}$ spectrum. 
His argument is based on the conjecture that the fluctuating velocity 
and magnetic fields tend to partially align at small scales, 
an idea that has had considerable numerical support 
\citep{Maron_Goldreich,Beresnyak_Lazarian1,Beresnyak_Lazarian2,Mason_Cattaneo_Boldyrev,Matthaeus_etal08}. 
The alignment weakens nonlinear interactions and alters the scalings. 
Another modification of the GS theory leading to an anisotropic 
$\kperp^{-3/2}$ spectrum was proposed by \citet{Gogoberidze07}, who 
assumed that MHD turbulence with a strong mean field is dominated 
by non-local interactions with the outer scale. However, in both arguments, 
the basic assumption that the turbulence is strong 
is retained. This is the main assumption that we make 
in this paper: the critical balance conjecture~\exref{crit_bal} is used below 
not as a scaling prescription but in a weaker sense of an ordering assumption, 
i.e., we simply take the wave propagation terms in the equations to be comparable to the 
nonlinear terms. It is not hard to show that 
the results derived in what follows remain valid whether or not 
the alignment is present. We note that observationally, only in the solar wind 
does one measure the spectra with sufficient accuracy to state that they 
are consistent with $\kperp^{-5/3}$ but {\em not} with $\kperp^{-3/2}$ 
(see \secref{sec_SW_Alfvenic}). \label{fn_Boldyrev}}

\subsection{MHD Turbulence with and without a Mean Field}
\label{sec_two_regimes}

In the discussion above, treating MHD turbulence as turbulence of Alfv\'enic fluctuations 
depended on assuming the presence of a mean (guide) field $B_0$ that is strong 
compared to the magnetic fluctuations, $\dB/B_0\sim u/v_A \ll1$. 
We will also need this assumption 
in the formal developments to follow (see \secref{sec_RMHDordering}, \secref{sec_params}). 
Is it legitimate to expect that such a spatially regular field will 
be generically present? 
\citet{Kraichnan} argued that in a generic situation 
in which all magnetic fields are produced by the turbulence 
itself via the dynamo effect, one could assume that 
the strongest field will be at the outer scale and 
that this field will play the role of an (approximately) 
uniform guide field for the Alfv\'en waves in the inertial range. 
Formally, this amounts to assuming that in the inertial range, 
\bea
\label{outer_scale_assumption}
{\dB\over B_0}\ll1,\quad\kpar\lf\ll1.
\eea
It is, however, by no means obvious that this should 
be true. When a strong mean field is imposed by some 
external mechanism, the turbulent 
motions cannot bend it significantly, so only small 
perturbations are possible and $\dB\ll B_0$. 
In contrast, without a strong imposed field, the 
energy density of the magnetic fluctuations is at most 
comparable to the kinetic-energy density of the plasma motions, 
which are then sufficiently energetic to randomly tangle the 
field, so $\dB\gg B_0$. 

In the weak-mean-field case, the dynamically strong stochastic 
magnetic field is a result of saturation of the {\em small-scale, or fluctuation, 
dynamo}---amplification of magnetic field due to random stretching 
by the turbulent motions \citep[see review by][]{SC_mhdbook}. The definitive 
theory of this saturated state remains to be discovered. Both physical 
arguments and numerical evidence \citep{SCTMM_stokes,YRS_exact} suggest 
that the magnetic field in this case is organized in folded flux sheets (or ribbons). 
The length of these folds is comparable to the outer scale, 
while the scale of the field-direction reversals transverse to the fold 
is determined by the dissipation physics: in MHD 
with isotropic viscosity and resistivity, 
it is the resistive scale.\footnote{\label{fn_mag_cutoff}
In weakly collisional astrophysical plasmas, such a description is not applicable: 
the field reversal scale is most probably determined by more complicated 
and as yet poorly understood kinetic plasma effects; below this scale, 
an Alfv\'enic turbulence of the kind discussed in this paper may 
exist \citep{SC_dpp05}.} 
Although Alfv\'en waves 
propagating along the folds may exist \citep{SCTMM_stokes,SC_mhdbook}, 
the presence of the small-scale direction reversals means that there is no 
scale-by-scale equipartition between the velocity and magnetic fields: 
while the magnetic energy is small-scale dominated due to the 
direction reversals,\footnote{See \citet{HBD_pre} for an alternative view. 
Note also that the numerical evidence cited above pertains to 
{\em forced} simulations. In {\em decaying} MHD turbulence simulations, 
the magnetic energy does indeed appear to be at the outer scale \citep{Biskamp_Mueller}, 
so one might expect an Alfv\'enic cascade deep in the inertial range.} 
the kinetic energy should be contained primarily at the outer scale, with some 
scaling law in the inertial range. 

Thus, at the current level of understanding we have to assume that 
there are two asymptotic regimes of MHD turbulence: 
anisotropic Alfv\'enic turbulence with $\dB\ll B_0$ and 
isotropic MHD turbulence with small-scale field reversals and $\dB\gg B_0$. 
In this paper, we shall only discuss the first regime. The origin 
of the mean field may be external (as, e.g., in the solar wind, where 
it is the field of the Sun) or due to some form of {\em mean-field dynamo} 
(rather than small-scale dynamo), as usually expected for galaxies 
\citep[see, e.g.,][]{Shukurov_review}. 

Note finally that the condition $\dB\ll B_0$ need not be satisfied 
at the outer scale and in fact is not satisfied in most space or astrophysical 
plasmas, where more commonly $\dB\sim B_0$ at the outer scale. 
This, however, is sufficient 
for the Kraichnan hypothesis to hold and for an Alf\'enic cascade to be set 
up, so at small scales (in the inertial range and beyond),
the assumptions \exref{outer_scale_assumption} are satisfied.  

\subsection{Kinetic Turbulence}
\label{sec_kinetic}

The GS theory of MHD turbulence (\secref{sec_GS}) 
allows us to make sense of the magnetized turbulence 
observed in cosmic plasmas exhibiting the same statistical scaling 
as turbulence in a neutral fluid (although the underlying 
dynamics are very different in these two cases!). 
However, there is an aspect of the observed astrophysical turbulence 
that undermines the applicability of any type of fluid description: 
in most cases, the inertial range where the Kolmogorov scaling 
holds extends to scales far below the mean free path deep into the 
collisionless regime. For example, 
in the case of the solar wind, the mean free path is close to 1~AU, 
so all scales are collisionless---an extreme case, which 
also happens to be the best studied, thanks 
to the possibility of in situ measurements (see \secref{sec_astro}). 

The proper way of treating such plasmas 
is using kinetic theory, not fluid equations. 
The basis for the application of the MHD fluid description to them has been 
the following well known result from the linear theory of plasma waves: 
while the fast, slow and entropy modes 
are damped at the mean-free-path scale both by collisional 
viscosity \citep[][see \secref{sec_visc_diss}]{Braginskii}
and by collisionless wave--particle interactions 
\citep[][see \secref{sec_barnes}]{Barnes}, the Alfv\'en waves 
are only damped at the ion gyroscale. It has, therefore, been assumed 
that the MHD description, inasmuch as it concerns the 
Alfv\'en-wave cascade, can be extended to the ion gyroscale, 
with the understanding that this cascade is decoupled from 
the damped cascades of the rest of the MHD modes. This approach
and its application to the turbulence in 
the ISM are best explained by \citet{Lithwick_Goldreich}. 

While the fluid description may be sufficient to understand 
the Alfv\'enic fluctuations in the inertial range, it is certainly 
inadequate for everything else: the compressive fluctuations in the inertial 
range and turbulence in the dissipation range (below the ion gyroscale), 
where power-law spectra are also detected \citep[e.g.,][see also \figref{fig_bale}]{Denskat_Beinroth_Neubauer,Leamon_etal98,Czaykowska_etal,Smith_etal06,Sahraoui_etal,Alexandrova_sw,Alexandrova_msheath}. The fundamental challenge that 
a comprehensive theory of astrophysical plasma turbulence must 
meet is to give the full account of how the turbulent 
fluctuation energy injected at the outer scale is cascaded 
to small scales and deposited into particle heat. 
We shall see (\secsand{sec_en_GK}{sec_heating}) 
that the familiar concept of an energy cascade can be generalized 
in the kinetic framework as the {\em kinetic cascade} of a single 
quantity that we call the {\em generalized energy} 
\citep[see also][and references therein]{SCDHHPQT_crete}. The small scales developed 
in the process are small scales both in the position and velocity space. 
The fundamental reason for this is the low collisionality of 
the plasma: since heating cannot ultimately be accomplished without 
collisions, large gradients in phase space are necessary for the 
collisions to be effective.

The idea of a generalized energy cascade in phase space as the engine of 
kinetic plasma turbulence is the central concept of this paper. 
In order to understand the physics of the kinetic cascade in various 
scale ranges, we derive in what follows a hierarchy of simplified, yet rigorous, 
reduced kinetic, fluid and hybrid descriptions. While the full kinetic theory 
of turbulence is very difficult to handle either analytically or 
numerically, the models we derive are much more tractable. 
For all, the regimes of applicability (scale/parameter ranges, underlying assumptions) 
are clearly stated. In each of these regimes, 
the kinetic cascade splits into several channels of energy 
transfer, some of them familiar (e.g., the Alfv\'enic cascade, \secsand{sec_AW}{sec_AW_coll}), 
others conceptually new (e.g., the kinetic cascade of collisionless compressive 
fluctuations, \secref{sec_colless}, or 
the entropy cascade, \secsdash{sec_ent_KAW}{sec_ent_els}). 

So as to introduce this theoretical framework in a way that is both analytically 
systematic and physically intelligible, let us first consider the characteristic 
scales that are relevant to the problem of astrophysical turbulence (\secref{sec_scales}). 
The models we derive are previewed in \secref{sec_models}, at the end 
of which the plan of further developments is given. 

\subsection{Scales in the Problem}
\label{sec_scales}

\begin{deluxetable}{llllll}
\tablewidth{0pt}
\tablecaption{Representative Parameters for Astrophysical Plasmas.\label{tab_scales}}
\tablehead{
\coltab{Parameter} & 
\coltab{Solar\\ wind\\ at\\ 1~AU\tablenotemark{(a)}} &
\coltab{Warm\\ionized\\ ISM\tablenotemark{(b)}} & 
\coltab{Accretion\\ flow~near\\ Sgr~A$^*$\tablenotemark{(c)}} &
\coltab{Galaxy\\ clusters\\ (cores)\tablenotemark{(d)}} 
} 
\startdata
$n_e=n_i$, cm$^{-3}$     & $30$          & $0.5$         & $10^6$           & $6\times10^{-2}$\\
$T_e$, K                 & $\sim T_i$\tablenotemark{(e)} & $8000$ & $10^{11}$   & $3\times10^7$\\
$T_i$, K                 & $5\times10^5$ & $8000$        & $\sim10^{12}$\tablenotemark{(f)} & ?\tablenotemark{(e)}\\
$B$, G                   & $10^{-4}$     & $10^{-6}$     & $30$             & $7\times10^{-6}$\\
$\beta_i$                & $5$           & $14$          & $4$              & $130$\\\\
$\vthi$, km/s            & $90$          & $10$          & $10^5$           & $700$\\
$v_A$, km/s              & $40$          & $3$           & $7\times10^4$    & $60$\\
$\uo$, km/s\tablenotemark{(f)} & $\sim10$   & $\sim10$      & $\sim10^4$    & $\sim10^2$\\\\
$\lf$, km\tablenotemark{(f)}   & $\sim10^5$ & $\sim10^{15}$ & $\sim10^8$    & $\sim10^{17}$\\
$(m_i/m_e)^{1/2}\mfp$, km  
                         & $10^{10}$     & $2\times10^8$ & $4\times10^{10}$ & $4\times10^{16}$\\
$\mfp$, km\tablenotemark{(g)} & $3\times10^8$ & $6\times10^6$ & $10^9$      & $10^{15}$\\
$\rho_i$, km             & $90$          & $1000$        & 0.4              & $10^4$\\
$\rho_e$, km             & $2$           & $30$          & 0.003            & $200$
\enddata
\tablenotetext{a}{Values for slow wind (mean flow speed $\Vsw=350$~km/s in this case) 
measured by Cluster spacecraft and taken from \citet{Bale_etal}, except 
the value of $T_e$, which they do not report, but which is expected 
to be of the same order as $T_i$ \citep{Newbury_etal}. 
Note that the data interval studied by \citet{Bale_etal} 
is slightly atypical, with $\beta_i$ higher than usual 
in the solar wind (the full range of $\beta_i$ variation in the 
solar wind is roughly between $0.1$ and $10$; 
see \citealt{Howes_etal2} for another, perhaps more typical, 
fiducial set of slow-wind parameters
and Appendix A of the review by 
\citealt{Bruno_Carbone} for slow- and fast-wind parameters measured by Helios 2). 
However, we use their parameter values as our representative example because the 
spectra they report show with particular clarity both the electric and magnetic 
fluctuations in both the inertial and dissipation ranges (see \figref{fig_bale}).
See further discussion in \secsand{sec_SW_ir}{sec_SW_dr}.}
\tablenotetext{b}{Typical values \citep[see, e.g.,][]{Norman_Ferrara,Ferriere_review}. 
See discussion in \secref{sec_ISM}.}
\tablenotetext{c}{Values based on observational constraints for the 
radio-emitting plasma around the Galactic Center (Sgr~A$^*$) as interpreted 
by \citet{Loeb_Waxman} \citep[see also][]{Quataert_SgrA}. 
See discussion in \secref{sec_disks}.}
\tablenotetext{d}{Values for the core region of the Hydra A 
cluster taken from \citet{Ensslin_Vogt_cores}; see \citealt{SC_dpp05} for 
a consistent set of numbers for the hot plasmas outside the cores. 
See discussion in \secref{sec_clusters}.}
\tablenotetext{e}{We assume $T_i\sim T_e$ for these estimates.} 
\tablenotetext{f}{Rough order-of-magnitude estimate.}
\tablenotetext{g}{Defined $\mfp=\vthi/\nui$, where $\nui$ is given by 
\eqref{nui_def}.} 
\end{deluxetable}

\pseudofigurewide{fig_validity_reduced}{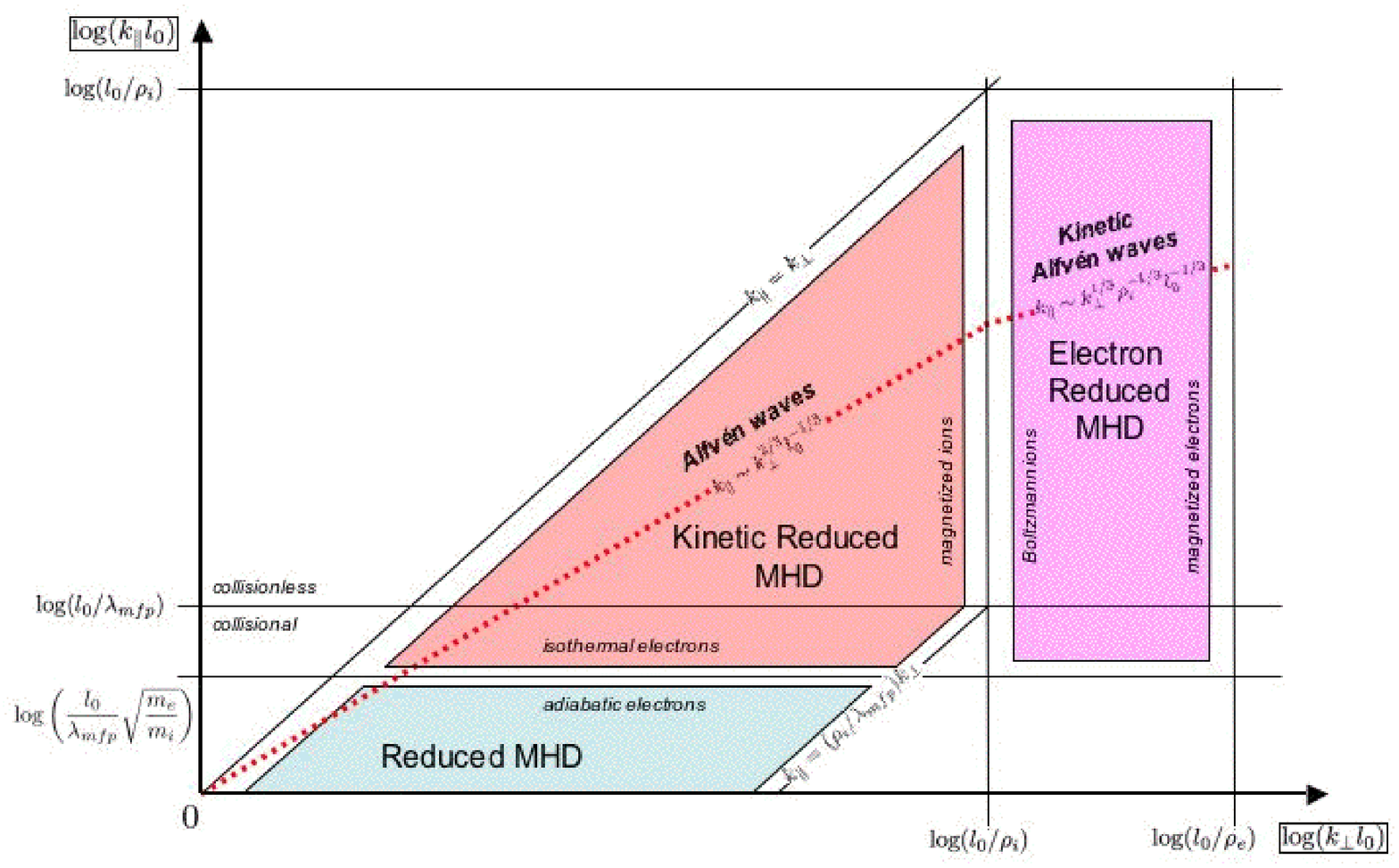}{f2.ps}{
Partition of the wavenumber space by characteristic scales. 
The wavenumbers are normalized by $\lo\sim v_A^3/\eps$, 
where $\eps$ is the total power input (see \secref{sec_GS}). 
Dotted line shows the path an Alfv\'en-wave cascade starting 
at the outer scale $\lf\sim\lo$ takes through the wavenumber space. 
We also show 
the regions of validity of the three tertiary approximations. 
They all require $\kpar\ll k_\perp$ (anisotropic fluctuations) 
and $\kpar\rho_i\ll1$ (i.e., $\kpar\vthi\ll\Omega_i$, low-frequency limit). 
Reduced MHD (RMHD, \secref{sec_RMHD}) is valid 
when $\kperp\rho_i \ll \kpar\mfp \ll(m_e/m_i)^{1/2}$ 
(strongly magnetized collisional limit, adiabatic electrons). 
The regions of validity of 
Kinetic Reduced MHD (KRMHD, \secref{sec_KRMHD}) 
and Electron Reduced MHD (ERMHD, \secref{sec_ERMHD}) 
lie within that of the isothermal electron/gyrokinetic ion 
approximation (\figref{fig_validity_isoth}) with the additional 
requirement that $\kperp\rho_i\ll\min(1, \kpar\mfp)$ 
(strongly magnetized ions) for KRMHD 
or $\kperp\rho_i\gg1$ (unmagnetized ions) for ERMHD. 
The collisional limit of KRMHD (\secref{sec_visc} and \apref{ap_visc}), 
$(m_e/m_i)^{1/2}\ll\kpar\mfp\ll1$, 
is similar to RMHD, except electrons are isothermal.
The dotted line is the scaling of $\kpar$ vs $\kperp$ 
from critical balance in both the Alfv\'en-wave 
[\secref{sec_GS}, \eqref{GS_aniso}] 
and kinetic-Alfv\'en-wave [\secref{sec_KAW_turb}, \eqref{KAW_aniso_scaling}] 
regimes.} 

\subsubsection{Outer Scale} 
\label{sec_outer_scale}

It is a generic feature of turbulent systems that energy is 
injected via some large-scale mechanism: ``large scale'' here 
means some scale (or a range of scales) 
comparable to the size of the system, depending on 
its global properties, and much larger than 
the microphysical scales at which energy is dissipated 
and converted into heat (\secref{sec_microscales}). Examples of large-scale stirring 
of turbulent fluctuations include 
the solar activity in the corona (launching Alfv\'en waves 
to produce turbulence in the solar wind); 
supernova explosions in the ISM 
\citep[e.g.,][]{Norman_Ferrara,Ferriere_review}; 
the magnetorotational instability in accretion disks 
\citep{Balbus_Hawley_review}; 
merger events, galaxy wakes and active galactic nuclei 
in galaxy clusters 
\citep[e.g.,][]{Subramanian_Shukurov_Haugen,Ensslin_Vogt_cores,Chandran_agns}. 
Since in this paper we are concerned with the local properties of astrophysical 
plasmas, let us simply assume that energy injection occurs at some 
characteristic {\em outer scale} $\lf$. All further considerations 
will apply to scales that are much smaller than $\lf$ and we 
will assume that the particular character of the energy injection 
does not matter at these small scales. 

In most astrophysical situations, one cannot assume that equilibrium 
quantities such as density, temperature, mean velocity and mean magnetic 
field are uniform at the outer scale. 
However, at scales much smaller 
than $\lf$, the gradients of the small-scale fluctuating fields are 
much larger than the outer-scale gradients (although the fluctuation 
amplitudes are much smaller; for the mean magnetic field, this assumption 
is discussed in some detail in \secref{sec_two_regimes}), 
so we may neglect the equilibrium gradients 
and consider the turbulence to be homogeneous. 
Specifically, this is a good assumption if $\kpar\lf\gg1$ [\eqref{outer_scale_assumption}], 
i.e., not only the perpendicular scales but also the much larger 
parallel ones are still shorter than the outer scale. 
Note that we cannot generally assume that the outer-scale energy 
injection is anisotropic, so the anisotropy is also the property 
of small scales only. 

\subsubsection{Microscales} 
\label{sec_microscales}
 
There are four microphysical scales that mark the transitions between 
distinct physical regimes: 

\paragraph{Electron diffusion scale.} At $\kpar\mfp(m_i/m_e)^{1/2}\gg1$, 
the electron response is isothermal (\secref{sec_dTe}, \apref{ap_isoth_els}). 
At $\kpar\mfp(m_i/m_e)^{1/2}\ll1$, it is adiabatic 
(\secref{sec_isoth_els}, \apref{ap_MHD}).\\ 

\paragraph{Mean free path.} At $\kpar\mfp\gg1$, the plasma 
is collisionless. In this regime, wave--particle interactions 
can damp compressive fluctuations 
via Barnes damping (\secref{sec_barnes}), so kinetic description 
becomes essential. At $\kpar\mfp\ll1$, the plasma is collisional and 
fluid-like (\secref{sec_visc}, \apsand{ap_Brag}{ap_visc}). 

\paragraph{Ion gyroscale.} At $\kperp\rho_i\ll1$, ions (as well as the 
electrons) are magnetized and the magnetic field is 
frozen into the ion flow (the $\vE\times\vB$ velocity field). 
At $\kperp\rho_i\sim1$, ions can exchange energy with electromagnetic 
fluctuations via wave--particle interactions 
(and ion heating eventually occurs via a kinetic ion-entropy cascade, see 
\secsdash{sec_ent_KAW}{sec_ent_no_KAW}). 
At $\kperp\rho_i\gg1$, the ions are unmagnetized 
and have a Boltzmann response (\secref{sec_ERMHD_eqns}). 
Note that the ion inertial scale $d_i= \rho_i/\sqrt{\beta_i}$ 
is comparable to the ion gyroscale unless the plasma beta  
$\beta_i=8\pi n_i T_i/B^2$ is very different from unity. 
In the theories developed below, $d_i$ does not play a special 
role except in the limit of $T_i\ll T_e$, which is not 
common in astrophysical plasmas 
(see further discussion in \secref{sec_transition} and \apref{ap_Hall}). 

\paragraph{Electron gyroscale.} 
At $\kperp\rho_e\ll1$, electrons are magnetized and 
the magnetic field is frozen into the electron flow 
(\secsref{sec_els}{sec_ERMHD}, \apref{ap_nongyro}).
At $\kperp\rho_e\sim1$, the electrons absorb 
the energy of the electromagnetic fluctuations 
via wave--particle interactions (leading to electron heating 
via a kinetic electron-entropy cascade, see \secref{sec_ent_els}).\\  

Typical values of these scales and of several other key 
parameters are given in \tabref{tab_scales}. In \figref{fig_validity_reduced}, 
we show how the wavenumber space, $(\kperp,\kpar)$, is divided by these 
scales into several domains, where the physics is different. 
Further partitioning of the wavenumber space results from comparing 
$\kperp\rho_i$ and $\kpar\mfp$ ($\kperp\rho_i\ll\kpar\mfp$ is 
the limit of strong magnetization, see \apref{ap_strongly_mag}) 
and, most importantly, from comparing parallel and perpendicular 
wavenumbers. As we explained above, observational and numerical 
evidence tells us that Alfv\'enic turbulence is anisotropic, 
$\kpar\ll\kperp$. In \figref{fig_validity_reduced}, 
we sketch the path the turbulent cascade is expected to take 
in the wavenumber space (we use the scalings of $\kpar$ with $\kperp$ 
that follow from the GS 
argument for the Alfv\'en waves and an analogous argument for 
the kinetic Alfv\'en waves, reviewed in \secsand{sec_GS}{sec_KAW_turb}, 
respectively). 

\subsection{Kinetic and Fluid Models} 
\label{sec_models}

What is the correct analytical description of 
the turbulent plasma fluctuations along the (presumed) path of the cascade? 
As we promised above, it is going to be possible to simplify 
the full kinetic theory substantially. These simplifications 
can be obtained in the form of a hierarchy of approximations 
and as these emerge, specific physical mechanisms that control 
the turbulent cascade in various physical regimes 
become more transparent. 

\paragraph{Gyrokinetics (\secref{sec_GK}).}
The starting point for these developments and the primary 
approximation in the hierarchy is {\em gyrokinetics}, 
a low-frequency kinetic theory resulting from 
averaging over the cyclotron motion of the particles. 
Gyrokinetics is appropriate 
for the study of subsonic plasma turbulence in virtually all 
astrophysically relevant parameter ranges \citep{Howes_etal}. 
For fluctuations at frequencies lower than the ion cyclotron 
frequency, $\omega\ll\Omega_i$, gyrokinetics can be systematically 
derived by making use of the following two assumptions, which 
also underpin the GS theory (\secref{sec_GS}): (a) anisotropy of the turbulence, so 
$\epsilon\sim \kpar/\kperp$ is used as the small parameter, 
and (b) strong interactions, i.e., the fluctuation amplitudes 
are assumed to be such that wave propagation and nonlinear interaction 
occur on comparable timescales: from \eqref{crit_bal}, 
$\uperp/v_A\sim\epsilon$. The first of these assumptions implies that 
fluctuations at Alfv\'enic frequencies satisfy $\omega\sim\kpar v_A\ll\Omega_i$
even when their perpendicular scale is such $\kperp\rho_i\sim1$. 
This makes gyrokinetics an ideal 
tool both for analytical theory and for numerical studies of 
astrophysical plasma turbulence; the numerical approaches are 
also made attractive by the long experience of gyrokinetic 
simulations accumulated in the fusion research and by the existence 
of publicly available gyrokinetic codes 
\citep{Kotschenreuther_Rewoldt_Tang,Jenko_etal,Candy_Waltz,Chen_Parker}.
A concise review of gyrokinetics is provided in \secref{sec_GK}
(see \citealt{Howes_etal} for a detailed derivation). 
The reader is urged to pay particular attention to 
\secsand{sec_en_GK}{sec_heating}, where the concept of 
the {\em kinetic cascade} of {\em generalized energy} is introduced 
and the particle heating in gyrokinetics is discussed 
(\apref{ap_inv} introduces additional conservation laws that 
arise in 2D and sometimes also in 3D). 
This establishes the conceptual framework in which most of the 
subsequent physical arguments are presented. 
The region of validity of gyrokinetics is illustrated in 
\figref{fig_validity_gk}: it covers virtually the entire 
path of the turbulent cascade, except the largest (outer) scales, 
where one cannot assume anisotropy.  
Note that the two-fluid theory, which is the starting 
point for the MHD theory (see \apref{ap_Brag}), 
is not a good description at collisionless scales. 
It is important to mention, however, that 
the formulation of gyrokinetics that we adopt, 
while appropriate for treating fluctuations at collisionless 
scales, does nevertheless require a certain (weak) degree of collisionality 
(see discussion in \secref{sec_order_coll} and an extended treatment 
of collisions in gyrokinetics in \apref{ap_coll}). 

\paragraph{Isothermal Electron Fluid (\secref{sec_els}).}
While gyrokinetics constitutes a significant simplification, 
it is still a fully kinetic description. Further progress towards 
simpler models is achieved by showing that, for parallel scales 
smaller than the electron diffusion scale, $\kpar\mfp\gg(m_e/m_i)^{1/2}$, 
and perpendicular scales larger than the electron gyroscale, $\kperp\rho_e\ll1$, 
the electrons are a magnetized isothermal fluid while ions  
must be treated (gyro)kinetically. This is the secondary approximation in 
our hierarchy, derived in \secref{sec_els} 
via an asymptotic expansion in $(m_e/m_i)^{1/2}$ 
(see also \apref{ap_el_eqns}). 
The plasma is described by the ion gyrokinetic equation and 
two fluid-like equations that contain electron dynamics---these 
are summarized in \secref{sec_els_sum}. 
The region of validity of this approximation is illustrated 
in \figref{fig_validity_isoth}: it does not capture the 
dissipative effects around the electron diffusion scale 
or the electron heating, but it remains uniformly valid as 
the cascade passes from collisional to collisionless scales 
and also as it crosses the ion gyroscale.\\ 

In order to elucidate the nature of the turbulence above and 
below the ion gyroscale, we derive two tertiary approximations, 
one of which is valid for $\kperp\rho_i\ll1$ (\secsand{sec_KRMHD}{sec_damping}) 
and the other for $\kperp\rho_i\gg1$ (\secref{sec_ERMHD}; see also \apref{ap_nongyro}, 
which gives a non-rigorous, non-gyrokinetic, but perhaps more intuitive, 
derivation of the results of \secsand{sec_els}{sec_ERMHD_eqns}).  

\paragraph{Kinetic Reduced MHD (\secsand{sec_KRMHD}{sec_damping}).} 
On scales above the ion gyroscale, known as the {\em ``inertial range''} 
we demonstrate that the decoupling 
of the Alfv\'en-wave cascade and its indifference to both collisional 
and collisionless damping are explicit and analytically provable 
properties. We show rigorously that the Alfv\'en-wave cascade is 
governed by a closed set of two fluid-like equations for the stream and flux 
functions---the Reduced Magnetohydrodynamics (RMHD)---independently 
of the collisionality 
(\secref{sec_AW} and \secref{sec_AW_coll}; the derivation of RMHD from MHD and 
its properties are presented in \secref{sec_RMHD}). 
The cascade proceeds via interaction of oppositely propagating wave packets 
and is decoupled from the density and magnetic-field-strength fluctuations 
(the ``compressive'' modes; in the collisional limit, these are the entropy and slow modes; 
see \secref{sec_visc} and \apref{ap_visc}). 
The latter are passively mixed by the Alfv\'en waves, but, unlike in 
the fluid (collisional) limit, this passive cascade is governed by a (simplified) 
kinetic equation for the ions (\secref{sec_sw}). 
Together with RMHD, it forms a hybrid fluid-kinetic description of 
magnetized turbulence in a weakly collisional plasma, which we call 
{\em Kinetic Reduced MHD (KRMHD)}. 
The KRMHD equations are summarized in \secref{sec_KRMHD_sum}. 
Their collisional and collisionless limits are 
explored in \secsand{sec_visc}{sec_colless}, respectively. 
Whereas the Alfv\'en waves are undamped in this approximation, 
the compressive fluctuations are subject to
damping both in the collisional 
(\citealt{Braginskii} viscous damping, \secref{sec_visc_diss}) and collisionless 
(\citealt{Barnes} damping, \secref{sec_barnes}) limits. 
In the collisionless limit, the compressive component of the turbulence 
is a simple example of an essentially kinetic turbulence, 
including such features as conservation of generalized energy 
despite collisionless damping and (parallel) phase mixing, 
possibly leading to ion heating (\secsdash{sec_inv_compr}{sec_en_compr}). 
How strongly the compressive fluctuations are damped 
depends on the parallel scale of these fluctuations. 
Since the ion kinetic equation turns out to be linear along the moving 
field lines associated with the Alfv\'en waves,  
the compressive fluctuations do not,  
in the absence of finite-gyroradius effects, develop 
small parallel scales and their cascade may be only weakly damped above 
the ion gyroscale---this is discussed~in~\secref{sec_par_cascade}. 

\paragraph{Electron Reduced MHD (\secref{sec_ERMHD}).}
At the ion gyroscale, the Alfv\'enic and the compressive 
cascades are no longer decoupled and their energy is 
partially damped via collisionless wave--particle interactions (\secref{sec_transition}). 
This part of the energy is channeled into ion heat. 
The rest of it is converted into a cascade of kinetic Alfv\'en waves 
(KAW). This cascade extends 
through what is known as the {\em ``dissipation range''} 
to the electron gyroscale, where its turn comes to be damped 
via wave--particle interaction and transferred into electron heat. 
The KAW turbulence is again anisotropic with $\kpar\ll\kperp$. 
It is governed by a pair of fluid-like equations, also derived from 
gyrokinetics. We call them {\em Electron Reduced MHD (ERMHD)}. 
In the high-beta limit, they coincide with the reduced (anisotropic) form 
of the previously known Electron MHD \citep{Kingsep_Chukbar_Yankov}.
The ERMHD equations are derived in \secref{sec_ERMHD_eqns} 
(see also \apref{ap_EMHD}) and the KAW cascade is considered 
in \secsdash{sec_KAW}{sec_KAW_turb}. 
The fate of the inertial-range energy collisionlessly damped 
at the ion gyroscale is investigated in \secsdash{sec_ent_KAW}{sec_superposed}; 
an analogous consideration for the KAW energy damped at the electron gyroscale 
is presented in \secref{sec_ent_els}. In these sections, 
we introduce the notion of the {\em entropy cascade}---a nonlinear phase-mixing process 
whereby the collisionless damping occurring at the ion and electron 
gyroscales is made irreversible and particles are heated. This part of 
the cascade is purely kinetic and its salient feature is 
the particle distribution functions developing small scales in 
the gyrokinetic phase space.
Note that besides deriving rigorous sets of equations for the dissipation-range 
turbulence, \secref{sec_ERMHD} also presents a number of Kolmogorov-style 
scaling predictions---both for the KAW cascade (\secref{sec_KAW_turb}) and 
for the entropy cascade (\secref{sec_KAW_scalings}, \secref{sec_electrost}, 
\secref{sec_mag}, \secref{sec_ent_els}). 

\paragraph{Hall Reduced MHD (\apref{ap_Hall}).}
The reduced (anisotropic) form of the popular Hall MHD system can be derived 
as a special limit of gyrokinetics ($\kperp\rho_i\ll1$, $T_i\ll T_e$, $\beta_i\ll1$). 
The resulting {\em Hall Reduced MHD (HRMHD)} 
equations are a convenient model for some purposes because they simultaneously capture 
the cold-ion, low-beta limits of both the KRMHD and ERMHD 
systems. However, they are usually not strictly applicable in space and astrophysical 
plasmas of interest, where ions are rarely cold and $\beta_i$ is not particularly low. 
The HRMHD equations are derived in \secref{ap_HRMHD}, the kinetic  
cascade of generalized energy in the Hall limit is discussed in \secref{ap_Hall_en}, 
and the circumstances 
under which the ion inertial and ion sound scales become important in 
theories of plasma turbulence are summarized in \secref{ap_Hall_sum}.
Theories of the dissipation-range turbulence based on Hall MHD are 
briefly discussed in \secref{sec_dr_alt}.\\

The regions of validity of the tertiary approximations---KRMHD and 
ERMHD---are illustrated in \figref{fig_validity_reduced}. 
In this figure, we also show the region of validity of the 
RMHD system derived from the standard compressible MHD equations 
by assuming anisotropy of the turbulence and strong interactions.
This derivation is the fluid analog of the derivation 
of gyrokinetics. We present it in \secref{sec_RMHD}, 
before embarking on the gyrokinetics-based path outlined above, 
in order to make a connection with the conventional 
MHD treatment and to demonstrate with particular simplicity 
how the assumption of anisotropy leads to a reduced fluid system 
in which the decoupling of the cascades of the Alfv\'en waves 
and of the compressive modes is manifest 
(\apref{ap_Brag} extends this derivation to \citealt{Braginskii} 
two-fluid equations in the limit of strong magnetization; 
it also works out rigorously the 
transition from the fluid limit to the KRMHD equations). 

The main formal developments of this paper are contained
in \secsdash{sec_GK}{sec_ERMHD}. The outline given above 
is meant to help the reader navigate these sections. 
In \secref{sec_astro}, we discuss at some length how our 
results apply to various astrophysical plasmas with weak 
collisionality: the solar wind and the magnetosheath, the ISM, accretion disks, 
and galaxy clusters (\secsand{sec_SW_ir}{sec_SW_dr} can also be read as 
an overall summary of the paper in light of the evidence available from 
space-plasma measurements). 
Finally, in \secref{sec_conc}, we provide a brief epilogue 
and make a few remarks about future directions of inquiry.

\section{Reduced MHD and the Decoupling of Turbulent Cascades}
\label{sec_RMHD}

Consider the equations of compressible MHD 
\bea
\label{MHD_rho}
{d\rho\over dt} &=& -\rho\vdel\cdot\vu,\\
\label{MHD_u}
\rho\,{d\vu\over dt} &=& -\vdel\lt(p+{B^2\over8\pi}\rt) + {\vB\cdot\vdel\vB\over4\pi},\\
\label{MHD_p}
{ds\over dt} &=& 0,\quad s={p\over\rho^\gamma},\quad \gamma={5\over3},\\
\label{MHD_B}
{d\vB\over dt} &=& \vB\cdot\vdel\vu - \vB\vdel\cdot\vu,
\eea
where $\rho$ is the mass density, $\vu$ velocity, $p$ pressure, $\vB$ magnetic field, 
$s$ the entropy density, and $d/dt=\dd/\dd t + \vu\cdot\vdel$
(the conditions under which these equations are 
valid are discussed in \apref{ap_Brag}). 
Consider a uniform static equilibrium with a straight mean field in the $z$ direction, 
so 
\bea
\rho = \rho_0 + \drho,\quad 
p = p_0 + \dpr,\quad 
\vB = B_0\vz + \dvB,
\eea 
where $\rho_0$, $p_0$, and $B_0$ are constants. 
In what follows, the subscripts $\parallel$ and $\perp$ will be used 
to denote the projections of fields, variables and gradients 
on the mean-field direction $\vz$ and onto the plane $(x,y)$ perpendicular 
to this direction, respectively. 

\subsection{RMHD Ordering}
\label{sec_RMHDordering}

As we explained in the Introduction, observational and numerical evidence 
makes it safe to assume that the turbulence in such a system will be anisotropic 
with $\kpar\ll\kperp$ (at scales smaller than the outer scale, $\kpar\lf\gg1$; 
see \secsand{sec_two_regimes}{sec_outer_scale}). 
Let us, therefore, introduce a small parameter 
$\epsilon\sim\kpar/\kperp$ and carry out a systematic expansion 
of \eqsdash{MHD_rho}{MHD_B} in $\epsilon$. In this expansion, the 
fluctuations are treated as small, but not arbitrarily so: 
in order to estimate their size, 
we shall adopt the critical-balance conjecture~\exref{crit_bal}, 
which is now treated {\em not} as a detailed scaling prescription but as an 
ordering assumption. This allows us to introduce the following ordering:
\bea
{\drho\over\rho_0}
\sim {\uperp\over v_A} \sim {\upar\over v_A}
\sim {\dpr\over p_0}
\sim {\dBperp\over B_0} \sim {\dBpar\over B_0} 
\sim {\kpar\over\kperp} 
\sim \epsilon, 
\label{RMHD_ordering}
\eea
where $v_A=B_0/\sqrt{4\pi\rho_0}$ is the Alfv\'en speed. Note that 
this means that we order the Mach number
\bea
\label{RMHD_Mach}
M\sim {u\over c_s} \sim {\epsilon\over\sqrt{\beta_i}},
\eea
where $c_s=(\gamma p_0/\rho_0)^{1/2}$ is the speed of sound and 
\bea
\label{beta_def}
\beta={8\pi p_0\over B_0^2}={2\over\gamma}{c_s^2\over v_A^2} 
\eea
is the plasma beta, which is ordered to be order unity in 
the $\epsilon$ expansion (subsidiary limits of high and 
low $\beta$ can be taken after the $\epsilon$ expansion 
is done; see \secref{sec_sw_fluid}). 
 
In \eqref{RMHD_ordering}, we made two auxiliary 
ordering assumptions: that the velocity and magnetic-field
fluctuations have the character of Alfv\'en and slow waves 
($\dBperp/B_0\sim\uperp/v_A$, $\dBpar/B_0\sim\upar/v_A$) 
and that the relative amplitudes of 
the Alfv\'en-wave-polarized fluctuations ($\dBperp/B_0$, $\uperp/v_A$),
slow-wave-polarized fluctuations ($\dBpar/B_0$, $\upar/v_A$)
and density/pressure/entropy fluctuations ($\drho/\rho_0$, $\dpr/p_0$) 
are all the same order. Strictly speaking, whether this is the case 
depends on the energy sources that drive the turbulence: as we shall  
see, if no slow waves (or entropy fluctuations) are launched, none will be present. 
However, in astrophysical contexts, the outer-scale energy input 
may be assumed random and, therefore, comparable power 
is injected into all types of fluctuations. 

We further assume that the characteristic frequency of the 
fluctuations is $\omega\sim\kpar v_A$ [\eqref{crit_bal}], meaning that 
the fast waves, for which $\omega\simeq\kperp(v_A^2+c_s^2)^{1/2}$, 
are ordered out. This restriction must be justified empirically. 
Observations of the solar-wind turbulence confirm that 
it is primarily Alfv\'enic \citep[see, e.g.,][]{Bale_etal} 
and that its compressive component is substantially 
pressure-balanced \citep[][see \eqref{MHD_pr_bal} below]{Roberts_prbal,Burlaga_etal_prbal,Marsch_Tu_prbal,Bavassano_etal_prbal}. 
A weak-turbulence calculation of compressible MHD turbulence 
in low-beta plasmas \citep{Chandran_fast_waves} 
suggests that only a small amount of energy is transferred from 
the fast waves to Alfv\'en waves with large $\kpar$. 
A similar conclusion emerges from numerical simulations 
\citep{Cho_Lazarian_low_beta,Cho_Lazarian_mnras}. 
As the fast waves are also expected to be subject to strong 
collisionless damping and/or to strong dissipation after they 
steepen into shocks, we eliminate them from our consideration of the 
problem and concentrate on low-frequency turbulence. 

\subsection{Alfv\'en Waves}
\label{sec_AW_fluid}

We start by observing that 
the Alfv\'en-wave-polarized fluctuations are two-dimensionally solenoidal: 
since, from \eqref{MHD_rho}, 
\bea
\vdel\cdot\vu = - {d\over dt}{\drho\over\rho_0} = O(\epsilon^2) 
\label{divu_eq}
\eea
and $\vdel\cdot\dvB=0$ exactly, 
separating the $O(\epsilon)$ part of these divergences gives 
$\vdperp\cdot\vuperp=0$ and $\vdperp\cdot\dvBperp=0$. To lowest order in 
the $\epsilon$ expansion, 
we may, therefore, express $\vuperp$ and $\dvBperp$ in terms of 
scalar stream (flux) functions:
\bea
\vuperp = \vz\times\vdperp\Phi,\qquad 
{\dvBperp\over\sqrt{4\pi\rho_0}} = \vz\times\vdperp\Psi.
\label{Phi_Psi_def}
\eea
Evolution equations for $\Phi$ and $\Psi$ 
are obtained by substituting the expressions \exref{Phi_Psi_def} into 
the perpendicular parts of the induction equation~\exref{MHD_B} and the 
momentum equation~\exref{MHD_u}---of the latter the curl is taken to 
annihilate the pressure term. Keeping only the terms of the lowest order, 
$O(\epsilon^2)$, we get 
\bea
\label{RMHD_Psi}
{\dd\Psi\over\dd t} + \lt\{\Phi,\Psi\rt\} &=& v_A{\dd\Phi\over\dd z},\\
\label{RMHD_Phi}
{\dd\over\dd t}\dperp^2\Phi + \lt\{\Phi,\dperp^2\Phi\rt\} 
&=& v_A\dpar\dperp^2\Psi + \lt\{\Psi,\dperp^2\Psi\rt\},
\eea
where $\lt\{\Phi,\Psi\rt\}=\vz\cdot(\vdperp\Phi\times\vdperp\Psi)$ 
and we have taken into account that, to lowest order,  
\bea
\label{dt_def}
{d\over dt} &=& {\dd\over\dd t} + \vuperp\cdot\vdperp={\dd\over\dd t} + \lt\{\Phi,\cdots\rt\},\\ 
\label{dpar_def}
\Dpar &=& \dpar + {\dvBperp\over B_0}\cdot\vdperp 
= \dpar + {1\over v_A}\lt\{\Psi,\cdots\rt\}.
\eea
Here $\vb=\vB/B_0$ is the unit vector along the perturbed field line. 

\Eqsdash{RMHD_Psi}{RMHD_Phi} are known as the Reduced Magnetohydrodynamics 
(RMHD). The first derivations of these equations (in the context of fusion plasmas) 
are due to \citet{Kadomtsev_Pogutse} and to \citet{Strauss76}. These were 
followed by many systematic derivations and generalizations employing various 
versions and refinements of the basic expansion, taking into account 
the non-Alfv\'enic modes (which we will do in \secref{sec_sw_fluid}), and 
including the effects of spatial gradients of equilibrium fields \citep[e.g.,][]{Strauss77,Montgomery,Hazeltine83,Zank_Matthaeus1,Kinney_McWilliams1,Bhattacharjee_Ng_Spangler,Kruger_Hegna_Callen}. 
A comparative review of these expansion schemes and their (often close) 
relationship to ours is outside the scope of this paper. 
One important point we wish to emphasize is that we do not assume the plasma 
beta [defined in \eqref{beta_def}] to be either large or small. 
 
\Eqsand{RMHD_Psi}{RMHD_Phi} form a closed set, meaning that the Alfv\'en-wave cascade 
decouples from the slow waves and density fluctuations.
It is to the turbulence described by \eqsdash{RMHD_Psi}{RMHD_Phi} 
that the GS theory outlined in \secref{sec_GS} applies.\footnote{The Alfv\'en-wave 
turbulence in the RMHD system has been studied by many authors. Some of the 
relevant numerical investigations are due to \citet{Kinney_McWilliams2}, 
\citet{Dmitruk_Gomez_Matthaeus}, \citet{Oughton_Dmitruk_Matthaeus}, 
\citet{Rappazzo_etal1,Rappazzo_etal2}, \citet{Perez_Boldyrev,Perez_Boldyrev_imb}. 
Analytical theory has mostly been confined to the weak-turbulence paradigm 
\citep{Ng_Bhattacharjee1,Ng_Bhattacharjee2,Bhattacharjee_Ng,Galtier_etal02,Lithwick_Goldreich_imb,Galtier_Chandran,Nazarenko}. 
We note that adopting the critical balance [\eqref{crit_bal}] as an ordering assumption 
for the expansion in $\kpar/\kperp$ does 
not preclude one from subsequently attempting a weak-turbulence approach: 
the latter should simply be treated as a subsidiary expansion. 
Indeed, implementing the anisotropy assumption 
on the level of MHD equations rather than simultaneously with the 
weak-turbulence closure \citep{Galtier_etal} significantly 
reduces the amount of algebra. One should, however, bear in mind 
that the weak-turbulence approximation always breaks down at some 
sufficiently small scale---namely, when $\kperp\sim (v_A/\uo)^2\kpar^2\lf$, 
where $\lf$ is the outer scale of the turbulence, $\uo$ velocity at the outer scale, 
and $\kpar$ the parallel wavenumber of the Alfv\'en waves (see \citealt{GS97} 
or the review by \citealt{SC_mhdbook}). 
Below this scale, interactions cannot be assumed weak.} 
In \secref{sec_AW}, we will show that \eqsand{RMHD_Psi}{RMHD_Phi} 
correctly describe inertial-range Alfv\'enic fluctuations even in a collisionless 
plasma, where the full MHD description [\eqsdash{MHD_rho}{MHD_B}] is not valid. 

\subsection{Elsasser Fields}
\label{sec_elsasser_AW}

The MHD equations~\exsdash{MHD_rho}{MHD_B} in the incompressible limit ($\rho=\const$) 
acquire a symmetric form if written in terms of the Elsasser fields 
${\bf z}^\pm=\vu\pm\dvB/\sqrt{4\pi\rho}$ \citep{Elsasser}. 
Let us demonstrate how this symmetry manifests itself in the reduced equations 
derived above. 

We introduce {\em Elsasser potentials} $\zeta^\pm=\Phi\pm\Psi$, 
so that ${\bf z}^\pm_\perp=\vz\times\vdperp\zeta^\pm$. For these potentials, 
\eqsdash{RMHD_Psi}{RMHD_Phi} become 
\bea
\nonumber
{\dd\over\dd t}\dperp^2\zeta^\pm \mp 
v_A\dpar\dperp^2\zeta^\pm 
&=& -{1\over2}\lt(\lt\{\zeta^+,\dperp^2\zeta^-\rt\}
+ \lt\{\zeta^-,\dperp^2\zeta^+\rt\}\rt.\\
&&\lt.\mp\dperp^2\lt\{\zeta^+,\zeta^-\rt\}\rt).\quad
\label{eq_zeta}
\eea
These equations show that the RMHD has a 
simple set of exact solutions: 
if $\zeta^-=0$ or $\zeta^+=0$, the nonlinear term vanishes 
and the other, non-zero, Elsasser potential is simply 
a fluctuation of arbitrary shape and magnitude 
propagating along the mean field at the Alfv\'en speed~$v_A$: 
$\zeta^\pm = f^\pm(x,y,z\mp v_A t)$. 
These solutions are finite-amplitude Alfv\'en-wave packets 
of arbitrary shape. 
Only counterpropagating such solutions can interact and thereby 
give rise to the Alfv\'en-wave cascade \citep{Kraichnan}. 
Note that these interactions are conservative in the sense 
that the ``$+$'' and ``$-$'' waves scatter off each other without 
exchanging energy. 

Note that the individual conservation of the ``$+$'' and ``$-$'' waves'
energies means that the energy fluxes associated with these waves 
need not be equal, so instead of a single Kolmogorov flux $\eps$ 
assumed in the scaling arguments reviewed in \secref{sec_GS}, 
we could have $\eps^+\neq\eps^-$. The GS theory can be generalized 
to this case of {\em imbalanced} Alfv\'enic cascades 
\citep{Lithwick_Goldreich_Sridhar,Beresnyak_Lazarian_imb,Chandran_imb}, 
but here we will focus on the balanced turbulence, $\eps^+\sim\eps^-$. 
If one considers the turbulence forced in a physical way (i.e., without 
forcing the magnetic field, which would break the flux conservation), 
the resulting cascade would always be balanced. In the real world, 
imbalanced Alfv\'enic fluxes are measured in the fast solar 
wind, where the influence of initial conditions in the solar 
atmosphere is more pronounced, while the slow-wind turbulence is 
approximately balanced (\citealt{Marsch_Tu_z}; see also reviews by
\citealt{Tu_Marsch_review,Bruno_Carbone} and references therein). 

\subsection{Slow Waves and the Entropy Mode}
\label{sec_sw_fluid}

In order to derive evolution equations for the remaining MHD modes, 
let us first revisit the perpendicular part of the momentum equation and 
use \eqref{RMHD_ordering} to order terms in it.
In the lowest order, $O(\epsilon)$, we get the pressure balance
\bea
\label{MHD_pr_bal}
\vdperp\lt(\dpr + {B_0\dBpar\over 4\pi}\rt) = 0 \quad\Rightarrow\quad
{\dpr\over p_0} = -\gamma\,{v_A^2\over c_s^2}{\dBpar\over B_0}.
\eea
Using \eqref{MHD_pr_bal} and the entropy equation \exref{MHD_p}, we get 
\bea
{d\ds\over dt} = 0,\quad 
{\ds\over s_0} = {\dpr\over p_0} - \gamma{\drho\over\rho_0} = 
- \gamma\lt({\drho\over\rho_0} + {v_A^2\over c_s^2}{\dBpar\over B_0}\rt),
\label{eq_ds}
\eea
where $s_0=p_0/\rho_0^\gamma$. 
Now, substituting \eqref{divu_eq} for $\vdel\cdot\vu$ 
in the parallel component of the induction equation~\exref{MHD_B}, 
we get
\bea
\label{eq1}
{d\over dt}\lt({\dBpar\over B_0} - {\drho\over\rho_0}\rt) - \Dpar\upar = 0.
\eea
Combining \eqsand{eq_ds}{eq1}, we obtain
\bea
\label{eq_drho}
{d\over dt}{\drho\over\rho_0} &=& - {1\over 1+ c_s^2/v_A^2}\,\Dpar\upar,\\
\label{eq_Bpar}
{d\over dt}{\dBpar\over B_0} &=& {1\over 1 + v_A^2/c_s^2}\,\Dpar\upar.
\eea
Finally, we take the parallel component of the momentum 
equation~\exref{MHD_u} and notice that, due to the pressure balance~\exref{MHD_pr_bal} 
and to the smallness of the parallel gradients, the pressure term is $O(\epsilon^3)$, 
while the inertial and tension terms are $O(\epsilon^2)$. Therefore, 
\bea
\label{eq_upar}
{d\upar\over dt} = v_A^2\Dpar{\dBpar\over B_0}.
\eea 

\Eqsdash{eq_Bpar}{eq_upar} describe the slow-wave-polarized fluctuations, 
while \eqref{eq_ds} describes the zero-frequency entropy mode, which is 
decoupled from the slow waves.\footnote{For other 
expansion schemes leading to reduced sets of equations for 
these ``compressive'' fluctuations see references in \secref{sec_AW_fluid}.
Note that the nature of the density fluctuations described above 
is distinct from the so called ``pseudosound'' density fluctuations that 
arise in the ``nearly incompressible'' MHD theories  
\citep{Montgomery_Brown_Matthaeus,Matthaeus_Brown,Matthaeus_etal91,Zank_Matthaeus2}.
The ``pseudosound'' is essentially the density response caused by the 
nonlinear pressure fluctuations calculated from the incompressibility 
constraint. The resulting density fluctuations are second order 
in Mach number and, therefore, order $\epsilon^2$ in our expansion 
[see \eqref{RMHD_Mach}]. The passive density fluctuations derived in this 
section are order $\epsilon$ and, therefore, supersede the ``pseudosound''
(see review by \citealt{Tu_Marsch_review} for a discussion of the relevant 
solar-wind evidence).} 
The nonlinearity in \eqsdash{eq_Bpar}{eq_upar} enters 
via the derivatives defined in \eqsdash{dt_def}{dpar_def} and is due solely to 
interactions with Alfv\'en waves. 
Thus, both the slow-wave and the entropy-mode 
cascades occur via passive scattering/mixing by Alfv\'en waves, in the course 
of which there is no energy exchange between the cascades.

Note that in the high-beta limit, $c_s\gg v_A$ [see \eqref{beta_def}], 
the entropy mode is dominated by density fluctuations [\eqref{eq_ds}, $c_s\gg v_A$], 
which also decouple from the slow-wave cascade [\eqref{eq_drho}, $c_s\gg v_A$]. 
and are passively mixed by the Alfv\'en-wave turbulence: 
\bea
{d\drho\over dt}=0.
\label{drho_cascade}
\eea 
The high-beta limit is equivalent to the incompressible 
approximation for the slow waves. 

In \secref{sec_sw}, we will derive a kinetic description for the inertial-range 
compressive fluctuations (density and magnetic-field strength), which is more generally valid 
in weakly collisional plasmas and which reduces to \eqsdash{eq_Bpar}{eq_upar} 
in the collisional limit (see \apref{ap_visc}). While these fluctuations 
will in general satisfy a kinetic equation, they will remain passive with respect 
to the Alfv\'en waves. 

\subsection{Elsasser Fields for the Slow Waves}
\label{sec_elsasser_SW}

The original \citet{Elsasser} symmetry was derived for incompressible 
MHD equations. However, for the ``compressive'' slow-wave fluctuations, 
we may introduce generalized Elsasser fields: 
\bea
\label{zpar_def}
\zpar^\pm = \upar\pm{\dBpar\over\sqrt{4\pi\rho_0}}\lt(1+{v_A^2\over c_s^2}\rt)^{1/2}.
\eea
Straightforwardly, the evolution equation for these fields~is
\bea
\nonumber
{\dd\zpar^\pm\over\dd t} &\mp& {v_A\over\sqrt{1+v_A^2/c_s^2}}{\dd\zpar^\pm\over\dd z}=\\ 
\nonumber
&-&{1\over2}\lt(1\mp{1\over\sqrt{1+v_A^2/c_s^2}}\rt)\bl\{\zeta^+,\zpar^\pm\br\}\\
&-&{1\over2}\lt(1\pm{1\over\sqrt{1+v_A^2/c_s^2}}\rt)\bl\{\zeta^-,\zpar^\pm\br\}.
\label{eq_zpar}
\eea

In the high-beta limit ($v_A\ll c_s$),  
the generalized Elsasser fields~\exref{zpar_def} become the 
parallel components of the conventional incompressible Elsasser fields. 
We see that only in this limit do the slow 
waves interact exclusively with the counterpropagating Alfv\'en 
waves, and so only in this limit does setting $\zeta^-=0$ or $\zeta^+=0$ 
gives rise to finite-amplitude slow-wave-packet solutions 
$\zpar^\pm = f^\pm(x,y,z\mp v_A t)$ analogous to the 
finite-amplitude Alfv\'en-wave packets discussed in 
\secref{sec_elsasser_AW}.\footnote{Obviously, setting {\em both} $\zeta^\pm=0$ does 
always enable these finite-amplitude slow-wave solutions. More 
non-trivially, such finite-amplitude solutions exist in the 
Lagrangian frame associated with the Alfv\'en waves---this is discussed 
in detail in \secref{sec_par_cascade}.} 
For general $\beta$, the phase speed of the slow waves is smaller than 
that of the Alfv\'en waves and, therefore, Alfv\'en waves can 
``catch up'' and interact with the slow waves that travel in the same 
direction. All of these interactions 
are of scattering type and involve no exchange of energy. 

\subsection{Scalings for Passive Fluctuations}
\label{sec_scalings_passive}

The scaling of the passively mixed scalar fields introduced above 
is slaved to the scaling of the Alfv\'enic fluctuations.
Consider for example the entropy mode [\eqref{eq_ds}]. 
As in Kolmogorov--Obukhov theory (see \secref{sec_K41}), 
one assumes a local-in-scale-space cascade of scalar variance 
and a constant flux $\epss$ of this variance. 
Then, analogously to \eqref{const_flux}, 
\bea
{\vthi^2\over s_0^2}{\dsl^2\over\taul}\sim\epss. 
\eea
Since the cascade time is 
$\taul^{-1}\sim\vuperp\cdot\vdperp\sim v_A/\lparl\sim\varepsilon/\upl^2$, 
\bea
{\dsl\over s_0}\sim\lt(\epss\over\eps\rt)^{1/2}{\upl\over\vthi}, 
\eea
so the scalar fluctuations have the same scaling as 
the turbulence that mixes them \citep{Obukhov,Corrsin}. 
In GS turbulence, the scalar-variance spectrum should, 
therefore, be $\kperp^{-5/3}$ \citep{Lithwick_Goldreich}. 
The same argument applies to all passive fields. 

It is the (presumably) passive electron-density spectrum 
that provides the main evidence of the $k^{-5/3}$ scaling 
in the interstellar turbulence 
\citep[][see further discussion in 
\secref{sec_el_den_ISM}]{Armstrong_Cordes_Rickett,Armstrong_Rickett_Spangler,Lazio_etal_review}. 
The explanation of this spectrum in terms of passive mixing 
of the entropy mode, originally proposed by \citet{Higdon}, 
was developed on the basis of the GS theory by \citet{Lithwick_Goldreich}. 
The turbulent cascade of the compressive fluctuations and the relevant solar-wind data 
is discussed further in \secref{sec_par_cascade}. In particular, 
it will emerge that the anisotropy of these fluctuations remains 
a non-trivial issue: is there an analog of the scaling relation~\exref{GS_aniso}?
The scaling argument outlined above does not invoke any assumptions 
about the relationship between the parallel and perpendicular 
scales of the compressive fluctuations (other than the assumption that 
they are anisotropic). \citet{Lithwick_Goldreich} argue that 
the parallel scales of the Alfv\'enic fluctuations will imprint 
themselves on the passively advected compressive ones, 
so \eqref{GS_aniso} holds for the latter as well. In \secref{sec_par_cascade}, 
we examine this conclusion in view of the solar-wind evidence 
and of the fact that the equations for the compressive modes 
become linear in the Lagrangian frame associated with the 
Alfv\'enic turbulence. 

\subsection{Five RMHD Cascades} 
\label{sec_RMHD_cascades}

Thus, the anisotropy and critical balance~\exref{crit_bal} 
taken as ordering assumptions lead to 
a neat decomposition of the MHD turbulent cascade into a decoupled 
Alfv\'en-wave cascade and cascades of slow waves and entropy fluctuations 
passively scattered/mixed by the Alfv\'en waves. More precisely, 
Eqs.~\exref{eq_ds}, \exref{eq_zeta} and \exref{eq_zpar} imply that, for 
arbitrary $\beta$, there are five conserved quantities:\footnote{Note that 
magnetic helicity of the perturbed field is not an invariant of RMHD, 
except in two dimensions (see \apref{ap_hel_RMHD}). 
In 2D, there is also conservation of the mean square flux, 
$\intr|\Psi|^2$ (see \apref{ap_Aparsq}).} 
\bea
\label{Wperp_def}
\Wperp^\pm &=& 
{1\over2}\intr\rho_0 |\vdperp\zeta^\pm|^2 
\qquad {\rm (Alfven~waves),}\\ 
\label{Wpar_def}
\Wpar^\pm &=& 
{1\over2}\intr\rho_0 |\zpar^\pm|^2 
\qquad\quad {\rm (slow~waves),}\\
\label{Ws_def}
\Ws &=& {1\over2}\intr{\ds^2\over s_0^2} 
\qquad\qquad\ \,{\rm (entropy~fluctuations).} 
\eea
$\Wperp^+$ and $\Wperp^-$ are always cascaded by interaction with each 
other, $\Ws$ is passively mixed by $\Wperp^+$ and $\Wperp^-$, 
$\Wpar^\pm$ are passively scattered by $\Wperp^\mp$ and, unless 
$\beta\gg1$, also by $\Wperp^\pm$. 

This is an example of splitting of the overall energy cascade 
into several channels (recovered as a particular case of 
the more general kinetic cascade in \apref{ap_en_RMHD})---a concept that 
will repeatedly arise in the kinetic treatment to follow. 

The decoupling of the slow- and Alfv\'en-wave cascades in 
MHD turbulence was studied in some detail 
and confirmed in direct numerical simulations 
by \citet[][for $\beta\gg1$]{Maron_Goldreich} 
and by \citet[][for a range of values of $\beta$]{Cho_Lazarian_low_beta,Cho_Lazarian_mnras}. 
The derivation given in \secsand{sec_AW_fluid}{sec_sw_fluid}
\citep[cf.][]{Lithwick_Goldreich} provides a straightforward theoretical basis 
for these results, assuming anisotropy of the turbulence 
(which was also confirmed in these numerical studies). 
 
It turns out that the decoupling of the Alfv\'en-wave cascade that 
we demonstrated above for the anisotropic MHD turbulence 
is a uniformly valid property of plasma turbulence 
at both collisional and collisionless scales 
and that this cascade is correctly described by the RMHD equations 
\exsdash{RMHD_Psi}{RMHD_Phi} all the way down to the ion gyroscale, 
while the fluctuations of density and magnetic-field strength do not satisfy 
simple fluid evolution equations anymore and require solving 
the kinetic equation. 
In order to prove this, we adopt a kinetic description 
and apply to it the same ordering (\secref{sec_RMHDordering}) 
as we used to reduce the MHD equations. The kinetic theory that 
emerges as a result is called gyrokinetics.

\section{Gyrokinetics}
\label{sec_GK}

The gyrokinetic 
formalism was first worked out for linear 
waves by \citet{Rutherford_Frieman} and by \citet{Taylor_Hastie} 
\cite[see also][]{Catto,Antonsen_Lane,Catto_Tang_Baldwin}
and subsequently extended to the nonlinear regime by \citet{Frieman_Chen}. 
Rigorous derivations of the gyrokinetic equation based on the 
Hamiltonian formalism were developed by \citet[][electrostatic]{Dubin_etal} 
and \citet[][electromagnetic]{Hahm_Lee_Brizard}. This approach 
is reviewed in \citet{Brizard_Hahm_review}. 
A more pedestrian, but perhaps also more transparent exposition 
of the gyrokinetics in a straight mean field 
can be found in \citet{Howes_etal}, who also provide a detailed explanation 
of the gyrokinetic ordering in the context of astrophysical plasma 
turbulence and a treatment of the linear waves and damping rates. 
Here we review only the main points so as to allow the reader 
to understand the present paper without referring elsewhere.

In general, a plasma is completely described by 
the distribution function $f_s(t,\vr,\vv)$---the probability 
density for a particle of species~$s$ ($=i,e$)  
to be found at the spatial position $\vr$ moving with velocity $\vv$. 
This function obeys the kinetic Vlasov--Landau (or Boltzmann) equation 
\bea
\label{Vlasov_eq}
{\dd f_s\over\dd t} + \vv\cdot\vdel f_s 
+ {\qs\over m_s}\(\vE + {\vv\times\vB\over c}\)\cdot{\dd f_s\over\dd\vv} 
= \({\dd f_s\over\dd t}\)_{\rm c}, 
\eea
where $\qs$ and $m_s$ are the particle's charge and mass, 
$c$ is the speed of light, and the right-hand side is 
the collision term (quadratic in $f$). 
The electric and magnetic fields are 
\bea
\label{E_B_def}
\vE = -\vdel\ephi - {1\over c}{\dd\vA\over \dd t},\quad
\vB = \vdel\times\vA.
\eea
The first equality is Faraday's law uncurled, 
the second the magnetic-field solenoidality condition;
we shall use the Coulomb gauge, $\vdel\cdot\vA=0$. 
The fields satisfy the Poisson and the Amp\`ere--Maxwell equations with the charge and 
current densities determined by $f_s(t,\vr,\vv)$: 
\bea
\label{Max_Poisson}
\vdel\cdot\vE &=& 4\pi\sum_s \qs n_s = 4\pi\sum_s \qs\int d^3\vv\,f_s,\\
\label{Max_Ampere}
\vdel\times\vB &-& {1\over c}{\dd \vE\over\dd t}
= {4\pi\over c}\,\vj = {4\pi\over c}\sum_s \qs \int d^3\vv\,\vv f_s.
\eea

\pseudofigurewide{fig_validity_gk}{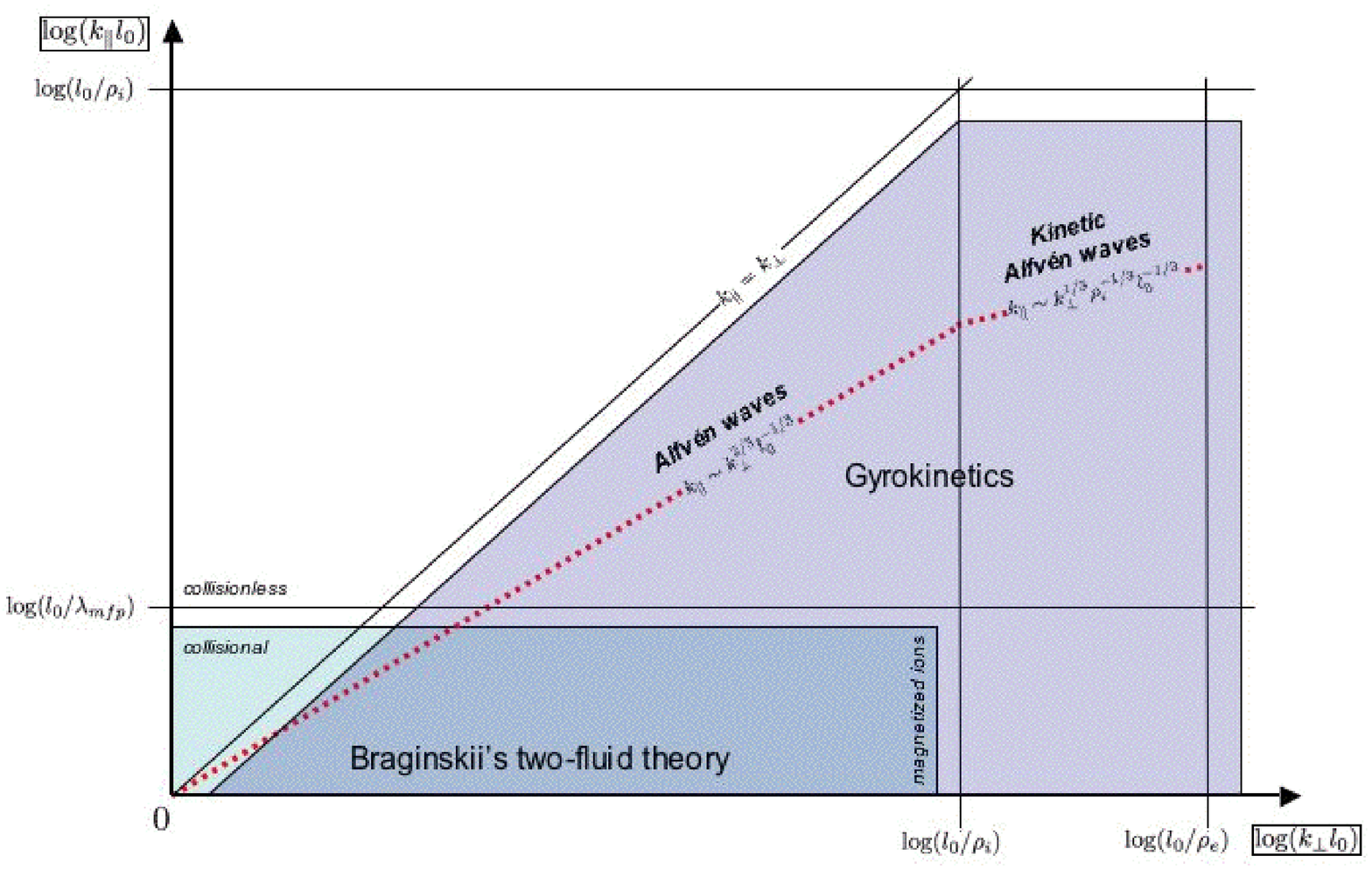}{f3.ps}{
Regions of validity in the wavenumber space 
of two primary approximations---the two-fluid (\apref{ap_two_fluid}) 
and gyrokinetic (\secref{sec_GK}). 
The gyrokinetic theory holds when $\kpar \ll \kperp$ and $\omega \ll \Omega_i$ 
[when $\kpar \ll \kperp < \rho_i^{-1}$, 
the second requirement is automatically satisfied for Alfv\'en, slow and entropy modes; 
see \eqref{omega_order}]. 
The two-fluid equations 
hold when $\kpar\mfp \ll 1$ (collisional limit) and $\kperp\rho_i \ll 1$ (magnetized plasma). 
Note that the gyrokinetic theory holds for all but the very largest (outer) scales, 
where anisotropy cannot be assumed.}

\subsection{Gyrokinetic Ordering and Dimensionless Parameters}
\label{sec_params}

As in \secref{sec_RMHD} we set up a static equilibrium with a uniform 
mean field, $\vB_0=B_0\vz$, $\vE_0=0$, assume that the perturbations 
will be anisotropic with $\kpar\ll\kperp$ 
(at scales smaller than the outer scale, $\kpar\lf\gg1$; 
see \secsand{sec_two_regimes}{sec_outer_scale}), 
and construct an expansion of the kinetic 
theory around this equilibrium with respect to the small parameter 
$\epsilon\sim\kpar/\kperp$. 
We adopt the ordering expressed by \eqsand{crit_bal}{RMHD_ordering}, i.e., 
we assume the perturbations to be strongly interacting Alfv\'en waves 
plus electron density and magnetic-field-strength fluctuations. 

Besides $\epsilon$, several other dimensionless parameters are present, 
all of which are formally considered to be of order unity in the gyrokinetic 
expansion: the electron--ion mass ratio $m_e/m_i$, the charge ratio 
\bea
Z={\qi/|\qe|}={\qi/e}
\label{Z_def}
\eea 
(for hydrogen, this is 1, which applies to most astrophysical plasmas 
of interest to us), the temperature ratio\footnote{It can be shown that 
equilibrium temperatures change on the timescale $\sim (\epsilon^2\omega)^{-1}$ 
\citep{Howes_etal}. On the other hand, from standard theory of collisional 
transport \citep[e.g.,][]{Helander_Sigmar}, 
the ion and electron temperatures equalize on the timescale 
$\sim \nuie^{-1}\sim (m_i/m_e)^{1/2}\nui^{-1}$ 
[see \eqref{nuie_def}]. Therefore, $\tau$ can depart from unity 
by an amount of order $\epsilon^2(\omega/\nui)(m_i/m_e)^{1/2}$. 
In our ordering scheme [\eqref{omega_vs_nu}], this is $O(\epsilon^2)$ and, therefore, 
we should simply set $\tau=1 + O(\epsilon^2)$. However, we shall carry 
the parameter $\tau$ because other ordering schemes 
are possible that permit arbitrary values of~$\tau$.
These are appropriate to plasmas with very 
weak collisions. For example, in the solar wind, $\tau$ appears 
to be order unity but not exactly 1 \citep{Newbury_etal}, while 
in accretion flows near the black hole, 
some models predict $\tau\gg1$ (see \secref{sec_disks}). 
\label{fn_temperatures}} 
\bea
\tau={T_i/ T_e},
\label{tau_def}
\eea 
and the plasma (ion) beta 
\bea
\beta_i={\vthi^2\over v_A^2}={8\pi n_iT_i\over B_0^2} 
= \beta\(1+{Z\over\tau}\)^{-1}, 
\label{betai_def}
\eea
where $\vthi=(2 T_i/m_i)^{1/2}$ is the ion thermal speed  
and the total $\beta$ was defined in \eqref{beta_def} based on 
the total pressure $p=n_iT_i + n_eT_e$. We shall occasionally 
also use the electron beta 
\bea
\beta_e={8\pi n_eT_e\over B_0^2}={Z\over \tau}\,\beta_i.
\eea 
The total beta is $\beta=\beta_i+\beta_e$. 

\subsubsection{Wavenumbers and Frequencies}
\label{sec_k_and_omega}

As we want our theory to be uniformly valid 
at all (perpendicular) scales above, at or below the ion gyroscale, 
we order
\bea
\label{kperp_order}
\kperp\rho_i \sim 1,
\eea
where $\rho_i=\vthi/\Omega_i$ is the ion gyroradius, 
$\Omega_i=\qi B_0/cm_i$ the ion cyclotron frequency. 
Note that 
\bea
\label{rho_ratio}
\rho_e = {Z\over\sqrt{\tau}}\sqrt{{m_e\over m_i}}\,\rho_i. 
\eea

Assuming Alfv\'enic frequencies implies 
\bea
{\omega\over\Omega_i}\sim{\kpar v_A\over\Omega_i}\sim 
{\kperp\rho_i\over\sqrt{\beta_i}}\,\epsilon.
\label{omega_order}
\eea
Thus, gyrokinetics is a low-frequency limit that averages over 
the timescales associated with the particle gyration. 
Because we have assumed that the fluctuations are 
anisotropic and have (by order of magnitude) Alfv\'enic frequencies, 
we see from \eqref{omega_order} that 
their frequency remains far below $\Omega_i$ at all scales, 
including the ion and even electron gyroscale---the gyrokinetics 
remains valid at all of these scales and the cyclotron-frequency effects 
are negligible \citep[cf.][]{Quataert_Gruzinov}. 

\subsubsection{Fluctuations} 
\label{sec_fluct}

\Eqref{crit_bal} allows us to order the fluctuations 
of the scalar potential: on the one hand, we have from \eqref{crit_bal} 
$\uperp\sim\epsilon v_A$; on the other hand, 
the plasma mass flow velocity is 
(to the lowest order) the $\vE\times\vB$ drift velocity of the ions,  
$\uperp\sim c\Eperp/B_0\sim c\kperp\ephi/B_0$, so
\bea
{e\ephi\over T_e} \sim {\tau\over Z}{1\over\kperp\rho_i\sqrt{\beta_i}}\,\epsilon.
\label{phi_order}
\eea
All other fluctuations (magnetic, density, parallel velocity) are ordered 
according to \eqref{RMHD_ordering}. 

Note that the ordering of the flow velocity dictated by \eqref{crit_bal}
means that we are considering the limit of small Mach numbers:
\bea
M\sim {u\over\vthi}\sim {\epsilon\over\sqrt{\beta_i}}.
\eea
This means that the gyrokinetic description in the form used below does 
not extend to large sonic flows that can be present in many astrophysical 
systems. It is, in principle, possible to extend the gyrokinetics 
to systems with sonic flows \citep[e.g., in the toroidal geometry; 
see][]{Artun_Tang,Sugama_Horton97}. However, we do not follow this route 
because such flows belong to the same class of non-universal outer-scale features 
as background density and temperature gradients, system-specific geometry 
etc.---these can all be ignored at small scales, where the turbulence should be
approximately homogeneous and subsonic 
(as long as $\kpar\lf\gg1$, see discussion in \secref{sec_outer_scale}). 

\subsubsection{Collisions}
\label{sec_order_coll}

Finally, we want our theory to be valid both in the collisional 
and the collisionless regimes, so we do not assume $\omega$ 
to be either smaller or larger than the (ion) collision frequency~$\nui$: 
\bea
\label{omega_vs_nu}
{\omega\over\nui}\sim {\kpar\mfp\over\sqrt{\beta_i}} \sim 1,
\eea
where $\mfp=\vthi/\nui$ is the ion mean free path 
\citep[this ordering can actually be inferred from equating the gyrokinetic entropy 
production terms to the collisional entropy production; 
see extended discussion in][]{Howes_etal}. 
Note that the ordering \exref{omega_vs_nu} holds on the understanding that we have ordered
$\kperp\rho_i\sim1$ [\eqref{kperp_order}] because the fluctuation frequency can depend 
on $\kperp\rho_i$ in the dissipation range (see \secref{sec_KAW}). 

Other collision rates are related to $\nui$ via a set 
of standard formulae \citep[see, e.g.,][]{Helander_Sigmar}, which will 
be useful in what follows:
\bea
\label{nue_def}
\nue &=& Z\nuee = {\tau^{3/2}\over Z^2}\sqrt{m_i\over m_e}\,\nui,\\
\label{nuie_def} 
\nuie &=& {8\over3\sqrt{\pi}}{\tau^{3/2}\over Z}\sqrt{m_e\over m_i}\,\nui,\\
\nui &=& {\sqrt{2}\pi Z^4 e^4 n_i \ln\Lambda\over m_i^{1/2} T_i^{3/2}},
\label{nui_def}
\eea
where $\ln\Lambda$ is the Coulomb logarithm and the numerical factor 
in the definition of $\nuie$ has been inserted for future notational 
convenience (see \apref{ap_Brag}). We always define 
\bea
\mfp={\vthi\over\nui},\quad 
\mfpe={\vthe\over\nue} = \lt({Z\over\tau}\rt)^2\mfp.
\eea

The ordering of the collision frequency expressed by \eqref{omega_vs_nu} 
means that collisions, while not dominant as in the fluid description 
(\apref{ap_Brag}), are still retained in the version of the 
gyrokinetic theory adopted by us. Their presence is required 
in order for us to be able to assume that the equilibrium distribution 
is Maxwellian [\eqref{fs_exp} below] and for the heating and entropy 
production to be treated correctly (\secsand{sec_en_GK}{sec_heating}). 
However, our ordering of collisions and of the fluctuation amplitudes (\secref{sec_fluct}) 
imposes certain limitations: thus, we cannot treat 
the class of nonlinear phenomena involving particle 
trapping by parallel-varying fluctuations,  
non-Maxwellian tails of particle distributions,  
plasma instabilities arising from the equilibrium pressure anisotropies 
(mirror, firehose) and their possible 
nonlinear evolution to large amplitudes (see discussion in 
\secref{sec_pressure_aniso}).\\ 

The region of validity of the gyrokinetic approximation in the 
wavenumber space is illustrated in \figref{fig_validity_gk}---it 
embraces all of the scales that are expected to be 
traversed by the anisotropic energy cascade (except 
the scales close to the outer scale). 

As we explained above, 
$m_e/m_i$, $\beta_i$, $\kperp\rho_i$ and $\kpar\mfp$ (or $\omega/\nui$) 
are assigned order unity in the gyrokinetic expansion. 
Subsidiary expansions in small $m_e/m_i$ (\secref{sec_els}) and in 
small or large values of the other three parameters (\secsdash{sec_KRMHD}{sec_ERMHD})
can be carried out at a later stage as long as their values 
are not so large or small as to interfere with the primary 
expansion in $\epsilon$. These expansions will yield simpler 
models of turbulence with more restricted domains of validity 
than gyrokinetics. 

\subsection{Gyrokinetic Equation}

Given the gyrokinetic ordering introduced above, the 
expansion of the distribution function up to first order in $\epsilon$ 
can be written as 
\bea
\label{fs_exp}
f_s(t,\vr,\vv) = \fMs(v) - {\qs\ephi(t,\vr)\over\Ts}\fMs(v) + \hs(t,\vR_s,\vperp,\vpar).
\eea
To zeroth order, it is a Maxwellian:\footnote{The use of isotropic equilibrium 
is a significant idealization---this is discussed in more detail 
in \secref{sec_pressure_aniso}.}
\bea
\fMs(v) = {\ns\over(\pi\vths^2)^{3/2}}\exp\(-{v^2\over\vths^2}\), \quad
\vths=\sqrt{2\Ts\over m_s},
\label{fMs_def}
\eea
with uniform density $\ns$ and temperature $\Ts$ and no mean flow. 
As will be explained in more detail in \secref{sec_heating}, 
$\fMs$ has a slow time dependence via the equilibrium temperature, 
$\Ts = \Ts(\epsilon^2 t)$. This reflects the slow heating of the 
plasma as the turbulent energy is dissipated. However, $\Ts$ can be 
treated as a constant with respect to the time dependence of the 
first-order distribution function (the timescale of the turbulent 
fluctuations). 
The first-order part of the distribution function is composed 
of the Boltzmann response [second term in \eqref{fs_exp}, ordered 
in \eqref{phi_order}] 
and the {\em gyrocenter distribution function}~$\hs$. 
The spatial dependence of the latter is expressed not by the particle 
position $\vr$ but by the position $\vR_s$ of the particle gyrocenter 
(or guiding center)---the center of the ring orbit that the particle follows 
in a strong guide field:  
\bea
\label{R_def}
\vR_s = \vr + {\vvperp\times\vz\over\Omega_s}. 
\eea
Thus, some of the velocity dependence of the distribution function 
is subsumed in the $\vR_s$ dependence of $\hs$. 
Explicitly, $\hs$ depends only on two velocity-space variables: 
it is customary in the gyrokinetic literature for these to be chosen as 
the particle energy $\varepsilon_s=m_sv^2/2$ 
and its first adiabatic invariant $\mu_s=m_s\vperp^2/2B_0$ 
(both conserved quantities to two lowest orders in the gyrokinetic expansion).
However, in a straight uniform guide field $B_0\vz$, 
the pair $(\vperp,\vpar)$ is a simpler choice, which will 
mostly be used in what follows (we shall sometimes find 
an alternative pair, $v$ and $\xi=\vpar/v$, useful, especially 
where collisions are concerned). It must be constantly 
kept in mind that derivatives of $\hs$ with respect 
to the velocity-space variables are taken at constant $\vR_s$, {\em not} 
at constant $\vr$. 

The function $\hs$ satisfies {\em the gyrokinetic equation:}
\beq
{\dd\hs\over\dd t} + \vpar{\dd\hs\over\dd z} +
{c\over B_0}\lt\{\avchi,\hs\rt\} = 
{\qs\fMs\over\Ts}\,{\dd\avchi\over\dd t} 
+ \dtcolls, 
\label{GK_eq}
\eeq
where 
\bea
\label{chi_def}
\chi(t,\vr,\vv) = \ephi - {\vpar\Apar\over c} 
- {\vvperp\cdot\vAperp\over c}, 
\eea
the Poisson brackets are defined in the usual way: 
\bea
\label{PB_def}
\{\avchi,\hs\} = \vz\cdot\({\dd\avchi\over\dd\vR_s}\times{\dd \hs\over\dd\vR_s}\),
\eea
and the ring average notation is introduced:
\bea
\label{ring_def}
\<\chi(t,\vr,\vv)\>_{\vR_s} = {1\over2\pi}\int_0^{2\pi}d\gktheta\,
\chi\(t,\vR_s-{\vvperp\times\vz\over\Omega_s},\vv\),
\eea
where $\gktheta$ is the angle in the velocity space 
taken in the plane perpendicular to the guide field $B_0\vz$. 
Note that, while $\chi$ is a function of $\vr$, 
its ring average is a function of $\vR_s$. Note also that 
the ring averages depend on the species index, as 
does the gyrocenter variable $\vR_s$. 
\Eqref{GK_eq} is derived 
by transforming the first-order kinetic equation to the 
gyrocenter variable~\exref{R_def} and ring averaging the result 
\cite[see][or the references given at the beginning of 
\secref{sec_GK}]{Howes_etal}. 
The ring-averaged collision integral 
$(\dd\hs/\dd t)_{\rm c}$ is discussed in \apref{ap_coll}. 

\subsection{Field Equations}
\label{sec_GK_field_eq}

To \eqref{GK_eq}, we must append the equations that determine 
the electromagnetic field, namely, the potentials 
$\ephi(t,\vr)$ and $\vA(t,\vr)$ that enter the expression 
for $\chi$ [\eqref{chi_def}]. 
In the non-relativistic limit ($\vthi\ll c$), 
these are the plasma quasi-neutrality constraint 
[which follows from the Poisson equation~\exref{Max_Poisson} 
to lowest order in $\vthi/c$]: 
\bea
\label{quasineut}
0 = \sum_s \qs\dn_s = 
\sum_s \qs\lt[-{\qs\ephi\over\Ts}\ns + \int d^3\vv\<\hs\>_\vr\rt]
\eea
and the parallel and perpendicular parts of Amp\`ere's 
law [\eqref{Max_Ampere} to lowest order in $\epsilon$ and in $\vthi/c$]:
\bea
\label{Amp_par}
\dperp^2\Apar &=& - {4\pi\over c}\,\jpar 
= -{4\pi\over c}\sum_s \qs\int d^3\vv\,\vpar\<\hs\>_\vr,\\
\nonumber
\dperp^2\dBpar &=& - {4\pi\over c}\,\vz\cdot\bl(\vdperp\times\vjperp\br)\\ 
&=& -{4\pi\over c}\,\vz\cdot\lt[\vdperp\times\sum_s \qs
\int d^3\vv\<\vvperp\hs\>_\vr\rt],
\label{Amp_perp}
\eea
where we have used $\dBpar = \vz\cdot\(\vdperp\times\vAperp\)$ 
and dropped the displacement current. 
Since field variables $\ephi$, $\Apar$ and $\dBpar$ are functions 
of the spatial variable $\vr$, not of the gyrocenter variable $\vR_s$, 
we had to determine the contribution from the gyrocenter distribution 
function $\hs$ to the charge distribution at fixed $\vr$ by 
performing a gyroaveraging operation dual to the ring average defined in 
\eqref{ring_def}:
\beq
\label{gyroavg_def}
\<\hs(t,\vR_s,\vperp,\vpar)\>_{\vr} = {1\over2\pi}\int_0^{2\pi}d\gktheta\,
\hs\(t,\vr+{\vvperp\times\vz\over\Omega_s},\vperp,\vpar\).
\eeq
In other words, the velocity-space integrals 
in \eqsdash{quasineut}{Amp_perp} are performed over $\hs$ 
at constant $\vr$, rather than constant $\vR_s$. 
If we Fourier transform $\hs$ in $\vR_s$, 
the gyroaveraging operation takes a simple mathematical form:
\bea
\nonumber
\<\hs\>_\vr &=& 
\sum_\vk \<e^{i\vk\cdot\vR_s}\>_\vr \hks(t,\vperp,\vpar)\\ 
\nonumber
&=& \sum_\vk e^{i\vk\cdot\vr} 
\lt<\exp\(i\vk\cdot{\vvperp\times\vz\over\Omega_s}\)\rt>_\vr
\hks(t,\vperp,\vpar) \\
&=&\sum_\vk e^{i\vk\cdot\vr}J_0(\kr_s)\hks(t,\vperp,\vpar), 
\label{int_h}
\eea
where $\kr_s=\kperp\vperp/\Omega_s$ and $J_0$ is a Bessel function 
that arose from the angle integral in the velocity space. 
In \eqref{Amp_perp}, an analogous calculation taking into account 
the angular dependence of $\vvperp$ leads to 
\beq
\label{dBpar_eq}
\dBpar = -{4\pi\over B_0}\sum_\vk e^{i\vk\cdot\vr} 
\sum_s\int d^3\vv\,m_s\vperp^2{J_1(\kr_s)\over\kr_s}\,\hks(t,\vperp,\vpar).
\eeq 

Note that \eqref{Amp_perp} [and, therefore, \eqref{dBpar_eq}]
is the gyrokinetic equivalent of 
the perpendicular pressure balance that appeared 
in \secref{sec_RMHD} [\eqref{MHD_pr_bal}]: 
\bea
\nonumber
\dperp^2{B_0\dBpar\over4\pi} 
= \vdperp\cdot\sum_s {\qs B_0\over c}\int d^3\vv\<{\vz\times\vvperp}\hs\>_\vr
\qquad\qquad\quad\\ 
\nonumber
= \vdperp\cdot\sum_s\Omega_s m_s\int d^3\vv\,{\dd\vvperp\over\dd\gktheta}\,
\hs\(t,\vr + {\vvperp\times\vz\over\Omega_s},\vperp,\vpar\)
\quad\ \ \\ 
= -\vdperp\vdperp:\sum_s\int d^3\vv\,m_s\<\vvperp\vvperp\,\hs\>_\vr 
= -\vdperp\vdperp:\delta{\bf P}_\perp,\quad
\label{GK_pr_bal}
\eea
where we have integrated by parts with respect to the gyroangle $\gktheta$ 
and used $\dd\vvperp/\dd\gktheta = \vz\times\vvperp$, $\dd^2\vvperp/\dd\gktheta^2 = -\vvperp$
\citep[cf.\ the Appendix of ][]{Roach_etal}. 

Once the fields are determined, they have to be substituted into 
$\chi$ [\eqref{chi_def}] and the result ring averaged [\eqref{ring_def}]. 
Again, we emphasize that $\ephi$, $\Apar$ and $\dBpar$ are functions of $\vr$, 
while $\avchi$ is a function of $\vR_s$. The transformation is 
accomplished via a calculation analogous to the one that led to 
\eqsand{int_h}{dBpar_eq}: 
\bea
\avchi &=& 
\sum_\vk e^{i\vk\cdot\vR_s}\avchik,\\ 
\avchik &=& J_0(\kr_s)\(\ephi_\vk - {\vpar\Apark\over c}\) 
+ {\Ts\over \qs}{2\vperp^2\over\vths^2}{J_1(\kr_s)\over\kr_s}{\dBpark\over B_0}.
\quad
\label{avchik_eq}
\eea
The last equation establishes a correspondence between the 
Fourier transforms of the fields with respect to $\vr$ 
and the Fourier transform of $\avchi$ with respect to $\vR_s$. 

\subsection{Generalized Energy and the Kinetic Cascade} 
\label{sec_en_GK} 

As promised in \secref{sec_kinetic}, the central unifying 
concept of this paper is now introduced. 

If we multiply the gyrokinetic equation \exref{GK_eq} 
by $\Ts\hs/\fMs$ and integrate over the velocities 
and gyrocenters, we find that the nonlinear 
term conserves the variance of $\hs$ and 
\bea
\nonumber
{d\over dt}\int d^3\vv\intRs{\Ts\hs^2\over2\fMs} 
= \int d^3\vv\intRs\qs\,{\dd\avchi\over\dd t}\,\hs\\
+ \int d^3\vv\intRs{\Ts\hs\over\fMs}\dtcolls.\quad
\label{hsq_eq}
\eea
Let us now sum this equation over all species. 
The first term on the right-hand side is
\bea
\nonumber
&&\!\!\!\!\!\!\!\!
\sum_s\qs\int d^3\vv\intRs{\dd\avchi\over\dd t}\,\hs\\
\nonumber
&=& \int d^3\vr\sum_s\qs\int d^3\vv\lt<{\dd\chi\over\dd t}\,\hs\rt>_\vr\\
\nonumber
&=& \int d^3\vr\lt[{\dd\ephi\over\dd t}\sum_s\qs\int d^3\vv\<\hs\>_\vr
- {1\over c}{\dd\vA\over\dd t}\cdot\sum_s\qs\int d^3\vv\<\vv\,\hs\>_\vr\rt]\\
&=& {d\over dt}\int d^3\vr\sum_s{\qs^2\ephi^2\ns\over2\Ts} + 
\intr\vE\cdot\vj,
\label{chi_term}
\eea
where we have used \eqref{quasineut} and Amp\`ere's law [\eqsdash{Amp_par}{Amp_perp}] 
to express the integrals of $\hs$. The second term on the right-hand side 
is the total work done on plasma per unit time. Using Faraday's law 
[\eqref{E_B_def}] and Amp\`ere's law [\eqref{Max_Ampere}], it can be written~as 
\bea
\intr\vE\cdot\vj = - {d\over dt}\intr{|\dvB|^2\over8\pi} + \Pa, 
\label{power}
\eea
where $\Pa\equiv-\intr\vE\cdot\vja$ is the total power injected into the system by 
the external energy sources (outer-scale stirring; 
in terms of the Kolmogorov energy flux $\eps$ used in the scaling arguments in 
\secref{sec_GS}, $\Pa=Vm_i\ni\eps$, where $V$ is the system volume).
Combining \eqsdash{hsq_eq}{power}, we find \citep{Howes_etal}
\bea
\nonumber
{dW\over dt}&\equiv&
{d\over dt}\int d^3\vr\lt[\sum_s\(\int d^3\vv\,{\Ts\<\hs^2\>_\vr\over2\fMs}
-{\qs^2\ephi^2\ns\over2\Ts}\)
+ {|\dvB|^2\over8\pi}\rt]\\
&=& \Pa + \sum_s\int d^3\vv
\intRs{\Ts\hs\over\fMs}\dtcolls.
\label{W_cons}
\eea
$W$ is a positive 
definite quantity---this becomes explicit if we use \eqref{quasineut} 
to express it in terms of the total perturbed distribution function 
$\dfs = -\qs\ephi\fMs/\Ts + \hs$ [see \eqref{fs_exp}]:
\bea
W = \int d^3\vr\lt(\sum_s\int d^3\vv\,{\Ts\dfs^2\over2\fMs}
+ {|\dvB|^2\over8\pi}\rt). 
\label{W_def}
\eea
We will refer to $W$ as the {\em generalized energy}.
We use this term to emphasize 
the role of $W$ as the cascaded quantity in gyrokinetic turbulence 
(see below). This quantity is, in fact, the gyrokinetic 
version of a collisionless kinetic invariant variously referred to as 
the {\em generalized grand canonical potential} \citep[see][who 
points out the fundamental role of this quantity in plasma 
turbulence simulations]{Hallatschek}
or {\em free energy} \citep[e.g.,][]{Fowler,Scott}. The non-magnetic part 
of $W$ is related to the perturbed entropy of the system 
\citep[][see discussion in \secref{sec_heating}]{Krommes_Hu,Sugama_etal,Howes_etal,SCDHHPQT_crete}.\footnote{
Note also that a quadratic form involving both the perturbed distribution 
function and the electromagnetic field appears, in a more general 
form than \eqref{W_def}, in the formulation of the energy principle for
the Kinetic MHD approximation 
\citep{Kruskal_Oberman,Kulsrud_KO,Kulsrud_Varenna}.
Regarding the relationship between Kinetic MHD and gyrokinetics, 
see footnote \ref{fn_KMHD}.} 

\Eqref{W_cons} is a conservation law of the generalized energy: 
$\Pa$ is the source and the second term on the right-hand side, 
which is negative definite, represents collisional dissipation. 
This suggests that we might think of kinetic plasma turbulence in 
terms of the generalized energy $W$ injected by the outer-scale 
stirring and dissipated by collisions. In order for the dissipation 
to be important, the collisional term in \eqref{W_cons} has 
to become comparable to $\Pa$. This can happen in two ways:

\begin{enumerate}

\item At collisional scales ($\kpar\mfp\sim1$) due to deviations 
of the perturbed distribution function from a local perturbed 
Maxwellian (see \secref{sec_visc} and \apref{ap_visc}); 

\item At collisionless scales ($\kpar\mfp\gg1$) due the 
development of small scales in the velocity space---large 
gradients in $\vpar$ (see \secref{sec_par_phase}) or 
$\vperp$ (which is accompanied by the development of small 
perpendicular scales in the position space; see 
\secref{sec_small_scales}). 

\end{enumerate}

Thus, the dissipation is only important at particular 
(small) scales, which are generally well separated  
from the outer scale. The generalized energy is transferred 
from the outer scale to the dissipation scales via 
a nonlinear cascade. We shall call it {\em the kinetic cascade.} 
It is analogous to the energy cascade in 
fluid or MHD turbulence, but a conceptually new feature 
is present: the small scales at which dissipation happens 
are small scales both in the velocity and position space. 
Whereas the large gradients in $\vpar$ are produced by 
the {\em linear} parallel phase mixing, whose role in the 
kinetic dissipation processes has been appreciated for some 
time \citep[][see \secref{sec_par_phase}]{Landau_damping,Hammett_Dorland_Perkins,Krommes_Hu,Krommes_df,Watanabe_Sugama04}, 
the emergence of large gradients in $\vperp$ is due to an 
essentially {\em nonlinear} phase mixing mechanism (\secref{sec_small_scales}). 
At spatial scales smaller than the ion gyroradius, this 
nonlinear perpendicular phase mixing turns out to be a faster 
and, therefore, presumably the dominant way of generating small-scale 
structure in the velocity space. It was anticipated 
in the development of gyrofluid moment hierarchies 
by \citet{Dorland_Hammett}. Here we treat it for the 
first time as a phase-space turbulent cascade: 
this is done in \secsand{sec_ent_KAW}{sec_ent_no_KAW}
\citep[see also][]{SCDHHPQT_crete}. 

In the sections that follow, we shall derive particular 
forms of $W$ for various limiting cases of the gyrokinetic 
theory (\secref{sec_en_els}, \secref{sec_en_KRMHD}, \secref{sec_en_compr},
\secref{sec_en_ERMHD}, \apsand{ap_en_RMHD}{ap_Hall_en}). 
We shall see that the kinetic cascade of $W$ is, indeed, 
a direct generalization of the more familiar fluid 
cascades (such as the RMHD cascades discussed in \secref{sec_RMHD})
and that $W$ contains the energy invariants of 
the fluid models in the appropriate limits. 
In these limits, the cascade of the generalized energy 
will split into several decoupled cascades, as it did 
in the case of RMHD (\secref{sec_RMHD_cascades}). 
Whenever one of the physically important 
scales (\secref{sec_microscales}) is crossed and a change 
of physical regime occurs, these cascades are mixed 
back together into the overall kinetic cascade of $W$, 
which can then be split in a different way as it emerges 
on the ``opposite side'' of the transition region 
in the scale space. The conversion of the Alfv\'enic cascade 
into the KAW cascade and the entropy cascade 
at $\kperp\rho_i\sim1$ is the 
most interesting example of such a transition, discussed 
in \secref{sec_ERMHD}. 

The generalized energy appears to be the only quadratic invariant 
of gyrokinetics in three dimensions; in two dimensions, many other 
invariants appear (see \apref{ap_inv}).

\subsection{Heating and Entropy} 
\label{sec_heating}

In a stationary state, all of the the turbulent power injected by the 
external stirring is dissipated and thus transferred into heat. 
Mathematically, this is expressed as a slow increase in the 
temperature of the Maxwellian equilibrium. In gyrokinetics, 
the heating timescale is ordered as $\sim (\epsilon^2\omega)^{-1}$. 

Even though the dissipation of turbulent 
fluctuations may be occurring ``collisionlessly'' 
at scales such that $\kpar\mfp\gg1$ (e.g., via wave--particle 
interaction at the ion gyroscale; \secref{sec_transition}), 
the resulting heating must ultimately be effected with the help of collisions. 
This is because heating is an irreversible process and 
it is a small amount of collisions that make 
``collisionless'' damping irreversible. In other words, 
slow heating of the Maxwellian equilibrium is 
equivalent to entropy production and Boltzmann's $H$-theorem
rigorously requires collisions to make this possible. 
Indeed, the total entropy of species $s$ is 
\bea
\nonumber
S_s&=&-\int d^3\vr\int d^3\vv\,f_s\ln f_s\\ 
&=& -\int d^3\vr\int d^3\vv\lt(\fMs\ln\fMs 
+ {\dfs^2\over 2\fMs}\rt) + O(\epsilon^3),
\label{ent_def}
\eea
where we took $\intr\dfs=0$. 
It is then not hard to show that 
\beq
{3\over2}\,V\ns\,{1\over\Ts}{d\Ts\over dt} = \overline{dS_s\over dt}
= -\overline{\int d^3\vv\intRs{\Ts\hs\over\fMs}\dtcolls}, 
\label{avg_heating} 
\eeq
where the overlines mean averaging over times longer than 
the characteristic time of the turbulent 
fluctuations $\sim\omega^{-1}$ but shorter than 
the typical heating time $\sim(\epsilon^2\omega)^{-1}$ 
(see \citealt{Howes_etal,SCDHHPQT_crete} for a detailed derivation of this 
and related results on heating in gyrokinetics; 
see also earlier discussions of the entropy production 
in gyrokinetics by \citealt{Krommes_Hu,Krommes_df,Sugama_etal}). 
We have omitted the term describing 
the interspecies collisional temperature equalization. 
Note that both sides of \eqref{avg_heating} 
are order~$\epsilon^2\omega$. 

If we now time average \eqref{W_cons} in a similar fashion, 
the left-hand side vanishes because it is a time derivative of 
a quantity fluctuating on the timescale $\sim\omega^{-1}$ 
and we confirm that the right-hand side of \eqref{avg_heating} 
is simply equal to the average power $\overline{\Pa}$ injected 
by external stirring. 
The import of \eqref{avg_heating} is that it tells us that 
heating can only be effected by collisions, while 
\eqref{W_cons} implies that the injected power 
gets to the collisional scales in velocity and position space 
by means of a kinetic cascade of generalized energy. 

The first term in the expression 
for the generalized energy \exref{W_def} is 
$-\sum_s\Ts\delta S_s$, where $\delta S_s$ is the perturbed 
entropy [see \eqref{ent_def}]. The second term in \eqref{W_def} is 
magnetic energy. Collisionless damping of electromagnetic 
fluctuations can be thought of as a redistribution of the generalized energy, 
transferring the electromagnetic energy into entropy fluctuations, 
while the total $W$ is conserved (a simple example of how that 
happens for collisionless compressive fluctuations in the inertial 
range is worked out in \secref{sec_inv_compr}). 

The contribution 
to the perturbed entropy from the gyrocenter distribution is the integral of 
$-\hs^2/2\fMs$, whose evolution 
equation \exref{hsq_eq} can be viewed as the gyrokinetic 
version of the $H$-theorem. The first term on the right-hand 
side of this equation represents the wave--particle interaction (collisionless damping). 
Under time average, it is related to the work done on plasma 
[\eqref{chi_term}] and hence to the average externally 
injected power $\overline{\Pa}$ 
via time-averaged \eqref{power}.\footnote{\label{fn_work_done}
Note that \eqref{power} is valid not only in the integral form but 
also individually for each wavenumber: indeed, using the Fourier-transformed 
Faraday and Amp\`ere's laws, we have
$\vE_\vk\cdot\vj_\vk^* + \vE_\vk^*\cdot\vj_\vk 
= \vE_\vk\cdot\vjak^* + \vE_\vk^*\cdot\vjak - (1/4\pi){\dd|\dvB_\vk|^2/\dd t}$.
In a stationary state, 
time averaging eliminates the time derivative of the 
magnetic-fluctuation energy, so 
$\overline{\vE_\vk\cdot\vj_\vk^* + \vE_\vk^*\cdot\vj_\vk} = 0$ 
at all $\vk$ except those corresponding to the outer scale, 
where the external energy injection occurs. This means that 
below the outer scale, the work done on one species balances 
the work done on the other. The wave--particle interaction 
term in the gyrokinetic equation is responsible for this energy exchange.}
In a stationary state, this is balanced by the second term
in the right-hand side of \eqref{hsq_eq}, 
which is the collisional-heating, or entropy-production, 
term that also appears in \eqref{avg_heating}. Thus, the generalized 
energy channeled by collisionless damping into entropy fluctuations
is eventually converted into heat by collisions. 
The sub-gyroscale 
entropy cascade, which brings the perturbed distribution function 
$\hs$ to collisional scales, will be discussed 
further in \secsand{sec_ent_KAW}{sec_ent_no_KAW} 
\citep[see also][]{SCDHHPQT_crete}.\\ 


This concludes a short primer on gyrokinetics 
necessary (and sufficient)
for adequate understanding of what is to follow. 
Formally, all further analytical derivations in this paper are simply  
subsidiary expansions of the gyrokinetics 
in the parameters we listed in \secref{sec_params}: 
in \secref{sec_els}, we expand in $(m_e/m_i)^{1/2}$, 
in \secref{sec_KRMHD} in $\kperp\rho_i$ 
(followed by further subsidiary expansions in large and small $\kpar\mfp$ 
in \secref{sec_damping}), and in \secref{sec_ERMHD} in $1/\kperp\rho_i$.

\section{Isothermal Electron Fluid}
\label{sec_els}

In this section, we carry out an expansion of the electron 
gyrokinetic equation in powers of $(m_e/m_i)^{1/2}\simeq0.02$ 
(for hydrogen plasma). 
In virtually all cases of interest, this expansion can be done 
while still considering $\sqrt{\beta_i}$, $\kperp\rho_i$, 
and $\kpar\mfp$ to be order unity.\footnote{One notable exception 
is the LAPD device at UCLA, where $\beta\sim10^{-4}-10^{-3}$ 
(due mostly to the electron pressure because the ions are cold, 
$\tau\sim0.1$, so $\beta_i\sim\beta_e/10$; see, e.g., 
\citealt{Morales_etal_LAPD,Carter_etal06}). 
This interferes with the mass-ratio expansion.} 
Note that the assumption $\kperp\rho_i\sim1$ together with 
\eqref{rho_ratio} mean that 
\bea
\kperp\rho_e\sim \kperp\rho_i(m_e/m_i)^{1/2} \ll 1,
\label{alphae_order}
\eea
i.e., the expansion in $(m_e/m_i)^{1/2}$ means also that we are 
considering scales larger than the electron gyroradius. 
The idea of such an expansion of the electron kinetic 
equation has been utilized many times 
in plasma physics literature. The mass-ratio expansion of the 
gyrokinetic equation in a form very similar to what is 
presented below is found in \citet{Snyder_Hammett}. 

The primary import of this section will be technical: we shall 
dispense with the electron gyrokinetic equation and thus 
prepare the necessary ground for further approximations. 
The main results are summarized in \secref{sec_els_sum}. 
A reader who is only interested in following qualitatively 
the major steps in the derivation may skip to this summary. 

\subsection{Ordering the Terms in the Kinetic Equation}
\label{IEF_ordering}

In view of \eqref{alphae_order}, $\kr_e\ll1$, so we 
can expand the Bessel functions arising 
from averaging over the electron ring motion: 
\bea
J_0(\kr_e)=1 - {1\over4}\,{\kr_e^2} + \cdots,\quad
{J_1(\kr_e)\over\kr_e}={1\over2}\(1 - {1\over8}\,\kr_e^2 + \cdots\).
\label{alphae_exp}
\eea 
Keeping only the lowest-order terms of the above expansions 
in \eqref{avchik_eq} for $\avchie$, 
then substituting this $\avchie$ and $\qe=-e$ in the electron 
gyrokinetic equation, we get the following kinetic equation 
for the electrons, accurate up 
to and including the first order in $(m_e/m_i)^{1/2}$ (or in $\kperp\rho_e$): 
\bea
\nonumber
\order{1}{{\dd\he\over\dd t}}
+ \order{0}{\vpar{\dd\he\over\dd z}} 
+ {c\over B_0}\biggl\{\order{1}{\ephi} 
- \order{0}{{\vpar\Apar\over c}}
- \order{1}{{\Te\over e}{\vperp^2\over\vthe^2}{\dBpar\over B_0}},\he\biggr\}\\ 
= -{e\fMe\over\Te}{\dd\over\dd t}
\biggl(\order{1}{\ephi} 
- \order{0}{{\vpar\Apar\over c}} 
- \order{1}{{\Te\over e}{\vperp^2\over\vthe^2}{\dBpar\over B_0}}\biggr)
+ \order{0}{\dtcolle}.
\label{he_eq}
\eea
Note that $\ephi$, $\Apar$, $\dBpar$ in \eqref{he_eq} are taken at $\vr=\vR_e$. 
We have indicated the lowest order to which each of the terms enters 
if compared with $\vpar\dd\he/\dd z$. 
In order to obtain these estimates, we have assumed that the physical ordering 
introduced in \secref{sec_params} 
holds with respect to the subsidiary expansion in $(m_e/m_i)^{1/2}$ as well 
as for the primary gyrokinetic expansion in $\epsilon$, so we can use 
\eqsand{crit_bal}{RMHD_ordering} to order terms with respect to $(m_e/m_i)^{1/2}$. 
We have also made use of \eqsref{rho_ratio}, \exref{phi_order}, 
and of the following three relations:   
\bea
{\kpar\vpar\over\omega}&\sim& {\vthe\over v_A}\sim 
\sqrt{\beta_i\over\tau}\sqrt{m_i\over m_e},\\
{(\vpar/c)\Apar\over\ephi}&\sim& {\vthe\dBperp\over c\kperp\ephi} 
\sim {1\over\kperp\rho_e}{\Te\over e\ephi}{\dBperp\over B_0}
\sim \sqrt{\beta_i\over\tau}\sqrt{m_i\over m_e},\quad\\
{\Te\over e\ephi}{\vperp^2\over\vthe^2}{\dBpar\over B_0} 
&\sim& {Z\over\tau}\,\kperp\rho_i\sqrt{\beta_i}.
\eea
The collision term is estimated to be zeroth order because 
[see \eqsref{omega_vs_nu}, \exref{nue_def}] 
\bea
{\nue\over\omega} \sim {\tau^{3/2}\sqrt{\beta_i}\over Z^2}\sqrt{m_i\over m_e}
{1\over\kpar\mfp}.
\label{ecoll_order}
\eea
The consequences of other possible orderings of the collision terms 
are discussed in \secref{sec_els_validity}. 
We remind the reader that all dimensionless parameters except $\kpar/\kperp\sim\epsilon$ 
and $(m_e/m_i)^{1/2}$ are held to be order unity. 

We now let $\he=\hezero + \heone + \dots$ and carry out the expansion 
to two lowest orders in $(m_e/m_i)^{1/2}$. 

\subsection{Zeroth Order}
\label{sec_els_zero}

To zeroth order, the electron kinetic equation is 
\bea
\vpar\Dpar\hezero = \vpar\,{e\fMe\over c\Te}{\dd\Apar\over\dd t} +
\({\dd\hezero\over\dd t}\)_{\rm c},
\label{hezero_eq}
\eea 
where we have assembled the terms in the left-hand side to take 
the form of the derivative of the distribution function along the 
perturbed magnetic field:
\bea
\Dpar = \dpar + {\dvBperp\over B_0}\cdot\vdel 
= \dpar - {1\over B_0}\lt\{\Apar,\cdots\rt\}. 
\eea 
We now multiply \eqref{hezero_eq} 
by $\hezero/\fMe$ and integrate over $\vv$ and $\vr$ 
(since we are only retaining lowest-order terms, the distinction 
between $\vr$ and $\vR_e$ does not matter here). 
Since $\vdel\cdot\vB=0$, the left-hand side vanishes 
(assuming that all perturbations are either periodic or 
vanish at the boundaries) and we get 
\beq
\int d^3\vr\int d^3\vv\, {\hezero\over\fMe}\({\dd\hezero\over\dd t}\)_{\rm c}
= - {e\ne\over c\Te}\int d^3\vr {\dd\Apar\over\dd t}\,\upare^{(0)} = 0. 
\label{coll_int_zero}
\eeq
The right-hand side of this equation is zero because 
the electron flow velocity is zero in the zeroth order, 
$\upare^{(0)} = (1/\ne)\int d^3\vv \vpar\hezero = 0$. 
This is a consequence of the parallel Amp\'ere's law [\eqref{Amp_par}], 
which can be written as follows
\bea
\label{upare_eq}
\upare = {c\over4\pi e\ne}\vdperp^2\Apar + \upari,
\eea
where
\bea
\upari = \sum_\vk e^{i\vk\cdot\vr}\intvi \vpar J_0(\kr_i)\hki. 
\label{upari_def}
\eea
The three terms in \eqref{upare_eq} can be estimated as follows
\bea
{\upare^{(0)}\over v_A} &\sim& {\epsilon\vthe\over v_A} \sim 
\sqrt{\beta_i\over\tau}\sqrt{m_i\over m_e}\,\epsilon,\\
{\upari\over v_A} &\sim& \epsilon,\\
{c\vdperp^2\Apar\over 4\pi e\ne v_A} &\sim& 
{\kperp\rho_i\over Z\sqrt{\beta_i}}\,\epsilon,
\eea
where we have used the fundamental ordering \exref{RMHD_ordering} 
of the slow waves ($\upari\sim \epsilon v_A$) and Alfv\'en waves 
($\dBperp\sim\epsilon B_0$). 
Thus, the two terms in the right-hand side of \eqref{upare_eq} are 
one order of $(m_e/m_i)^{1/2}$ smaller than $\upare^{(0)}$, which 
means that to zeroth order, the parallel Amp\`ere's law is 
$\upare^{(0)}=0$. 

The collision operator in \eqref{coll_int_zero} contains 
electron--electron and electron--ion collisions. To lowest order 
in $(m_e/m_i)^{1/2}$, the electron--ion collision operator is simply 
the pitch-angle scattering operator [see \eqref{Cgk_ei} in 
\apref{ap_coll} and recall that $\upari$ is first order]. 
Therefore, we may then rewrite \eqref{coll_int_zero} 
as follows
\bea
\nonumber
&&\int d^3\vr\int d^3\vv\, {\hezero\over\fMe}\,\dC_{ee}[\hezero]\\ 
&&- \int d^3\vr\int d^3\vv\, {\nuDei(v)\over\fMe}{1-\xi^2\over2}
\({\dd\hezero\over\dd\xi}\)^2 = 0.
\eea 
Both terms in this expression are negative definite and must, therefore, 
vanish individually. This implies that $\hezero$ must be a perturbed 
Maxwellian distribution with zero mean velocity \citep[this follows from the 
proof of Boltzmann's H theorem; see, e.g.,][]{Longmire_book}, i.e.,  
the full electron distribution function to zeroth order in the mass-ratio expansion 
is [see \eqref{fs_exp}]: 
\bea
f_e = \fMe + {e\ephi\over\Te} + \hezero = 
{n_e\over\(2\pi T_e/m_e\)^{3/2}}\exp\(-{m_e v^2\over2 T_e}\),
\eea 
where $n_e=\ne+\dne$, $T_e=\Te + \dTe$. Expanding around the unperturbed 
Maxwellian $\fMe$, we get
\bea
\hezero = \lt[{\dne\over\ne} - {e\ephi\over\Te} + 
\({v^2\over\vthe^2} - {3\over2}\){\dTe\over\Te}\rt]\fMe, 
\label{hezero_formula_temp}
\eea
where the fields are taken at $\vr=\vR_e$. 
Now substitute this solution back into \eqref{hezero_eq}. 
The collision term vanishes and the remaining equation must be 
satisfied at all values of $v$. This gives 
\bea
\label{Apar_eq}
{1\over c}{\dd\Apar\over\dd t} + \Dpar\ephi &=& \Dpar{\Te\over e}{\dne\over\ne},\\
\label{dTe_const}
\Dpar{\dTe\over\Te} &=& 0. 
\eea
The collision term is neglected in \eqref{Apar_eq} because, 
for $\hezero$ given by \eqref{hezero_formula_temp}, 
it vanishes to zeroth order. 

\subsection{Flux Conservation}
\label{sec_flux}

\Eqref{Apar_eq} implies that the magnetic flux is conserved and magnetic-field lines 
cannot be broken to lowest order in the mass-ratio expansion. 
Indeed, we may follow \citet{Cowley_thesis} 
and argue that the left-hand side of \eqref{Apar_eq} is 
minus the projection of the electric 
field on the total magnetic field [see \eqref{E_B_def}], 
so we have 
\bea
\vE\cdot\vb = -\vb\cdot\vdel\({\Te\over e}{\dne\over\ne}\);
\eea 
hence the total electric field is 
\bea
\vE = \(\unity - \vb\vb\)\cdot\(\vE + \vdel{\Te\over e}{\dne\over\ne}\) 
- \vdel{\Te\over e}{\dne\over\ne}
\eea   
and Faraday's law becomes
\bea
\label{Far_law}
{\dd\vB\over\dd t} &=& -c\vdel\times\vE
= \vdel\times\(\vueff\times\vB\),\\
\vueff &=& {c\over B^2}\(\vE + \vdel{\Te\over e}{\dne\over\ne}\)\times\vB,
\eea 
i.e., the magnetic field lines are frozen into the velocity field $\vueff$. 
In \apref{ap_el_eqns}, we show that 
this effective velocity is the part of the electron 
flow velocity $\vu_e$ perpendicular to the total magnetic field $\vB$
[see \eqref{uperpe_eq}]. 

The flux conservation is broken in the higher orders of the mass-ratio 
expansion. In the first order, Ohmic resistivity formally enters in \eqref{Apar_eq} 
(unless collisions are even weaker than assumed so far; if they 
are downgraded one order as is done in \secref{sec_weaker_colls}, resistivity 
enters in the second order). In the second order, the electron inertia 
and the finiteness of the electron gyroradius also lead to unfreezing 
of the flux. This can be seen formally by keeping second-order terms 
in \eqref{he_eq}, multiplying it by $\vpar$ and integrating over velocities.
The relative importance of these flux unfreezing mechanisms is evaluated 
in \secref{sec_unfreezing}. 

\subsection{Isothermal Electrons}
\label{sec_dTe}

\Eqref{dTe_const} mandates that the perturbed electron temperature 
must remain constant along the perturbed field lines. Strictly speaking, 
this does not preclude $\dTe$ varying across the field lines. However, 
we shall now assume $\dTe=\const$ (has no spatial variation), which is justified, 
e.g., if the field lines are stochastic.
Assuming that no spatially uniform perturbations exist, we may set $\dTe=0$. 
\Eqref{hezero_formula_temp} then reduces to 
\bea
\hezero = \({\dne\over\ne} - {e\ephi\over\Te}\)\fMe(v),
\label{hezero_formula}
\eea
or, using \eqref{fs_exp}, 
\bea
\dfe = {\dne\over\ne}\,\fMe(v). 
\label{dfe_formula}
\eea
Hence follows the equation of state for isothermal electrons:
\bea
\dpe = {\Te\dne}. 
\label{dpe_eq}
\eea 

\subsection{First Order}
\label{sec_els_first}

We now integrate \eqref{he_eq} over the velocity space 
and retain the lowest (first) order terms only. 
Using \eqref{hezero_formula}, we get 
\bea
\nonumber
{\dd\over\dd t}\({\dne\over\ne}-{\dBpar\over B_0}\)
+ {c\over B_0}\lt\{\ephi,{\dne\over\ne}-{\dBpar\over B_0}\rt\}\qquad\qquad&&\\ 
+\ {\dd\upare\over\dd z} - {1\over B_0}\lt\{\Apar,\upare\rt\} 
+ {c\Te\over eB_0}\lt\{{\dne\over\ne},{\dBpar\over B_0}\rt\} &=& 0,\qquad 
\label{dne_eq}
\eea
where the parallel electron velocity is first order: 
\bea
\label{upare_def}
\upare = \upare^{(1)}={1\over\ne}\int d^3\vv\,\vpar\heone.
\eea
The velocity-space integral of the collision term does not enter 
because it is subdominant by at least one factor of $(m_e/m_i)^{1/2}$: 
indeed, as shown in \apref{ap_int_coll}, the velocity integration leads to an extra 
factor of $\kperp^2\rho_e^2$, so that 
\bea
\nonumber
{1\over\ne}\int d^3\vv \({\dd\he\over\dd t}\)_{\rm c}
&\sim&\nue\kperp^2\rho_e^2\,{\dne\over\ne}\\
&\sim&\sqrt{\tau\,{m_e\over m_i}} \kperp^2\rho_i^2\nui\,{\dne\over\ne},
\eea
where we have used \eqsand{rho_ratio}{nue_def}. The collision term is 
subdominant because of the ordering of the ion collision frequency
given by \eqref{omega_vs_nu}. 

\subsection{Field Equations}
\label{sec_els_fields}

Using \eqref{hezero_formula} and $\qi=Ze$, $\ne=Z\ni$, $\Te=\Ti/\tau$, 
we derive from the quasi-neutrality 
equation~\exref{quasineut} [see also \eqref{int_h}] 
\bea
{\dne\over\ne} = {\dni\over\ni} = -{Ze\ephi\over\Ti} + 
\sum_\vk e^{i\vk\cdot\vr}\intvi J_0(\kr_i)\hki,
\label{quasineut2} 
\eea
and, from the perpendicular part of Amp\`ere's law 
[\eqref{dBpar_eq}, using also \eqref{quasineut2}], 
\bea
\nonumber
{\dBpar\over B_0} &=& {\beta_i\over2}\Biggl\{\(1+{Z\over\tau}\){Ze\ephi\over\Ti}
-\sum_\vk e^{i\vk\cdot\vr}\Biggr.\\
&&\times\Biggl.\intvi\lt[{Z\over\tau}\,J_0(\kr_i) 
+ {2\vperp^2\over\vthi^2}{J_1(\kr_i)\over\kr_i}\rt]\hki\Biggr\}.
\label{dBpar_eq2}
\eea
The parallel electron velocity, $\upare$, is determined 
from the parallel part of Amp\`ere's law, \eqref{upare_eq}.  

The ion distribution function $\hi$ that enters these equations 
has to be determined by solving the ion gyrokinetic equation: 
\eqref{GK_eq} with $s=i$. 


\subsection{Generalized Energy} 
\label{sec_en_els} 

The generalized energy (\secref{sec_en_GK})
for the case of isothermal electrons is calculated by substituting 
\eqref{dfe_formula} into~\eqref{W_def}: 
\beq
W = \int d^3\vr\lt(\int d^3\vv\,{\Ti\dfi^2\over2\fMi} + 
{\ne\Te\over2}{\dne^2\over\ne^2}
+ {|\dvB|^2\over8\pi}\rt),
\label{W_els} 
\eeq
where $\dfi=\hi - \lt(Ze\ephi/\Ti\rt)\fMi$ [see \eqref{fs_exp}].

\subsection{Validity of the Mass-Ratio Expansion}
\label{sec_els_validity}

\pseudofigurewide{fig_validity_isoth}{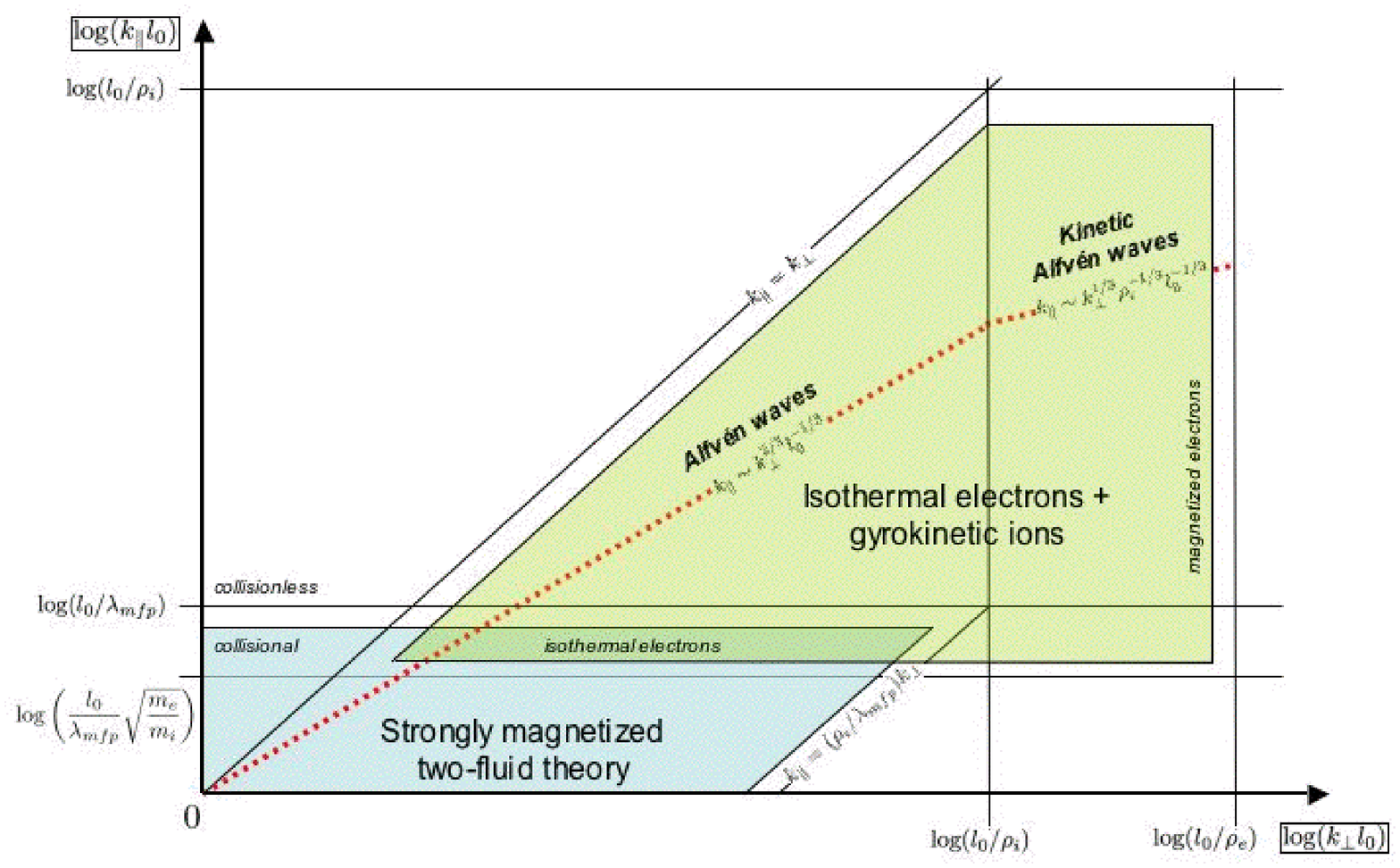}{f4.ps}{
Region of validity in the wavenumber space 
of the secondary approximation---isothermal electrons 
and gyrokinetic ions (\secref{sec_els}). 
It is the region of validity of the gyrokinetic approximation 
(\figref{fig_validity_gk}) further circumscribed by two conditions: 
$\kpar\mfp \gg (m_e/m_i)^{1/2}$ (isothermal electrons) 
and $k_\perp\rho_e \ll 1$ (magnetized electrons). 
The region of validity of the strongly magnetized 
two-fluid theory (\apref{ap_strongly_mag})
is also shown. It is the same as for the full two-fluid 
theory plus the additional constraint $\kperp\rho_i\ll\kpar\mfp$. 
The region of validity of MHD (or one-fluid theory) is the subset 
of this with $\kpar\mfp\ll(m_e/m_i)^{1/2}$ (adiabatic electrons).}

Let us examine the range of spatial scales in which 
the equations derived above are valid. 
In carrying out the expansion in $(m_e/m_i)^{1/2}$, we ordered 
$\kperp\rho_i\sim1$ [\eqref{alphae_order}] and $\kpar\mfp\sim1$ [\eqref{ecoll_order}]. 
Formally, this means that 
the perpendicular and parallel wavelengths of the perturbations 
must not be so small or so large as to interfere with the mass ratio 
expansion. We now discuss the four conditions that this requirement 
leads to and whether any of them can be violated without destroying 
the validity of the equations derived above. 

\subsubsection{$\kperp\rho_i\ll(m_i/m_e)^{1/2}$.} 
\label{sec_mag_els}

This is equivalent to demanding 
that $\kperp\rho_e\ll1$, a condition that was, indeed, essential 
for the expansion to hold [\eqref{alphae_exp}]. 
This is not a serious limitation because electrons can be considered 
well magnetized at virtually all scales of interest for 
astrophysical applications. However, we do forfeit 
the detailed information about some important electron physics 
at $\kperp\rho_e\sim1$: for example such effects as wave damping at the 
electron gyroscale and the electron heating (although 
the total amount of the electron heating can be deduced by 
subtracting the ion heating from the total energy input). 
The breaking of the flux conservation (resistivity) is also 
an effect that requires incorporation of the finite electron 
gyroscale physics. 

\subsubsection{$\kperp\rho_i\gg(m_e/m_i)^{1/2}$.} 

If this condition is 
broken, the small-$\kperp\rho_i$ expansion, carried out in 
\secref{sec_KRMHD}, must, formally speaking, precede the mass-ratio 
expansion. However, it turns out that the small-$\kperp\rho_i$ 
expansion commutes with the mass-ratio expansion 
\citep[][see also footnote~\ref{fn_KMHD}]{SCD_kiev}, so 
we may use the equations derived in \secsdash{sec_els_zero}{sec_els_fields} 
when $\kperp\rho_i\lesssim(m_e/m_i)^{1/2}$. 

\subsubsection{$\kpar\mfp\ll(m_i/m_e)^{1/2}$.} 
\label{sec_weaker_colls}

Let us consider what happens if this condition is broken and 
$\kpar\mfp\gtrsim(m_i/m_e)^{1/2}$. 
In this case, the collisions become even weaker and the expansion 
procedure must be modified. Namely, 
the collision term picks up one extra order of $(m_e/m_i)^{1/2}$, 
so it is first order in \eqref{he_eq}. To zeroth order, the electron kinetic 
equation no longer contains collisions: instead of \eqref{hezero_eq}, we have
\bea
\vpar\Dpar\hezero = \vpar\,{e\fMe\over c\Te}{\dd\Apar\over\dd t}. 
\label{hezero_eq_collless}
\eea 
We may seek the solution of this equation in the form 
$\hezero=H(t,\vR_e)\fMe + \tH$, where $H(t,\vR_e)$ is an unknown 
function to be determined and $\tH$ is the homogeneous solution satisfying 
\bea
\Dpar\tH = 0, 
\eea
i.e., $\tH$ must be constant along the perturbed magnetic field. 
This is a generalization of \eqref{dTe_const}. Again assuming stochastic field 
lines, we conclude that $\tH$ is independent of space. If we rule out 
spatially uniform perturbations, we may set $\tH=0$. 
The unknown function $H(t,\vR_e)$ is readily expressed in terms 
of $\dne$ and $\ephi$:
\bea
{\dne\over\ne} = {e\ephi\over\Te} + {1\over\ne}\int d^3\vv\hezero
\quad\Rightarrow\quad
H = {\dne\over\ne} - {e\ephi\over\Te}, 
\eea
so $\hezero$ is again given by \eqref{hezero_formula}, so 
the equations derived in \secsdash{sec_els_zero}{sec_els_fields} are unaltered. 
Thus, the mass-ratio expansion remains valid at $\kpar\mfp\gtrsim(m_i/m_e)^{1/2}$. 

\subsubsection{$\kpar\mfp\gg(m_e/m_i)^{1/2}$.} 
\label{sec_isoth_els}

If the parallel wavelength 
of the fluctuations is so long that this is violated, 
$\kpar\mfp\lesssim(m_e/m_i)^{1/2}$, 
the collision term in \eqref{he_eq} is minus first order. 
This is the lowest-order term in the equation. Setting it to zero obliges 
$\hezero$ to be a perturbed Maxwellian again given by \eqref{hezero_formula_temp}. 
Instead of \eqref{hezero_eq}, the zeroth-order kinetic equation is
\bea
\vpar\Dpar\hezero = \vpar\,{e\fMe\over c\Te}{\dd\Apar\over\dd t} +
\({\dd\heone\over\dd t}\)_{\rm c}.
\label{heone_eq}
\eea 
Now the collision term in this order contains $\heone$, which can be determined 
from \eqref{heone_eq} by inverting the collision operator. This sets up a 
perturbation theory that in due course leads to 
the Reduced MHD version of the general MHD equations---this 
is what was considered in \secref{sec_RMHD}. 
\Eqref{dTe_const} no 
longer needs to hold, so the electrons are not isothermal. 
In this true one-fluid limit, both electrons and ions are adiabatic 
with equal temperatures [see \eqref{equal_temperatures} below]. 
The collisional transport terms in this limit 
(parallel and perpendicular resistivity, viscosity, heat fluxes, etc.) 
were calculated [starting not from gyrokinetics but from the 
general Vlasov--Landau equation \exref{Vlasov_eq}] 
in exhaustive detail by \citet{Braginskii}. 
His results and the way RMHD emerges from them are 
reviewed in \apref{ap_Brag}. 

In physical terms, the electrons can no longer be 
isothermal if the parallel electron diffusion time becomes longer than 
the characteristic time of the fluctuations (the Alfv\'en time): 
\beq
{1\over\vthe\mfp\kpar^2} \gtrsim {1\over\kpar v_A} 
\quad\Leftrightarrow\quad
\kpar\mfp \lesssim {1\over\sqrt{\beta_i}}\sqrt{m_e\over m_i}.
\eeq
Furthermore, under a similar condition, electron and ion temperatures 
must equalize: this happens if the ion--electron collision time is shorter 
than the Alfv\'en time, 
\bea
{1\over\nuie}\lesssim {1\over\kpar v_A}
\quad\Leftrightarrow\quad
\kpar\mfp \lesssim \sqrt{\beta_i}\sqrt{m_e\over m_i}\quad
\label{equal_temperatures}
\eea 
(see \citealt{Lithwick_Goldreich} for a discussion of these 
conditions in application to the ISM). 

\subsection{Summary} 
\label{sec_els_sum}

The original gyrokinetic description introduced in \secref{sec_GK} was 
a system of two kinetic equations [\eqref{GK_eq}] that evolved 
the electron and ion distribution functions $\he$, $\hi$ 
and three field equations [\eqsdash{quasineut}{Amp_perp}] 
that related $\ephi$, $\Apar$ and $\dBpar$ to $\he$ and $\hi$. 
In this section, we have taken advantage of the smallness of the electron 
mass to treat the electrons as an isothermal magnetized fluid, 
while ions remained fully gyrokinetic. 

In mathematical terms, we solved the electron kinetic equation 
and replaced the gyrokinetics with a simpler closed system of 
equations that evolve 6 unknown functions: 
$\ephi$, $\Apar$, $\dBpar$, $\dne$, $\upare$ and $\hi$. These satisfy 
two fluid-like evolution equations~\exsand{Apar_eq}{dne_eq}, 
three integral relations~\exref{quasineut2}, \exref{dBpar_eq2}, 
and \exref{upare_eq} which involve $\hi$, and the kinetic equation~\exref{GK_eq} 
for $\hi$. The system is simpler because the full electron distribution 
function has been replaced by two scalar fields $\dne$ and $\upare$. 
We now summarize this new system of equations:
denoting $\kr_i={\kperp\vperp/\Omega_i}$, we have
\bea
\label{Apar_eq_sum}
{1\over c}{\dd\Apar\over\dd t} + \Dpar\ephi 
&=& \Dpar{\Te\over e}{\dne\over\ne},\\ 
{d\over d t}\({\dne\over\ne}-{\dBpar\over B_0}\) +\Dpar\upare 
&=& -{c\Te\over eB_0}\lt\{{\dne\over\ne},{\dBpar\over B_0}\rt\}, 
\label{dne_eq_sum}
\eea
\bea
\label{quasineut_sum}
{\dne\over\ne} &=& -{Ze\ephi\over\Ti} + 
\sum_\vk e^{i\vk\cdot\vr}\intvi J_0(\kr_i)\hki,\\
\label{upare_eq_sum}
\upare &=& {c\over4\pi e\ne}\vdperp^2\Apar 
+ \sum_\vk e^{i\vk\cdot\vr}\intvi \vpar J_0(\kr_i)\hki,\quad\\
\nonumber
{\dBpar\over B_0} &=& {\beta_i\over2}\Biggl\{\(1+{Z\over\tau}\){Ze\ephi\over\Ti}
-\sum_\vk e^{i\vk\cdot\vr}\Biggr.\\
&&\times\Biggl.\intvi\lt[{Z\over\tau}\,J_0(\kr_i) 
+ {2\vperp^2\over\vthi^2}{J_1(\kr_i)\over\kr_i}\rt]\hki\Biggr\},
\label{dBpar_eq_sum}
\eea 
and \eqref{GK_eq} for $s=i$ and ion--ion collisions only:
\beq
{\dd\hi\over\dd t} + \vpar{\dd\hi\over\dd z} + 
{c\over B_0}\lt\{\avchii,\hi\rt\} =
{Ze\over\Ti}{\dd\avchii\over\dd t}\,\fMi 
+ \lt<\dC_{ii}[\hi]\rt>_{\vR_i}, 
\label{GK_ions_sum}
\eeq
where $\lt<\dC_{ii}[\dots]\rt>_{\vR_i}$ is the gyrokinetic 
ion--ion collision operator (see \apref{ap_coll}) 
and the ion--electron collisions have been neglected 
to lowest order in $(m_e/m_i)^{1/2}$ [see \eqref{nuie_def}]. 
Note that \eqsdash{Apar_eq_sum}{dne_eq_sum} have been written 
in a compact form, where 
\bea
{d\over dt} = {\dd\over\dd t} + \vu_E\cdot\vdel 
= {\dd\over\dd t} + {c\over B_0}\lt\{\ephi,\cdots\rt\}
\label{def_ddt}
\eea
is the convective derivative with respect to 
the $\vE\times\vB$ drift velocity, $\vu_E=-c{\vdperp\ephi\times\vz/B_0}$, 
and  
\bea
\Dpar = \dpar + {\dvBperp\over B_0}\cdot\vdel 
= \dpar - {1\over B_0}\lt\{\Apar,\cdots\rt\}
\label{def_Bdgrad}
\eea 
is the gradient along the total magnetic field 
(mean field plus perturbation). 

The generalized energy conserved 
by \eqsdash{Apar_eq_sum}{GK_ions_sum} is 
given by \eqref{W_els}. 

It is worth observing that 
the left-hand side of \eqref{Apar_eq_sum} is simply minus the 
component of the electric field along the total magnetic 
field [see \eqref{E_B_def}]. This was used in \secref{sec_flux} to prove 
that the magnetic flux described by \eqref{Apar_eq_sum} 
is exactly conserved (see \secref{sec_unfreezing} for a discussion 
of scales at which this conservation is broken). 
\Eqref{Apar_eq_sum} is the projection of the 
generalized Ohm's law onto the total magnetic field---the right-hand 
side of this equation is the so-called thermoelectric term. 
This is discussed in more detail in \apref{ap_el_eqns}, where 
we also show that \eqref{dne_eq_sum} is  
the parallel part of Faraday's law and give a qualitative 
non-gyrokinetic derivation of \eqsdash{Apar_eq_sum}{dne_eq_sum}. 

We will refer to \eqsdash{Apar_eq_sum}{GK_ions_sum} 
as {\em the equations of isothermal electron fluid.}
They are valid in a broad range 
of scales: the only constraints are that 
$\kpar\ll\kperp$ (gyrokinetic ordering, \secref{sec_params}),
$\kperp\rho_e\ll1$ (electrons are magnetized, \secref{sec_mag_els}) 
and $\kpar\mfp\gg(m_e/m_i)^{1/2}$ (electrons are isothermal, \secref{sec_isoth_els}). 
The region of validity of \eqsdash{Apar_eq_sum}{GK_ions_sum} in the 
wavenumber space is illustrated in \figref{fig_validity_isoth}. 
A particular advantage of this hybrid fluid-kinetic system is 
that it is uniformly valid across the transition from magnetized 
to unmagnetized ions (i.e., from $\kperp\rho_i\ll1$ to $\kperp\rho_i\gg1$). 


\section{Turbulence in the Inertial Range: Kinetic RMHD} 
\label{sec_KRMHD}

Our goal in this section is to derive a reduced set of equations that 
describe the magnetized plasma in the limit of small $\kperp\rho_i$. 
Before we proceed with an expansion in $\kperp\rho_i$, 
we need to make a formal technical step, the usefulness of which 
will become clear shortly. A reader with no patience for this or 
any of the subsequent technical developments may skip to 
the summary at the end of this section (\secref{sec_KRMHD_sum}). 

\subsection{A Technical Step} 

Let us formally split the ion gyrocenter distribution function 
into two parts:
\bea
\nonumber
\hi &=& {Ze\over\Ti}\lt\<\ephi-{\vvperp\cdot\vAperp\over c}\rt\>_{\vR_i}\fMi + \gi\\ 
&=& \sum_\vk e^{i\vk\cdot\vR_i}\lt[J_0(\kr_i){Ze\ephi_\vk\over \Ti} 
+ {2\vperp^2\over\vthi^2}{J_1(\kr_i)\over\kr_i}{\dBpark\over B_0}\rt]\fMi 
+ \gi.\quad\ 
\label{g_ansatz}
\eea
Then $\gi$ satisfies the following equation, obtained by substituting 
\eqref{g_ansatz} and the expression for $\dd\Apar/\dd t$ that follows 
from \eqref{Apar_eq_sum} into the ion gyrokinetic equation~\exref{GK_ions_sum}:
\bea
\nonumber
\order{0}{{\dd\gi\over\dd t} + \vpar{\dd\gi\over\dd z} + {c\over B_0}\lt\{\avchii,\gi\rt\} 
- \lt<\dC_{ii}[\gi]\rt>_{\vR_i}} = \\ 
\nonumber
-{Ze\over\Ti}\,\vpar\lt\<\order{1}{{1\over B_0}\lt\{\Apar,\ephi-\<\ephi\>_{\vR_i}\rt\}}\rt.\\
\nonumber
+ \lt. \order{0}{\Dpar\({\Te\over e}{\dne\over\ne}-
\lt\<{\vvperp\cdot\vAperp\over c}\rt\>_{\vR_i}\)}
\rt\>{\vphantom{\Biggr>}}_{\vR_i}\,\fMi\\ 
+ {Ze\over\Ti}\lt<\dC_{ii}\biggl[\biggl\<\order{1}{\ephi}
-\order{0}{{\vvperp\cdot\vAperp\over c}}\biggr\>{\vphantom{\biggr>}}_{\vR_i}\,\fMi\biggr]\rt>{\vphantom{\Biggr>}}_{\vR_i}\,.
\label{g_eq}
\eea
In the above equation, we have used compact notation 
in writing out the nonlinear terms: e.g., 
$\lt<\lt\{\Apar,\ephi-\<\ephi\>_{\vR_i}\rt\}\rt\>_{\vR_i}
= \lt<\lt\{\Apar(\vr),\ephi(\vr)\rt\}\rt\>_{\vR_i}
-\lt\{\<\Apar\>_{\vR_i},\<\ephi\>_{\vR_i}\rt\}$,
where the first Poisson bracket involves derivatives 
with respect to $\vr$ and the second with respect to $\vR_i$. 

The field equations~\exsdash{quasineut_sum}{dBpar_eq_sum} rewritten 
in terms of~$\gi$~are
\bea
\nonumber
\order{0}{{\dnek\over\ne}} - \order{0}{\Gamma_1(\krsq_i){\dBpark\over B_0}} 
+ \order{1}{\bl[1-\Gamma_0(\krsq_i)\br]{Ze\ephi_\vk\over\Ti}}\\ 
= \order{0}{\intvi J_0(\kr_i)\gki},
\label{dnek_from_g}
\eea
\beq
\order{0}{\uparek} + \order{1}{{c\over4\pi e\ne}\,\kperp^2\Apark} =  
\order{0}{\intvi \vpar J_0(\kr_i)\gki} = \uparik,
\label{uparek_from_g}
\eeq
\bea
\nonumber
\order{0}{{Z\over\tau}{\dnek\over\ne}} +
\order{0}{\lt[\Gamma_2(\krsq_i)+{2\over\beta_i}\rt]{\dBpark\over B_0}} 
- \order{1}{\bl[1-\Gamma_1(\krsq_i)\br]{Ze\ephi_\vk\over\Ti}}\\
= -\order{0}{\intvi {2\vperp^2\over\vthi^2}{J_1(\kr_i)\over\kr_i}\gki},
\label{dBpark_from_g} 
\eea
where $\kr_i=\kperp\vperp/\Omega_i$, $\krsq_i=\kperp^2\rho_i^2/2$ and we have defined 
\bea
\nonumber
\Gamma_0(\krsq_i) &=& \intvi\lt[J_0(\kr_i)\rt]^2\fMi\\ 
\label{G0_def}
&=& I_0(\krsq_i)\,e^{-\krsq_i} = 1-\krsq_i + \cdots,\\
\nonumber
\Gamma_1(\krsq_i) &=& \intvi{2\vperp^2\over\vthi^2}J_0(\kr_i)
{J_1(\kr_i)\over\kr_i}\fMi = -\Gamma_0'(\krsq_i)\\ 
\label{G1_def}
&=& \lt[I_0(\krsq_i)-I_1(\krsq_i)\rt]\,e^{-\krsq_i} 
= 1 - {3\over2}\,\krsq_i + \cdots,\\
\Gamma_2(\krsq_i) &=& \intvi\lt[{2\vperp^2\over\vthi^2}
{J_1(\kr_i)\over\kr_i}\rt]^2\fMi = 2\Gamma_1(\krsq_i). 
\eea
Underneath each term in \eqsdash{g_eq}{dBpark_from_g}, we have indicated 
the lowest order in $\kperp\rho_i$ to which this term enters. 

\subsection{Subsidiary Ordering in $\kperp\rho_i$} 
\label{sec_sub_order}

In order to carry out a subsidiary expansion in small $\kperp\rho_i$, 
we must order all terms in \eqsdash{Apar_eq}{dne_eq} and \exsdash{g_eq}{dBpark_from_g} 
with respect to $\kperp\rho_i$. 
Let us again assume, like we did when expanding the electron equation 
(\secref{sec_els}), that the ordering introduced for 
the gyrokinetics in \secref{sec_params} holds also 
for the subsidiary expansion in $\kperp\rho_i$. 
First note that, in view of \eqref{phi_order}, 
we must regard $Ze\ephi/\Ti$ to be minus first order:
\bea
{Ze\ephi\over\Ti}\sim {\epsilon\over\kperp\rho_i\sqrt{\beta_i}}.
\label{order_phi2}
\eea 
Also, as $\dBperp/B_0\sim\epsilon$ [\eqref{RMHD_ordering}], 
\bea
{(\vpar/c)\Apar\over\ephi}\sim {\vthi\dBperp\over c\kperp\ephi} 
\sim {1\over\kperp\rho_i}{\Ti\over Ze\ephi}{\dBperp\over B_0}
\sim \sqrt{\beta_i},
\eea
so $\ephi$ and $(\vpar/c)\Apar$ are same order. 

Since $\upar=\upari$ (electrons do not contribute 
to the mass flow), assuming that slow waves and Alfv\'en waves 
have comparable energies implies $\upari\sim\uperp$. 
As $\upari$ is determined by the second equality in \eqref{uparek_from_g}, 
we can order~$\gi$ [using \eqref{RMHD_ordering}]:
\bea
\label{order_g}
{\gi\over\fMi}\sim {\upar\over\vthi}\sim{\uperp\over\vthi} 
\sim{\epsilon\over\sqrt{\beta_i}},
\eea
so $\gi$ is zeroth order in $\kperp\rho_i$.  
Similarly, $\dne/\ne\sim\dBpar/B_0\sim\epsilon$ are zeroth order in $\kperp\rho_i$---this 
follows directly from \eqref{RMHD_ordering}. 

Together with \eqref{crit_bal}, the above considerations 
allow us to order all terms in our equations. 
The ordering of the collision term involving $\ephi$ 
is explained in \apref{ap_ii}.

\subsection{Alfv\'en Waves: Kinetic Derivation of RMHD}
\label{sec_AW}

We shall now show that the RMHD equations~\exsdash{RMHD_Psi}{RMHD_Phi} 
hold in this approximation. There is a simple correspondence 
between the stream and flux functions defined in \eqref{Phi_Psi_def} 
and the electromagnetic potentials $\ephi$ and $\Apar$: 
\bea
\Phi = {c\over B_0}\,\ephi,\quad
\Psi = -{\Apar\over\sqrt{4\pi m_i\ni}}.
\label{Phi_Psi_def2}
\eea
The first of these definitions says that the perpendicular 
flow velocity $\vuperp$ is the $\vE\times\vB$ drift velocity; 
the second definition is the standard MHD relation between the 
magnetic flux function and the parallel component of the vector 
potential. 

\subsubsection{Derivation of \eqref{RMHD_Psi}}
\label{sec_RMHD_Psi_deriv}

Deriving \eqref{RMHD_Psi} is straightforward: 
in \eqref{Apar_eq}, we retain only the lowest---minus 
first---order terms (those that contain $\ephi$ and $\Apar$).
The result is 
\bea
{\dd\Apar\over\dd t} + c{\dd\ephi\over\dd z} - {c\over B_0}\lt\{\Apar,\ephi\rt\} = 0. 
\label{RMHD_Apar}
\eea 
Using \eqref{Phi_Psi_def2} and the definition of the 
Alfv\'en speed, $v_A=B_0/\sqrt{4\pi m_i\ni}$, we get \eqref{RMHD_Psi}. 
By the argument of \secref{sec_flux}, \eqref{RMHD_Apar} 
expresses the fact that that magnetic-field lines are frozen 
into the $\vE\times\vB$ velocity field, which is the 
mean flow velocity associated with the Alfv\'en waves 
(see \secref{sec_AW_coll}). 

\subsubsection{Derivation of \eqref{RMHD_Phi}}
\label{sec_RMHD_Phi_deriv}

As we are about to see, in order to derive \eqref{RMHD_Phi}, 
we have to separate the first-order part of the $\kperp\rho_i$ expansion. 
The easiest way to achieve this, is to integrate \eqref{g_eq} over the 
velocity space (keeping $\vr$ constant) 
and expand the resulting equation in small $\kperp\rho_i$. 
Using \eqsand{dnek_from_g}{uparek_from_g} to express 
the velocity-space integrals of $\gi$, we get 
\bea
\nonumber
\order{1}{{\dd\over\dd t}\bl[1-\Gamma_0(\krsq_i)\br]{Ze\ephi_\vk\over\Ti}} 
&+& \order{0}{{\dd\over\dd t}\biggl[{\dnek\over\ne} - \Gamma_1(\krsq_i){\dBpark\over B_0}\biggr]}\\
\nonumber
&+& \dpar\biggl(\order{0}{\uparek} + 
\order{1}{{c\over4\pi e\ne}\,\kperp^2\Apark}\biggr)\\
\nonumber
&+& \order{0}{{c\over B_0}\intvi J_0(\kr_i)\lt\{\avchii,\gi\rt\}_\vk}\\
\nonumber
= \intvi J_0(\kr_i)&&\hspace{-0.75em}
\Biggl<\dC_{ii}\biggl[{Ze\over\Ti}\biggr\<\order{3}{\ephi}
-\order{2}{{\vvperp\cdot\vAperp\over c}}\biggl\>{\vphantom{\biggr>}}_{\vR_i}\!\!\!\fMi
\biggr.\Biggr.\\ 
&+& \Biggl.\biggr. \order{2}{\gi} \biggr]\Biggr>{\vphantom{\Biggr>}}_{\vR_i,\vk}\,.
\label{dni_eq}
\eea
Underneath each term, the lowest order in $\kperp\rho_i$ to which it enters is shown. 
We see that terms containing $\ephi$ are all first order, so 
it is up to this order that we shall retain terms. 
The collision term integrated over the velocity space 
picks up two extra orders of $\kperp\rho_i$ (see \apref{ap_int_coll}), 
so it is second order and can, therefore, be dropped. 
As a consequence of quasi-neutrality, the zeroth-order part of the above equation 
exactly coincides with \eqref{dne_eq}, i.e, $\dni/\ni=\dne/\ne$ satisfy the same equation. 
Indeed, neglecting second-order terms (but not first-order ones!), 
the nonlinear term in \eqref{dni_eq} (the last term on the left-hand side)~is 
\bea
\nonumber
{c\over B_0}\lt\{\ephi,\intvi\gi\rt\}
- {1\over B_0}\lt\{\Apar,\intvi \vpar\gi\rt\}\\ 
+\ {c\Ti\over Ze B_0}\lt\{{\dBpar\over B_0},\intvi {\vperp^2\over\vthi^2}\gi\rt\}, 
\eea
and, using \eqsdash{dnek_from_g}{dBpark_from_g} to express 
velocity-space integrals of $\gi$ in the above expression, 
we find that the zeroth-order 
part of the nonlinearity is the same as the nonlinearity in \eqref{dne_eq}, 
while the first-order part is 
\beq
-{c\over B_0}\lt\{\ephi,{1\over2}\,\rho_i^2\dperp^2{Ze\ephi\over\Ti}\rt\} 
+ {1\over B_0}\lt\{\Apar,{c\over 4\pi e\ne}\,\dperp^2\Apar\rt\},
\eeq
where we have used the expansion~\exref{G0_def} of $\Gamma_0(\krsq_i)$ 
and converted it back into $x$~space.

Thus, if we subtract \eqref{dne_eq} from \eqref{dni_eq}, 
the remainder is first order and reads
\bea
\nonumber
{\dd\over\dd t}{1\over2}\,\rho_i^2\dperp^2{Ze\ephi\over\Ti} +
{c\over B_0}\lt\{\ephi,{1\over2}\,\rho_i^2\dperp^2{Ze\ephi\over\Ti}\rt\} 
\qquad\quad&&\\ 
+\ \dpar{c\over 4\pi e\ne}\dperp^2\Apar
- {1\over B_0}\lt\{\Apar,{c\over 4\pi e\ne}\,\dperp^2\Apar\rt\} &=& 0.\quad
\label{RMHD_phi}
\eea
Multiplying \eqref{RMHD_phi} by $2\Ti/Ze\rho_i^2$ and using 
\eqref{Phi_Psi_def2}, we get the second RMHD equation~\exref{RMHD_Phi}. 

We have established that the Alfv\'en-wave component of the turbulence is 
decoupled and fully described by the RMHD equations \exsand{RMHD_Psi}{RMHD_Phi}. 
This result is the same as that in \secref{sec_AW_fluid} but 
now we have proven that collisions do not affect the Alfv\'en waves and that 
a fluid-like description only requires $\kperp\rho_i\ll1$ to be valid. 

\subsection{Why Alfv\'en Waves Ignore Collisions}
\label{sec_AW_coll}

Let us write explicitly the distribution function of the ion gyrocenters 
[\eqref{g_ansatz}] to two lowest orders in $\kperp\rho_i$:
\bea
\hi = {Ze\over\Ti}\<\ephi\>_{\vR_i}\fMi + 
{\vperp^2\over\vthi^2}{\dBpar\over B_0}\,\fMi + \gi + \cdots,
\label{hi_KRMHD}
\eea 
where, up to corrections of order $\kperp^2\rho_i^2$, 
the ring-averaged scalar potential is 
$\<\ephi\>_{\vR_i} = \ephi(\vR_i)$, the scalar potential 
taken at the position of the ion gyrocenter. 
Note that in \eqref{hi_KRMHD}, the first term is minus first 
order in $\kperp\rho_i$ [see \eqref{order_phi2}], the second 
and third terms are zeroth order [\eqref{order_g}], and 
all terms of first and higher orders are omitted. 
In order to compute the full ion distribution function 
given by \eqref{fs_exp}, we have to convert $\hi$ to the 
$\vr$ space. Keeping terms up to zeroth order, we get 
\bea
\nonumber
{Ze\over\Ti}\<\ephi\>_{\vR_i} \simeq 
{Ze\over\Ti}\ephi(\vR_i) &=& 
{Ze\over\Ti}\lt[\ephi(\vr) + {\vvperp\times\vz\over\Omega_i}\cdot\vdel\ephi(\vr) 
+ \cdots\rt]\\
&=& {Ze\over\Ti}\,\ephi(\vr) + {2\vvperp\cdot\vu_E\over\vthi^2} + \dots,
\label{phi_exp}
\eea
where $\vu_E = -c\vdel\ephi(\vr)\times\vz/B_0$, the $\vE\times\vB$ drift 
velocity. Substituting \eqref{phi_exp} into \eqref{hi_KRMHD} and 
then \eqref{hi_KRMHD} into \eqref{fs_exp}, we find
\bea
f_i = \fMi + {2\vvperp\cdot\vu_E\over\vthi^2}\,\fMi 
+ {\vperp^2\over\vthi^2}{\dBpar\over B_0}\,\fMi + \gi + \cdots.
\label{fi_KRMHD}
\eea
The first two terms can be combined into a Maxwellian with 
mean perpendicular flow velocity $\vuperp=\vu_E$. These are the terms 
responsible for the Alfv\'en waves. The remaining terms, 
which we shall denote $\tdfi$, are the perturbation of the Maxwellian 
in the moving frame of the Alfv\'en waves---they describe  
the passive (compressive) component of the turbulence (see \secref{sec_sw}). 
Thus, the ion distribution function is 
\bea
f_i = {\ni\over (\pi\vthi^2)^{3/2}}\,\exp\lt[-{(\vvperp-\vu_E)^2 + \vpar^2\over\vthi}\rt] 
+ \tdfi.
\label{fi_AW}
\eea

This sheds some light on the indifference of Alfv\'en waves 
to collisions: Alfv\'enic perturbations do not change the 
Maxwellian character of the ion distribution. 
Unlike in a neutral fluid or gas, where viscosity 
arises when particles transport the local mean momentum 
a distance $\sim\mfp$, the particles in a magnetized plasma 
instantaneously take on the local $\vE\times\vB$ velocity
(they take a cyclotron period to adjust, so, roughly speaking, 
$\rho_i$ plays the role of the mean free path). 
Thus, there is no memory of the mean perpendicular motion 
and, therefore, no perpendicular momentum transport.

Some readers may find it illuminating to notice that 
\eqref{RMHD_phi} can be interpreted as stating simply 
$\vdel\cdot\vj=0$: the first two terms represent 
the divergence of the polarization current, 
which is perpendicular to the magnetic field;\footnote{The polarization-drift 
velocity is formally higher order than $\vu_E$ in the gyrokinetic expansion. 
However, since $\vu_E$ does not produce any current, the lowest-order 
contribution to the perpendicular current comes from the polarization 
drift. The higher-order contributions to the gyrocenter distribution 
function did not need to be calculated explicitly because 
the information about the polarization charge is effectively carried 
by the quasi-neutrality condition \exref{quasineut}. We do not belabor 
this point because, in our approach, the notion of polarization 
charge is only ever brought in for interpretative purposes, but is 
not needed to carry out calculations. 
For further qualitative discussion of the role of 
the polarization charge and polarization drift in gyrokinetics, 
we refer the reader to \citealt{Krommes_lectures} and references therein.} 
the last two terms are $\Dpar\jpar$. No contribution to the current arises 
from the collisional term in \eqref{dni_eq} 
as ion--ion collisions cause no particle transport to lowest 
order in $\kperp\rho_i$. 

\subsection{Compressive Fluctuations} 
\label{sec_sw}

The equations that describe the density ($\dne$)
and magnetic-field-strength ($\dBpar$) 
fluctuations follow immediately from \eqsdash{g_eq}{dBpark_from_g} 
if only zeroth-order terms are kept. In these equations, terms that involve 
$\ephi$ and $\Apar$ also contain factors $\sim\kperp^2\rho_i^2$ and are, 
therefore, first-order [with the exception of the nonlinearity on the left-hand side 
of \eqref{g_eq}]. The fact that $\lt<\dC_{ii}[\<\ephi\>_{\vR_i}\fMi]\rt>_{\vR_i}$ in \eqref{g_eq} 
is first order is proved in \apref{ap_ii}. Dropping these terms along with 
all other contributions of order higher than zeroth and 
making use of \eqref{avchik_eq} to write out $\<\chi\>_{\vR_i}$, 
we find that \eqref{g_eq} takes the form 
\bea
\nonumber
{d\gi\over dt} + \vpar\,\Dpar\lt[\gi+
\lt({Z\over\tau}{\dne\over\ne} + {\vperp^2\over\vthi^2}{\dBpar\over B_0}\rt)\fMi\rt] &&\\
= \lt<\dC_{ii}\lt[\gi+{\vperp^2\over\vthi^2}{\dBpar\over B_0}\fMi\rt]\rt>_{\vR_i},&&
\label{sw_g}
\eea
where we have used definitions~\exsdash{def_ddt}{def_Bdgrad} of the convective 
time derivative $d/dt$ and the total gradient along the magnetic field $\Dpar$ 
to write our equation in a compact form. Note that, in view of the correspondence 
between $\Phi$, $\Psi$ and $\ephi$, $\Apar$ [\eqref{Phi_Psi_def2}], 
these nonlinear derivatives are the same as those defined in \eqsdash{dt_def}{dpar_def}. 
The collision term in the right-hand side of the above equation 
is the zeroth-order limit of the gyrokinetic ion--ion collision operator: a useful 
model form of it is given in \apref{ap_ss} [\eqref{Cgk_lowest}]. 

To zeroth order, \eqsdash{dnek_from_g}{dBpark_from_g} are
\bea
\label{sw_n}
{\dne\over\ne} - {\dBpar\over B_0} &=& \intvi\gi,\\
\label{sw_upar}
\upar &=& \intvi \vpar\gi,\\
\label{sw_Bpar}
{Z\over\tau}{\dne\over\ne} + 2\(1+{1\over\beta_i}\){\dBpar\over B_0} 
&=& - \intvi {\vperp^2\over\vthi^2}\,\gi.
\eea
Note that $\upar$ is not an independent quantity---it can be computed from 
the ion distribution but is not needed for the determination of the latter. 

\Eqsdash{sw_g}{sw_Bpar} evolve the ion distribution function $\gi$, 
the ``slow-wave quantities'' $\upar$, $\dBpar$, and the 
density fluctuations~$\dne$. 
The nonlinearities in \eqref{sw_g}, contained in $d/dt$ 
and $\Dpar$, involve the Alfv\'en-wave quantities 
$\Phi$ and $\Psi$ (or, equivalently, $\ephi$ and $\Apar$) determined 
separately and independently by the RMHD equations~\exsdash{RMHD_Psi}{RMHD_Phi}. 
The situation is qualitatively similar to that in MHD (\secref{sec_sw_fluid}), 
except now a kinetic description is necessary---\eqsdash{sw_g}{sw_Bpar} 
replace \eqsdash{eq_drho}{eq_upar}---and the nonlinear scattering/mixing 
of the slow waves and the entropy mode by the Alfv\'en waves takes the form of 
passive advection of the distribution function~$\gi$. 
The density and magnetic-field-strength fluctuations 
are velocity-space moments of~$\gi$. 

Another way to understand the passive nature of the 
compressive component of the turbulence discussed above 
is to think of it as the perturbation of 
a local Maxwellian equilibrium associated 
with the Alfv\'en waves. 
Indeed, in \secref{sec_AW_coll}, 
we split the full ion distribution function [\eqref{fi_AW}] 
into such a local Maxwellian and its perturbation 
\bea
\tdfi = \gi + {\vperp^2\over\vthi^2}{\dBpar\over B_0}\,\fMi. 
\label{dfi_def}
\eea
It is this perturbation that contains all the information about 
the compressive component; 
the second term in the above expression enforces to lowest 
order the conservation of the first adiabatic invariant 
$\mu_i=m_i\vperp^2/2B$. 
In terms of the function \exref{dfi_def}, \eqsdash{sw_g}{sw_Bpar} take 
a somewhat more compact form \citep[cf.][]{SCD_kiev}: 
\bea
\nonumber
{d\over dt}\lt(\tdfi - {\vperp^2\over\vthi^2}{\dBpar\over B_0}\,\fMi\rt) 
+ \vpar\Dpar\lt(\tdfi + {Z\over\tau}{\dne\over\ne}\,\fMi\rt)&&\\ 
= \lt<\dC_{ii}\lt[\tdfi\rt]\rt>_{\vR_i},&&
\label{KRMHD_dfi}
\eea
\bea
\label{KRMHD_dfi_n}
{\dne\over\ne} &=& \intvi\tdfi,\\
\label{KRMHD_dfi_Bpar}
{\dBpar\over B_0} &=& 
- {\beta_i\over2}\intvi\lt({Z\over\tau} + {\vperp^2\over\vthi^2}\rt)\tdfi.
\eea

\pseudofigureone{fig_cascade_channels}{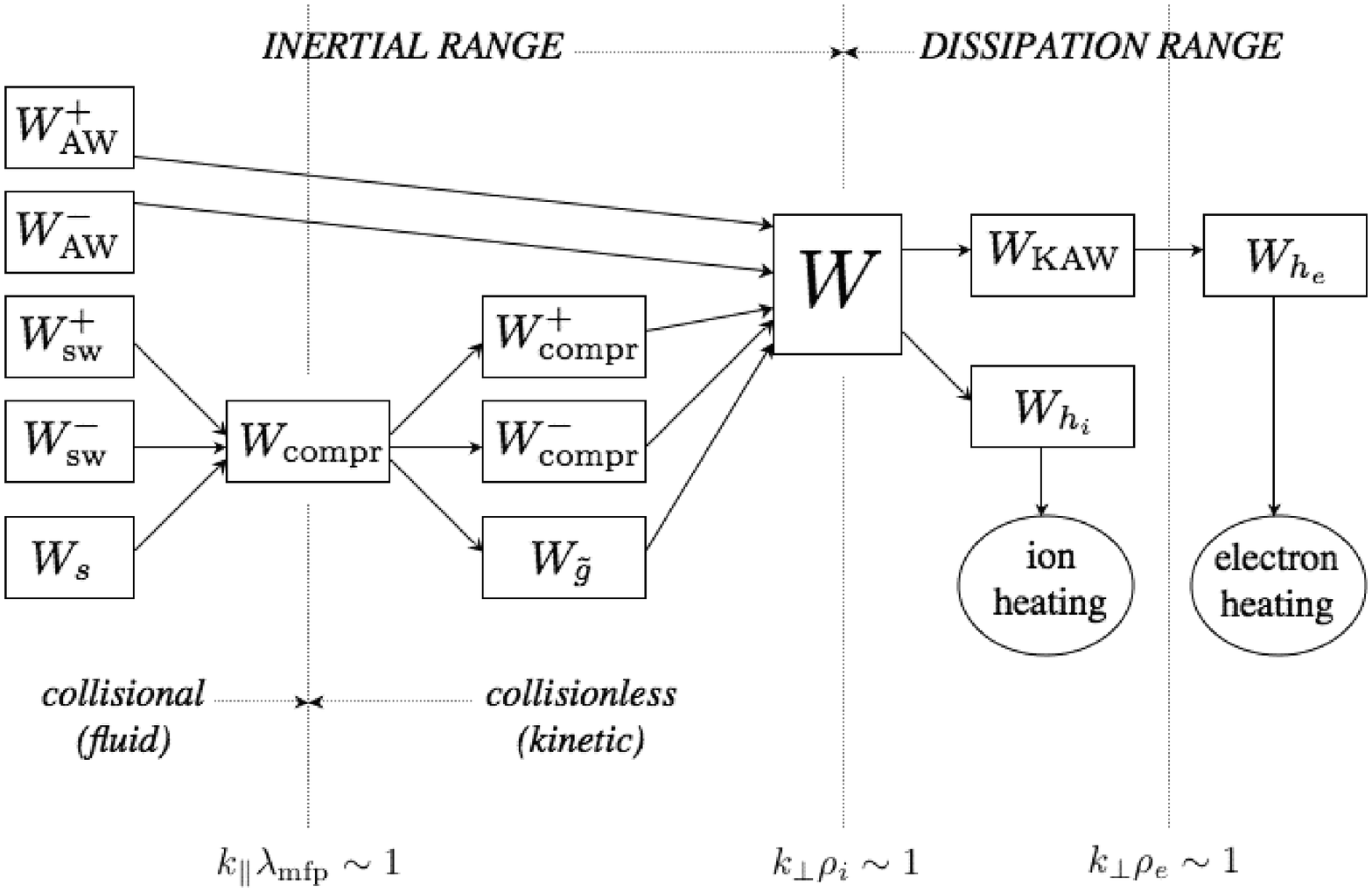}{f5.ps}{Channels 
of the kinetic cascade of generalized energy (\secref{sec_en_GK}) from large 
to small scales: see \secref{sec_RMHD_cascades} and \apref{ap_en_RMHD} (inertial range, collisional regime), 
\secsand{sec_en_KRMHD}{sec_en_compr} (inertial range, collisionless regime), 
\secref{sec_en_ERMHD} and \secref{sec_ent_els} (dissipation range). Note that 
some ion heating probably also results from the collisional and collisionless 
damping of the compressive fluctuations in the inertial range (see \secsand{sec_visc_diss}{sec_par_phase}).}

\subsection{Generalized Energy: Three KRMHD Cascades} 
\label{sec_en_KRMHD} 

The generalized energy (\secref{sec_en_GK}) 
in the limit $\kperp\rho_i\ll1$ is 
calculated by substituting into \eqref{W_els} 
the perturbed ion distribution function
$\dfi = 2\vvperp\cdot\vu_E\fMi/\vthi^2 + \tdfi$ 
[see \eqsand{fi_KRMHD}{dfi_def}].  
After performing velocity integration, we get 
\bea
\nonumber
W &=& \int d^3\vr
\lt[{m_i\ni u_E^2\over2} + {\dBperp^2\over8\pi}
\rt.\\
\nonumber
&&+\ \lt.{\ni\Ti\over2} 
\lt({Z\over\tau}{\dne^2\over\ne^2} +
{2\over\beta_i}{\dBpar^2\over B_0^2} 
+ \intvi{\tdfi^2\over\fMi}\rt)\rt]\\
&=& W_{\rm AW} + \Wcompr.
\label{W_KRMHD}
\eea
We see that the kinetic energy of the Alfv\'enic fluctuations 
has emerged from the ion-entropy part of the generalized energy.
The first two terms in \eqref{W_KRMHD} are the total 
(kinetic plus magnetic) energy of the Alfv\'en waves,  
denoted $W_{\rm AW}$. As we learned from \secref{sec_AW}, 
it cascades independently of the rest of the generalized 
energy, $\Wcompr$, which contains the 
compressive component of the turbulence (\secref{sec_sw}) 
and is the invariant conserved by \eqsdash{KRMHD_dfi}{KRMHD_dfi_Bpar}.

In terms of the potentials used 
in our discussion of RMHD in \secref{sec_RMHD}, we have 
\bea
\nonumber
W_{\rm AW} &=& \intr{m_i\ni\over2}\lt(|\vdperp\Phi|^2 + |\vdperp\Psi|^2\rt)\\
\nonumber
&=&\intr{m_i\ni\over2}\lt(|\vdperp\zeta^+|^2 + |\vdperp\zeta^-|^2\rt)\\ 
&=& \Wperp^+ + \Wperp^-
\label{W_AW}
\eea
where $\Wperp^+$ and $\Wperp^-$ are the energies of the 
``$+$'' and ``$-$'' waves [\eqref{Wperp_def}], which, as we know 
from \secref{sec_elsasser_AW}, cascade by scattering off each 
other but without exchanging energy. 

Thus, the kinetic cascade in the limit $\kperp\rho_i\ll1$ is 
split, independently of the collisionality, into three 
cascades: of $\Wperp^+$, $\Wperp^-$ and $\Wcompr$. 
The compressive cascade is, in fact, split into three 
independent cascades---the splitting is different 
in the collisional limit (\apref{ap_en_RMHD}) and 
in the collisionless one (\secref{sec_en_compr}). 
\Figref{fig_cascade_channels} schematically summarizes 
both the splitting of the kinetic cascade that we have 
worked out so far and the upcoming developments.

\subsection{Summary}
\label{sec_KRMHD_sum}

In \secref{sec_els}, gyrokinetics was reduced to a hybrid fluid-kinetic 
system by means of an expansion in the electron mass, which was valid for 
$\kperp\rho_e\ll1$. In this section, we have further restricted the scale 
range by taking $\kperp\rho_i\ll1$ and as a result have been able to 
achieve a further reduction in the complexity of the kinetic theory 
describing the turbulent cascades. The reduced theory derived here 
evolves 5 unknown functions: $\Phi$, $\Psi$, $\dBpar$, $\dne$ and $\gi$. 
The stream and flux functions, $\Phi$ and $\Psi$ 
are related to the fluid quantities (perpendicular velocity and 
magnetic field perturbations) via \eqref{Phi_Psi_def} and 
to the electromagnetic potentials $\ephi$, $\Apar$ via \eqref{Phi_Psi_def2}. 
They satisfy a closed system of equations, \eqsdash{RMHD_Psi}{RMHD_Phi}, 
which describe the decoupled cascade of Alfv\'en waves. These are the 
same equations that arise from the MHD approximations, but we have now 
proven that their validity does not depend on the assumption of high 
collisionality (the fluid limit) and extends to scales well below the mean 
free path, but above the ion gyroscale. The physical reasons for this 
are explained in \secref{sec_AW_coll}. The density and 
magnetic-field-strength fluctuations (the ``compressive'' fluctuations, 
or the slow waves and the entropy mode in the MHD limit) now 
require a kinetic description in terms of the ion distribution function 
$\gi$ [or $\tdfi$, \eqref{dfi_def}], 
evolved by the kinetic equation \exref{sw_g} [or \eqref{KRMHD_dfi}]. 
The kinetic equation contains $\dne$ and $\dBpar$, which are, in turn 
calculated in terms of the velocity-space integrals of $\gi$ via 
\eqsand{sw_n}{sw_Bpar} [or \eqsand{KRMHD_dfi_n}{KRMHD_dfi_Bpar}]. 
The nonlinear evolution 
(turbulent cascade) of $\gi$, $\dBpar$ and $\dne$ is due solely 
to passive advection of $\gi$ by the Alfv\'en-wave turbulence. 

Let us summarize the new set of equations: 
\bea
\label{RMHD_Psi_sum}
{\dd\Psi\over\dd t} &=& v_A\Dpar\Phi,\\
\label{RMHD_Phi_sum}
{d\over dt}\dperp^2\Phi &=& v_A\Dpar\dperp^2\Psi,\\
\nonumber
{d\gi\over dt} + \vpar\,\Dpar\lt[\phantfrac\gi\rt.&+&\lt. 
\lt({Z\over\tau}{\dne\over\ne} + {\vperp^2\over\vthi^2}{\dBpar\over B_0}\rt)\fMi\rt]\\
&=& \lt<\dC_{ii}\lt[\gi+{\vperp^2\over\vthi^2}{\dBpar\over B_0}\fMi\rt]\rt>_{\vR_i},
\label{sw_g_sum}
\eea
\beq
{\dne\over\ne} = -\lt[{Z\over\tau} + 2\(1 + {1\over\beta_i}\)\rt]^{-1}\!\!\!
\intvi\lt[{\vperp^2\over\vthi^2} - 2\(1+{1\over\beta_i}\)\rt]\gi, 
\label{sw_n_sum}
\eeq
\beq
{\dBpar\over B_0} = -\lt[{Z\over\tau} + 2\(1 + {1\over\beta_i}\)\rt]^{-1}\!\!\!
\intvi\({\vperp^2\over\vthi^2} + {Z\over\tau}\)\gi,
\label{sw_Bpar_sum}
\eeq
where
\bea
{d\over dt} = {\dd\over\dd t} + \lt\{\Phi,\cdots\rt\},\quad
\Dpar = \dpar + {1\over v_A}\lt\{\Psi,\cdots\rt\}.
\label{dd_def_sum}
\eea
An explicit form of the collision term in the right-hand side of 
\eqref{sw_g_sum} is provided in \apref{ap_ss} [\eqref{Cgk_lowest}]. 

The generalized energy conserved by \eqsdash{RMHD_Psi_sum}{sw_Bpar_sum} 
is given by \eqref{W_KRMHD}. 
The kinetic cascade is split, the Alfv\'enic cascade 
proceeding independently of the compressive one (see \figref{fig_cascade_channels}). 

The decoupling of the Alfv\'enic cascade is manifested 
by \eqsdash{RMHD_Psi_sum}{RMHD_Phi_sum} 
forming a closed subset. As already noted in \secref{sec_els_sum}, 
\eqref{RMHD_Psi_sum} is the component of Ohm's law 
along the total magnetic field, $\vB\cdot\vE=0$. 
\Eqref{RMHD_Phi_sum} can be interpreted as the evolution 
equation for the vorticity of the perpendicular plasma flow 
velocity, which is the $\vE\times\vB$ drift velocity. 

We shall refer to the system of equations~\exsdash{RMHD_Psi_sum}{sw_Bpar_sum} 
as {\em Kinetic Reduced Magnetohydrodynamics (KRMHD)}.\footnote{\label{fn_KMHD} The term 
is introduced by analogy with a popular fluid-kinetic system known as Kinetic MHD, 
or KMHD \citep[see][]{Kulsrud_Varenna,Kulsrud_HPP}. KMHD is derived for 
magnetized plasmas ($\rho_i\ll\mfp$) under the assumption that 
$k\rho_s\ll1$ and $\omega\ll\Omega_s$ but without assuming either 
strong anisotropy ($\kpar\ll\kperp$) or small fluctuations 
($|\dvB|\ll B_0$). The KRMHD equations~\exsdash{RMHD_Psi_sum}{sw_Bpar_sum} 
can be recovered from KMHD by applying to it the GK-RMHD ordering
[\eqref{RMHD_ordering} and \secref{sec_params}] 
and an expansion in $(m_e/m_i)^{1/2}$ \citep{SCD_kiev}. This means that 
the $\kperp\rho_i$ expansion (\secref{sec_KRMHD}), which for KMHD is 
the primary expansion, commutes with the gyrokinetic expansion 
(\secref{sec_GK}) and the $(m_e/m_i)^{1/2}$ expansion (\secref{sec_els}), 
both of which preceded it in this paper.} 
It is a hybrid fluid-kinetic description of low-frequency turbulence in strongly magnetized 
weakly collisional plasma that is uniformly valid at all scales satisfying 
$\kperp\rho_i\ll\min(1, \kpar\mfp)$ 
(ions are strongly magnetized)\footnote{The condition $\kperp\rho_i\ll\kpar\mfp$ must be 
satisfied because in our estimates of the collision terms (\apref{ap_ii}) 
we took $\kperp\rho_i\ll1$ while assuming that $\kpar\mfp\sim1$.\label{fn_str_mag}} 
and $\kpar\mfp\gg(m_e/m_i)^{1/2}$ 
(electrons are isothermal), as illustrated in \figref{fig_validity_reduced}. 
Therefore, it smoothly connects the collisional and collisionless 
regimes and is the appropriate theory for the study of the 
turbulent cascades in the inertial range. 
The KRMHD equations generalize rather straightforwardly to 
plasmas that are so collisionless that one cannot assume a 
Maxwellian equilibrium distribution function 
\citep{Chen_etal_KRMHD}---a situation that is relevant in 
some of the solar-wind measurements (see further discussion in \secref{sec_pressure_aniso}).

KRMHD describe what happens to the turbulent  
cascade at or below the ion gyroscale---we shall move on to these scales 
in \secref{sec_ERMHD}, but first we would like to discuss the 
turbulent cascades of density and magnetic-field-strength fluctuations 
and their damping by collisional and collisionless mechanisms. 

\section{Compressive Fluctuations in the Inertial Range} 
\label{sec_damping}

Here we first derive the nonlinear equations that govern the evolution 
of the compressive (density and magnetic-field-strength) fluctuations in the 
collisional ($\kpar\mfp\ll1$, \secref{sec_visc} and \apref{ap_visc}) 
and collisionless ($\kpar\mfp\gg1$, \secref{sec_colless}) 
limits, discuss the linear damping that these fluctuations 
undergo in the two limits and work out the form the generalized energy 
takes for compressive fluctuations (which is particularly interesting 
in the collisionless limit, \secsdash{sec_inv_compr}{sec_en_compr}). 
As in previous sections, an impatient 
reader may skip to \secref{sec_par_cascade} where the results of 
the previous two subsections are summarized and the implications 
for the structure of the turbulent cascades of the density 
and field-strength fluctuations are discussed. 

\subsection{Collisional Regime}
\label{sec_visc}

\subsubsection{Equations}

In the collisional regime, $\kpar\mfp\ll1$, the fluid limit is recovered 
by expanding \eqsdash{RMHD_Psi_sum}{sw_Bpar_sum} in small $\kpar\mfp$. 
The calculation that is necessary to achieve this is done in \apref{ap_visc} 
(see also \apref{ap_isoth_els}). 
The result is a closed set of three fluid equations that evolve $\dBpar$, $\dne$ and $\upar$:
\bea
\label{coll_dBpar}
{d\over dt}{\dBpar\over B_0} &=& \Dpar\upar + {d\over dt}{\dne\over\ne},\\
\label{coll_upar}
{d\upar\over dt} &=& v_A^2\Dpar{\dBpar\over B_0} + \nupar\Dpar\lt(\Dpar\upar\rt),\\
\label{coll_dTi}
{d\over dt}{\dTi\over\Ti} &=& {2\over3}{d\over dt}{\dne\over\ne} 
+ \kappar\Dpar\lt(\Dpar{\dTi\over\Ti}\rt),
\eea
where 
\beq
\lt(1+ {Z\over\tau}\rt){\dne\over\ne} = - {\dTi\over\Ti} 
- {2\over\beta_i}\lt({\dBpar\over B_0} + 
{1\over 3v_A^2}
\nupar\Dpar\upar\rt), 
\label{coll_pr_bal}
\eeq
and $\nupar$ and $\kappar$ are the coefficients of parallel ion viscosity and 
thermal diffusivity, respectively. The viscous and thermal diffusion 
are anisotropic because plasma is magnetized, $\mfp\gg\rho_i$ \citep{Braginskii}. 
The method of calculation of $\nupar$ and $\kappar$ is explained in 
\apref{ap_transport}. Here we shall ignore numerical prefactors of order 
unity and give order-of-magnitude values for these coefficients:
\bea
\label{nupar_kappar_est}
\nupar\sim\kappar\sim {\vthi^2\over\nui}\sim\vthi\mfp.
\eea

If we set $\nupar=\kappar=0$, \eqsdash{coll_dBpar}{coll_pr_bal} 
are the same as the RMHD equations of \secref{sec_RMHD}
with the sound speed defined~as 
\bea
\label{cs_def}
c_s = v_A\sqrt{{\beta_i\over2}\({Z\over\tau} + {5\over3}\)} 
= \sqrt{{Z\Te\over m_i} + {5\over3}{\Ti\over m_i}}.
\eea
This is the natural definition of $c_s$ for 
the case of adiabatic ions, whose specific heat ratio is $\gamma_i=5/3$, 
and isothermal electrons, whose specific heat ratio is $\gamma_e=1$ 
[because $\dpe=\Te\dne$; see \eqref{dpe_eq}]. Note that 
\eqref{coll_pr_bal} is equivalent to the pressure balance [\eqref{MHD_pr_bal} 
of \secref{sec_RMHD}] with $p=n_i T_i + n_e T_e$ and $\dpe=\Te\dne$. 

As in \secref{sec_RMHD}, 
the fluctuations described by \eqsdash{coll_dBpar}{coll_pr_bal} 
separate into the zero-frequency entropy mode and the left- and right-propagating 
slow waves with 
\bea
\label{sw_disp_rln}
\omega = \pm {\kpar v_A\over \sqrt{1+v_A^2/c_s^2}}
\eea
[see \eqref{eq_zpar}]. 
All three are cascaded independently of each other 
via nonlinear interaction with the Alfv\'en waves.
In \apref{ap_en_RMHD}, we show that 
the generalized energy $\Wcompr$ for this system, given 
in \secref{sec_en_KRMHD}, splits into the three familiar 
invariants $\Wpar^+$, $\Wpar^-$, and $\Ws$, defined by 
\eqsdash{Wpar_def}{Ws_def} (see \figref{fig_cascade_channels}). 

\subsubsection{Dissipation}
\label{sec_visc_diss}

The diffusion terms add dissipation to the equations. 
Because diffusion occurs along the field lines of the total magnetic field 
(mean field plus perturbation), the diffusive terms are nonlinear 
and the dissipation process also involves interaction with the Alfv\'en waves. 
We can estimate the characteristic parallel scale at which the diffusion 
terms become important by balancing the nonlinear cascade time and the 
typical diffusion time: 
\bea
\kpar v_A \sim \vthi\mfp\kpar^2
\quad\Leftrightarrow\quad
\kpar\mfp\sim {1/\sqrt{\beta_i}},
\label{coll_cutoff}
\eea
where we have used \eqref{nupar_kappar_est}. 

Technically speaking, the cutoff given by \eqref{coll_cutoff} 
always lies in the range of $\kpar$ that is outside the region of validity 
of the small-$\kpar\mfp$ expansion adopted in the derivation of 
\eqsdash{coll_dBpar}{coll_dTi}. 
In fact, in the low-beta limit, the collisional cutoff 
falls manifestly in the collisionless scale range, i.e., the collisional 
(fluid) approximation breaks down before the slow-wave and entropy cascades 
are damped and one must use the collisionless (kinetic) limit to calculate 
the damping (see \secref{sec_barnes}). The situation is different 
in the high-beta limit: in this case, the expansion in small 
$\kpar\mfp$ can be reformulated as an expansion in small $1/\sqrt{\beta_i}$
and the cutoff falls within the range of validity of the fluid approximation. 
\Eqsdash{coll_dBpar}{coll_dTi} in this limit are
\bea
\label{coll_dBpar_inc}
{d\over dt}{\dBpar\over B_0} &=& \Dpar\upar,\\
\label{coll_upar_inc}
{d\upar\over dt} &=& v_A^2\Dpar{\dBpar\over B_0} + \nupar\Dpar\lt(\Dpar\upar\rt),\\
\label{coll_dne_inc}
{d\over dt}{\dne\over\ne} &=& {1+Z/\tau\over 5/3 + Z/\tau}\,\kappar\Dpar\lt(\Dpar{\dne\over\ne}\rt).
\eea
As in \secref{sec_RMHD} [\eqref{drho_cascade}], the density fluctuations 
[\eqref{coll_dne_inc}] have decoupled from the slow waves 
[\eqsdash{coll_dBpar_inc}{coll_upar_inc}]. The former are damped by 
thermal diffusion, the latter by viscosity. The corresponding linear dispersion 
relations are 
\bea
\label{disp_rln_n}
\omega &=& -i{1+Z/\tau\over 5/3 + Z/\tau}\,\kappar\kpar^2,\\
\omega &=& \pm\kpar v_A\sqrt{1-\({\nupar\kpar\over2v_A}\)^2} 
- i\,{\nupar\kpar^2\over2}. 
\label{disp_rln_sw_coll}
\eea
\Eqref{disp_rln_n} describes strong diffusive damping of the density fluctuations. 
The slow-wave dispersion relation~\exref{disp_rln_sw_coll} has two distinct regimes: 

\begin{enumerate}

\item When $\kpar<2v_A/\nupar$, it describes 
viscously damped slow waves. In particular, 
in the limit $\kpar\mfp\ll1/\sqrt{\beta_i}$, we have 
\bea
\omega \simeq \pm\kpar v_A - i\,{\nupar\kpar^2\over2}.
\eea 

\item For $\kpar>2v_A/\nupar$, both solutions become purely imaginary, so the slow waves 
are converted into aperiodic decaying fluctuations. 
The stronger-damped (diffusive) branch has $\omega\simeq-i\nupar\kpar^2$, 
the weaker-damped one has 
\bea
\omega \simeq -i\,{v_A^2\over\nupar} \sim -{i\over\beta_i}{\vthi\over\mfp} 
\sim -{i\over\sqrt{\beta_i}}{v_A\over\mfp}.
\label{weaker_branch} 
\eea
This damping effect is called viscous relaxation. 
It is valid until $\kpar\mfp\sim1$, where it is replaced by the collisionless 
damping discussed in \secref{sec_barnes} [see \eqref{damping_high_beta}]. 

\end{enumerate}

The viscous and thermal-diffusive dissipation mechanisms described above
lead, in the limits where they are efficient, to ion heating via the standard fluid 
(collisional) route, involving the development of small parallel scales 
in the position space, but not in velocity space (see \secsand{sec_en_GK}{sec_heating}).   

\subsection{Collisionless Regime} 
\label{sec_colless}

\subsubsection{Equations}
\label{sec_colless_eqns}

In the collisionless regime, $\kpar\mfp\gg1$, the collision integral in the 
right-hand side of the kinetic equation \exref{sw_g_sum} can be neglected. 
The $\vperp$ dependence can then be integrated out of \eqref{sw_g_sum}. 
Indeed, let us introduce the following two auxiliary functions:
\bea
\nonumber
\Gn(\vpar) &=& -\lt[{Z\over\tau} + 2\(1 + {1\over\beta_i}\)\rt]^{-1}\\
\label{Gn_def}
&& \times\,{2\pi\over\ni}\int_0^\infty d\vperp\,\vperp
\lt[{\vperp^2\over\vthi^2} - 2\(1+{1\over\beta_i}\)\rt]\gi,\\ 
\nonumber
\GB(\vpar) &=& -\lt[{Z\over\tau} + 2\(1 + {1\over\beta_i}\)\rt]^{-1}\\
&& \times\,{2\pi\over\ni}\int_0^\infty d\vperp\,\vperp
\({\vperp^2\over\vthi^2} + {Z\over\tau}\)\gi. 
\label{GB_def}
\eea
In terms of these functions, 
\bea
\label{nB_from_GnB}
{\dne\over\ne} = \intvpar\Gn,\quad
{\dBpar\over B_0} = \intvpar\GB 
\eea
and \eqref{sw_g_sum} reduces to the following 
two coupled one-dimensional kinetic equations 
\bea
\nonumber
{d\Gn\over dt} &+& \vpar\Dpar\Gn 
= -\lt[{Z\over\tau} + 2\(1 + {1\over\beta_i}\)\rt]^{-1}\vpar\FM(\vpar)\\
&&\times\Dpar\lt[{Z\over\tau}\(1+{2\over\beta_i}\){\dne\over\ne} 
+ {2\over\beta_i}{\dBpar\over B_0}\rt],\\
\nonumber
{d\GB\over dt} &+& \vpar\Dpar\GB 
= \lt[{Z\over\tau} + 2\(1 + {1\over\beta_i}\)\rt]^{-1}\vpar\FM(\vpar)\\
&&\times\Dpar\lt[{Z\over\tau}\(1+{Z\over\tau}\){\dne\over\ne}
+ \(2 + {Z\over\tau}\){\dBpar\over B_0}\rt],
\eea
where $\FM(\vpar) = (1/\sqrt{\pi}\vthi)\exp(-\vpar^2/\vthi^2)$ is a 
one-dimensional Maxwellian. 
This system can be diagonalized, so it splits into two decoupled 
equations 
\beq
\label{Gpm_eq}
{dG^\pm\over dt} + \vpar\Dpar G^\pm = {\vpar\FM(\vpar)\over\Lambda^\pm}\,\Dpar 
\int_{-\infty}^{+\infty}d\vpar'\,G^\pm(\vpar'), 
\eeq
where 
\bea
\Lambda^\pm = -{\tau\over Z} + {1\over\beta_i} \pm 
\sqrt{\(1+{\tau\over Z}\)^2 + {1\over\beta_i^2}}
\eea
and we have introduced a new pair of functions 
\bea
\label{Gpm_def}
G^+ = \GB + {1\over\sigma}\(1+{Z\over\tau}\)\Gn,\quad
G^- = \Gn + {1\over\sigma}{\tau\over Z}{2\over\beta_i}\GB,
\eea
where
\bea
\label{sigma_def}
\sigma = 1 + {\tau\over Z} + {1\over\beta_i} 
+ \sqrt{\(1+{\tau\over Z}\)^2 + {1\over\beta_i^2}}. 
\eea
\Eqref{Gpm_eq} describes two decoupled kinetic cascades, which we 
will discuss in greater detail in \secsdash{sec_inv_compr}{sec_en_compr}. 

\subsubsection{Collisionless Damping}
\label{sec_barnes}

Fluctuations described by 
\eqref{Gpm_eq} are subject to collisionless damping. Indeed, let us linearize 
\eqref{Gpm_eq}, Fourier transform in time and space, 
divide through by $-i(\omega-\kpar\vpar)$, and integrate over $\vpar$. 
This gives the following dispersion relation 
(the ``$-$'' branch is for $\tGn$, the ``$+$'' branch for $\tGB$) 
\bea
\label{disp_rln_colless}
\zeta_iZ\(\zeta_i\) = \Lambda^\pm - 1,
\eea 
where $\zeta_i=\omega/|\kpar|\vthi=\omega/|\kpar|v_A\sqrt{\beta_i}$ and 
we have used the plasma dispersion function \citep{Fried_Conte}
\bea
\label{def_Z}
Z\(\zeta_i\) = {1\over\sqrt{\pi}}\int_{-\infty}^\infty dx\,{e^{-x^2}\over x-\zeta_i}
\eea
(the integration is along the Landau contour). 
This function is not to be confused with the ion charge parameter $Z=q_i/e$. 

Formally, \eqref{disp_rln_colless} has an infinite number of solutions.
When $\beta_i\sim1$, they are all 
strongly damped with damping rates ${\rm Im}(\omega)\sim |\kpar|\vthi\sim|\kpar|v_A$, 
so the damping time is comparable to the characteristic timescale on 
which the Alfv\'en waves cause these fluctuations to cascade to smaller 
scales. 

It is interesting to consider the high- and low-beta limits. 


\paragraph{High-Beta Limit.} 
\pseudofigureone{fig_gamma_plot}{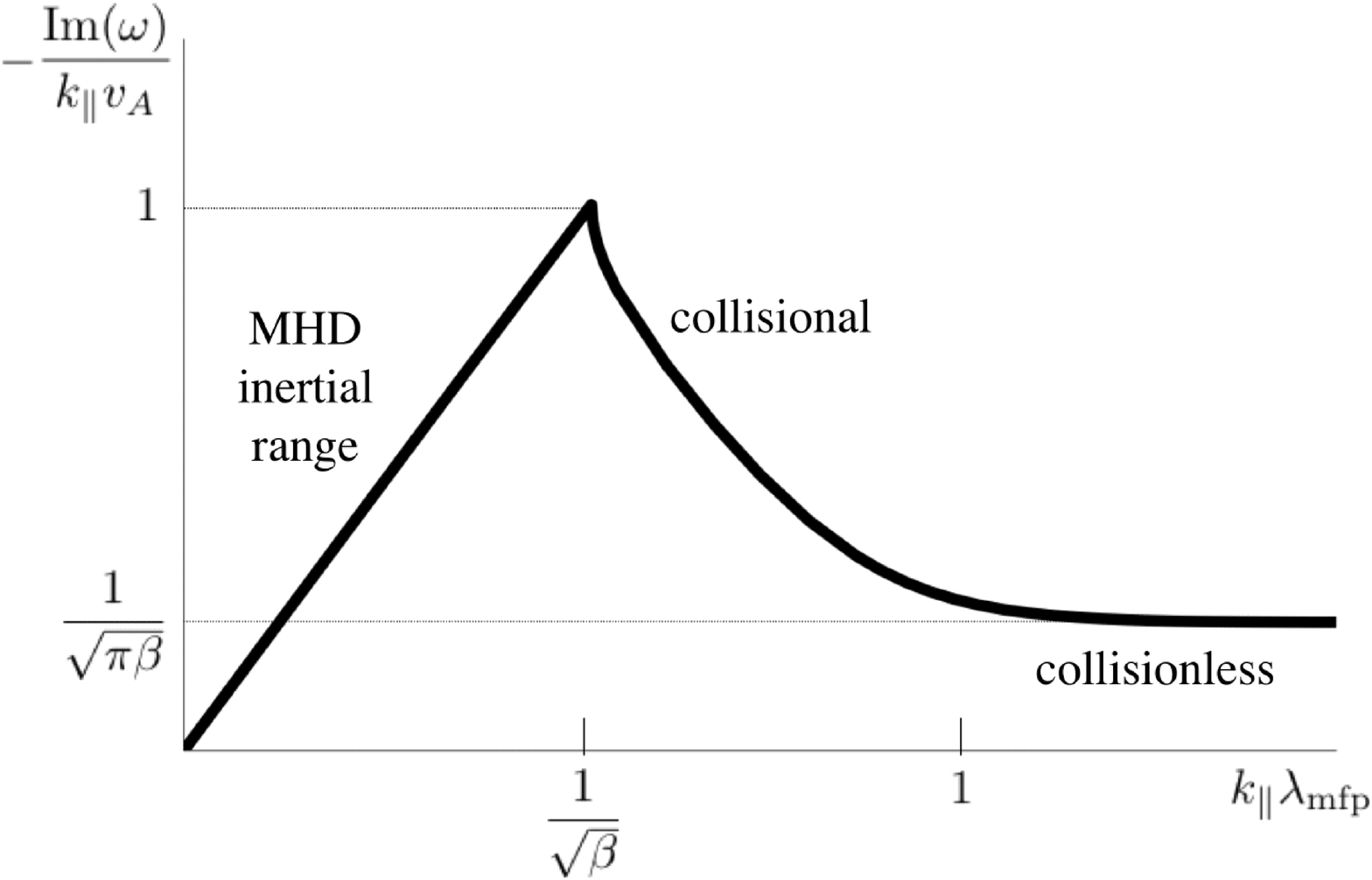}{f6.ps}{Schematic 
log-log plot (artist's impression) of the 
ratio of the damping rate of magnetic-field-strength fluctuations 
to the Alfv\'en frequency $\kpar v_A$ 
in the high-beta limit [see \eqsand{disp_rln_sw_coll}{damping_high_beta}]. 
In \citet{Barnes_etal}, this plot is reproduced via a direct numerical solution 
of the linearized ion gyrokinetic equation with collisions.}

When $\beta_i\gg1$, we have in \eqref{disp_rln_colless}
\bea
\label{tGn_high_beta}
\Lambda^- -1 &\simeq& -2\lt(1 + {\tau\over Z}\rt),
\qquad \tGn \simeq \Gn,\\ 
\label{tGB_high_beta}
\Lambda^+ -1 &\simeq& {1\over\beta_i}, 
\qquad\qquad\quad\ \ \,\tGB \simeq \GB + {1\over2}{Z\over\tau}\,\Gn.
\eea
The ``$-$'' branch corresponds to the density fluctuations. 
The solution of \eqref{disp_rln_colless} has ${\rm Im}(\zeta_i)\sim 1$, 
so these fluctuations are strongly damped:
\bea
\omega\sim-i|\kpar|v_A\sqrt{\beta_i}.
\eea
The damping rate is much greater than the Alfv\'enic rate $\kpar v_A$ 
of the nonlinear cascade. 
In contrast, for the ``$+$'' branch, the damping rate is small: 
it can be obtained by expanding $Z(\zeta_i)=i\sqrt{\pi}+O\(\zeta_i\)$, 
which gives\footnote{This is the gyrokinetic limit ($\kpar/\kperp\ll1$) 
of the more general damping effect known in astrophysics as the 
\citet{Barnes} damping and in plasma physics as transit-time damping. 
We remind the reader that our approach was to carry out the gyrokinetic 
expansion (in small $\kpar/\kperp$) first, and then take the high-beta 
limit as a subsidiary expansion. A more standard approach in the linear 
theory of plasma waves is to take the limit of high $\beta_i$ while 
treating $\kpar/\kperp$ as an arbitrary quantity. 
A detailed calculation of the damping rates done in this way 
can be found in \citet{Foote_Kulsrud}.} 
\bea
\omega = - i\,{|\kpar|\vthi\over\sqrt{\pi}\beta_i}
= - i\,{|\kpar|v_A\over\sqrt{\pi\beta_i}}.  
\label{damping_high_beta}
\eea
Since $\Gn$ is strongly damped, \eqref{tGB_high_beta} 
implies $\tGB\simeq\GB$, i.e., the fluctuations that are 
damped at the rate \exref{damping_high_beta} are predominantly 
of the magnetic-field strength. 
The damping rate is a constant (independent of $\kpar$) 
small fraction $\sim1/\sqrt{\beta_i}$ of the Alfv\'enic cascade rate. 

In \figref{fig_gamma_plot}, we give a schematic plot 
of the damping rate of the magnetic-field-strength fluctuations 
(slow waves) connecting the fluid and kinetic limits for $\beta_i\gg1$. 


\paragraph{Low-Beta Limit.} When $\beta_i\ll1$, we have 
\bea
\label{tGn_low_beta}
\Lambda^- -1 &\simeq& -\lt(1+{\tau\over Z}\rt),
\qquad \tGn \simeq \Gn + {\tau\over Z}\,\GB,\\ 
\label{tGB_low_beta}
\Lambda^+ -1 &\simeq& {2\over\beta_i}, 
\qquad\qquad\quad\,\tGB \simeq \GB.
\eea
For the ``$-$'' branch, we again have ${\rm Im}(\zeta_i)\sim1$, so 
\bea
\omega\sim-i|\kpar|v_A\sqrt{\beta_i},
\eea
which now is much smaller than the Alfv\'enic cascade rate $\kpar v_A$. 
For the ``$+$'' branch (predominantly the field-strength fluctuations), 
we seek a solution with $\zeta=-i\tilde\zeta_i$ and $\tilde\zeta_i\gg1$. 
Then \eqref{disp_rln_colless} becomes 
$\zeta_iZ(\zeta_i) \simeq 2\sqrt{\pi}\,\tilde\zeta_i\exp(\tilde\zeta_i) = 2/\beta_i$. 
Up to logarithmically small corrections, this gives
$\tilde\zeta_i\simeq\sqrt{|\ln\beta_i|}$, whence 
\bea
\omega\sim-i|\kpar|v_A\sqrt{\beta_i|\ln\beta_i|}.
\eea
While this damping rate is slightly greater than that of the 
``$-$'' branch, it is still much smaller than the Alfv\'enic 
cascade rate. 

\subsubsection{Collisionless Invariants}
\label{sec_inv_compr}

\Eqref{Gpm_eq} obeys a conservation law, which is very easy to derive. 
Multiplying \eqref{Gpm_eq} by $G^\pm/\FM$ and integrating over 
space and velocities and performing integration by parts in the 
right-hand side, we get 
\bea
\nonumber
{d\over dt}\intr\intvpar {(G^\pm)^2\over2\FM}
\qquad\qquad\qquad\qquad\qquad\qquad\\ 
= -{1\over\Lambda^\pm}\intr\lt(\intvpar G^\pm\rt)
\Dpar\intvpar\vpar G^\pm.\quad
\label{Gsq_eq}
\eea
On the other hand, integrating \eqref{Gpm_eq} over $\vpar$ gives
\bea
{d\over dt}\intvpar G^\pm = - \Dpar\intvpar\vpar G^\pm. 
\eea
Using this to express the right-hand side of \eqref{Gsq_eq} as 
a full time derivative, we find
\bea
{d\Wcompr^\pm\over dt} = 0,
\eea
where the two invariants are
\beq
\Wcompr^\pm = \intr{\ni\Ti\over2}\lt[\intvpar{(G^\pm)^2\over\FM}
- {1\over\Lambda^\pm}\lt(\intvpar G^\pm\rt)^2\rt].
\label{Wcompr_def}
\eeq

It is useful (and always possible) to split 
\bea
G^\pm = \FM\intvpar G^\pm + \tilde G^\pm,
\eea 
where $\intvpar\tilde G^\pm=0$ by construction. Then 
\bea
\nonumber
\Wcompr^\pm &=& \intr {\ni\Ti\over2}\lt[
\intvpar{(\tilde G^\pm)^2\over\FM} \rt.\\
&&+ \lt.\lt(1- {1\over\Lambda^\pm}\rt)\lt(\intvpar G^\pm\rt)^2\rt].
\label{Wcompr_positive}
\eea
Written in this form, the two invariants $\Wcompr^\pm$ are manifestly 
positive definite quantities because $\Lambda^+>1$ and $\Lambda^-<0$.  
The invariants regulate the two decoupled kinetic 
cascades of compressive fluctuations in the collisionless regime. 
The collisionless damping derived in \secref{sec_barnes} 
leads to exponential decay of the density and field-strength 
fluctuations, or, equivalently, of $\intvpar G^\pm$, while conserving 
$\Wcompr^\pm$. This means that the damping is merely a redistribution 
of the conserved quantity $\Wcompr^\pm$: the first term in 
\eqref{Wcompr_positive} grows to compensate for the decay of the second. 

\subsubsection{Linear Parallel Phase Mixing}
\label{sec_par_phase}

In dynamical terms, how does the kinetic system \eqref{Gpm_eq} arrange
for the integral of the distribution function $G^\pm(\vpar)$ to decay 
while allowing its norm to grow? This is a very well known phenomenon of (linear) 
phase mixing \citep{Landau_damping,Hammett_Dorland_Perkins,Krommes_Hu,Krommes_df,Watanabe_Sugama04}. 
To put it in simple terms, the solution of the linearized \eqref{Gpm_eq} consists 
of the inhomogeneous part, which contains the collisionless damping 
and the homogeneous part (solution of the left-hand side $=0$) given by 
$G^\pm \propto e^{-i\kpar\vpar t}$, the so-called ballistic response
(this is also the nonlinear solution if $t$ and $\kpar$ are interpreted 
as Lagrangian variables in the frame of the Alfv\'en waves; see \secref{sec_par_cascade}). 
As time goes on, this part of the solution becomes increasingly oscillatory 
in $\vpar$, so its velocity integral tends to zero, while its amplitude 
does not decay. It is such ballistic contributions that make up the 
$\tilde G^\pm$ term in \eqref{Wcompr_positive}. 

As the velocity gradient of $\tilde G^\pm$ increases with time, 
$\dd\tilde G^\pm/\dd\vpar \sim \kpar t G^\pm$, at some point 
it can become sufficiently large to activate the collision integral  
[the right-hand side of \eqref{sw_g_sum}], which has so far been neglected. 
This way the collisionless damping of compressive fluctuations 
can be turned into ion heating---a simple example of a more general principle 
of how electromagnetic fluctuation energy is transferred into heat 
via the entropy part of the generalized energy (\secref{sec_heating}). 
Indeed, we will prove in \secref{sec_en_compr} that the invariants 
$\Wcompr^\pm$ are constituent parts of the overall generalized energy 
functional for the compressive fluctuations, so their cascade to small 
scales in phase space is part of the overall kinetic cascade introduced 
in \secref{sec_en_GK}. 

It is not entirely clear how efficient 
is the parallel-phase-mixing route to ion heating
and, therefore, whether the collisionlessly damped energy of 
compressive fluctuations ends up in the ion heat or rather 
reaches the ion gyroscale and couples back to the Alfv\'enic 
component of the turbulence (\secref{sec_transition}). 
The answer to this question will depend on whether compressive 
fluctuations can develop large $\kpar$---a non-trivial issue 
further discussed in \secref{sec_par_cascade}. 

\subsubsection{Generalized Energy: Three Collisionless Cascades}
\label{sec_en_compr}

We will now show how the generalized energy for compressive fluctuations 
in the collisionless regime incorporates the two 
invariants derived in \secref{sec_inv_compr}. 

Rewriting the compressive part of the KRMHD generalized 
energy [\eqref{W_KRMHD}] in terms of the function $\gi$ [see \eqref{dfi_def}], we get
\bea
\nonumber
\Wcompr = {\ni\Ti\over2}\intr\lt\{\intvi{\gi^2\over\fMi}\rt.
\qquad\qquad\\ 
\lt.+\, {Z\over\tau}\lt({\dne\over\ne} - {\dBpar\over B_0}\rt)^2
- \lt[{Z\over\tau} + 2\lt(1+{1\over\beta_i}\rt)\rt]{\dBpar^2\over B_0^2}\rt\}.
\label{Wcompr_g}
\eea
Using \eqsand{nB_from_GnB}{Gpm_def}, we can express $\dne$ and $\dBpar$ in terms 
of $\intvpar G^\pm$ as follows
\bea
\label{dne_Gpm}
{\dne\over\ne} &=& {1\over\kappa}\lt(\sigma \intvpar G^- 
- {\tau\over Z}{2\over\beta_i} \intvpar G^+\rt),\\
\label{dBpar_Gpm}
{\dBpar\over B_0} &=& {1\over\kappa}\lt[\sigma \intvpar G^+ 
- \lt(1+{Z\over\tau}\rt) \intvpar G^-\rt],
\eea
where $\sigma$ was defined in \eqref{sigma_def} and 
\bea
\label{kappa_def}
\kappa = \sqrt{\(1+{\tau\over Z}\)^2 + {1\over\beta_i^2}}. 
\eea
In order to express $\gi$ in terms of $G^\pm$, we have to reconstruct 
the $\vperp$ dependence of $g$, which we integrated out at the beginning 
of \secref{sec_colless_eqns}. 

Let us represent the distribution function as follows
\bea
\gi = {\ni\over\pi\vthi^2}\,e^{-x}\hat\gi(x,\vpar),\quad
\hat\gi(x,\vpar) = \sum_{l=0}^\infty L_l(x) G_l(\vpar),
\eea
where $x=\vperp^2/\vthi^2$ and we have expanded $\hat\gi$ in 
Laguerre polynomials $L_l(x)=(e^x/l!)(d^l/dx^l)x^le^{-x}$. Since  
Laguerre polynomials are orthogonal, the first term in \eqref{Wcompr_g} 
splits into a sum of ``energies'' associated with the expansion coefficients:
\bea
\label{gsq_exp}
\intvi {\gi^2\over\fMi} = \sum_{l=0}^\infty\intvpar {G_l^2\over\FM}.
\eea
The expansion coefficients are determined via the Laguerre transform:
\bea
G_l(\vpar) = \int_0^\infty dx\, e^{-x} L_l(x)\hat\gi(x,\vpar).
\eea 
As $L_0=1$ and $L_1=1-x$, it is easy to see that $\dne$ and $\dBpar$ 
can be expressed as linear combinations of $\intvpar G_0$ and $\intvpar G_1$
[see \eqsdash{Gn_def}{nB_from_GnB}]. 
Using \eqsref{Gn_def}, \exref{GB_def}, and \exref{Gpm_def}, we can show that 
\bea
\label{G0_eq}
G_0 &=& -{1\over\kappa}\lt[\lt(\sigma - {2\over\beta_i}\rt)\Lambda^+ G^+ 
+ {Z\over\tau}\lt(\sigma - 1 - {\tau\over Z}\rt)\Lambda^- G^-\rt],\quad\\
\label{G1_eq}
G_1 &=& {1\over\kappa}\lt[\sigma\Lambda^+ G^+ 
- \lt(1+{Z\over\tau}\rt)\Lambda^- G^-\rt],
\eea
where $G^\pm$ satisfy \eqref{Gpm_eq}. As follows from \eqref{sw_g_sum} (neglecting 
the collision integral), all higher-order expansion coefficients 
satisfy a simple homogeneous equation:
\bea
{d G_l\over dt} + \vpar\Dpar G_l &=& 0,\quad l>1.
\eea
Thus, the distribution function can be explicitly written 
in terms of $G^\pm$: 
\bea
\gi = \lt[G_0(\vpar) + \lt(1-{\vperp^2\over\vthi^2}\rt)G_1(\vpar)\rt]
{\ni\over\pi\vthi^2}\,e^{-\vperp^2/\vthi^2} + \tilde\gi,
\eea
where $G_0$ and $G_1$ are given by \eqsdash{G0_eq}{G1_eq} and 
$\tilde\gi$ comprises the rest of the Laguerre expansion (all $G_l$ with 
$l>1$), i.e., it is the homogeneous solution of \eqref{sw_g_sum} that 
does not contribute to either density or magnetic-field strength: 
\bea
{d\tilde\gi\over dt} + \vpar\Dpar\tilde\gi = 0,\ 
\intv\,\tilde\gi = 0,\ 
\intv\,{\vperp^2\over\vthi^2}\,\tilde\gi = 0.
\label{gt_eq}
\eea 

Now substituting \eqsand{G0_eq}{G1_eq} into \eqref{gsq_exp} and then 
substituting the result and \eqsdash{dne_Gpm}{dBpar_Gpm} into 
\eqref{Wcompr_g}, we find after some straightforward manipulations
\bea
\nonumber
\Wcompr &=& \intr\intv {\Ti\tilde\gi^2\over2\fMi}\\ 
\nonumber
&& +\,\,4\lt[1+{1\over\kappa}\lt(1+{\tau\over Z}\rt)\rt]
(\Lambda^+)^2\Wcompr^+\\
&& +\,\, 2\,{Z^2\over\tau^2}\lt(1 + {1\over\kappa}{1\over\beta_i}\rt)
(\Lambda^-)^2\Wcompr^-,
\label{Wcompr_colless}
\eea
where $\kappa$ is defined by \eqref{kappa_def} and 
$\Wcompr^\pm$ are the two independent invariants that we derived in 
\secref{sec_inv_compr}. 
Thus, the generalized energy for compressive fluctuations splits into three 
independently cascading parts: $\Wcompr^\pm$ associated with the density 
and magnetic-field-strength fluctuations and a purely kinetic part given 
by the first term in \eqref{Wcompr_colless} (see \figref{fig_cascade_channels}). 
The dynamical evolution of this purely kinetic component is described by 
\eqref{gt_eq}---it is a passively mixed, undamped ballistic-type mode. 

All three cascade channels lead to small perpendicular spatial scales 
via passive mixing by the Alfv\'enic turbulence and also 
to small scales in $\vpar$ via the parallel phase mixing process 
discussed in \secref{sec_par_phase} (note that $\tilde\gi$ 
is subject to this process as well).

\subsection{Parallel and Perpendicular Cascades}
\label{sec_par_cascade} 

Let us return to the kinetic equation \exref{sw_g_sum}
and transform it to the Lagrangian 
frame associated with the velocity field $\vuperp=\vz\times\vdperp\Phi$ 
of the Alfv\'en waves: $(t,\vr) \to (t, \vr_0)$, where 
\bea
\vr(t,\vr_0) = \vr_0 + \int_0^t dt'\vuperp(t',\vr(t',\vr_0)). 
\label{r_Lagr}
\eea
In this frame, the convective derivative $d/dt$ defined in \eqref{dd_def_sum} 
turns into $\dd/\dd t$, while the parallel spatial gradient $\Dpar$ 
can be calculated by employing the Cauchy solution for the perturbed magnetic 
field $\dvBperp=\vz\times\vdperp\Psi$: 
\bea
\vb(t,\vr) = \vz + {\dvBperp(t,\vr)\over B_0}
= \vb(0,\vr_0)\cdot\vdel_0\vr,
\label{b_sln}
\eea
where $\vr$ is given by \eqref{r_Lagr} and $\vdel_0=\dd/\dd\vr_0$. 
Then 
\bea
\Dpar = \vb(0,\vr_0)\cdot\bl(\vdel_0\vr\br)\cdot\vdel 
= \vb(0,\vr_0)\cdot\vdel_0 = {\dd\over\dd s_0},
\eea
where $s_0$ is the arc length along the perturbed magnetic 
field taken at $t=0$ [if $\dvBperp(0,\vr_0)=0$, $s_0=z_0$]. 
Thus, in the Lagrangian frame associated with the Alfv\'enic 
component of the turbulence, \eqref{sw_g_sum} is linear. 
This means that, if the effect of finite ion gyroradius 
is neglected, the KRMHD system does not give rise to a cascade 
of density and magnetic-field-strength fluctuations to smaller 
scales along the moving (perturbed) field lines, i.e., 
$\Dpar\dne$ and $\Dpar\dBpar$ do not increase. In contrast, 
there is a perpendicular cascade (cascade in $\kperp$): 
the perpendicular wandering of field lines due to 
the Alfv\'enic turbulence causes passive mixing of 
$\dne$ and $\dBpar$ in the direction transverse to the 
magnetic field (see \secref{sec_scalings_passive} for 
a quick recapitulation of the standard scaling argument 
on the passive cascade that leads to a $\kperp^{-5/3}$ 
in the perpendicular direction). \Figref{fig_cartoon} illustrates 
this situation.\footnote{\label{fn_par_cascade} 
Note that effectively, 
there is also a cascade in $\kpar$ if the latter is measured along 
the unperturbed field---more precisely, a cascade in $k_z$.  
This is due to the perpendicular deformation of the 
perturbed magnetic field by the Alfv\'en-wave turbulence: 
since $\vdperp$ grows while $\vb\cdot\vdel$ remains the same, 
we have from \eqref{def_Bdgrad} $\dd/\dd z \simeq -(\dvBperp/B_0)\cdot\vdperp$.} 

\pseudofigureone{fig_cartoon}{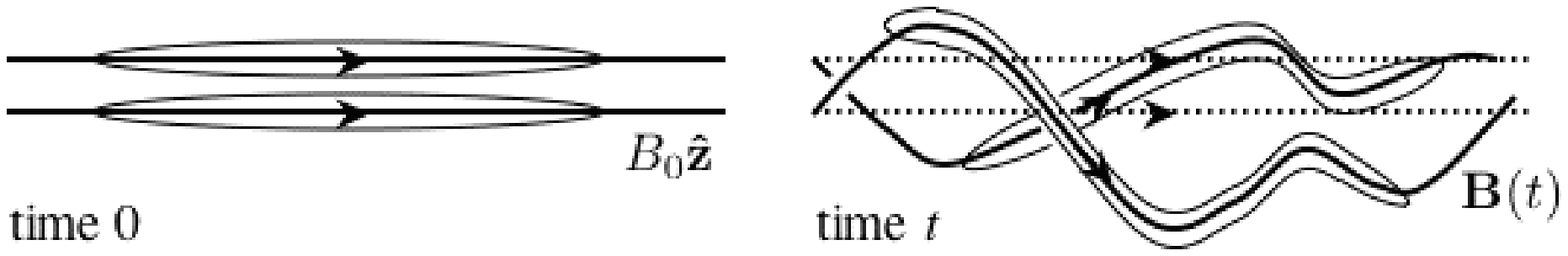}{f7.ps}{Lagrangian mixing of passive 
fields: fluctuations develop small scales across, but not along the exact field lines.}

We emphasize that this lack of nonlinear refinement of the 
scale of $\dne$ and $\dBpar$ along the moving field lines 
is a particular property of the compressive component of the 
turbulence, not shared by the Alfv\'en waves. Indeed, 
unlike \eqref{sw_g_sum}, the RMHD equations \exsdash{RMHD_Psi_sum}{RMHD_Phi_sum}, 
do not reduce to a linear form under the Lagrangian transformation \exref{r_Lagr}, 
so the Alfv\'en waves should develop small scales both across and 
along the perturbed magnetic field. 

Whether the density and magnetic-field-strength fluctuations 
develop small scales along the magnetic field has direct physical and 
observational consequences. Damping of these fluctuations, 
both in the collisional and collisionless regimes, 
discussed in \secsand{sec_visc_diss}{sec_barnes}, 
respectively, depends precisely on their scale along 
the perturbed field: indeed, the linear results derived 
there are exact in the Lagrangian frame \exref{r_Lagr}. 
To summarize these results, the damping rate of $\dne$ 
and $\dBpar$ at $\beta_i\sim 1$~is 
\bea
\gamma &\sim& \vthi\mfp\kparo^2,\quad 
\kparo\mfp\ll1,\\
\gamma &\sim& \vthi\kparo,\quad\qquad 
\kparo\mfp\gg1,
\eea
where $\kparo\sim\Dpar$ is the wavenumber along the perturbed 
field (i.e., if there is no parallel cascade, the wavenumber 
of the large-scale stirring). 

Whether this damping cuts off the cascades 
of $\dne$ and $\dBpar$ depends on the relative magnitudes of the damping 
rate $\gamma$ for a given $\kperp$ and the characteristic rate 
at which the Alfv\'en waves cause $\dne$ and $\dBpar$ 
to cascade to higher $\kperp$. This rate is 
$\omega_A\sim\kparA v_A$, where $\kparA$ is the parallel wave 
number of the Alfv\'en waves that have the same $\kperp$. 
Since the Alfv\'en waves do have a parallel cascade, 
assuming scale-by-scale critical balance \exref{crit_bal} 
leads to [\eqref{GS_aniso}] 
\bea
\label{kpar_vs_kperp}
\kparA\sim\kperp^{2/3}\lo^{-1/3}.
\eea 
If, in contrast to the Alfv\'en waves, $\dne$ and $\dBpar$ have no parallel 
cascade, $\kparo$ does not grow with $\kperp$, so, for large enough $\kperp$, 
$\kparo\ll\kparA$ and $\gamma\ll\omega_A$. 
This means that, despite the damping, 
the density and field-strength fluctuations 
should have perpendicular cascades extending to the ion gyroscale. 

The validity of the argument at the beginning of this section 
that ruled out the parallel cascade of $\dne$ and $\dBpar$ 
is not quite as obvious as it might appear. 
\citet{Lithwick_Goldreich} argued that
the dissipation of $\dne$ and $\dBpar$ at the ion gyroscale 
would cause these fluctuations to become uncorrelated at the 
same parallel scales as the Alfv\'enic fluctuations by which 
they are mixed, i.e., $\kparo\sim\kparA$. The damping rate 
then becomes comparable to the cascade rate, cutting off the 
cascades of density and field-strength fluctuations 
at $\kpar\mfp\sim1$. The corresponding perpendicular 
cutoff wavenumber is [see \eqref{kpar_vs_kperp}]
\bea
\label{kperp_LG}
\kperp \sim \lo^{1/2}\mfp^{-3/2}. 
\eea

Asymptotically speaking, in a weakly collisional plasma, 
this cutoff is far above the ion gyroscale, $\kperp\rho_i\ll1$. 
However, the relatively small value of $\mfp$ in the 
warm ISM, which was the main focus of \citealt{Lithwick_Goldreich}, 
meant that the numerical value of the perpendicular cutoff 
scale given by \eqref{kperp_LG} was, in fact, quite close 
both to the ion gyroscale (see \tabref{tab_scales}) 
and to the observational estimates for the inner scale 
of the electron-density fluctuations in the ISM 
\citep{Spangler_Gwinn,Armstrong_Rickett_Spangler}. 
Thus, it was not possible to tell whether \eqref{kperp_LG}, 
rather than $\kperp\sim\rho_i^{-1}$, represented the 
correct prediction. 

The situation is rather different 
in the nearly collisionless case of the solar wind, 
where the cutoff given by \eqref{kperp_LG} would mean 
that very little density or field-strength fluctuations 
should be detected above the ion gyroscale. Observations do not support  
such a conclusion: the density fluctuations appear to follow 
a $k^{-5/3}$ law at all scales larger than a few times $\rho_i$ 
\citep{Lovelace_etal,Woo_Armstrong,Celnikier_etal83,Celnikier_etal87,Coles_Harmon,Marsch_Tu_compr,Coles_etal},
consistently with the expected behavior of an undamped
passive scalar field (see \secref{sec_scalings_passive}). 
An extended range of $k^{-5/3}$ scaling above the ion 
gyroscale is also observed for the fluctuations of the magnetic-field strength 
\citep{Marsch_Tu_compr,Bershadskii_Sreeni_Bpar,Hnat_Chapman_Rowlands2,Alexandrova_sw}. 

These observational facts suggest that the cutoff formula \exref{kperp_LG} 
does not apply. This does not, however, 
conclusively vitiate the \citet{Lithwick_Goldreich} theory. 
Heuristically, their argument is plausible, although it is, perhaps, 
useful to note that in order for the effect of the perpendicular 
dissipation terms, not present in 
the KRMHD equations \exsdash{sw_g_sum}{sw_Bpar_sum}, 
to be felt, the density and field-strength 
fluctuations should reach the ion gyroscale in the first place. 
Quantitatively, the failure of the compressive fluctuations 
in the solar wind to be damped could still be 
consistent with the \citet{Lithwick_Goldreich} theory because 
of the relative weakness of the collisionless damping, 
especially at low beta (\secref{sec_barnes})---the explanation they themselves favor. 
The way to check observationally whether this explanation 
suffices would be to make a comparative study of the 
compressive fluctuations for solar-wind 
data with different values of $\beta_i$. If the strength 
of the damping is the decisive factor, one should always see 
cascades of both $\dne$ and $\dBpar$ at low $\beta_i$, 
no cascades at $\beta_i\sim1$, and a cascade of $\dBpar$ 
but not $\dne$ at high $\beta_i$ (in this limit, the damping of the density 
fluctuations is strong, of the field-strength weak; see \secref{sec_barnes}). 
If, on the other hand, the parallel cascade of the compressive 
fluctuations is intrinsically inefficient, very little 
$\beta_i$ dependence is expected and a perpendicular 
cascade should be seen in all cases. 

Obviously, an even more direct  observational 
(or numerical) test would be the detection or non-detection 
of near-perfect alignment of the density and field-strength 
structures with the moving field lines ({\em not} with 
the mean magnetic field---see footnote~\ref{fn_par_cascade}), 
but it is not clear how to measure this reliably.
It is interesting, in this context, that 
in near-the-Sun measurements, the density fluctuations 
are reported to have the form of highly anisotropic 
filaments aligned with the magnetic field 
\citep{Armstrong_etal_aniso,Grall_etal_aniso,Woo_Habbal}. 
Another intriguing piece of observational evidence is the discovery 
that the local structure of the magnetic-field-strength and 
density fluctuations at 1~AU is, in a certain sense, correlated 
with the solar cycle \citep{Kiyani_etal,Hnat_etal,Wicks_Chapman_Dendy}---this 
suggests a dependence on initial 
conditions that is absent in the Alfv\'enic fluctuations and that 
presumably should also disappear in the compressive fluctuations
if the latter are fully mixed both in the perpendicular and parallel
directions. 

\section{Turbulence in the Dissipation Range: Electron RMHD and the Entropy Cascade}
\label{sec_ERMHD}

\pseudofigurewide{fig_omegas}{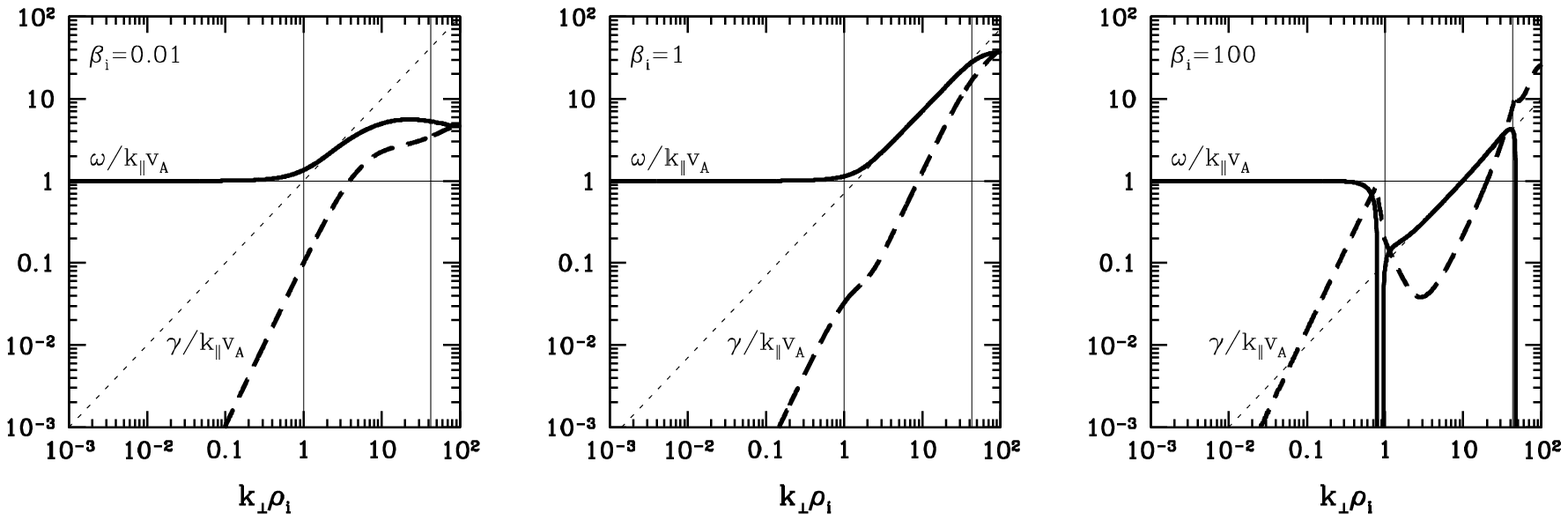}{f8.ps}{Numerical solutions of the 
linear gyrokinetic dispersion relation \citep[for a detailed treatment of the linear 
theory, see][]{Howes_etal} showing 
the transition from the Alfv\'en wave to KAW between the inertial range 
($\kperp\rho_i\ll1$) and the dissipation range ($\kperp\rho_i\gtrsim1$). 
We show three cases: 
low beta ($\beta_i=0.01$), $\beta_i=1$, and high beta ($\beta_i=100$). In 
all three cases, $\tau=1$ and $Z=1$.
Bold solid lines show the real frequency $\omega$, 
bold dashed lines the damping rate $\gamma$, 
both normalized by $\kpar v_A$ (in gyrokinetics, $\omega/\kpar v_A$ and 
$\gamma/\kpar v_A$ are functions of $\kperp$ only). 
Dotted lines show the asymptotic KAW solution \exref{omega_KAW}. 
Horizontal solid line shows the Alfv\'en wave $\omega=\kpar v_A$. 
Vertical solid lines show $\kperp\rho_i=1$ and $\kperp\rho_e=1$.
Note that the damping can be considered strong if the characteristic 
decay time is comparable or shorter than the wave period, i.e., 
$\gamma/\omega\gtrsim 1/2\pi$. Thus, in these plots, the damping 
at $\kperp\rho_i\sim1$ is relatively weak 
for $\beta_i=1$, relatively strong for low beta and very strong for high beta.} 

\subsection{Transition at the Ion Gyroscale}
\label{sec_transition}

The validity of the theory discussed in 
\secsand{sec_KRMHD}{sec_damping} breaks down when $\kperp\rho_i\sim1$. 
As the ion gyroscale is approached, the Alfv\'en waves are 
no longer decoupled from the rest of the plasma dynamics. 
All modes now contain perturbations of density and magnetic-field strength 
and can be collisionlessly damped. Because 
of the low-frequency nature of the Alfv\'en-wave cascade, 
$\omega\ll\Omega_i$ even at $\kperp\rho_i\sim1$ [\eqref{omega_order}], so 
the ion cyclotron resonance ($\omega-\kpar\vpar=\pm\Omega_i$) is not important, 
while the Landau one ($\omega=\kpar\vpar$) is. 
The linear theory of this collisionless damping in the gyrokinetic 
approximation is worked out in detail in \citet{Howes_etal} 
\citep[see also][]{Gary_Borovsky}. 
\Figref{fig_omegas} shows the solutions of their dispersion relation 
that illustrate how the Alfv\'en wave becomes a dispersive 
{\em kinetic Alfv\'en wave (KAW)} (see \secref{sec_KAW}) and collisionless 
damping becomes important as the ion gyroscale is reached. 

We stress that this transition occurs at the ion gyroscale, not 
at the ion inertial scale $d_i=\rho_i/\sqrt{\beta_i}$ 
(except in the limit of cold ions, $\tau=\Ti/\Te\ll1$; see \apref{ap_Hall}). 
This statement is true even when $\beta_i$ is not order unity, 
as illustrated in \figref{fig_omegas}: for the three cases plotted there, 
$\kperp d_i=1$ corresponds to $\kperp\rho_i=0.1$, $1$ and $10$ for 
$\beta_i=0.01$, $1$ and $100$, respectively, but there is no trace 
of the ion inertial scale in the solutions of the linear dispersion relation. 
Nonlinearly, in the limit $\beta_i\ll1$, we may consider the scales $\kperp d_i\sim1$ 
and expand the gyrokinetics in $\kperp\rho_i=\kperp d_i\sqrt{\beta_i}\ll1$ in a way similar 
to how it was done in \secref{sec_KRMHD} and obtain precisely the same 
results: Alfv\'enic fluctuations described by the RMHD equations and 
compressive fluctuations passively advected by them and satisfying 
the reduced kinetic equation derived in \secref{sec_sw}. 
Thus, even though $d_i\gg\rho_i$ at low beta, there is 
no change in the nature of the turbulent cascade 
until $\kperp\rho_i\sim1$ is reached.

The nonlinear theory of what happens at $\kperp\rho_i\sim1$ 
is very poorly understood. It is, however, possible to make progress 
by examining what kind of fluctuations emerge on the other side 
of the transition, at $\kperp\rho_i\gg1$. As we will demonstrate below, 
it turns out that another turbulent cascade---this time of KAW---is possible 
in this so-called {\em dissipation range}. It can transfer the 
energy of KAW-like fluctuations down to the electron gyroscale, 
where electron Landau damping becomes important (see \citealt{Howes_etal}). 
Some observational evidence of KAW is, indeed, available 
in the solar wind and the magnetosphere \citep[][see further discussion in 
\secref{sec_dr_spectra}]{Bale_etal,Grison_etal}. 
Below we derive the equations that describe KAW-like fluctuations in the 
scale range $\kperp\rho_i\gg1$, $\kperp\rho_e\ll1$ (\secref{sec_ERMHD_eqns}) and 
work out a Kolmogorov-style scaling theory for this cascade (\secref{sec_KAW_turb}). 

Because of the presence of the collisionless damping at the ion 
gyroscale, only a certain fraction of the turbulent power arriving 
there from the inertial range is converted into the KAW cascade, 
while the rest is Landau-damped. The damping leads to the heating of the ions, 
but the process of depositing the collisionlessly damped 
fluctuation energy into the ion heat is non-trivial because, as we 
explained in \secref{sec_heating}, collisions do need to play a role 
in order for true heating to occur. As we explained in \secref{sec_heating}
and will see specifically for the dissipation range in \secref{sec_en_ERMHD}, 
the electromagnetic-fluctuation energy does not disappear as a result of  
the Landau damping but is converted into ion entropy fluctuations, 
while the generalized energy is conserved.
Collisions are then accessed and ion heating achieved 
via a purely kinetic phenomenon: the ion entropy cascade in phase space
(nonlinear phase mixing), for which a theory is developed 
in \secsand{sec_ent_KAW}{sec_ent_no_KAW}.
A similar process of conversion of the KAW energy 
into electron entropy fluctuations and then electron heat 
is treated in \secref{sec_ent_els}. 

\Figref{fig_cascade_channels} illustrates the routes energy 
takes from the ion gyroscale towards heating. 
Crucially, it is at $\kperp\rho_i\sim1$ that it is decided 
how much energy would eventually go into the ions and how much 
into electrons.\footnote{Some of the energy of compressive fluctuations 
may go into ion heat via collisional (\secref{sec_visc_diss}) or collisionless 
(\secref{sec_barnes})
damping of these fluctuations in the inertial range. Whether this is a significant 
ion heating mechanism depends on the efficiency of the parallel cascade 
(see \secsand{sec_par_phase}{sec_par_cascade}).} 
How this distribution of energy depends on 
plasma parameters ($\beta_i$ and $\Ti/\Te$) is an open theoretical 
question\footnote{How much energy is converted into ion entropy 
fluctuations in the process of a {\em nonlinear} turbulent cascade 
is not necessarily directly related to the strength of the 
{\em linear} collisionless damping.} 
of considerable astrophysical interest: e.g., the efficiency 
of ion heating is a key unknown in the theory of advection-dominated accretion 
flows \citep[][see discussion in \secref{sec_disks}]{Quataert_Gruzinov} 
and of the solar corona \citep[e.g.,][]{Cranmer_vanBallegooijen}; 
we will also see in \secref{sec_superposed} that it may determine 
the form of the observed dissipation-range spectra in space plasmas. 

A short summary of this section is given in \secref{sec_ERMHD_sum}. 

\subsection{Equations of Electron Reduced MHD}
\label{sec_ERMHD_eqns}

The derivation is straightforward: when $\kr_i\sim\kperp\rho_i\gg1$, 
all Bessel functions in \eqsdash{quasineut_sum}{dBpar_eq_sum} are small,
so the integrals of the ion distribution function 
vanish and \eqsdash{quasineut_sum}{dBpar_eq_sum} become 
\bea
\label{EMHD_dne} 
{\dne\over\ne} &=& -{Ze\ephi\over\Ti} = 
- {2\over\sqrt{\beta_i}}{\Phi\over\rho_i v_A},\\
\label{EMHD_upar}
\upare &=& {c\over4\pi e\ne}\dperp^2\Apar 
= - {\rho_i\dperp^2\Psi\over\sqrt{\beta_i}},\qquad
\upari = 0,\\
\label{EMHD_dBpar}
{\dBpar\over B_0} &=& {\beta_i\over2}\(1+{Z\over\tau}\){Ze\ephi\over\Ti} 
= \sqrt{\beta_i}\(1+{Z\over\tau}\){\Phi\over\rho_i v_A},
\eea
where we used the definitions~\exref{Phi_Psi_def2} of the stream and flux 
functions $\Phi$ and $\Psi$. 

These equations are a reflection of the fact that, 
for $\kperp\rho_i\gg1$, the ion response is effectively 
purely Boltzmann, with the gyrokinetic part $\hi$ contributing 
nothing to the fields or flows [see \eqref{fs_exp} with $\hi$ omitted; 
$\hi$ does, however, play an important role in the energy balance and ion heating, 
as explained in \secsdash{sec_en_ERMHD}{sec_ent_no_KAW} below]. 
The Boltzmann response for ion density is expressed by \eqref{EMHD_dne}. 
\Eqref{EMHD_upar} states that the parallel ion flow velocity 
can be neglected. 
Finally, \eqref{EMHD_dBpar} expresses the pressure balance for 
Boltzmann (and, therefore, isothermal) electrons [\eqref{dpe_eq}] 
and ions: if we write 
\beq
{B_0\dBpar\over4\pi} = -\dpi - \dpe = 
-\Ti\dni - \Te\dne,
\eeq
it follows that 
\beq
{\dBpar\over B_0} = -{\beta_i\over2}\lt(1+{Z\over\tau}\rt){\dne\over\ne},
\label{dBpar_via_dne}
\eeq
which, combined with \eqref{EMHD_dne}, gives \eqref{EMHD_dBpar}. 
We remind the reader that the perpendicular Amp\`ere's law, from which 
\eqref{EMHD_dBpar} was derived [\eqref{dBpar_eq} via \eqref{dBpar_eq_sum}] 
is, in gyrokinetics, indeed equivalent to the statement of perpendicular 
pressure balance (see \secref{sec_GK_field_eq}). 
 
Substituting \eqsdash{EMHD_dne}{EMHD_dBpar} into \eqsdash{Apar_eq_sum}{dne_eq_sum}, 
we obtain the following closed system of equations
\bea
\label{EMHD_Psi}
{\dd\Psi\over\dd t} &=& 
v_A\(1+{Z\over\tau}\)\Dpar\Phi,\\
\label{EMHD_Phi}
{\dd\Phi\over\dd t} &=& 
-{v_A\over2 + \beta_i\(1+Z/\tau\)}\,\Dpar\(\rho_i^2\dperp^2\Psi\).
\eea
Note that, using \eqref{EMHD_dBpar}, \eqsand{EMHD_Psi}{EMHD_Phi}
can be recast as two coupled evolution equations for the perpendicular 
and parallel components of the perturbed magnetic field, respectively
[\eqsref{ERMHD_eqns_ap} in \apref{ap_EMHD}]. 

We shall refer to \eqsdash{EMHD_Psi}{EMHD_Phi} as {\em Electron Reduced MHD (ERMHD)}. 
They are related to the Electron Magnetohydrodynamics 
(EMHD)---a fluid-like approximation that evolves the magnetic field only 
and arises if one assumes that the magnetic 
field is frozen into the electron flow velocity $\vu_e$, while the 
ions are immobile, $\vu_i=0$ \citep{Kingsep_Chukbar_Yankov}:
\bea
\label{EMHD_eq}
{\dd\vB\over\dd t} = -{c\over4\pi e\ne}\vdel\times\lt[\lt(\vdel\times\vB\rt)\times\vB\rt].
\eea 
As explained in \apref{ap_EMHD}, the result of applying the 
RMHD/gyrokinetic ordering (\secsand{sec_RMHDordering}{sec_params}) 
to \eqref{EMHD_eq}, where $\vB=B_0\vz + \dvB$ and 
\bea
\label{dB_decomp}
{\dvB\over B_0} = {1\over v_A}\vz\times\vdperp\Psi + \vz\,{\dBpar\over B_0},
\eea
coincides with our \eqsdash{EMHD_Psi}{EMHD_Phi} 
in the effectively incompressible limits of $\beta_i\gg1$ or $\beta_e=\beta_iZ/\tau\gg1$. 
When betas are arbitrary, density fluctuations cannot be neglected compared 
to the magnetic-field-strength fluctuations [\eqref{dBpar_via_dne}] and
give rise to perpendicular ion flows with $\vdel\cdot\vu_i\neq0$. 
Thus, our ERMHD system constitutes the appropriate generalization 
of EMHD for low-frequency anisotropic fluctuations without the assumption 
of incompressibility. 

A (more tenuous) relationship also exists between our 
ERMHD system and the so-called Hall MHD, which, like EMHD, is based on 
the magnetic field being frozen into the electron flow, but includes the 
ion motion via the standard MHD momentum equation [\eqref{MHD_u}]. 
Strictly speaking, Hall MHD can only be used in the limit of 
cold ions, $\tau=\Ti/\Te\ll1$ \citep[see, e.g.,][and \apref{ap_Hall}]{Ito_etal,Hirose_etal}, 
in which case it can be shown to reduce to \eqsdash{EMHD_Psi}{EMHD_Phi} 
in the appropriate small-scale limit (\apref{ap_Hall}). 
Although $\tau\ll1$ is not a natural assumption for most 
space and astrophysical plasmas, Hall MHD has, due to its simplicity, 
been a popular theoretical paradigm in the studies of space and 
astrophysical plasma turbulence (see \secref{sec_dr_alt}). 
We have therefore devoted \apref{ap_Hall}  
to showing how this approximation fits into the theoretical 
framework proposed here: namely, we derive 
the anisotropic low-frequency version of the Hall MHD 
approximation from gyrokinetics under the assumption $\tau\ll1$ 
and discuss the role of the ion inertial and ion sound scales, which 
acquire physical significance in this limit. 
However, outside this Appendix, we assume $\tau\sim1$ everywhere and 
shall not use Hall MHD. 

The validity of the ERMHD equations as a model for plasma dynamics in 
the dissipation range is further discussed in \secref{sec_validity_ERMHD}. 

\subsection{Kinetic Alfv\'en Waves}
\label{sec_KAW}

The linear modes supported by ERMHD are kinetic Alfv\'en 
waves (KAW) with frequencies 
\bea
\omega_\vk = \pm\sqrt{1+Z/\tau\over2+\beta_i\(1+Z/\tau\)}\,
\kperp\rho_i\kpar v_A.
\label{omega_KAW}
\eea
This dispersion relation is illustrated in \figref{fig_omegas}: 
note that the transition from Alfv\'en waves to dispersive KAW 
always occurs at $\kperp\rho_i\sim1$, even when 
$\beta_i\ll1$ or $\beta_i\gg1$. 
In the latter case, there is a sharp frequency jump at the transition 
(accompanied by very strong ion Landau damping). 

The eigenfunctions corresponding to the two waves with frequencies 
\exref{omega_KAW} are
\beq
\label{KAW_ef}
\Theta^\pm_\vk = \sqrt{\(1+{Z\over\tau}\)\lt[2+\beta_i\(1+{Z\over\tau}\)\rt]}
\,{\Phi_\vk\over\rho_i} \mp 
\kperp\Psi_\vk.
\eeq
Using \eqsand{dB_decomp}{EMHD_dBpar}, 
the perturbed magnetic-field vector can be expressed as follows
\beq
\label{dB_KAW}
{\dvB_\vk\over B_0}  
= -i\vz\times{\vkperp\over\kperp}{\Theta^+_\vk - \Theta^-_\vk\over 2v_A}
+\vz\,\sqrt{1+Z/\tau\over2+\beta_i\(1+Z/\tau\)}\,
{\Theta^+_\vk + \Theta^-_\vk\over 2v_A}, 
\eeq
so, for a single ``$+$'' or ``$-$'' wave (corresponding to 
$\Theta^-_\vk=0$ or $\Theta^+_\vk=0$, respectively), 
$\dvB_\vk$ rotates in the plane perpendicular to 
the wave vector $\vkperp$ 
clockwise with respect to the latter, while the wave 
propagates parallel or antiparallel to the guide field
(\figref{fig_kaw}). 

The waves are elliptically right-hand polarized. 
Indeed, using \eqref{EMHD_dBpar}, the perpendicular 
electric field is:
\bea
\nonumber
\vEperpk &=& -i\vkperp\ephi + {i\omega_\vk\over c}\,\vAperpk\\
&=& \lt[-i\vkperp + \vz\times\vkperp{\omega_\vk\over\Omega_i}
{\beta_i\over\kperp^2\rho_i^2}\lt(1+{Z\over\tau}\rt)\rt]\ephi
\label{Eperp_KAW}
\eea
\citep[cf.][]{Gary_KAW,Hollweg_KAW}. 
The second term is small in the gyrokinetic expansion, 
so this is a very elongated ellipse (\figref{fig_kaw}). 

\pseudofigureone{fig_kaw}{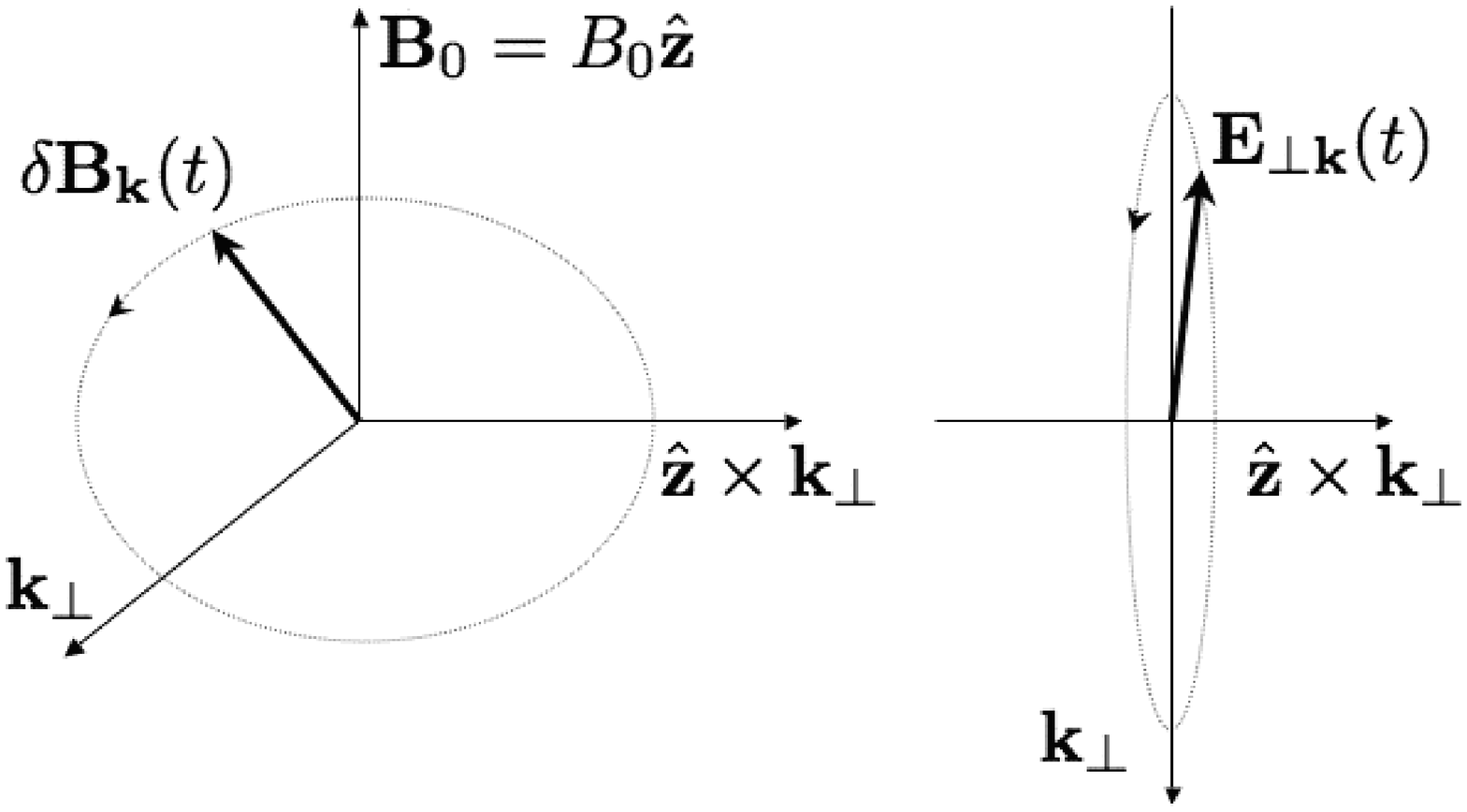}{f9.ps}{Polarization of the 
kinetic Alfv\'en wave, see \eqsand{dB_KAW}{Eperp_KAW}.}

\subsection{Finite-Amplitude Kinetic Alfv\'en Waves}
\label{sec_KAW_nlin}

As we are about to argue for a critically balanced KAW turbulence 
in a fashion analogous to the GS theory for the 
Alfv\'en waves (\secref{sec_GS}), it is a natural question 
to ask how similar the nonlinear properties of a putative 
KAW cascade will be to an Alfv\'en-wave cascade. 
As in the case of Alfv\'en waves, there are two 
counterpropagating linear modes [\eqsand{omega_KAW}{KAW_ef}], 
and it turns out that certain superpositions of these modes 
(KAW packets) are also exact {\em nonlinear} solutions 
of \eqsdash{EMHD_Psi}{EMHD_Phi}. Let us show that this is the case. 

We might look for the nonlinear solutions of \eqsdash{EMHD_Psi}{EMHD_Phi} 
by requiring that the nonlinear terms vanish.
Since $\Dpar={\dd/\dd z} + (1/v_A)\{\Psi,\cdots\}$, this gives 
\bea
\label{eq_c1}
\{\Psi,\Phi\} = 0 &\quad\Rightarrow\quad& \Psi = c_1\Phi,\\
\label{eq_c2}
\{\Psi,\rho_i^2\dperp^2\Psi\} = 0 &\quad\Rightarrow\quad& \rho_i^2\dperp^2\Psi = c_2\Psi,
\eea
where $c_1$ and $c_2$ are constants. Whether such solutions 
are possible is determined by substituting \eqsand{eq_c1}{eq_c2} 
into \eqsand{EMHD_Psi}{EMHD_Phi} and demanding that the two 
resulting {\em linear} equations be consistent with each other
(both equations now just evolve $\Psi$). This is achieved if\footnote{Formally speaking, 
$c_1$ and $c_2$ can depend on $t$ and $z$. If this is allowed, we still 
recover \eqref{c1c2_constraint}, but in addition to it, we get the 
evolution equation $c_1\dd c_1/\dd t = v_A(1+Z/\tau)\dd c_1/\dd z$. 
This allows $c_1=\const$, but there are, of course, other solutions. 
We shall not consider them here.} 
\bea
\label{c1c2_constraint}
c_1^2 = -{1\over c_2}\lt(1+{Z\over\tau}\rt)\lt[2+\beta_i\lt(1+{Z\over\tau}\rt)\rt],
\eea
so real solutions exist if $c_2<0$. In particular, wave packets 
consisting of KAW given by one of the linear 
eigenmodes \exref{KAW_ef} with an arbitrary shape in $z$ but 
confined to a single shell $|\vkperp|=\kperp=\const$, 
satisfy \eqsdash{eq_c1}{c1c2_constraint} with $c_2=-\kperp^2\rho_i^2$. 
This outcome is, in fact, only mildly non-trivial: 
in gyrokinetics, the Poisson bracket nonlinearity [\eqref{PB_def}] 
vanishes for any monochromatic (in $\vkperp$) mode because the Poisson bracket of two modes 
with wavenumbers $\vkperp$ and $\vkperp'$ is $\propto\vz\cdot(\vkperp\times\vkperp')$. 
Therefore, any monochromatic solution of 
the linearized equations is also an exact nonlinear solution. 
As we have shown above, 
a superposition of monochromatic KAW that have a fixed $\kperp$, or, 
somewhat more generally, satisfy \eqref{eq_c2} with a fixed $c_2$, 
is still an exact solution. 

Note that a similar procedure 
applied to the RMHD equations \exsdash{RMHD_Psi}{RMHD_Phi} 
returns the Elsasser solutions: perturbations of 
arbitrary shape that satisfy $\Phi=\pm\Psi$. 
The physical difference between these finite-amplitude Alfven-wave 
packets and the finite-amplitude KAW packets discussed above 
is that nonlinear interactions can occur not just between 
counterpropagating KAW but also between copropagating ones---a natural 
conclusion because KAW are dispersive (their group velocity along 
the guide field is $\propto v_A\kperp\rho_i$), so copropagating waves with 
different $\kperp$ can ``catch up'' with each other and 
interact.\footnote{The calculation above is analogous to the calculation by 
\citet{Mahajan_Krishan} for incompressible Hall MHD 
(i.e., essentially, the high-$\beta_e$ limit of the equations 
discussed in \apref{ap_Hall}), but the result is more general 
in the sense that it holds at arbitrary ion and electron betas. 
The Mahajan--Krishan solution in the EMHD limit 
amounts to noticing that \eqref{EMHD_eq} becomes linear 
for force-free (Beltrami) magnetic perturbations, 
$\vdel\times\dvB=\lambda\dvB.$ 
Substituting \eqref{dB_decomp} into this equation and using 
\eqref{EMHD_dBpar}, we see that the force-free equation 
is equivalent to \eqsdash{eq_c1}{c1c2_constraint} if $c_2=-\lambda^2$ 
and the incompressible limit 
($\beta_i\gg1$ or $\beta_e=\beta_iZ/\tau\gg1$) is taken.} 

\subsection{Scalings for KAW Turbulence}
\label{sec_KAW_turb}

A scaling theory for the turbulence described by \eqsdash{EMHD_dne}{EMHD_Phi} 
can be constructed along the same lines as the GS theory for 
the Alfv\'en-wave turbulence (\secref{sec_GS}). Namely, we shall 
assume that the turbulence below the ion gyroscale 
consists of KAW-like fluctuations with $\kpar\ll\kperp$ \citep{Quataert_Gruzinov} 
and that the interactions between them are critically balanced
\citep{Cho_Lazarian_EMHD}, i.e., that the propagation time and nonlinear 
interaction time are comparable at every scale. 
We stress that none of these assumptions are, strictly speaking, 
inevitable\footnote{In fact, the EMHD turbulence was thought to be weak by several 
authors, who predicted a $k^{-2}$ spectrum of magnetic energy assuming 
isotropy \citep{Goldreich_Reisenegger} or $\kperp^{-5/2}$ for the anisotropic case 
\citep{Voitenko2,Galtier_Bhattacharjee,Galtier_HMHD}.} 
(and, in fact, neither were they inevitable in the case of Alfv\'en waves). 
Since we have derived \eqsdash{EMHD_Psi}{EMHD_Phi} 
from gyrokinetics, the anisotropy of the fluctuations 
described by these equations is hard-wired, but it is not 
guaranteed that the actual physical cascade below the ion gyroscale 
is indeed anisotropic, although analysis of solar-wind 
measurements does seem to indicate that at least a significant fraction 
of it is \citep[see][]{Leamon_etal98,Hamilton_etal}. 
Numerical simulations based on \eqref{EMHD_eq} 
\citep{Biskamp_etal_EMHD1,Biskamp_etal_EMHD2,Ghosh_etal,Ng_etal_EMHD,Cho_Lazarian_EMHD,Shaikh_Zank05} 
have revealed that the spectrum of magnetic fluctuations 
scales as $\kperp^{-7/3}$, the outcome consistent with 
the assumptions stated above. Let us outline the argument that leads to this scaling. 

First assume that the fluctuations are KAW-like and that 
$\Theta^+$ and $\Theta^-$ [\eqref{KAW_ef}] have similar scaling. 
This implies 
\bea
\Psil\sim\sqrt{1+\beta_i}\,{\lambda\over\rho_i}\,{\Phil}
\label{KAW_fluct}
\eea
(for the purposes of scaling arguments and order-of-magnitude 
estimates, we set $Z/\tau=1$, but keep the $\beta_i$ dependence 
so low- and high-beta limits could be recovered if necessary). 
The fact that fixed-$\kperp$ KAW packets, 
which satisfy \eqref{KAW_fluct} with $\lambda=1/\kperp$, 
are exact nonlinear solutions of the ERMHD equations (\secref{sec_KAW_nlin}) 
lends some credence to this assumption. 

Assuming scale-space locality of interactions implies a constant-flux KAW cascade: 
analogously to \eqref{const_flux}, 
\bea
{(\Psil/\lambda)^2\over\tKAW} \sim 
{(1+\beta_i)(\Phil/\rho_i)^2\over\tKAW} \sim
 \epsB = \const,
\label{const_flux_KAW}
\eea
where $\tKAW$ is the cascade time and $\epsB$ is the KAW energy 
flux proportional to the fraction of the total flux $\eps$ 
(or the total turbulent power $\overline{\Pa}$; see \secref{sec_en_GK}) 
that was converted into the KAW cascade at the ion gyroscale. 

Using \eqsdash{EMHD_Psi}{EMHD_Phi} and 
\eqref{KAW_fluct}, it is not hard to see that the characteristic 
nonlinear decorrelation time is $\lambda^2/\Phil$. 
If the turbulence is strong, then this time 
is comparable to the inverse KAW frequency [\eqref{omega_KAW}] 
scale by scale and we may assume the cascade time 
is comparable to either:
\bea
\tKAW\sim {\lambda^2\over\Phil} \sim 
{1\over\sqrt{1+\beta_i}}{\rho_i\over\lambda}{v_A\over\lparl}.
\label{crit_bal_KAW}
\eea
In other words, this says that 
$\dd/\dd z\sim (\dvBperp/B_0)\cdot\vdperp$ 
and so $\dBpl/B_0\sim\lambda/\lparl$
(note that the last relation confirms that our scaling arguments 
do not violate the gyrokinetic ordering; 
see \secsand{sec_RMHDordering}{sec_params}). 
\Eqref{crit_bal_KAW} is the critical-balance assumption
for KAW. As in the case of the Alfv\'en waves (\secref{sec_GS}), 
we might argue physically that the critical balance is set up 
because the parallel correlation length $\lparl$ 
is determined by the condition that a wave can propagate the distance 
$\lparl$ in one nonlinear decorrelation time corresponding to the perpendicular 
correlation length $\lambda$. 

Combining \eqsand{const_flux_KAW}{crit_bal_KAW}, we get 
the desired scaling relations for the KAW turbulence: 
\bea
\label{KAW_Phi_scaling}
\Phil &\sim& \({\epsB\over\eps}\)^{1/3}
{v_A\over(1+\beta_i)^{1/3}}\,\lo^{-1/3}\rho_i^{2/3}\lambda^{2/3},\\
\label{KAW_aniso_scaling}
\lparl &\sim& \({\eps\over\epsB}\)^{1/3}
{\lo^{1/3}\rho_i^{1/3}\lambda^{1/3}\over(1+\beta_i)^{1/6}},
\eea
where $\lo=v_A^3/\eps$, as in \secref{sec_GS}. 
The first of these scaling relations is equivalent to a $\kperp^{-7/3}$ 
spectrum of magnetic energy, the second quantifies the anisotropy 
(which is stronger than for the GS turbulence). 
Both scalings were confirmed in the numerical simulations of 
\citet{Cho_Lazarian_EMHD}---it is their detection 
of the scaling \exref{KAW_aniso_scaling} that makes a 
particularly strong case that KAW turbulence is not weak 
and that the critical balance hypothesis applies. 

For KAW-like fluctuations, the density [\eqref{EMHD_dne}] and  
magnetic field [\eqsand{EMHD_dBpar}{KAW_ef}] have the same 
spectrum as the scalar potential, i.e., $\kperp^{-7/3}$, 
while the electric field $E\sim\kperp\ephi$ has a $\kperp^{-1/3}$ 
spectrum. The solar-wind fluctuation spectra reported by \citet{Bale_etal} 
indeed are consistent with a transition to KAW turbulence 
around the ion gyroscale: $k^{-5/3}$ magnetic and electric-field 
power spectra at $k\rho_i\ll1$ are replaced, for $k\rho_i\gtrsim1$, 
with what appears to be consistent with 
a $k^{-7/3}$ scaling for the magnetic-field spectrum 
and a $k^{-1/3}$ for the electric one (see \figref{fig_bale}). 
A similar result is recovered in 
fully gyrokinetic simulations with $\beta_i=1$, $\tau=1$ \citep{Howes_etal3}. 
However, not all solar-wind observations are quite 
as straightforwardly supportive of the notion of the
KAW cascade and much steeper magnetic-fluctuation spectra 
have also been reported \citep[e.g.,][]{Denskat_Beinroth_Neubauer,Leamon_etal98,Smith_etal06}. 
Possible reasons for this will emerge in \secsand{sec_validity_ERMHD}{sec_superposed} 
and the solar-wind data are further discussed in \secsand{sec_dr_spectra}{sec_dr_variability}. 

\subsection{Validity of the Electron RMHD and the Effect of Electron Landau Damping}
\label{sec_validity_ERMHD}

The ERMHD equations derived in \secref{sec_ERMHD} are valid 
provided $\kperp\rho_i\gg1$ and also 
provided it is sufficient to use the leading order in 
the mass-ratio expansion (isothermal electrons; see \secref{sec_els}). 
In particular, this means that the electron Landau damping is neglected. 
Asymptotically speaking, this is a rigorous limit, but one must
be cautious in applying it to real plasmas.
Since the width of the scale range where $\kperp\rho_i\gg1$ and $\kperp\rho_e\ll1$ 
is only $\sim(m_i/m_e)^{1/2}\simeq43$, for some values of the plasma parameters 
($\Ti/\Te$ and $\beta_i$) there may not be a very broad interval of scales 
where the electron Landau damping is truly negligible. 
Consider, for example, the low-beta limit, $\beta_i\ll1$.  
In this limit, the KAW frequency 
is $\omega\sim\kperp\rho_i\kpar v_A$ [\eqref{omega_KAW}].  
The electron Landau damping becomes important when $\omega\sim\kpar\vthe$, 
or $\kperp\rho_e\sim\sqrt{\beta_i}\ll1$, so the ERMHD approximation breaks 
down and, consequently, the KAW cascade, if any, 
should be interrupted well before the electron gyroscale is reached.
\Figref{fig_omegas} shows the solution of the full gyrokinetic 
dispersion relation \citep{Howes_etal} for small, unity and large $\beta_i$. 
One can judge for which scales and how well (or how badly) the ERMHD approximation holds 
from the precision with which the exact frequency follows the asymptotic 
solution \eqref{omega_KAW} and from the relative strength of the damping compared 
to the real frequency of the waves.

Non-negligible electron Landau damping may affect turbulence spectra because 
one can no longer assume a constant flux of KAW energy as we did in \secref{sec_KAW_turb}. 
To evaluate the consequences of this effect, \citet{Howes_etal2} constructed a simple model 
of spectral energy transfer and concluded that Landau damping leads to steepening 
of the KAW spectra---one of several possible reasons for steep dissipation-range 
spectra observed in space plasmas (see also \secref{sec_superposed}). 

\subsection{Unfreezing of Flux}
\label{sec_unfreezing}

As ERMHD is a limit of the isothermal-electron-fluid system (\secref{sec_els}), 
the magnetic-field lines remain unbroken (see \secref{sec_flux}).
Within the orderings employed above 
(small mass ratio, $\nui\sim\omega$, $\beta_i\sim1$, $\tau\sim1$), 
the flux unfreezes only in the vicinity of the electron gyroscale. 
It is interesting to evaluate somewhat more precisely the scale 
at which this happens as a function of plasma parameters. 

Physically, there are three kinds of mechanisms by which the flux conservation 
is broken: electron inertia, the effects of finite electron gyroradius, and Ohmic 
resistivity. Let us take the $\vpar$ moment of the electron gyrokinetic equation 
[\eqref{GK_eq}, $s=e$, integration at constant $\vr$] and use \eqref{EMHD_upar} 
to evaluate the inertial term in the resulting parallel electron momentum equation:
\bea
{cm_e\over e}{\dd\upare\over\dd t} = {\dd\over\dd t}\,d_e^2\dperp^2\Apar,
\eea 
where $d_e=\rho_e/\sqrt{\beta_e}$ is the electron inertial scale and $\beta_e= Z\beta_i/\tau$. 
Comparing this with the $\dd\Apar/\dd t$ term in the right-hand side of the electron 
momentum equation, 
we see that the electron inertia becomes important when $\kperp\rho_e\sim\sqrt{\beta_e}$. 
The finite-gyroradius effects enter when $\kperp\rho_e\sim1$. 
Thus, at low $\beta_e$, the electron inertia becomes important 
above the electron gyroscale, whereas at high $\beta_e$, the 
finite-gyroradius effects enter first. 
Finally, the Ohmic resistivity comes from the collision term 
(see \apref{ap_ei}):
\beq
{cm_e\over e}\intve\vpar\dtcolle\sim {cm_e\over e}\nue\upare
\sim \nue \kperp^2 d_e^2\Apar.
\eeq
Thus, resistivity starts to act when $\kperp d_e\sim (\omega/\nue)^{1/2}$. 
Using the KAW frequency [\eqref{omega_KAW}] to estimate $\omega$ and 
assuming that $\tau$ is not small, we get 
\bea
\kperp\rho_e \sim \kpar\mfp\sqrt{\beta_i\over 1+\beta_i}{Z^2\over\tau^2}.
\eea
Thus, the resistive scale can only be larger the electron gyroscale if the plasma
is collisional ($\kpar\mfp\ll1$) and/or 
electrons are much colder than ions ($\tau\gg1$) 
and/or $\beta_i\ll1$. Note if only the last of these conditions 
is satisfied, the electron inertia still 
becomes important at larger scales than resistivity. 

\subsection{Generalized Energy: KAW and Entropy Cascades} 
\label{sec_en_ERMHD} 

The generalized energy (\secref{sec_en_GK}) 
in the limit $\kperp\rho_i\gg1$ 
is calculated by substituting \eqsand{EMHD_dne}{EMHD_dBpar} 
into \eqref{W_els}: 
\bea
\nonumber
W &=& \int d^3\vr\lt\{\int d^3\vv\,{\Ti\<\hi^2\>_\vr\over2\fMi} 
+ {\dBperp^2\over8\pi} \rt.\\
\nonumber 
&&\lt.+\ {\ni\Ti\over2}\(1+{Z\over\tau}\)\lt[1+{\beta_i\over2}\(1+{Z\over\tau}\)\rt]
\lt({Ze\ephi\over\Ti}\rt)^2\rt\}\\
&=& \Whi + W_{\rm KAW}.
\label{W_ERMHD}
\eea
Here the first term, $\Whi$, is the total variance of $\hi$, which is 
proportional to minus the entropy of the ion gyrocenter distribution 
(see \secref{sec_heating}) and 
whose cascade to collisional scales will be discussed in 
\secsand{sec_ent_KAW}{sec_ent_no_KAW}. 
The remaining two terms are the independently cascaded KAW energy: 
\bea
\nonumber
W_{\rm KAW} &=& \intr{m_i\ni\over2}\Biggl\{|\vdperp\Psi|^2 \Biggr.\\
\nonumber
&&\Biggl.
+\ \(1+{Z\over\tau}\)\lt[1+{\beta_i\over2}\(1+{Z\over\tau}\)\rt]
{\Phi^2\over\rho_i^2}\Biggr\}\\
&=& \intr{m_i\ni\over2}\lt(|\Theta^+|^2 + |\Theta^-|^2\rt).
\eea
Although we can write $W_{\rm KAW}$ as the sum of the energies of the 
``$+$'' and ``$-$'' linear KAW eigenmodes [\eqref{KAW_ef}], 
which are also exact nonlinear solutions (\secref{sec_KAW_nlin}), 
the two do not cascade independently and can exchange energy. 
Note that the ERMHD equations also conserve $\intr\Psi\Phi$, 
which is readily interpreted as the helicity of the perturbed 
magnetic field (see \apref{ap_hel_els}). However, it does not 
affect the KAW cascade discussed in \secref{sec_KAW_turb} 
because it can be argued to have a tendency to cascade inversely 
(\apref{ap_inv_scalings}). 

Comparing the way the generalized energy is split 
above and below the ion gyroscale
(see \secref{sec_en_KRMHD} for the $\kperp\rho_i\ll1$ limit), 
we interpret what happens at the $\kperp\rho_i\sim1$ 
transition as a redistribution of the power that arrived 
from large scales between a cascade of KAW and a cascade 
of the (minus) gyrocenter entropy in the phase space 
(see \figref{fig_cascade_channels}). 
The latter cascade is the way in which the energy 
diverted from the electromagnetic fluctuations by 
the collisionless damping (wave--particle interaction) 
can be transferred to the collisional scales and 
deposited into heat (\secref{sec_transition}). 
The concept of entropy cascade 
as the key agent in the heating of the plasma was introduced 
in \secref{sec_heating}, where we promised a more detailed 
discussion later on. We now proceed to this discussion. 

\subsection{Entropy Cascade}
\label{sec_ent_KAW}

The ion-gyrocenter distribution function $\hi$ 
satisfies the ion gyrokinetic equation \exref{GK_ions_sum}, 
where ion--electron collisions are neglected under the 
mass-ratio expansion. At $\kperp\rho_i\gg1$, 
the dominant contribution to $\avchii$ comes from 
the electromagnetic fluctuations associated with KAW 
turbulence. Since the KAW cascade is decoupled from 
the entropy cascade, $\hi$ is a passive tracer of the 
ring-averaged KAW turbulence in phase space. 
Expanding the Bessel functions in the expression for $\avchiik$ 
[$\kr_i\gg1$ in \eqref{avchik_eq} with $s=i$] 
and making use of \eqsdash{EMHD_upar}{EMHD_dBpar} 
and of the KAW scaling $\Psi\sim\Phi/\kperp\rho_i$ 
[\eqref{KAW_ef}], it is not hard to show that 
\bea
{Ze\over\Ti}\,\avchiik \simeq
{Ze\over\Ti}\,\<\ephi\>_{\vR_i,\vk}
= {2\over\sqrt{\beta_i}}{J_0(\kr_i)\Phi_\vk\over\rho_i v_A},
\label{avchi_KAW}
\eea
where 
\bea
J_0(a_i)\simeq\sqrt{2\over\pi\kr_i}\,\cos\(\kr_i-{\pi\over 4}\),\quad
\kr_i=\kperp\rho_i\,{\vperp\over\vthi},
\label{Bessel_osc} 
\eea
so $\hi$ satisfies [\eqref{GK_ions_sum}]
\beq
{\dd\hi\over\dd t} + \vpar{\dd\hi\over\dd z} + 
\lt\{\<\Phi\>_{\vR_i},\hi\rt\} =
{2\over\sqrt{\beta_i}\,\rho_i v_A}
{\dd\<\Phi\>_{\vR_i}\over\dd t}\,\fMi 
+ \lt<\dC_{ii}[\hi]\rt>_{\vR_i} 
\label{hi_eq_KAW}
\eeq
with the conservation law [\eqref{hsq_eq}, $s=i$] 
\bea
\nonumber
{1\over\Ti}{d\Whi\over dt} &\equiv& 
{d\over dt}\int d^3\vv\intRi{\hi^2\over2\fMi}\\ 
\nonumber
&=& {2\over\sqrt{\beta_i}\,\rho_i v_A}\int d^3\vv\intRi{\dd\<\Phi\>_{\vR_i}\over\dd t}\,\hi\\
&&+\ \int d^3\vv\intRi{\hi\lt<\dC_{ii}[\hi]\rt>_{\vR_i}\over\fMi}.
\qquad
\label{hisq_eq}
\eea

\subsubsection{Nonlinear Perpendicular Phase Mixing}
\label{sec_small_scales}

The wave--particle interaction term (the first term on the 
right hand sides of these two equations) will shortly be seen 
to be subdominant at $\kperp\rho_i\gg1$. 
It represents the source of the invariant $\Whi$ due to the 
collisionless damping at the ion gyroscale of some fraction 
of the energy arriving from the inertial range. 
In a stationary turbulent state, we should have 
$\overline{d\Whi/dt} = 0$ and this source should be balanced 
on average by the (negative definite) collisional 
dissipation term (~=~heating; see \secref{sec_heating}). 
This balance can only be achieved if $\hi$ 
develops small scales in the velocity space and carries 
the generalized energy, or, in this case, entropy, 
to scales in the phase space at which collisions are important. 
A quick way to see this is by recalling that the collision 
operator has two velocity derivatives and can only balance 
the terms on the left-hand side of \eqref{hi_eq_KAW}~if 
\bea
\nui\vthi^2\({\dd\over\dd\vv}\)^2\sim \omega
\quad\Rightarrow\quad
{\delta v\over\vthi}\sim\Biggl({\nui\over\omega}\Biggr)^{1/2},
\label{dv_small}
\eea
where $\omega$ is the characteristic frequency of the fluctuations 
of $\hi$. If $\nui\ll\omega$, $\delta v/\vthi\ll1$. 
This is certainly true for $\kperp\rho_i\sim1$: 
taking $\omega\sim\kpar v_A$ and using 
$\kpar\mfp\gg1$ (which is the appropriate limit 
at and below the ion gyroscale for most of the plasmas of interest; 
cf.\ footnote~\ref{fn_str_mag}), 
we have $\nui/\omega\sim\sqrt{\beta_i}/\kpar\mfp\ll1$. 

\pseudofigureone{fig_gyroorbits}{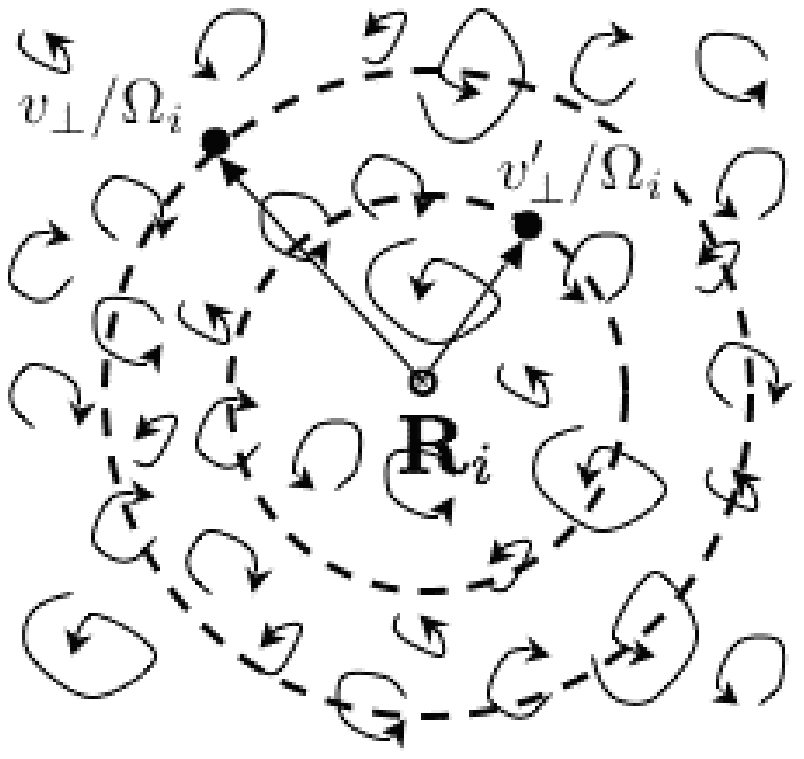}{f10.ps}{Nonlinear 
perpendicular phase-mixing mechanism: the gyrocenter distribution 
function at $\vR_i$ of particles with velocities $\vperp$ and $\vperp'$ 
is mixed by turbulent fluctuations of the potential $\Phi$ 
($\vE\times\vB$ flows) averaged over particle orbits separated by a 
distance greater than the correlation length of~$\Phi$.}

The condition \exref{dv_small} 
means that the collision rate can be arbitrarily 
small---this will always be compensated by the sufficiently 
fine velocity-space structure of the distribution function
to produce a finite amount of entropy production (heating) 
independent of $\nui$ in the limit $\nui\to +0$. 
The situation bears some resemblance to the emergence of 
small spatial scales in neutral-fluid turbulence with arbitrarily 
small but non-zero viscosity \citep{K41}. The analogy is not 
perfect, however, because the ion gyrokinetic equation \exref{hi_eq_KAW} 
does not contain a nonlinear interaction term that would explicitly 
cause a cascade in the velocity space. Instead, the 
(ring-averaged) KAW turbulence mixes $\hi$ in the gyrocenter space 
via the nonlinear term in \eqref{hi_eq_KAW}, so $\hi$ will 
have small-scale structure in $\vR_i$ on characteristic scales 
much smaller than $\rho_i$. Let us assume that the dominant 
nonlinear effect is a local interaction of the small-scale 
fluctuations of $\hi$ with the similarly 
small-scale component of $\<\Phi\>_{\vR_i}$. 
Since ring averaging 
is involved and $\kperp\rho_i$ is large, the values 
of $\<\Phi\>_{\vR_i}$ corresponding to two velocities $\vv$ and $\vv'$ 
will come from spatially decorrelated electromagnetic fluctuations 
if $\kperp\vperp/\Omega_i$ and $\kperp\vperp'/\Omega_i$ 
[the argument of the Bessel function in \eqref{avchi_KAW}] 
differ by order unity, i.e., for 
\bea
{\delta v_\perp\over\vthi}={|\vperp-\vperp'|\over\vthi}\sim 
{1\over\kperp\rho_i}
\label{dv_vs_kperp}
\eea
(see \figref{fig_gyroorbits}). 
This relation gives a correspondence between the decorrelation scales 
of $\hi$ in the position and velocity space. 
Combining \eqsand{dv_vs_kperp}{dv_small}, we see that there is a 
collisional cutoff scale determined by 
$\kperp\rho_i\sim(\omega/\nui)^{1/2}\gg1$.\footnote{Another 
source of small-scale spatial smoothing 
comes from the perpendicular gyrocenter-diffusion terms 
$\sim-\nui(v/\vthi)^2\kperp^2\rho_i^2\hki$ that arise in 
the ring-averaged collision operators, e.g., the second 
term in the model operator \exref{Cgk_formula}. 
These terms again enforce a cutoff wavenumber such that 
$\kperp\rho_i\sim(\omega/\nui)^{1/2}\gg1$.}
The cutoff scale is much smaller than the ion gyroscale. 
In the range between these scales, collisional dissipation is small. 
The ion entropy fluctuations are transferred across this scale 
range by means of a cascade, for which we will construct 
a scaling theory in \secref{sec_KAW_scalings} (and, for the 
case without the background KAW turbulence, in \secref{sec_ent_no_KAW}). 

It is important to emphasize that no matter how small the 
collisional cutoff scale is, all of the generalized energy 
channeled into the entropy cascade at the ion gyroscale 
eventually reaches it and is converted into heat. 
Note that the rate at which this happens is in general 
amplitude-dependent because the process is nonlinear, 
although we will argue in \secref{sec_par_with_KAW} (see also 
\secref{sec_par_no_KAW}) that the nonlinear cascade time 
and the parallel linear propagation (particle streaming) time 
are related by a critical-balance-like condition (we will 
also argue there that the linear parallel phase mixing, which 
can generate small scales in $\vpar$, is a less efficient process 
than the nonlinear perpendicular one discussed above). 

It is interesting to note the connection between the entropy 
cascade and certain aspects of the gyrofluid closure 
formalism developed by \citet{Dorland_Hammett}. In their theory, 
the emergence of small scales in $\vperp$ manifested itself as the 
growth of high-order $\vperp$ moments of the gyrocenter distribution function.
They correctly identified this effect as a consequence of the nonlinear 
perpendicular phase mixing of the gyrocenter distribution function caused by 
a perpendicular-velocity-space spread in the ring-averaged $\vE\times\vB$ velocities 
(given by $\<\vu_E\>_{\vR_i} = \vz\times\vdel\<\Phi\>_{\vR_i}$ in our notation) 
arising at and below the ion gyroscale. 

\subsubsection{Scalings}
\label{sec_KAW_scalings}

Since entropy is a conserved quantity, we will follow 
the well trodden Kolmogorov path, assume locality 
of interactions in scale space and constant entropy flux, 
and conclude, analogously to \eqref{const_flux},  
\bea
{\vthi^8\over\ni^2}{\hl^2\over\th}\sim\epsh=\const,
\label{const_flux_h}
\eea
where $\epsh$ is the entropy flux proportional to the fraction of the 
total turbulent power $\eps$ (or $\overline{\Pa}$; see \secref{sec_en_GK}) 
that was diverted into the entropy cascade at the ion gyroscale,  
and $\th$ is the cascade time that we now need to find. 

By the critical-balance assumption, the decorrelation time of the 
electromagnetic fluctuations in KAW turbulence 
is comparable at each scale to the KAW period at that scale 
and to the nonlinear interaction time [\eqref{crit_bal_KAW}]: 
\beq
\tKAW\sim {\lambda^2\over\Phil} 
\sim \({\eps\over\epsB}\)^{1/3}(1+\beta_i)^{1/3} 
{\lo^{1/3}\rho_i^{-2/3}\lambda^{4/3}\over v_A}.
\label{tKAW_scaling}
\eeq
The characteristic time associated with 
the nonlinear term in \eqref{hi_eq_KAW} is longer than 
$\tKAW$ by a factor of $(\rho_i/\lambda)^{1/2}$ due 
to the ring averaging, which reduces the strength of the 
nonlinear interaction. This weakness of the nonlinearity
makes it possible to develop a systematic analytical
theory of the entropy cascade \citep{SC_entropy}. 
It is also possible to estimate the 
cascade time $\th$ via a more qualitative argument analogous 
to that first devised by \citet{Kraichnan} for the weak turbulence 
of Alfv\'en waves: during each KAW correlation 
time $\tKAW$, the nonlinearity changes 
the amplitude of $\hi$ by only a small 
amount: 
\bea
\label{Delta_h}
\Delta\hl \sim (\lambda/\rho_i)^{1/2}\hl \ll \hl;
\eea 
these changes accumulate with time 
as a random walk, so after time $t$, the cumulative 
change in amplitude is $\Delta\hl (t/\tKAW)^{1/2}$;
finally, the cascade time $t=\th$ is the time after which 
the cumulative change in amplitude is comparable to the 
amplitude itself, which gives, using \eqref{tKAW_scaling}, 
\beq
\th \sim {\rho_i\over\lambda}\,\tKAW
\sim \({\eps\over\epsB}\)^{1/3}(1+\beta_i)^{1/3} 
{\lo^{1/3}\rho_i^{1/3}\lambda^{1/3}\over v_A}.
\label{tau_h}
\eeq  
Substituting this into \eqref{const_flux_h}, we get 
\beq
\hl \sim {\ni\over\vthi^3}\Biggl({\epsh\over\eps}\Biggr)^{1/2}
\({\eps\over\epsB}\)^{1/6}
{(1+\beta_i)^{1/6}\over\sqrt{\beta_i}}\,
\lo^{-1/3}\rho_i^{1/6}\lambda^{1/6}, 
\label{hi_scaling_KAW}
\eeq
which corresponds to a $\kperp^{-4/3}$ spectrum of entropy. 

In the argument presented above, we assumed that the scaling 
of $\hi$ was determined by the nonlinear mixing of $\hi$ by the 
ring-averaged KAW fluctuations rather than by 
the wave--particle interaction term on 
the right-hand side of \eqref{hi_eq_KAW}. 
We can now confirm the validity of this assumption. 
The change in amplitude of $\hi$ in one KAW correlation time 
$\tKAW$ due to the wave--particle interaction term is 
\bea
\nonumber
\Delta\hl &\sim& {\ni\over\vthi^3}\({\lambda\over\rho_i}\)^{1/2}
{\Phil\over\sqrt{\beta_i}\,\rho_i v_A}\\ 
&\sim& {\ni\over\vthi^3}\({\epsB\over\eps}\)^{1/3}
{1\over\sqrt{\beta_i}\,(1+\beta_i)^{1/3}}\,
\lo^{-1/3}\rho_i^{-5/6}\lambda^{7/6},\qquad 
\label{h_steep}
\eea
where we have used \eqref{KAW_Phi_scaling}. 
Comparing this with \eqref{Delta_h} and using \eqref{hi_scaling_KAW}, 
we see that $\Delta\hl$ in \eqref{h_steep} is a factor 
of $(\lambda/\rho_i)^{1/2}$ smaller than $\Delta\hl$ due 
to the nonlinear mixing. 

\subsubsection{Phase-Space Cutoff}
\label{sec_cutoff}

To work out the cutoff scales both in the position and velocity 
space, we use \eqsand{dv_small}{dv_vs_kperp}: in \eqref{dv_small}, 
$\omega\sim 1/\th$, where $\th$ is the characteristic 
decorrelation time of $\hi$ given by \eqref{tau_h}; 
using \eqref{dv_vs_kperp}, we find the cutoffs: 
\beq
{\delta v_\perp\over\vthi}\sim 
{1\over\kperp\rho_i}\sim 
(\nui\taur)^{3/5} = \Do^{-3/5},
\label{cutoffs_KAW}
\eeq
where $\taur$ is the cascade time [\eqref{tau_h}]  
taken at $\lambda=\rho_i$. By a recently established 
convention, the dimensionless number $\Do=1/\nui\taur$ is 
called the Dorland number. It plays the role 
of Reynolds number for kinetic turbulence, measuring the 
scale separation between the ion gyroscale and the 
collisional dissipation scale \citep{SCDHHPQT_crete,Tatsuno_etal1,Tatsuno_etal2}. 

\subsubsection{Parallel Phase Mixing}
\label{sec_par_with_KAW}

Another assumption, which was made implicitly, was that 
the parallel phase mixing due to the second term on the 
left-hand side of \eqref{hi_eq_KAW} could be ignored. 
This requires justification, especially because it is 
with this ``ballistic'' term that one traditionally associates 
the emergence of small-scale structure in the velocity 
space \citep[e.g.,][]{Krommes_Hu,Krommes_df,Watanabe_Sugama04}. 
The effect of the parallel phase mixing is to produce 
small scales in velocity space $\delta\vpar\sim1/\kpar t$. 
Let us assume that the KAW turbulence imparts its parallel 
decorrelation scale to $\hi$ and 
use the scaling relation \exref{KAW_aniso_scaling} 
to estimate $\kpar\sim\lparl^{-1}$. Then, after one cascade time $\th$ 
[\eqref{tau_h}], $\hi$ is decorrelated on the parallel velocity 
scales
\bea
{\delta\vpar\over\vthi}\sim {\lparl\over\vthi\th}
\sim {1\over\sqrt{\beta_i(1+\beta_i)}}\sim1. 
\label{vpar_KAW}
\eea 
We conclude that the nonlinear 
perpendicular phase mixing [\eqref{cutoffs_KAW}] is 
more efficient than the linear parallel one. 
Note that up to a $\beta_i$-dependent factor \eqref{vpar_KAW} 
is equivalent to a critical-balance-like assumption for 
$\hi$ in the sense that the propagation time is
comparable to the cascade time, or
$\kpar\vpar\sim \th^{-1}$ [see \eqref{hi_eq_KAW}]. 

\subsection{Entropy Cascade in the Absence of KAW Turbulence}
\label{sec_ent_no_KAW}

It is not currently known how one might determine analytically 
what fraction of the turbulent power arriving from the inertial 
range to the ion gyroscale is channeled into the KAW cascade and 
what fraction is dissipated via the kinetic ion-entropy cascade 
introduced in \secref{sec_ent_KAW} (perhaps it can only be determined 
by direct numerical simulations). It is certainly a fact 
that in many solar-wind measurements, the relatively 
shallow magnetic-energy spectra associated with the 
KAW cascade (\secref{sec_KAW_turb}) fail to appear and 
much steeper spectra are detected 
\citep[close to $k^{-4}$; see][]{Leamon_etal98,Smith_etal06}. 
In view of this evidence, it is 
interesting to ask what would be the nature of electromagnetic 
fluctuations below the ion gyroscale if the KAW cascade failed 
to be launched, i.e., if all (or most) of the turbulent power 
were directed into the entropy cascade (i.e., if $W\simeq\Whi$ 
in \secref{sec_en_ERMHD}). 

\subsubsection{Equations}
\label{sec_eqs_no_KAW}

It is again possible to derive a closed set of equations 
for all fluctuating quantities. 

Let us assume (and verify a posteriori; \secref{sec_mag}) 
that the characteristic frequency of such fluctuations is much 
lower than the KAW frequency [\eqref{omega_KAW}] so 
that the first term in \eqref{Apar_eq_sum} is small 
and the equation reduces to the balance of the other 
two terms. This gives 
\bea
{\dne\over\ne} = {e\ephi\over\Te},
\label{Boltzmann_els}
\eea
meaning that the electrons are purely Boltzmann 
[$\he=0$ to lowest order; see \eqref{hezero_formula}]. 
Then, from \eqref{quasineut_sum}, 
\beq
{Ze\ephi\over\Ti} \equiv {2\Phi\over\rho_i\vthi}
= \(1+{\tau\over Z}\)^{-1}\sum_\vk e^{i\vk\cdot\vr}
\intvi J_0(\kr_i)\hki 
\label{phi_from_h}
\eeq

Using \eqref{phi_from_h}, we find from \eqref{dBpar_eq_sum} 
that the field-strength fluctuations are 
\beq
\label{Bpar_no_KAW}
{\dBpar\over B_0} = -{\beta_i\over2}\sum_\vk e^{i\vk\cdot\vr}
\intvi {2\vperp^2\over\vthi^2}{J_1(\kr_i)\over\kr_i}\,\hki,
\eeq
which is smaller than $Ze\ephi/\Ti$ by a 
factor of $\beta_i/\kperp\rho_i$. 

Therefore, we can neglect $\dBpar/B_0$ compared to $\dne/\ne$ 
in \eqref{dne_eq_sum}. Using \eqref{Boltzmann_els}, we get 
what is physically the electron continuity equation:
\bea
\label{Apar_no_KAW}
{\dd\over\dd t}{e\ephi\over\Te} + \Dpar\({c\over4\pi e\ne}\,\vdperp^2\Apar 
+ \upari\) = 0,\\
\upari = \sum_\vk e^{i\vk\cdot\vr}\intvi \vpar J_0(\kr_i)\hki.
\eea
Note that in terms of the stream and flux functions, 
\eqref{Apar_no_KAW} takes the form
\bea
\label{Psi_no_KAW}
{\dd\over\dd z}\,\rho_i^2\vdperp^2\Psi = \sqrt{\beta_i}\lt(
{2\tau\over Z}{1\over\vthi}{\dd\Phi\over\dd t} 
+ \rho_i\,{\dd\upari\over\dd z}\rt),
\eea
where we have approximated $\Dpar\simeq\dd/\dd z$, 
which will, indeed, be shown to be correct in \secref{sec_mag}. 

Together with the ion gyrokinetic equation, which determines $\hi$, 
\eqsdash{Boltzmann_els}{Apar_no_KAW} form a closed set. 
They describe low-frequency fluctuations of the density 
and electromagnetic field due solely to the 
presence of fluctuations of $\hi$ below the ion gyroscale. 

It follows from \eqref{Bpar_no_KAW} that $\dBpar/B_0$ 
contributes subdominantly to $\avchii$ 
[\eqref{avchik_eq} with $s=i$ and $\kr_i\gg1$]. 
It will be verified a posteriori (\secref{sec_mag}) 
that the same is true for $\Apar$. Therefore, 
\eqsand{avchi_KAW}{hi_eq_KAW} continue to hold, 
as in the case with KAW. 
This means that \eqsand{hi_eq_KAW}{phi_from_h} form 
a closed subset. Thus the kinetic ion-entropy cascade 
is self-regulating in the sense that 
$\hi$ is no longer passive (as it was in the presence of KAW turbulence; 
\secref{sec_ent_KAW}) but is mixed by the ring-averaged 
``electrostatic'' fluctuations of the 
scalar potential, which themselves are produced by $\hi$ 
according to \eqref{phi_from_h}. 

The magnetic fluctuations are passive and determined 
by the electrostatic and entropy fluctuations via 
\eqsand{Bpar_no_KAW}{Apar_no_KAW}. 

\subsubsection{Scalings}
\label{sec_electrost}

From \eqref{phi_from_h}, we can establish a correspondence 
between $\Phil$ and $\hl$ (the electrostatic fluctuations 
and the fluctuations of the ion-gyrocenter distribution 
function):  
\beq
\Phil\sim {\rho_i\vthi}\({\lambda\over\rho_i}\)^{1/2}
{\hl\vthi^3\over\ni}\({\delta\vperp\over\vthi}\)^{1/2}
\sim {\vthi^4\over\ni}\,\hl\lambda,
\label{phi_from_h_scaling}
\eeq
where the factor of $(\lambda/\rho_i)^{1/2}$ comes from 
the Bessel function [\eqref{Bessel_osc}] and the factor of 
$(\delta\vperp/\vthi)^{1/2}$ results from the 
$\vperp$ integration of the oscillatory factor 
in the Bessel function times $\hi$, which decorrelates on small 
scales in the velocity space and, therefore, 
its integral accumulates in a random-walk-like fashion. 
The velocity-space scales are related to the spatial scales via 
\eqref{dv_vs_kperp}, which was arrived at by an argument 
not specific to KAW-like fluctuations and, therefore, 
continues to hold. 

Using \eqref{phi_from_h_scaling}, we find that the wave--particle 
interaction term in the right-hand side of \eqref{hi_eq_KAW} 
is subdominant: comparing it with $\dd\hi/\dd t$ shows that 
it is smaller by a factor of $(\lambda/\rho_i)^{3/2}\ll1$. 
Therefore, it is the nonlinear term in \eqref{hi_eq_KAW} that 
controls the scalings of $\hl$ and $\Phil$. 

We now assume again the scale-space locality and constancy of 
the entropy flux, so \eqref{const_flux_h} holds. 
The cascade (decorrelation) time is 
equal to the characteristic time associated with the 
nonlinear term in \eqref{hi_eq_KAW}: 
$\th\sim (\rho_i/\lambda)^{1/2}\lambda^2/\Phil$. 
Substituting this into \eqref{const_flux_h} and using 
\eqref{phi_from_h_scaling}, we arrive at the desired scaling 
relations for the entropy cascade \citep{SCDHHPQT_crete}:
\bea
\label{hi_scaling_no_KAW}
\hl &\sim& {\ni\over\vthi^3}\({\epsh\over\eps}\)^{1/3}
{1\over\sqrt{\beta_i}}\,
\lo^{-1/3}\rho_i^{1/6}\lambda^{1/6},\\
\label{phi_scaling_no_KAW}
\Phil &\sim& \({\epsh\over\eps}\)^{1/3}
{\vthi\over\sqrt{\beta_i}}\,\lo^{-1/3}\rho_i^{1/6}\lambda^{7/6},\\
\label{tau_scaling_no_KAW}
\th &\sim& \({\eps\over\epsh}\)^{1/3}
{\sqrt{\beta_i}\over\vthi}\,
\lo^{1/3}\rho_i^{1/3}\lambda^{1/3},
\eea
where $\lo=v_A^3/\eps$, as in \secref{sec_GS}. 
Note that since the existence of this cascade depends 
on it not being overwhelmed by the KAW fluctuations, 
we should have $\epsB\ll\eps$ and 
$\epsh=\eps-\epsB\approx\eps$. 

The scaling for the ion-gyrocenter distribution 
function, \eqref{hi_scaling_no_KAW}, implies a $\kperp^{-4/3}$
spectrum---the same as for the KAW turbulence [\eqref{hi_scaling_KAW}]. 
The scaling for the the cascade time, \eqref{tau_scaling_no_KAW}, 
is also similar to that for the KAW turbulence [\eqref{tau_h}]. 
Therefore the velocity- and gyrocenter-space cutoffs 
are still given by \eqref{cutoffs_KAW}, where $\taur$ 
is now given by \eqref{tau_scaling_no_KAW} taken at $\lambda=\rho_i$. 

A new feature is the scaling of the scalar potential, 
given by \eqref{phi_scaling_no_KAW}, which corresponds to 
a $\kperp^{-10/3}$ spectrum (unlike the KAW spectrum, 
\secref{sec_KAW_turb}). This is a measurable prediction for the 
electrostatic fluctuations: the implied electric-field spectrum
is $\kperp^{-4/3}$. From \eqref{Boltzmann_els}, we also conclude 
that the density fluctuations should have the same spectrum 
as the scalar potential, $\kperp^{-10/3}$---another measurable 
prediction. 

The scalings derived above for the spectra of the ion distribution 
function and of the scalar potential have been confirmed in 
the numerical simulations by \citet{Tatsuno_etal1,Tatsuno_etal2}, who 
studied decaying electrostatic gyrokinetic turbulence in 
two spatial dimensions. They also found velocity-space scalings in accord 
with \eqref{dv_vs_kperp} (using a spectral representation of the correlation 
functions in the $\vperp$ space based on the Hankel transform of the 
distribution function; see \citealt{Plunk_etal}). 

\subsubsection{Parallel Cascade and Parallel Phase Mixing} 
\label{sec_par_no_KAW}

We have again ignored the ballistic term 
(the second on the left-hand side) in \eqref{hi_eq_KAW}. 
We will estimate the 
efficiency of the parallel spatial cascade of the ion entropy 
and of the associated parallel phase mixing
by making a conjecture analogous to the critical balance: 
assuming that any two perpendicular planes only remain correlated 
provided particles can stream between them in one nonlinear 
decorrelation time (cf.\ \secsand{sec_GS}{sec_par_with_KAW}), 
we conclude that the parallel 
particle-streaming frequency $\kpar\vpar$ should be comparable 
at each scale to the inverse nonlinear time $\th^{-1}$, so  
\bea
\kpar\vthi\th\sim1. 
\label{kin_crit_bal}
\eea
As we explained in \secref{sec_par_with_KAW}, 
the parallel scales in the velocity space generated 
via the ballistic term are related to the parallel 
wavenumbers by $\delta\vpar\sim1/\kpar t$. 
From \eqref{kin_crit_bal}, we find that after one 
cascade time $\th$, the typical parallel 
velocity scale is $\delta\vpar/\vthi\sim 1$, 
so the parallel phase mixing is again much less efficient 
than the perpendicular one. 

Note that \eqref{kin_crit_bal} combined with \eqref{tau_scaling_no_KAW} 
means that the anisotropy is again characterized by the scaling 
relation $\kpar\sim\kperp^{1/3}$, similarly to the case of KAW 
turbulence [see \eqref{KAW_aniso_scaling} and \secref{sec_par_with_KAW}]. 

\subsubsection{Scalings for the Magnetic Fluctuations}
\label{sec_mag}

The scaling law for the fluctuations of the magnetic-field strength 
follows immediately from \eqsand{Bpar_no_KAW}{phi_scaling_no_KAW}: 
\bea
{\dBparl\over B_0}\sim\beta_i\,{\lambda\over\rho_i}
{\Phil\over\rho_i\vthi}\sim 
\sqrt{\beta_i}\,\lo^{-1/3}\rho_i^{-11/6}\lambda^{13/6},
\label{Bpar_scaling_no_KAW}
\eea
whence the spectrum of these fluctuations is $\kperp^{-16/3}$. 

The scaling of $\Apar$ (the perpendicular magnetic fluctuations)
depends on the relation between $\kpar$ and $\kperp$. 
Indeed, the ratio between the first and the third terms on the 
left-hand side of \eqref{Apar_no_KAW} [or, equivalently, 
between the first and second terms on the right-hand side 
of \eqref{Psi_no_KAW}] is $\sim\(\kpar\vthi\th\)^{-1}$. 
For a critically balanced cascade, this makes the two 
terms comparable [\eqref{kin_crit_bal}]. 
Using the first term to work out the scaling for the 
perpendicular magnetic fluctuations, we get, using 
\eqref{phi_scaling_no_KAW}, 
\beq
{\dBpl\over B_0}\sim {1\over\lambda}{\Psil\over v_A}
\sim \beta_i\,{\lambda\over\rho_i}{\Phil\over\rho_i\vthi}
\sim \sqrt{\beta_i}\,\lo^{-1/3}\rho_i^{-11/6}\lambda^{13/6},
\label{Bperp_scaling_no_KAW}
\eeq
which is the same scaling as for $\dBpar/B_0$ [\eqref{Bpar_scaling_no_KAW}]. 

Using \eqref{Bperp_scaling_no_KAW} together with 
\eqsand{phi_scaling_no_KAW}{tau_scaling_no_KAW}, it is now 
straightforward to confirm the three assumptions made in 
\secref{sec_eqs_no_KAW} that we promised to verify a posteriori: 

\begin{enumerate}

\item In \eqref{Apar_eq_sum}, $\dd\Apar/\dd t\ll c\Dpar\ephi$, so
\eqref{Boltzmann_els} holds (the electrons remain Boltzmann). 
This means that no KAW can be excited by the cascade.

\item $\dBperp/B_0\ll\kpar/\kperp$, so $\Dpar\simeq\dd/\dd z$ in 
\eqref{Apar_no_KAW}. This means that field lines are not 
significantly perturbed. 

\item In the expression for $\avchii$ [\eqref{avchik_eq}], 
$\vpar\Apar/c\ll\ephi$, so \eqref{hi_eq_KAW} holds. 
This means that the electrostatic fluctuations 
dominate the cascade. 

\end{enumerate}

\subsection{Cascades Superposed?}
\label{sec_superposed} 

The spectra of magnetic fluctuations obtained in \secref{sec_mag} are 
very steep---steeper, in fact, than those normally observed 
in the dissipation range of the solar wind (\secref{sec_dr_variability}). 
One might speculate that the observed spectra 
may be due to a superposition of the two cascades 
realizable below the ion gyroscale: a high-frequency 
cascade of KAW (\secref{sec_KAW_turb}) and a low-frequency 
cascade of electrostatic fluctuations due to the 
ion entropy fluctuations (\secref{sec_ent_no_KAW}). 
Such a superposition could happen if the power going 
into the KAW cascade is relatively small, $\epsB\ll\eps$. 
One then expects an electrostatic cascade to be set up 
just below the ion gyroscale with the KAW cascade 
superseding it deeper into the dissipation range. 
Comparing \eqsand{KAW_Phi_scaling}{phi_scaling_no_KAW}, 
we can estimate the position of the spectral break: 
\bea
\label{kperp_break}
\kperp\rho_i \sim \lt(\eps/\epsB\rt)^{2/3}. 
\eea
Since $\rho_i/\rho_e\sim (\tau m_i/m_e)^{1/2}/Z$ is not 
a very large number, the dissipation range is not very wide. 
It is then conceivable that the observed spectra 
are not true power laws but simply non-asymptotic 
superpositions of the electrostatic and KAW spectra 
with the observed range of ``effective'' spectral 
exponents due to varying values of the spectral 
break \exref{kperp_break} between the two 
cascades.\footnote{Several alternative 
theories that aim to explain the dissipation-range spectra 
exist: see \secref{sec_dr_alt}.}

The value of $\epsB/\eps$ specific to any particular 
set of parameters ($\beta_i$, $\tau$, etc.) 
is set by what happens at $\kperp\rho_i\sim1$
(\secref{sec_transition}; see \secref{sec_dr_transition}, 
\secref{sec_dr_variability}, and \secref{sec_disks} for further discussion). 

\subsection{Below the Electron Gyroscale: The Last Cascade} 
\label{sec_ent_els}

Finally, let us consider what happens when $\kperp\rho_e\gg1$. 
At these scales, we have to return to the full gyrokinetic 
system of equations. The quasi-neutrality [\eqref{quasineut}], 
parallel [\eqref{Amp_par}] and perpendicular [\eqref{dBpar_eq}] 
Amp\`ere's law become
\bea
\label{phi_from_he}
{e\ephi\over\Te} = -\lt(1+{Z\over\tau}\rt)^{-1}\sum_\vk e^{i\vk\cdot\vr}
\intve J_0(\kr_e)\hke,\\ 
\label{Apar_from_he}
{c\over4\pi e\ne}\,\dperp^2\Apar = \sum_\vk e^{i\vk\cdot\vr}
\intve \vpar J_0(\kr_e)\hke,\\
\label{Bpar_from_he}
{\dBpar\over B_0} = -{\beta_e\over2}\sum_\vk e^{i\vk\cdot\vr}
\intve {2\vperp^2\over\vthe^2}{J_1(\kr_e)\over\kr_e}\,\hke,
\eea
where $\beta_e=\beta_iZ/\tau$. 
We have discarded the velocity integrals of $\hi$ both 
because the gyroaveraging makes them subdominant in powers of 
$(m_e/m_i)^{1/2}$ and because the fluctuations of $\hi$ are 
damped by collisions [assuming the collisional cutoff given 
by \eqref{cutoffs_KAW} lies above the electron gyroscale]. 
To \eqsdash{phi_from_he}{Bpar_from_he}, we must append the 
gyrokinetic equation for $\he$ [\eqref{GK_eq} with $s=e$], 
thus closing the system. 

The type of turbulence described by these equations is very 
similar to that discussed in \secref{sec_ent_no_KAW}. 
It is easy to show from \eqsdash{phi_from_he}{Bpar_from_he} that 
\bea
{\dBperp\over B_0}\sim{\dBpar\over B_0}\sim 
{\beta_e\over\kperp\rho_e}\,{e\ephi\over\Te}.
\eea
Hence the magnetic fluctuations are subdominant in 
the expression for $\avchie$ [\eqref{avchik_eq} with $s=e$
and $\kr_e\gg1$], so $\avchie\simeq\<\ephi\>_{\vR_e}$. 
The electron gyrokinetic equation then is 
\bea
{\dd\he\over\dd t} + \vpar\,{\dd\he\over\dd z} + 
{c\over B_0}\lt\{\<\ephi\>_{\vR_e},\he\rt\} = \dtcolle,
\label{he_eq_electrost}
\eea
where the wave--particle interaction term in the right-hand side 
has been dropped because it can be shown to be small via the 
same argument as in \secref{sec_electrost}. 

Together with \eqref{phi_from_he}, \eqref{he_eq_electrost} 
describes the kinetic cascade of electron entropy from the 
electron gyroscale down to the scale at which electron 
collisions can dissipate it into heat. This cascade the 
result of collisionless damping of KAW at $\kperp\rho_e\sim1$, 
whereby the power in the KAW cascade is converted into 
the electron-entropy fluctuations: indeed, in the limit 
$\kperp\rho_e\gg1$, the generalized energy is simply 
\bea
W = \int d^3\vv\intRe{\Te\he^2\over2\fMe} = \Whe 
\label{W_last}
\eea
(see \figref{fig_cascade_channels}).

The same scaling arguments as in \secref{sec_electrost} apply 
and scaling relations analogous to \eqsdash{hi_scaling_no_KAW}{tau_scaling_no_KAW}, 
and \exref{Bpar_scaling_no_KAW} duly follow:
\bea
\hle &\sim& {\ne\over\vthe^3}
\({\epsB\over\eps}\)^{1/3}\lt({1\over\beta_e}{m_e\over m_i}\rt)^{1/2}
\lo^{-1/3}\rho_e^{1/6}\lambda^{1/6},\\
\Phil &\sim& \({\epsB\over\eps}\)^{1/3}\lt({1\over\beta_e}{m_e\over m_i}\rt)^{1/2}
\vthe\,\lo^{-1/3}\rho_e^{1/6}\lambda^{7/6},\\
\label{tau_scaling_els}
\th &\sim& \({\eps\over\epsB}\)^{1/3}\lt({\beta_e}{m_i\over m_e}\rt)^{1/2}
{\lo^{1/3}\rho_e^{1/3}\lambda^{1/3}\over\vthe},\\
{\dBl\over B_0}&\sim& \({\epsB\over\eps}\)^{1/3}\lt(\beta_e{m_e\over m_i}\rt)^{1/2}
\lo^{-1/3}\rho_e^{-11/6}\lambda^{13/6},
\eea
where $\lo=v_A^3/\eps$, as in \secref{sec_GS}. 
The formula for the collisional cutoffs in the 
wavenumber and velocity space is analogous to \eqref{cutoffs_KAW}:
\bea
{\delta v_\perp\over\vthi}\sim 
{1\over\kperp\rho_i}\sim 
(\nue\taure)^{3/5}, 
\label{cutoffs_els}
\eea 
where $\taure$ is the cascade time \exref{tau_scaling_els} 
taken at $\lambda=\rho_e$.

\subsection{Validity of Gyrokinetics in the Dissipation Range}
\label{sec_gk_validity}

As the kinetic cascade takes the (generalized) energy to ever 
smaller scales, the frequency $\omega$ of the fluctuations increases. 
In applying the gyrokinetic theory, one must be mindful of 
the need for this frequency to stay smaller than $\Omega_i$. 
Using the scaling formulae for the characteristic times of 
the fluctuations derived above [\eqsref{tKAW_scaling}, 
\exref{tau_scaling_no_KAW} and \exref{tau_scaling_els}], 
we can determine the conditions for $\omega\ll\Omega_i$. 
Thus, for the gyrokinetic theory to be valid 
everywhere in the inertial range, we must have
\bea
\kperp\rho_i\ll \beta_i^{3/4} \({\lo\over\rho_i}\)^{1/2}
\eea
at all scales down to $\kperp\rho_i\sim1$, i.e., 
$\rho_i/\lo\ll\beta_i^{3/2}$, not a very stringent condition. 

Below the ion gyroscale, the KAW cascade (\secref{sec_KAW_turb}) remains 
in the gyrokinetic regime as long as 
\beq
\kperp\rho_i\ll \({\eps\over\epsB}\)^{1/4}\beta_i^{3/8}(1+\beta_i)^{1/4}
\({\lo\over\rho_i}\)^{1/4}
\label{KAW_validity}
\eeq
(we are assuming $T_i/T_e\sim1$ everywhere). 
The condition for this still to be true at the electron 
gyroscale~is 
\beq
{\rho_i\over\lo}\ll {\eps\over\epsB}\,\beta_i^{3/2}(1+\beta_i)
\({m_e\over m_i}\)^2. 
\eeq
The ion entropy fluctuations passively mixed by the KAW 
turbulence (\secref{sec_ent_KAW}) 
satisfy \eqref{KAW_validity} at all scales down to 
the ion collisional cutoff [\eqref{cutoffs_KAW}] if 
\beq
{\mfp\over\lo}\ll \({\eps\over\epsB}\)^{3/4}\beta_i^{9/8}(1+\beta_i)^{3/4} 
\({\rho_i\over\lo}\)^{1/4}.
\label{ent_validity}
\eeq
Note that the condition for the ion collisional cutoff 
to lie above the electron gyroscale is 
\beq
{\mfp\over\lo}\ll \({\eps\over\epsB}\)^{1/3}\sqrt{\beta_i}(1+\beta_i)^{1/3}
\({m_i\over m_e}\)^{5/6}\({\rho_i\over\lo}\)^{2/3}.
\label{cutoff_KAW_above}
\eeq
In the absence of KAW turbulence, 
the pure ion-entropy cascade (\secref{sec_ent_no_KAW}) 
remains gyrokinetic~for 
\beq
\kperp\rho_i\ll \beta_i^{3/2}{\lo\over\rho_i}.
\eeq 
This is valid at all scales down to the ion collisional cutoff 
provided ${\mfp/\lo}\ll \beta_i^3(\lo/\rho_i)$, 
an extremely weak condition, which is always satisfied.
This is because the ion-entropy fluctuations in this case 
have much lower frequencies than in the KAW regime. 
The ion collisional cutoff lies above the electron gyroscale 
if, similarly to \eqref{cutoff_KAW_above}, 
\beq
{\mfp\over\lo}\ll\sqrt{\beta_i}\({m_i\over m_e}\)^{5/6}
\({\rho_i\over\lo}\)^{2/3}.
\label{cutoff_no_KAW_above}
\eeq

If the condition \exref{cutoff_KAW_above} is satisfied, 
all fluctuations of the ion distribution function are 
damped out above the electron gyroscale. 
This means that below this scale, we only need 
the electron gyrokinetic equation to be valid, i.e., 
$\omega\ll\Omega_e$. The electron-entropy cascade 
(\secref{sec_ent_els}), whose characteristic timescale 
is given by \eqref{tau_scaling_els}, satisfies this condition for 
\beq
\kperp\rho_e\ll \({\eps\over\epsB}\) \beta_e^{3/2}
\({m_i\over m_e}\)^{3/2}{\lo\over\rho_e}.
\eeq
This is valid at all scales down to the electron collisional 
cutoff [\eqref{cutoffs_els}] provided 
$\mfpe/\lo\ll(\eps/\epsB)^2\beta_e^3(m_i/m_e)^3(\lo/\rho_e)$, 
which is always satisfied. 

Within the formal expansion we have adopted 
($\kperp\rho_i\sim1$ and $\kpar\mfp\sim\sqrt{\beta_i}$), 
it is not hard to see that 
$\mfp/\lo\sim\epsilon^2$ and $\rho_i/\lo\sim\epsilon^3$. 
Since all other parameters ($m_e/m_i$, $\beta_i$, $\beta_e$ etc.) 
are order unity with respect to $\epsilon$, 
all of the above conditions for the validity of the 
gyrokinetics are asymptotically correct by construction. 
However, in application to real astrophysical plasmas, 
one should always check whether this construction holds. 
For example, substituting the relevant parameters for the solar 
wind shows that the gyrokinetic approximation is, in fact, 
likely to start breaking down somewhere 
between the ion and electron gyroscales 
\citep{Howes_etal2}.\footnote{See this paper also for a set of numerical tests of 
the validity of gyrokinetics in the dissipation range, a 
linear theory of the conversion of KAW into 
ion-cyclotron-damped Bernstein waves, and a discussion of the 
potential (un)importance of ion cyclotron damping for the dissipation of 
turbulence.} 
This releases a variety of high-frequency wave modes, 
which may be participating in the turbulent cascade 
around and below the electron gyroscale 
(see, e.g., the recent detailed observations of these 
scales in the magnetosheath by \citealt{Mangeney_etal06,Lacombe_etal06}
or the early measurements of high-frequency fluctuations 
in the solar wind by \citealt{Denskat_Beinroth_Neubauer,Coroniti_etal}). 

\subsection{Summary} 
\label{sec_ERMHD_sum}

In this section, we have analyzed the turbulence in the dissipation 
range, which turned out to have many more essentially kinetic features 
than the inertial range.

At the ion gyroscale, $\kperp\rho_i\sim1$, the kinetic cascade rearranged 
itself into two distinct components: part of the (generalized) energy 
arriving from the inertial range was collisionlessly damped, giving 
rise to a purely kinetic cascade of ion-entropy fluctuations, the rest 
was converted into a cascade of Kinetic Alfv\'en Waves (KAW) 
(\figref{fig_cascade_channels}; see \secsand{sec_transition}{sec_en_ERMHD}). 

The KAW cascade is described by two fluid-like equations for 
two scalar functions, the magnetic flux function 
$\Psi=-\Apar/\sqrt{4\pi m_i\ni}$ and the scalar 
potential, expressed, for continuity with the results of \secref{sec_KRMHD}, 
in terms of the function $\Phi=(c/B_0)\ephi$. The equations 
are (see \secref{sec_ERMHD_eqns})
\bea
\label{ERMHD_Psi_sum}
{\dd\Psi\over\dd t} &=& 
v_A\(1+{Z\over\tau}\)\Dpar\Phi,\\
\label{ERMHD_Phi_sum}
{\dd\Phi\over\dd t} &=& 
-{v_A\over2 + \beta_i\(1+Z/\tau\)}\,\Dpar\(\rho_i^2\dperp^2\Psi\),
\eea
where $\Dpar=\dd/\dd z + (1/v_A)\{\Psi,\cdots\}$.
The density and magnetic-field-strength fluctuations are 
directly related to the scalar potential:
\bea
{\dne\over\ne} = -{2\over\sqrt{\beta_i}}{\Phi\over\rho_i v_A},\quad
{\dBpar\over B_0} = \sqrt{\beta_i}\lt(1+{Z\over\tau}\rt){\Phi\over\rho_i v_A}.
\label{ERMHD_nB_sum}
\eea
We call \eqsdash{ERMHD_Psi_sum}{ERMHD_nB_sum} the
{\em Electron Reduced Magnetohydrodynamics (ERMHD)}. 

The ion-entropy cascade is described by the ion gyrokinetic 
equation:
\beq
{\dd\hi\over\dd t} + \vpar{\dd\hi\over\dd z} + 
\lt\{\<\Phi\>_{\vR_i},\hi\rt\} = \lt<\dC_{ii}[\hi]\rt>_{\vR_i}. 
\label{ERMHD_hi_sum}
\eeq
The ion distribution function is mixed by the ring-averaged 
scalar potential and undergoes a cascade both in the velocity 
and gyrocenter space---this phase-space cascade is essential 
for the conversion of the turbulent energy into the ion heat, 
which can ultimately only be done by collisions 
(see \secref{sec_ent_KAW}). 

If the KAW cascade is strong (its power $\epsB$ is an order-unity 
fraction of the total injected turbulent power $\eps$), it determines 
$\Phi$ in \eqref{ERMHD_hi_sum}, so the ion-entropy 
cascade is passive with respect to the KAW turbulence. 
\Eqsdash{ERMHD_Psi_sum}{ERMHD_Phi_sum} and \exref{ERMHD_hi_sum} 
form a closed system that determines the three functions 
$\Phi$, $\Psi$, $\hi$, of which the latter is slaved to the 
first two. One can also compute $\dne$ and $\dBpar$, which 
are proportional to $\Phi$ [\eqref{ERMHD_nB_sum}]. The generalized 
energy conserved by these equations is given by \eqref{W_ERMHD}. 

If the KAW cascade is weak ($\epsB\ll\eps$), the ion-entropy 
cascade dominates the turbulence in the dissipation range 
and drives low-frequency mostly electrostatic fluctuations, 
with a subdominant magnetic component. These are given by 
the following relations (see \secref{sec_ent_no_KAW})
\bea
\label{Phi_sum}
\Phi &=& {\rho_i\vthi\over 2(1+\tau/Z)}
\sum_\vk e^{i\vk\cdot\vr} \intvi J_0(\kr_i)\hki,\\ 
\label{dne_sum}
{\dne\over\ne} &=& {2Z\over\tau} 
{\Phi\over\rho_i\vthi},\\
\nonumber
\Psi &=& \rho_i\sqrt{\beta_i}
\sum_\vk e^{i\vk\cdot\vr}\times\\ 
\label{Psi_sum}
&&\intvi \lt({1\over1+Z/\tau}{i\over\kpar}{\dd\over\dd t} - \vpar\rt)
{J_0(\kr_i)\over\kperp^2\rho_i^2}\,\hki,\\
{\dBpar\over B_0} &=& -{\beta_i\over2}\sum_\vk e^{i\vk\cdot\vr}
\intvi {2\vperp^2\over\vthi^2}{J_1(\kr_i)\over\kr_i}\,\hki,
\label{Bpar_sum}
\eea
where $\kr_i={\kperp\vperp/\Omega_i}$,
\Eqsand{ERMHD_hi_sum}{Phi_sum} form a closed system for $\Phi$ 
and $\hi$. The rest of the fields, 
namely $\dne$, $\Psi$ and $\dBpar$, are slaved to $\hi$ via 
\eqsdash{dne_sum}{Bpar_sum}. 

The fluid and kinetic models summarized above are valid 
between the ion and electron gyroscales. Below the electron 
gyroscale, the collisionless damping of the KAW cascade 
converts it into a cascade of electron entropy, similar 
in nature to the ion-entropy cascade (\secref{sec_ent_els}). 

The KAW cascade and the low-frequency turbulence 
associated with the ion-entropy cascade have distinct scaling 
behaviors. For the KAW cascade, the spectra of the 
electric, density and magnetic fluctuations are (\secref{sec_KAW_turb})
\beq
E_E(\kperp)\propto \kperp^{-1/3},\quad
E_n(\kperp)\propto \kperp^{-7/3},\quad
E_B(\kperp)\propto \kperp^{-7/3}.
\eeq
For the ion- and electron-entropy cascades 
(\secsand{sec_ent_KAW}{sec_ent_els}), 
\beq
E_E(\kperp)\propto \kperp^{-4/3},\quad
E_n(\kperp)\propto \kperp^{-10/3},\quad
E_B(\kperp)\propto \kperp^{-16/3}.
\eeq
We argued in \secref{sec_superposed} that the observed 
spectra in the dissipation range of the solar wind 
could be the result of a superposition of these two cascades, 
although a number of alternative theories exist (\secref{sec_dr_alt}). 

\section{Discussion of Astrophysical Applications}
\label{sec_astro}

We have so far only occasionally referred to some 
relevant observational evidence for space and astrophysical plasmas. 
We now discuss in more detail how the theoretical framework laid out above applies 
to real plasma turbulence in space. 

Although we will discuss the interstellar medium, accretion disks and galaxy 
clusters towards the end of this section, the most rewarding source 
of observational information about 
plasma turbulence in astrophysical conditions is the solar wind 
and the magnetosheath because only there direct in situ measurements 
of all the interesting quantities are possible.
Measurements of the fluctuating magnetic and velocity 
fields in the solar wind 
have been available since the 1960s \citep{Coleman} and a vast 
literature now exists on their spectra, anisotropy, 
Alfv\'enic character and many other aspects
(a short recent review is \citealt{Horbury_etal_review}; 
two long ones are \citealt{Tu_Marsch_review,Bruno_Carbone}). 
It is not our aim here to provide a comprehensive 
survey of what is known about plasma turbulence in the solar 
wind. Instead, we shall limit our discussion to a few points that 
we consider important in light of the theoretical framework 
proposed in this paper.\footnote{An extended quantitative 
discussion of the applicability of the gyrokinetic theory 
to the turbulence in the slow solar wind was given by 
\citet{Howes_etal2}.} As we do this, we shall provide 
copious references to the main body of the paper, so this section 
can be read as a data-oriented guide to it, aimed 
both at a thorough reader who has arrived here after going through 
the preceding sections and an impatient one who has skipped to this 
one hoping to find out whether there is anything of ``practical'' use in the 
theoretical developments above. 

\subsection{Inertial-Range Turbulence in the Solar Wind}
\label{sec_SW_ir}

In the inertial range, i.e., for $\kperp\rho_i\ll1$, 
the solar-wind turbulence should be described 
by the reduced hybrid fluid-kinetic theory derived in \secref{sec_KRMHD}
(KRMHD). Its applicability hinges on three key assumptions: 
(i) the turbulence is Alfv\'enic, i.e., consists 
of small ($\dB/B_0\ll1$) low-frequency ($\omega\sim\kpar v_A\ll\Omega_i$) 
perturbations of an ambient mean magnetic field and corresponding 
velocity fluctuations; 
(ii) it is strongly anisotropic, $\kperp\gg\kpar$; 
(iii) the equilibrium distribution can be approximated or, at least, 
reasonably modeled 
by a Maxwellian without loss of essential physics (this will be discussed 
in \secref{sec_pressure_aniso}). 
If these assumptions are satisfied, KRMHD (summarized in \secref{sec_KRMHD_sum}) 
is a rigorous set of dynamical equations for the inertial range, a set of 
Kolmogorov-style scaling predictions for the Alfv\'enic component of the 
turbulence can be produced (the GS theory, reviewed in \secref{sec_GS}), 
while to the compressive fluctuations, the considerations of \secref{sec_damping} apply. 
So let us examine the observational evidence. 

\subsubsection{Alfv\'enic Nature of the Turbulence} 
\label{sec_SW_Alfvenic}

The presence of Alfv\'en waves in the solar wind was reported  
already the early works of \citet{Unti_Neugebauer} and 
\citet{Belcher_Davis}. Alfv\'en waves are detected 
already at very low frequencies (large scales)---and, 
at these low frequencies, have a $k^{-1}$ spectrum.\footnote{Inferred 
from the frequency spectrum $f^{-1}$ via the 
\citet{Taylor_hyp} hypothesis, $f\sim \vk\cdot\vVsw$, 
where $\vVsw$ is the mean velocity at which the wind blows past 
the spacecraft. The Taylor hypothesis is a good assumption for the solar 
wind because $\Vsw$ ($\sim800$~km/s in the fast wind, 
$\sim300$~km/s in the slow wind) is highly supersonic, super-Alfv\'enic 
and far exceeds the fluctuating velocities.}
This spectrum corresponds to a uniform distribution of scales/frequencies 
of waves launched by the coronal activity of the Sun. 
Nonlinear interaction of these waves gives rise to an Alfv\'enic 
turbulent cascade of the type that was discussed above. 
The effective outer scale of this cascade can be detected as 
a spectral break where the $k^{-1}$ scaling steepens to 
the Kolmogorov slope $k^{-5/3}$ (see \citealt{Bavassano_etal,Marsch_Tu_z,Horbury_etal96}
for fast-wind results on the spectral break; for a discussion of the effective outer 
scale in the slow wind at 1~AU, see \citealt{Howes_etal2}). 
The particular scale at which this happens increases with the distance 
from the Sun \citep{Bavassano_etal}, reflecting 
the more developed state of the turbulence at later stages 
of evolution. At 1~AU, the outer scale is roughly in the 
range of $10^5-10^6$~km; the $k^{-5/3}$ range extends down to 
scales/frequencies that correspond to a few times 
the ion gyroradius ($10^2-10^3$~km; see \tabref{tab_scales}). 

The range between the outer scale (the spectral break) and the ion 
gyroscale is the inertial range. In this range, $\dB/B_0$ decreases 
with scale because of the steep negative spectral slope. 
Therefore, the assumption of small fluctuations, $\dB/B_0\ll1$, 
while not necessarily true at the outer scale, is 
increasingly better satisfied further into the inertial 
range (cf.\ \secref{sec_two_regimes}). 

Are these fluctuations Alfv\'enic? In a plasma such as the solar wind, 
they ought to be because, as showed in \secref{sec_AW}, 
for $\kperp\rho_i\ll1$, these fluctuations are rigorously 
described by the RMHD equations. The magnetic flux is frozen into the ion 
motions, so displacing a parcel of plasma should produce a matching 
(Alfv\'enic) perturbation of the magnetic field line and vice versa:
in an Alfv\'en wave, $\vuperp = \pm\dvBperp/\sqrt{4\pi m_i\ni}$.
The strongest confirmation that this is indeed true for the inertial-range fluctuations 
in the solar wind was achieved by \citet{Bale_etal}, who 
compared the spectra of electric and magnetic fluctuations
and found that they both scale as $k^{-5/3}$ and 
follow each other with remarkable precision (see \figref{fig_bale}). 
The electric field is a very good measure of the perpendicular 
velocity field because, for $\kperp\rho_i\ll1$, the 
plasma velocity is the $\vE\times\vB$ drift velocity, 
$\vuperp = c\vE\times\vz/B_0$ (see \secref{sec_AW_coll}). 

This picture of agreement between basic theory and observations 
is upset in a disturbing fashion by an extraordinary recent result by 
\citet{Chapman_Hnat,Podesta_Roberts_Goldstein} and J.~E.~Borovsky (2008, private communication), 
who claim different spectral indices for velocity and 
magnetic fluctuations---$k^{-3/2}$ and $k^{-5/3}$, respectively. 
This result is puzzling because if it is asymptotically correct 
in the inertial range, it implies either $\uperp\gg\dBperp$ or $\uperp\ll\dBperp$ 
and it is not clear how perpendicular velocity fluctuations in a near-ideal plasma 
could fail to produce Alfv\'enic displacements and, therefore, perpendicular 
magnetic field fluctuations with matching energies. 
Plausible explanations may be either that the 
velocity field in these measurements is polluted by a 
non-Alfv\'enic component parallel to the magnetic field 
(although data analysis by \citealt{Chapman_Hnat} does not support this)
or that the flattening of the velocity spectrum is 
due to some form of a finite-gyroradius effect or 
even an energy injection into the velocity fluctuations 
at scales approaching the ion gyroscale (e.g., from 
the pressure-anisotropy-driven instabilities, \secref{sec_pressure_aniso}). 

\subsubsection{Energy Spectrum}

How solid is the statement that the observed spectrum has a $k^{-5/3}$ scaling?
In individual measurements of the magnetic-energy spectra, very high accuracy is claimed 
for this scaling: the measured spectral exponent 
is between 1.6 and 1.7; agreement with Kolmogorov value 1.67 is often 
reported to be within a few percent \citep[see, e.g.,][]{Horbury_etal96,Leamon_etal98,Bale_etal,Narita_Glassmeier_Treumann,Alexandrova_sw,Horbury_etal_aniso}). 
There is a somewhat wider scatter of spectral indices 
if one considers large sets of measurement intervals \citep{Smith_etal06}, but 
overall, the observational evidence does not appear to be consistent with a $\kperp^{-3/2}$ 
spectrum consistently found in the MHD simulations with a strong 
mean field \citep{Maron_Goldreich,Mueller_Biskamp_Grappin,Mason_Cattaneo_Boldyrev2,Perez_Boldyrev,Perez_Boldyrev_imb,Beresnyak_Lazarian2} 
and defended on theoretical grounds in the recent modifications 
of the GS theory by \citet{Boldyrev_spectrum2} and 
by \citet{Gogoberidze07} (see footnote \ref{fn_Boldyrev}). 
This discrepancy between observations and simulations 
remains an unresolved theoretical issue. 
It is probably best addressed by numerical 
modeling of the RMHD equations (\secref{sec_AW_fluid}) 
and by a detailed comparison of the structure of the 
Alfv\'enic fluctuations in such simulations and in the solar wind. 

\subsubsection{Anisotropy}
\label{sec_ir_aniso}

Building up evidence for anisotropy of turbulent fluctuations 
has progressed from merely detecting their elongation along 
the magnetic field \citep{Belcher_Davis}---to fitting data 
to an ad hoc model mixing a 2D perpendicular 
and a 1D parallel (``slab'') turbulent components in some 
proportion\footnote{These techniques originate from the 
view of MHD turbulence as a superposition of a 2D turbulence 
and an admixture of Alfv\'en waves \citep{Fyfe_Joyce_Montgomery,Montgomery_Turner}. 
As we discussed in \secref{sec_GS}, we consider the 
\citet{GS95,GS97} view of a critically balanced Alfv\'enic cascade
to be better physically justified.} 
\citep{Matthaeus_Goldstein_Roberts,Bieber_etal,Dasso_etal,Hamilton_etal}---to 
formal systematic unbiased analyses showing the persistent presence 
of anisotropy at all scales \citep{Bigazzi_etal,SorrisoValvo_etal}---
to direct measurements of three-dimensional correlation functions 
\citep{Osman_Horbury}---and 
finally to computing spectral exponents at fixed angles 
between $\vk$ and $\vB_0$ \citep{Horbury_etal_aniso}. 
The latter authors appear to have achieved the first direct 
quantitative confirmation of the GS theory by demonstrating that 
the magnetic-energy spectrum scales as $\kperp^{-5/3}$ 
in wavenumbers perpendicular to the mean field 
and as $\kpar^{-2}$ in wavenumbers parallel to it
[consistent with the first scaling relation in \eqref{GS_scaling}]. 
This is the closest that observations have got to confirming 
the GS relation $\kpar\sim\kperp^{2/3}$ 
[see \eqref{GS_aniso}] in a real astrophysical turbulent plasma. 

\subsubsection{Compressive Fluctuations}

According to the theory developed in \secref{sec_KRMHD}, 
the density and magnetic-field-strength fluctuations
are passive, energetically decoupled from and mixed by the Alfv\'enic cascade 
(\secref{sec_sw}; these are slow and entropy modes in the collisional 
MHD limit---see \secsand{sec_sw_fluid}{sec_visc}). 
These fluctuations are expected to be pressure-balanced, as 
expressed by \eqref{MHD_pr_bal} or, more generally in gyrokinetics, 
by \eqref{GK_pr_bal}. There is, indeed, strong evidence that 
magnetic and thermal pressures in the solar wind are anticorrelated, 
although there are some indications of the presence of compressive, 
fast-wave-like fluctuations as well 
\citep{Roberts_prbal,Burlaga_etal_prbal,Marsch_Tu_prbal,Bavassano_etal_prbal}. 

Measurements of density and field-strength fluctuations done 
by a variety of different methods both at 1~AU \citep{Celnikier_etal83,Celnikier_etal87,Marsch_Tu_compr,Bershadskii_Sreeni_Bpar,Hnat_Chapman_Rowlands2,Kellogg_Horbury,Alexandrova_sw} 
and near the Sun \citep{Lovelace_etal,Woo_Armstrong,Coles_Harmon,Coles_etal} 
show fluctuation levels of order 10\% and 
spectra that appear to have a $k^{-5/3}$ scaling 
above scales of order $10^2-10^3$~km, which approximately 
corresponds to the ion gyroscale.
The Kolmogorov value of the spectral exponent is, as in the 
case of Alfv\'enic fluctuations, measured quite accurately 
in individual cases \citep[$1.67\pm0.03$ in][]{Celnikier_etal87}.
Interestingly, the higher-order structure function exponents 
measured for the magnetic-field strength show that it is a more 
intermittent quantity than the velocity or the vector magnetic 
field (i.e., than the Alfv\'enic fluctuations) and that the scaling exponents 
are quantitatively very close to the values found for passive scalars 
in neutral fluids \citep{Bershadskii_Sreeni_Bpar,Bruno_etal}. 
One might argue that this lends some support to 
the theoretical expectation of passive 
magnetic-field-strength fluctuations. 
 
Considering that in the collisionless regime 
these fluctuations are supposed to be subject to 
strong kinetic damping (\secref{sec_barnes}), 
the presence of well-developed Kolmogorov-like 
and apparently undamped turbulent spectra is more surprising 
than has perhaps been publicly acknowledged. 
An extended discussion of this issue was given in \secref{sec_par_cascade}. 
Without the inclusion of the dissipation effects 
associated with the finite ion gyroscale, the passive 
cascade of the density and field strength is purely 
perpendicular to the (exact) local magnetic field and does not 
lead to any scale refinement along the field. This 
implies highly anisotropic field-aligned structures, 
whose length is determined by the initial conditions (i.e., 
conditions in the corona). The kinetic damping is inefficient 
for such fluctuations. While 
this would seem to explain the presence of fully fledged 
power-law spectra, it is not entirely obvious that 
the parallel cascade is really absent once dissipation 
is taken into account \citep{Lithwick_Goldreich}, so 
the issue is not yet settled. 
This said, we note that there is plenty of evidence 
of a high degree of anisotropy and field alignment of the 
density microstructure in the inner solar wind and 
outer corona \citep[e.g.,][]{Armstrong_etal_aniso,Grall_etal_aniso,Woo_Habbal}. 
There is also evidence that the local structure 
of the compressive fluctuations at 1~AU is correlated with 
the coronal activity, implying some form of memory 
of initial conditions \citep{Kiyani_etal,Hnat_etal,Wicks_Chapman_Dendy}. 

We note, finally, that whether compressive fluctuations in the inertial 
range can develop short parallel scales should also tell us 
how much ion heating can result from their damping (see \secref{sec_par_phase}). 

\subsection{Dissipation-Range Turbulence in the Solar Wind and the Magnetosheath}
\label{sec_SW_dr}

At scales approaching the ion gyroscale, $\kperp\rho_i\sim1$, 
effects associated with the finite extent of ion gyroorbits 
start to matter. Observationally, this transition manifests itself as a clear break 
in the spectrum of magnetic fluctuations, with the inertial-range 
$k^{-5/3}$ scaling replaced by a steeper slope (see \figref{fig_bale}). 
While the electrons at these scales can be treated as an isothermal fluid 
(as long as we are considering fluctuations above the electron gyroscale, 
$\kperp\rho_e\ll1$; see \secref{sec_els}), the fully gyrokinetic 
description (\secref{sec_GK}) has to be adopted for the ions. 
It is, indeed, to understand plasma dynamics at and around $\kperp\rho_i\sim1$ 
that gyrokinetics was first designed in fusion plasma theory 
\citep{Frieman_Chen,Brizard_Hahm_review}. 
In order for gyrokinetics and further dissipation-range approximations 
that follow from it (\secref{sec_ERMHD}) to be a credible approach 
in the solar wind and other space plasmas, it has to be established 
that fluctuations at and below the ion 
gyroscale are still strongly anisotropic, $\kpar\ll\kperp$. 
If that is the case, then their frequencies ($\omega\sim\kpar v_A\kperp\rho_i$, 
see \secref{sec_KAW}) will still be smaller than the cyclotron frequency 
in at least a part of the ``dissipation range''\footnote{This term, 
customary in the space-physics literature, is somewhat of a misnomer 
because, as we have seen in \secref{sec_ERMHD}, rich dissipationless 
turbulent dynamics are present in this range alongside what is normally thought 
of as dissipation.}---the range of scales $\kperp\rho_i\gtrsim1$ 
(see \secref{sec_gk_validity}).

Note that additional information about the dissipation-range turbulence 
can be extracted from the measurements in the magnetosheath---while scales 
above the ion gyroscale are probably non-universal there, the dissipation 
range appears to display universal behavior, mostly similar to the solar 
wind \citep[see, e.g.,][]{Alexandrova_review}. This complements the observational 
picture emerging from the solar-wind data and allows us to learn more 
as fluctuation amplitudes in the magnetosheath are larger and much smaller 
scales can be probed than in the solar wind 
\citep{Mangeney_etal06,Lacombe_etal06,Alexandrova_msheath}. 

\subsubsection{Anisotropy}
\label{sec_dr_aniso}

We know with a fair degree of certainty that the fluctuations that cascade down 
to the ion gyroscale from the inertial range are strongly anisotropic 
(\secref{sec_ir_aniso}). While it appears likely that the anisotropy 
persists at $\kperp\rho_i\sim1$, it is extremely important to have a clear 
verdict on this assumption from solar wind measurements. 
While \citet{Leamon_etal98} and, more recently, \citet{Hamilton_etal} 
did present some evidence that magnetic 
fluctuations in the solar wind 
have a degree of anisotropy below the ion gyroscale, 
no definitive study similar to \citet{Horbury_etal_aniso} or 
\citet{Bigazzi_etal,SorrisoValvo_etal} exists as yet. 
In the magnetosheath, where the dissipation-range scales are easier to 
measure than in the solar wind, recent analysis by \citet{Sahraoui_etal,Alexandrova_msheath}
does show evidence of strong anisotropy. 

Besides confirming the presence of the anisotropy, it would be 
interesting to study its scaling characteristics: e.g., check the 
scaling prediction $\kpar\sim\kperp^{1/3}$ [\eqref{KAW_aniso_scaling}; 
see also \secsand{sec_par_with_KAW}{sec_par_no_KAW}] 
in a similar fashion as the GS relation $\kpar\sim\kperp^{2/3}$ [\eqref{GS_aniso}]
was corroborated by \citet{Horbury_etal_aniso}. 

In this paper, we have proceeded on the assumption that the anisotropy, 
and, therefore, low frequencies ($\omega\ll\Omega_i$) do characterize 
fluctuations in the dissipation range---or, at least, that the low-frequency 
anisotropic fluctuations are a significant energy cascade channel 
and can be considered decoupled from any possible high-frequency dynamics.

\subsubsection{Transition at the Ion Gyroscale: 
Collisionless Damping and Heating}
\label{sec_dr_transition}

If the fluctuations at the ion gyroscale have $\kpar\ll\kperp$ and 
$\omega\ll\Omega_i$ (\secref{sec_dr_aniso}), 
they are not subject to the cyclotron resonance ($\omega-\kpar\vpar=\pm\Omega_i$), 
but are subject to the Landau one ($\omega=\kpar\vpar$).
Alfv\'enic fluctuations at the ion gyroscale are no longer decoupled 
from the compressive fluctuations and can be Landau-damped 
(\secref{sec_transition}). 
It seems plausible that it is the inflow of energy from the 
Alfv\'enic cascade that accounts for a pronounced 
local flattening of the spectrum of density fluctuations 
in the solar wind observed just above the ion gyroscale \citep{Woo_Armstrong,Celnikier_etal83,Celnikier_etal87,Coles_Harmon,Marsch_Tu_compr,Coles_etal,Kellogg_Horbury}.\footnote{\citet{Celnikier_etal87}  
proposed that the flattening might be a $k^{-1}$ spectrum 
analogous to Batchelor's spectrum of passive scalar variance in the 
viscous-convective range. We think this analogy cannot apply 
because density is not passive at or below the ion gyroscale.} 

In energetic terms, Landau damping amounts to a redistribution 
of generalized energy from electromagnetic fluctuations to 
entropy fluctuations (\secsref{sec_en_GK}{sec_en_ERMHD}). This gives rise to the entropy 
cascade, ultimately transferring the Landau-damped energy into ion heat
(\secref{sec_heating}, \secsand{sec_ent_KAW}{sec_ent_no_KAW}). However, 
only part of the inertial-range cascade is so damped because an alternative, 
electron, cascade channel exists: the kinetic Alfv\'en waves 
(\secsdash{sec_ERMHD_eqns}{sec_en_ERMHD}). The energy transferred into 
the KAW-like fluctuations can cascade to the electron gyroscale, where 
it is Landau damped on electrons, converting first into the 
electron entropy cascade and then electron heat (\secref{sec_ent_els}). 

Thus, the transition at the ion gyroscale ultimately decides 
in what proportion the turbulent energy arriving from the inertial 
range is distributed between 
the ion and electron heat. How the fraction of power going into either 
depends on parameters---$\beta_i$, $T_i/T_e$, amplitudes, \dots---is a key 
unanswered question both in space and astrophysical (see, e.g., \secref{sec_disks}) 
plasmas. Gyrokinetics appears to be an ideal tool for addressing this question 
both analytically and numerically \citep{Howes_etal3}. Within the 
framework outlined in this paper, the minimal model 
appropriate for studying the transition at the ion gyroscale
is the system of equations for isothermal electrons and gyrokinetic ions 
derived in \secref{sec_els} (it is summarized in \secref{sec_els_sum}). 

\subsubsection{Ion Gyroscale vs.\ Ion Inertial Scale}
\label{sec_dr_scale}

It is often assumed in the space physics literature that 
it is at the ion inertial scale, $d_i=\rho_i/\sqrt{\beta_i}$, rather than at 
the ion gyroscale $\rho_i$ that the spectral 
break between the inertial and dissipation range occurs. 
The distinction between $d_i$ and $\rho_i$ becomes noticeable 
when $\beta_i$ is significantly different from unity, 
a relatively rare occurrence in the solar 
wind. While some attempts to determine at which of these two scales 
a spectral break between the inertial and dissipation ranges occurs 
have produced claims that $d_i$ is a more likely candidate
\citep{Smith_etal01}, more comprehensive studies of the available data 
sets conclude basically that it is hard to tell 
\citep{Leamon_etal00,Markovskii_Vasquez_Smith}. 

In the gyrokinetic approach advocated in this paper, the ion inertial 
scale does not play a special role (see \secref{sec_transition}). 
The only parameter regime in which $d_i$ does appear as a special scale 
is $T_i\ll T_e$ (``cold ions''), when the Hall MHD approximation can be derived 
in a systematic way (see \apref{ap_Hall}). This, however, is not the right 
limit for the solar wind or most other astrophysical plasmas of interest 
because ions are rarely cold. Hall MHD is discussed further in \secref{sec_dr_alt} 
and \apref{ap_Hall}. 

\subsubsection{KAW Turbulence}
\label{sec_dr_spectra}

If gyrokinetics is valid at scales $\kperp\rho_i\gtrsim1$ (i.e., if 
$\kpar\ll\kperp$, $\omega\ll\Omega_i$ and it is acceptable to at least 
model the equilibrium distribution as a Maxwellian; see \secref{sec_pressure_aniso}), 
the electromagnetic fluctuations below the ion gyroscale will 
be described by the fluid approximation that we derived in \secref{sec_ERMHD_eqns} 
and referred to ERMHD. The wave solutions of this system of equations are the kinetic 
Alfv\'en waves (\secsdash{sec_KAW}{sec_KAW_nlin}) and it is possible to argue for 
a GS-style critically balanced cascade of KAW-like electromagnetic fluctuations 
(\secref{sec_KAW_turb}) between the ion and electron gyroscales 
(Landau damped on electrons at $\kperp\rho_e\sim1$;
the expression for the KAW damping rate in the gyrokinetic limit is given in 
\citealt{Howes_etal}; see also \figref{fig_omegas}).

Individual KAW have, indeed, been detected in space plasmas \citep[e.g.,][]{Grison_etal}. 
What about KAW turbulence? How does one tell whether any particular spectral slope 
one is measuring corresponds to the KAW cascade or fits some alternative scheme 
for the dissipation-range turbulence (\secref{sec_dr_alt})? It appears to be a sensible 
program to look for specific relationships between different fields predicted 
by theory (\secref{sec_ERMHD_eqns}) and for the corresponding spectral slopes 
and scaling relations for the anisotropy (\secref{sec_KAW_turb}). 
This means that simultaneous measurements 
of magnetic, electric, density and magnetic-field-strength fluctuations are needed. 

For the solar wind, the spectra of electric and magnetic 
fluctuations below the ion gyroscale reported by 
\citet{Bale_etal} are consistent with the 
$k^{-1/3}$ and $k^{-7/3}$ scalings predicted for 
an anisotropic critically balanced KAW cascade 
(\secref{sec_KAW_turb}; see \figref{fig_bale} for 
theoretical scaling fits superimposed on a 
plot taken from \citealt{Bale_etal}; note, however, that 
\citealt{Bale_etal} themselves interpreted their data 
in a somewhat different way and that their resolution was 
in any case not sufficient to be sure of the scalings). 
They were also able to check that their fluctuations satisfied 
the KAW dispersion relation---for critically balanced fluctuations, 
this is, indeed, plausible. 
Magnetic-fluctuation spectra recently reported by 
\citet{Alexandrova_sw} are only slightly steeper than 
the theoretical $k^{-7/3}$ KAW spectrum. These authors also 
find a significant amount of magnetic-field-strength fluctuations 
in the dissipation range, with a spectrum that follows the same 
scaling---this is again consistent with the theoretical picture 
of KAW turbulence [see \eqref{EMHD_dBpar}]. Measurements reported 
by \citet{Czaykowska_etal,Alexandrova_msheath} for the magnetosheath
appear to present a similar picture. 

The density spectra measured by \citet{Celnikier_etal83,Celnikier_etal87} 
steepen below the ion gyroscale following the flattened segment around 
$\kperp\rho_i\sim1$ (discussed in \secref{sec_dr_transition}). 
For a KAW cascade, the density spectrum should be $k^{-7/3}$ (\secref{sec_KAW_turb}); 
without KAW, $k^{-10/3}$ (\secref{sec_electrost}). 
The slope observed in the papers cited above appears to be somewhat shallower 
even than $k^{-2}$ (cf.\ a similar result by \citealt{Spangler_Gwinn} for the 
ISM; see \secref{sec_el_den_ISM}), 
but, given imperfect resolution, neither seriously in contradiction 
with the prediction based on the 
KAW cascade, nor sufficient to corroborate it. Unfortunately, we have 
not found published simultaneous measurements of density- and 
magnetic- or electric-fluctuation spectra.

\subsubsection{Variability of the Spectral Slope}
\label{sec_dr_variability}

While many measurements consistent with the KAW picture can be found, 
there are also many in which the spectra are much steeper 
\citep{Denskat_Beinroth_Neubauer,Leamon_etal98}. 
Analysis of a large set of measurements of the magnetic-fluctuation 
spectra in the dissipation range of the solar wind 
reveals a wide spread in the spectral indices: 
roughly between $-1$ and $-4$ \citep{Smith_etal06}. 
There is evidence of a weak 
positive correlation between steeper dissipation-range spectra 
and higher ion temperatures \citep{Leamon_etal98} 
or higher cascade rates calculated from 
the inertial range \citep{Smith_etal06}. This suggests 
that a larger amount of ion heating may correspond 
to a fully or partially suppressed KAW cascade, which is 
in line with our view of the ion heating and the KAW 
cascade as the two competing channels of the overall 
kinetic cascade (\secref{sec_en_ERMHD}). With a 
weakened KAW cascade, all or part of 
the dissipation range would be dominated by the ion entropy 
cascade---a purely kinetic phenomenon manifested 
by predominantly electrostatic fluctuations and very 
steep magnetic-energy spectra (\secref{sec_ent_no_KAW}). 
This might account both for the steepness of the observed spectra
and for the spread in their indices (\secref{sec_superposed}), 
although many other theories exist (see \secref{sec_dr_alt}). 

While we may thus have a plausible argument, 
this is not yet a satisfactory quantitative 
theory that would allow us to predict when the KAW cascade is present 
and when it is not or what dissipation-range spectrum should 
be expected for given values of the solar-wind parameters 
($\beta_i$, $T_i/T_e$, etc.). 
Resolution of this issue again appears to hinge on 
the question of how much turbulent power is diverted 
into the ion entropy cascade (equivalently, 
into ion heat) at the ion gyroscale (see \secref{sec_dr_transition}). 

\subsubsection{Alternative Theories of the Dissipation Range} 
\label{sec_dr_alt}

A number of alternative theories and models have been put forward 
to explain the observed spectral slopes (and their variability) in the 
dissipation range. It is not our aim to review or critique them all in detail, 
but perhaps it is useful to provide a few brief comments 
about some of them in light of the theoretical framework 
constructed in this paper. 

This entire theoretical framework hinges 
on adopting gyrokinetics as a valid description or, at least, 
a sensible model that does not miss any significant channels of 
energy cascade and dissipation. While we obviously believe this to 
be the right approach, it is worth spelling out what effects are 
left out ``by construction.'' 

\paragraph{Parallel Alfv\'en-wave cascade and ion cyclotron damping.} The use 
of gyrokinetics assumes 
that fluctuations stay anisotropic at all scales, $\kpar\ll\kperp$, 
and, therefore, $\omega\ll\Omega_i$, so the cyclotron resonances 
are ordered out. However, 
if one insists on routing the Alfv\'en-wave energy into a parallel cascade, 
e.g., by forcibly setting $\kperp=0$, it is possible to construct 
a weak turbulence theory in which it is dissipated by the 
ion cyclotron damping \citep{Yoon_Fang}. 
Numerical simulations of 3D MHD turbulence do not support the possibility 
of a parallel Alfv\'en-wave cascade \citep{Shebalin_Matthaeus_Montgomery,Oughton_Priest_Matthaeus,CV_aniso,Maron_Goldreich,CLV_aniso,Mueller_Biskamp_Grappin}. 
Solar-wind evidence that the perpendicular 
cascade dominates is quite strong for the inertial range (\secref{sec_ir_aniso}) 
and less so for the dissipation range (\secref{sec_dr_aniso}). 
While, as stated in \secref{sec_dr_aniso}, one cannot yet definitely claim that 
observations tell us that $\omega\ll\Omega_i$ at $\kperp\rho_i\sim1$, 
it has been argued that observations do not appear to be consistent with 
cyclotron damping being the main mechanism for the dissipation of the 
inertial-range Alfv\'enic turbulence at the ion gyroscale 
\citep{Leamon_etal98,Leamon_etal00,Smith_etal01}. 
Ion-cyclotron resonance could conceivably be reached 
somewhere in the dissipation range (see \secref{sec_gk_validity}). 
At this point gyrokinetics will formally break down, 
although, as argued by \citet[][see their \S\,3.6]{Howes_etal2}, 
this does not necessarily mean that ion cyclotron damping will 
become the dominant dissipation channel for the turbulence. 

\paragraph{Parallel whistler cascade.} A parallel magnetosonic/whistler 
cascade eventually damped by the electron cyclotron resonance \citep{Stawicki_Gary_Li}
is also excluded in the construction of gyrokinetics. The whistler cascade has been 
given some consideration in the Hall MHD approximation 
(further discussed at the end of this section). 
Both weak-turbulence theory \citep{Galtier_HMHD} 
and 3D numerical simulations \citep{Cho_Lazarian_EMHD} 
concluded that, like in MHD, the turbulent cascade is highly anisotropic, 
with perpendicular energy transfer dominating over the parallel one.\footnote{It 
is possible to produce a parallel cascade artificially by running 
1D simulations \citep{Matthaeus_Servidio_Dmitruk_comment}.} 
The same conclusion appears to have been reached in recent 2D kinetic PIC simulations 
by \citet{Gary_etal_PIC,Saito_etal_PIC}. 
Thus, the turbulence again seems to be driven into 
the gyrokinetically accessible regime.\\ 

While theory and numerical simulations appear to make arguing in favor of 
a parallel cascade and cyclotron heating difficult, there exists some 
observational evidence in support of 
them, especially for the near-Sun solar wind 
\citep[e.g.,][]{Harmon_Coles05}. 
Thus, the presence or relative importance of the cyclotron 
heating in the solar wind and, more generally, the mechanism(s) responsible for 
the observed perpendicular ion heating \citep{Marsch_etal83} 
remain a largely open problem. 
Besides the theories mentioned above, many other ideas have been proposed, 
some of which attempted to reconcile the dominance of the 
low-frequency perpendicular cascade with the possibility of cyclotron heating 
(e.g., \citealt{Chandran_fast_waves,Markovskii_etal}; see \citealt{Hollweg_review}
for a concise recent review of the problem). 

\paragraph{Mirror cascade.} \citet{Sahraoui_etal} analyzed a set of 
Cluster multi-spacecraft measurements in the magnetosheath and reported  
a broad power-law ($\sim k^{-8/3}$) spectrum of mirror structures at and below 
the ion gyroscale. They claim that these are {\em not} KAW-like fluctuations because 
their frequency is zero in the plasma frame. Although these structures are highly 
anisotropic with $\kpar\ll\kperp$, they cannot be described 
by the gyrokinetic theory in its present form because $\dBpar/B_0$ is 
very large ($\sim40\%$, occasionally reaching unity) and because the 
particle trapping by fluctuations, which is likely to be important in the nonlinear 
physics of the mirror instability \citep{Kivelson_Southwood_nlin,Pokhotelov_etal_mirror,Rincon_etal_mirror}, is ordered out in gyrokinetics. Thus, if a ``mirror cascade'' exists, 
it is not captured in our description. More generally, 
the effect of the pressure-anisotropy-driven instabilities 
on the turbulence in the dissipation range is a wide open area, 
requiring further analytical effort (see \secref{sec_pressure_aniso}).\\ 

If $\kpar\ll\kperp$, $\omega\ll\Omega_i$, 
and $\dB/B_0\ll1$ are accepted for the dissipation range 
and plasma instabilities at the ion gyroscale (\secref{sec_pressure_aniso}) 
are ignored, the formal gyrokinetic theory and its asymptotic consequences 
derived above should hold. There are two essential features of the linear 
physics at and below the ion gyroscale that must play some role: 
the collisionless (Landau) damping and the dispersive nature of the 
wave solutions (see \figref{fig_omegas} and \secref{sec_KAW}; cf., e.g., 
\citealt{Leamon_etal99,Stawicki_Gary_Li}). Both of these features have been employed 
to explain the spectral break at the ion gyroscale and the spectral slopes 
below it. 

\paragraph{Landau damping and instrumental effects.} In most of our discussion, 
(\secref{sec_ERMHD}, \secsdash{sec_dr_spectra}{sec_dr_variability}), 
we effectively assumed that the Landau damping is only important at 
$\kperp\rho_i\sim1$ and $\kperp\rho_e\sim1$, but not in between, 
so we could talk about asymptotic scalings and dissipationless cascades. 
However, as was noted in \secref{sec_validity_ERMHD},  
a properly asymptotic scaling behavior in the dissipation range is probably 
impossible in nature because the scale separation between the ion and electron gyroscales 
is only about $(m_i/m_e)^{1/2}\simeq43$. 
In particular, there is not always a wide scale interval 
where the kinetic damping is negligibly small (especially at low $\beta_i$; 
see \figref{fig_omegas}; cf.\ \citealt{Leamon_etal99}). 
\citet{Howes_etal2} proposed a model 
of how the presence of damping combined with instrumental effects 
(a resolution floor) could lead to measured spectra that look 
like power laws steeper than $k^{-7/3}$, with the effective spectral 
exponent depending on plasma parameters (we refer the reader to that 
paper for a discussion of how this compares with previous models of 
a similar kind, e.g., \citealt{Li_Gary_Stawicki}). A key physical 
assumption of theirs and similar models is that the amount of power drained 
from the Alfv\'en-wave and KAW cascades into the ion heat 
is set by the strength of the {\em linear} damping. Whether this is 
justified is not yet clear. 

\paragraph{Hall and Electron MHD.} If Landau damping is deemed unimportant 
in some part of the dissipation range \citep[which can be true in some regimes; 
see \figref{fig_omegas} and][]{Howes_etal,Howes_etal2,Howes_etal3}
and the wave dispersion is considered to be the salient feature, 
it might appear that a fluid, rather 
than kinetic, description should be sufficient. 
Hall MHD \citep{Mahajan_Yoshida} or its $k d_i\gg1$ limit the Electron MHD 
\citep{Kingsep_Chukbar_Yankov} have been embraced by many authors as such a description, 
suitable both for analytical arguments 
\citep{Goldreich_Reisenegger,Krishan_Mahajan,Gogoberidze05,Galtier_Bhattacharjee,Galtier_HMHD,Alexandrova_sw} 
and numerical simulations \citep{Biskamp_etal_EMHD1,Biskamp_etal_EMHD2,Ghosh_etal,Ng_etal_EMHD,Cho_Lazarian_EMHD,Shaikh_Zank05,Galtier_Buchlin,Matthaeus_Servidio_Dmitruk_comment}.

To what extent does this constitute an approach {\em alternative} to (and better than?) 
gyrokinetics \citep[as suggested, e.g., by][]{Matthaeus_Servidio_Dmitruk_comment}? 
For fluctuations with $\kpar\ll\kperp$, Hall MHD is merely a particular limit of 
gyrokinetics: $\beta_i\ll1$ and $T_i/T_e\ll1$ (cold-ion limit; see \apref{ap_Hall}). 
If $\kpar$ is not small compared 
to $\kperp$, then the gyrokinetics is not valid, while Hall MHD continues to describe 
the cold-ion limit correctly \citep[e.g.,][]{Ito_etal,Hirose_etal}, capturing in particular 
the whistler branch of the dispersion relation. However, as we have already 
mentioned above, the dominance of the perpendicular energy transfer ($\kpar\ll\kperp$) 
is supported both by weak-turbulence theory for Hall MHD \citep{Galtier_HMHD} 
and by 3D numerical simulations of the Electron MHD \citep{Cho_Lazarian_EMHD}. 

Thus, the gyrokinetic theory and its rigorous limits, such as ERMHD (\secref{sec_ERMHD_eqns}), 
supersede Hall MHD for anisotropic turbulence. Since ions are generally 
not cold in the solar wind (or any other plasma discussed here), 
Hall MHD is not formally a relevant approximation. It also entirely 
misses the kinetic damping and the associated entropy cascade channel 
leading to particle heating (\secref{sec_transition}, \secsand{sec_ent_KAW}{sec_ent_no_KAW}). 
However, Hall MHD does capture the Alfv\'en waves becoming dispersive 
and numerical simulations of it do show a spectral break, 
although, technically speaking, at the wrong scale 
($d_i$ instead of $\rho_i$; see \secref{sec_transition}). 
Although Hall MHD cannot be rigorously used as quantitative theory 
of the spectral break and the associated change in the nature 
of the turbulent cascade, 
the Hall MHD equations in the limit $k d_i\gg1$ are mathematically 
similar to our ERMHD equations (see \secref{sec_ERMHD_eqns} and \apref{ap_Hall}) 
to within constant 
coefficients probably not essential for qualitative models of turbulence. 
Therefore, results of numerical simulations of Hall and Electron MHD cited above 
are directly useful for understanding the KAW cascade---and, indeed, 
in the limit $k d_i\gg1$, $k d_e\ll1$, they 
are mostly consistent with the scaling arguments of \secref{sec_KAW_turb}. 


\paragraph{Alfv\'en vortices.} Finally we mention an argument pertaining 
to the dissipation-range spectra that is not based on energy cascades at all.  
Based on the evidence of Alfv\'en vortices in the magnetosheath, 
\citet{Alexandrova_review} speculated that steep power-law spectra observed 
in the dissipation range at least in some cases could reflect the geometry 
of the ion-gyroscale structures rather than a local energy cascade. 
If Alfv\'en vortices are a common feature, this possibility cannot 
be excluded. However, the resulting geometrical spectra are quite steep ($k^{-4}$ and steeper), 
so they can become important only if the KAW cascade is weak or suppressed---somewhat 
similarly to the steep spectra associated with the entropy cascade (\secref{sec_superposed}).  

\subsection{Is Equilibrium Distribution Isotropic and Maxwellian?}
\label{sec_pressure_aniso}

In rigorous theoretical terms, the weakest point of this paper 
is the use of a Maxwellian equilibrium. 
Formally, this is only justified when the collisions are weak but 
not too weak: we ordered the collision frequency  
as similar to the fluctuation frequency [\eqref{omega_vs_nu}]. 
This degree of collisionality is sufficient 
to prove that a Maxwellian equilibrium 
distribution $\fMs(v)$ does indeed emerge in the lowest 
order of the gyrokinetic expansion \citep{Howes_etal}. 
This argument works well for plasmas such as the ISM (\secref{sec_ISM}), 
where collisions are weak ($\mfp\gg\rho_i$) but non-negligible 
($\mfp\ll\lf$). In space plasmas, the mean free path is 
of the order of 1~AU---the distance between the Sun and 
the Earth (see \tabref{tab_scales}). 
Strictly speaking, in so highly collisionless a plasma, 
the equilibrium distribution does not have to be either 
Maxwellian or isotropic. 

The conservation of the first adiabatic 
invariant, $\mu=\vperp^2/2B$, suggests that temperature anisotropy 
with respect to the magnetic-field direction ($\Tperp\neq\Tpar$) may 
exist. When the relative anisotropy is larger than (roughly) $1/\beta_i$, 
it triggers several very fast growing 
plasma instabilities: most prominently the firehose ($\Tperp<\Tpar$) 
and mirror ($\Tperp>\Tpar$) modes \citep[e.g.,][]{Gary_etal76}. 
Their growth rates peak around the ion gyroscale, thus giving 
rise to additional energy injection at $\kperp\rho_i\sim1$. 

No definitive analytical theory of how these fluctuations 
saturate, cascade and affect the equilibrium distribution 
has been proposed. It appears to be a reasonable expectation 
that the fluctuations resulting from temperature anisotropy 
will saturate by limiting this anisotropy. This 
idea has some support in solar-wind observations: 
while the degree of anisotropy of the core particle distribution 
functions varies considerably between data sets, the observed 
anisotropies do seem to populate the part of the parameter 
plane $(\Tperp/\Tpar,\beta_i)$ circumscribed in a rather precise way
by the marginal stability boundaries for the mirror and firehose 
\citep{Gary_etal_ACE,Kasper_Lazarus_Gary,Marsch_Ao_Tu,Hellinger_etal,Matteini_etal}.\footnote{Note that \citet{Kellogg_etal06} measure the electric-field fluctuations 
in the ion-cyclotron frequency range, estimate the resulting 
velocity-space diffusion and argue that it is sufficient to 
isotropize the ion distribution}

If we want to study turbulence in data sets that do not lie too close 
to these stability boundaries, assuming an isotropic Maxwellian equilibrium 
distribution [\eqref{fs_exp}] is probably an acceptable simplification, 
although not an entirely rigorous one. Further theoretical work 
is clearly possible on this subject: thus, it is not a problem to 
formulate gyrokinetics with an arbitrary equilibrium distribution 
\citep{Frieman_Chen} and starting from that, once can  
generalize the results of this paper \citep[for the KRMHD system, 
\secref{sec_KRMHD}, this has been done by][]{Chen_etal_KRMHD}. 
Treating the instabilities themselves might prove more difficult, 
requiring the gyrokinetic ordering to be modified and the expansion 
carried to higher orders to incorporate features that are not captured 
by gyrokinetics, e.g., short parallel scales \citep{Rosin_etal_firehose}, 
particle trapping \citep{Pokhotelov_etal_mirror,Rincon_etal_mirror}, 
or nonlinear finite-gyroradius effects \citep{Califano_etal_mirror}. 
Note that the theory of the dissipation-range turbulence 
will probably need to be modified to account for the additional energy injection 
from the instabilities and for the (yet unclear) way in which this energy 
makes its way to dissipation and into heat. 

Besides the anisotropies, the particle distribution functions in the 
solar wind (especially the electron one) exhibit non-Maxwellian suprathermal 
tails \citep[see][and references therein]{Maksimovic_etal,Marsch_review}. 
These contain small ($\sim5\%$ of the total density) populations of 
energetic particles. Both the origin of these particles and their effect 
on turbulence have to be modeled kinetically. Again, it is possible 
to formulate gyrokinetics for general equilibrium distributions of this 
kind and examine the interaction between them and the turbulent 
fluctuations, but we leave such a theory outside the scope of this paper. 

Thus, much remains to be done to incorporate realistic equilibrium distribution 
functions into the gyrokinetic description of the solar wind plasma. 
In the meanwhile, we believe that 
the gyrokinetic theory based on a Maxwellian equilibrium distribution 
as presented in this paper, while idealized 
and imperfect, is nevertheless a step forward in the analytical treatment 
of the space-plasma turbulence compared to the fluid descriptions that 
have prevailed thus far. 

\subsection{Interstellar Medium}
\label{sec_ISM}

While the solar wind is unmatched by other astrophysical 
plasmas in the level of detail with which turbulence in it 
can be measured, the interstellar medium (ISM) also offers 
an observer a number of ways of diagnosing 
plasma turbulence, which, in the case of the ISM, is thought to 
be primarily excited by supernova explosions \citep{Norman_Ferrara}. 
The accuracy and resolution of this analysis are due to 
improve rapidly thanks to many new observatories, 
e.g., LOFAR,\footnote{http://www.lofar.org} 
Planck \citep{EWVS_bologna}, and, in more distant 
future, the SKA \citep{Lazio_etal_review}. 

The ISM is a spatially inhomogeneous environment consisting 
of several phases that have different temperatures, 
densities and degrees of ionization 
\citep{Ferriere_review}.\footnote{And, therefore, different 
degrees of importance of the neutral particles 
and the associated ambipolar damping effects---these 
will not be discussed here; see \citealt{Lithwick_Goldreich}.} 
We will use the Warm ISM phase 
(see \tabref{tab_scales}) as our fiducial interstellar 
plasma and discuss briefly what is known about 
the two main observationally accessible quantities---the electron 
density and magnetic fields---and how this information fits 
into the theoretical framework proposed here. 

\subsubsection{Electron Density Fluctuations} 
\label{sec_el_den_ISM}

The electron-density fluctuations inferred from the interstellar 
scintillation measurements appear to have a spectrum with an 
exponent $\simeq-1.7$, consistent with the Kolmogorov scaling 
(\citealt{Armstrong_Cordes_Rickett,Armstrong_Rickett_Spangler,Lazio_etal_review}; see, however, 
dissenting evidence by \citealt{Smirnova_Gwinn_Shishov}, who 
claim a spectral exponent closer to $-1.5$). This holds 
over about 5 decades of scales: $\lambda\in(10^5,10^{10})$~km. 
Other observational evidence at larger and smaller scales 
supports the case for this presumed inertial range to be 
extended over as many as 12 decades: $\lambda\in(10^2,10^{15})$~km, 
a fine example of scale separation that prompted an impressed 
astrophysicist to dub the density scaling 
``The Great Power Law in the Sky.'' 
The upper cutoff here is consistent with the estimates 
of the supernova scale of order $100$~pc---presumably 
the outer scale of the turbulence \citep{Norman_Ferrara}  
and also roughly the scale height of the galactic disk 
(obviously the upper bound on the validity 
of any homogeneous model of the ISM turbulence). 
The lower cutoff is an estimate for the inner scale 
below which the logarithmic slope of the density spectrum 
steepens to about $-2$ \citep{Spangler_Gwinn}. 

\citet{Higdon} was the first to realize that the electron-density 
fluctuations in the ISM could be attributed to a cascade of 
a passive tracer mixed by the ambient turbulence
(the MHD entropy mode; see \secref{sec_scalings_passive}). 
This idea was brought to maturity by 
\citet{Lithwick_Goldreich}, who studied the passive cascades 
of the slow and entropy modes in the framework of the GS theory 
\citep[see also][]{Maron_Goldreich}. 
If the turbulence is assumed anisotropic, 
as in the GS theory, the passive nature of the density fluctuations 
with respect to the decoupled Alfv\'en-wave cascade becomes a 
rigorous result both in MHD (\secref{sec_sw_fluid}) 
and, as we showed above, in the more general gyrokinetic description 
appropriate for weakly collisional plasmas (\secref{sec_sw}). 
Anisotropy of the electron-density fluctuations in the ISM is, 
indeed, observationally supported \citep[][see also \citealt{Lazio_etal_review}
for a concise discussion]{Wilkinson_Narayan_Spencer,Trotter_Moran_Rodriguez,Rickett_etal_aniso,DennettThorpe_deBruyn,Heyer_etal}, 
although detailed scale-by-scale measurements are not 
currently possible. 

If the underlying Alfv\'en-wave turbulence in the ISM 
has a $\kperp^{-5/3}$ spectrum, as predicted by GS, 
so should the electron density (see \secref{sec_scalings_passive}). 
As we discussed in 
\secref{sec_par_cascade}, the physical nature of the 
inner scale for the density fluctuations depends on whether 
they have a cascade in $\kpar$ and are efficiently damped 
when $\kpar\mfp\sim1$ or fail to develop small parallel scales 
and can, therefore, reach $\kperp\rho_i\sim1$. 
The observationally estimated inner scale is 
consistent with the ion gyroscale, $\rho_i\sim10^3$~km 
(see \tabref{tab_scales}; 
note that the ion inertial scale $d_i= \rho_i/\sqrt{\beta_i}$
is similar to $\rho_i$ at the moderate values of $\beta_i$ 
characteristic of the ISM---see further discussion 
of the (ir)relevance of $d_i$ in \secref{sec_transition}, 
\secref{sec_dr_scale} and \apref{ap_Hall}). 
However, since the mean free path in the ISM is not 
huge (\tabref{tab_scales}), 
it is not possible to distinguish this from the 
perpendicular cutoff $\kperp^{-1}\sim\mfp^{3/2}\lf^{-1/2}\sim500$~km 
implied by the parallel cutoff at $\kpar\mfp\sim1$ 
[see \eqref{kperp_LG}], as advocated by 
\citet{Lithwick_Goldreich}. 
Note that the relatively short mean free path means 
that much of the scale range spanned by the Great Power 
Law in the Sky is, in fact, well described by the 
MHD approximation either with adiabatic 
(\secref{sec_RMHD}) or isothermal 
(\secref{sec_visc} and \apref{ap_visc}) electrons. 

Below the ion gyroscale, the $-2$ spectral exponent 
reported by \citet{Spangler_Gwinn} is measured sufficiently 
imprecisely to be consistent with the $-7/3$ expected 
for the density fluctuations in the KAW cascade 
(\secref{sec_KAW_turb}). However, given the high degree of 
uncertainty about what happens in this ``dissipation range'' 
even in the much better resolved case of the solar wind 
(\secref{sec_SW_dr}), it would probably be wise 
to reserve judgment until better data are available.

\subsubsection{Magnetic Fluctuations} 

The second main observable type of turbulent fluctuations in the ISM 
are the magnetic fluctuations, accessible indirectly via the 
measurements of the Faraday rotation of the polarization 
angle of the pulsar light travelling through the ISM. 
The structure function of the 
rotation measure (RM) should have the Kolmogorov slope of $2/3$ if the 
magnetic fluctuations are due to Alfv\'enic turbulence 
described by the GS theory. There is a considerable uncertainty 
in interpreting the available data, primarily due to insufficient 
spatial resolution (rarely better than a few parsec). 
Structure function slopes consistent with $2/3$ have been 
reported \citep{Minter_Spangler}, but, depending on where 
one looks, shallower structure functions that seem to steepen at scales 
of a few parsec are also observed \citep{Haverkorn_etal_ApJ}. 

A recent study by \citet{Haverkorn_etal_arms} detected an 
interesting trend: the RM structure functions computed for 
regions that lie in the galactic spiral arms 
are nearly perfectly flat down to the resolution limit, 
while in the interarm regions, they have detectable 
slopes (although these are mostly shallower that $2/3$). 
Observations of magnetic fields in 
external galaxies also reveal a marked difference in 
the magnetic-field structure between arms and interarms: 
the spatially regular (mean) fields are stronger in 
the interarms, while in the arms, the stochastic fields 
dominate \citep{Beck_structure}. 
This qualitative difference between the magnetic-field 
structure in the arms and interarms has been attributed 
to smaller effective outer scale in the arms 
\citep[$\sim1$~pc, compared to $\sim10^2$~pc in the interarms; 
see][]{Haverkorn_etal_arms2} or to the turbulence in 
the arms and interarms belonging to the two distinct asymptotic 
regimes described in \secref{sec_two_regimes}: 
closer to the anisotropic Alfv\'enic turbulence 
with a strong mean field in the interarms and 
to the isotropic saturated state of small-scale dynamo 
in the arms \citep{SCD_kiev}. 

\subsection{Accretion Disks}
\label{sec_disks}

Accretion of plasma onto a central black hole or neutron star is
responsible for many of the most energetic phenomena observed in
astrophysics (see, e.g., \citealt{Narayan_Quataert_Sci} for a review).  
It is now believed that a linear instability of differentially rotating
plasmas---the magnetorotational instability (MRI)---amplifies 
magnetic fields and gives rise to MHD turbulence in astrophysical
disks \citep{Balbus_Hawley_review}. Magnetic stresses due to this turbulence 
transport angular momentum, allowing plasma to accrete. 
The MRI converts the gravitational potential energy of the inflowing 
plasma into turbulence at the outer scale that is comparable to the scale 
height of the disk. This energy is then cascaded to small scales 
and dissipated into heat---powering the radiation that we see from accretion flows. 
Fluid MHD simulations show that the MRI-generated turbulence in disks 
is subsonic and has $\beta\sim10-100$. 
Thus, on scales much smaller than the scale height of the disk, 
homogeneous turbulence in the parameter regimes considered in this paper 
is a valid idealization and the kinetic models developed above 
should represent a step forward compared to the purely fluid approach. 

Turbulence is not yet directly observable in disks, so models 
of turbulence are mostly used to produce testable 
predictions of observable properties of disks such as their 
X-ray and radio emission. One of the best observed cases is the 
(presumed) accretion flow onto the 
black hole coincident with the radio source Sgr~A$^*$ in the 
center of our Galaxy \citep[see review by][]{Quataert_SgrA}. 

Depending on the rate of heating and cooling in the inflowing plasma
(which in turn depend on accretion rate and other properties of the
system under consideration), there are different models that 
describe the physical properties of accretion flows onto a central object. 
In one class of
models, a geometrically thin optically thick accretion disk \citep{Shakura_Sunyaev}, 
the inflowing plasma is cold and dense and well described as an MHD fluid. 
When applied to Sgr~A$^*$, these models produce a prediction 
for its total luminosity that is several orders of magnitude larger than observed. 
Another class of models, which appears to be more consistent with the 
observed properties of Sgr~A$^*$, 
is called radiatively inefficient accretion flows (RIAFs; see 
\citealt{Rees_etal,Narayan_Yi} and review by \citealt{Quataert_SgrA} of the 
applications and observational constraints in Sgr~A$^*$). 
In these models, the inflowing plasma near the black hole 
is believed to adopt a
two-temperature configuration, with the ions ($T_i\sim10^{11}-10^{12}$~K)
hotter than the electrons ($T_e\sim10^9-10^{11}$~K).\footnote{It is 
partly with this application in mind that we carried the general 
temperature ratio in our calculations; see footnote \ref{fn_temperatures}.} 
The electron and ion thermodynamics decouple because the densities are 
so low that the temperature equalization time $\sim\nuie^{-1}$ is 
longer than the time for the plasma to flow into the black hole. Thus,
like the solar wind, RIAFs are macroscopically collisionless plasmas
(see \tabref{tab_scales} for plasma parameters in the Galactic center; 
note that these parameters are so extreme that 
the gyrokinetic description, while probably better than the fluid one, 
cannot be expected to be rigorously valid; at the very least, it needs to 
be reformulated in a relativistic form). 
At the high temperatures appropriate to RIAFs, electrons radiate energy 
much more efficiently than the ions (by virtue of their much smaller mass)
and are, therefore, expected to contribute dominantly to the observed 
emission, while the thermal energy of the ions is swallowed by the 
black hole. Since the plasma is collisionless, the electron heating 
by turbulence largely determines the thermodynamics 
of the electrons and thus the observable properties of RIAFs. 
The question of which fraction of the turbulent energy 
goes into ion and which into electron heating is, therefore, crucial 
for understanding accretion flows---and the answer to this question 
depends on the detailed properties of the small-scale kinetic turbulence
\citep[e.g.,][]{Quataert_Gruzinov,Sharma_etal07}, as well as on the 
linear properties of the collisionless MRI \citep{Quataert_Dorland_Hammett,Sharma_Hammett_Quataert}.

Since all of the turbulent power coming down the cascade must be 
dissipated into either ion or electron heat, it is really 
the amount of generalized energy diverted at the ion gyroscale 
into the ion entropy cascade (\secsdash{sec_en_ERMHD}{sec_ent_KAW}) that decides 
how much energy is left to heat the electrons via the KAW cascade
(\secsdash{sec_ERMHD_eqns}{sec_KAW_turb}, \secref{sec_ent_els}). 
Again, as in the case of the 
solar wind (\secsand{sec_dr_transition}{sec_dr_variability}), the transition around 
the ion gyroscale from the Alfv\'enic turbulence at $\kperp\rho_i\ll1$ 
to the KAW turbulence at $\kperp\rho_i\gg1$ emerges as a key 
unsolved problem. 

\subsection{Galaxy Clusters}
\label{sec_clusters}

Galaxy clusters are the largest plasma objects in the Universe. 
Like the other examples discussed above, 
the intracluster plasma is in the weakly collisional 
regime (see \tabref{tab_scales}). 
Fluctuations of electron density, temperature and of magnetic 
fields are measured in clusters by X-ray and radio 
observatories, but the resolution 
is only just enough to claim that a fairly broad 
scale range of fluctuations exists \citep{Schuecker_etal,Vogt_Ensslin2}. 
No power-law scalings have yet been established beyond reasonable doubt. 

What fundamentally hampers quantitative modeling of turbulence 
and related effects in clusters 
is that we do not have a definite theory of the basic 
properties of the intracluster medium: its (effective) viscosity, 
magnetic diffusivity or thermal conductivity. In a weakly 
collisional and strongly magnetized plasma, all of these 
depend on the structure of the magnetic field \citep{Braginskii},
which is shaped by the turbulence. If (or at scales where) 
a reasonable {\em a priori} assumption can be made about 
the field structure, further analytical progress is possible: 
thus, the theoretical models presented in this paper 
assume that the magnetic field is a sum of 
a slowly varying in space ``mean field'' and small 
low-frequency perturbations ($\dB\ll B_0$). 

In fact, 
since clusters do not have mean fields of any magnitude that could 
be considered dynamically significant, but do have stochastic fields, 
the outer-scale MHD turbulence in clusters falls into the weak-mean-field 
category (see \secref{sec_two_regimes}). The magnetic field should be 
highly filamentary, organized in long folded direction-reversing 
structures. It is not currently known what determines the reversal 
scale.\footnote{See \citet{SC_dpp05} for a detailed presentation 
of our views on the interplay between turbulence, magnetic field 
and plasma effects in cluster; for further discussions and disagreements, 
see \citet{Ensslin_Vogt_cores,Subramanian_Shukurov_Haugen,Brunetti_Lazarian}.} 
Observations, while tentatively confirming the existence of 
very long filaments \citep{Clarke_Ensslin}, suggest that 
the reversal scale is much larger than the ion 
gyroscale: thus, the magnetic-energy spectrum for the Hydra A cluster 
core reported by \citet{Vogt_Ensslin2} peaks at around $1$~kpc, compared 
to $\rho_i\sim10^5$~km. Below this scale, 
an Alfv\'en-wave cascade should exist (as is, indeed, suggested 
by Vogt \& En{\ss}lin's spectrum being roughly consistent with $k^{-5/3}$ 
at scales below the peak). As these scales are collisionless 
($\mfp\sim100$~pc in the cores and $\sim10$~kpc in the bulk 
of the clusters), it is to this turbulence that 
the theory developed in this paper should be applicable. 

Another complication exists, similar 
to that discussed in \secref{sec_pressure_aniso}: 
pressure anisotropies could 
give rise to fast plasma instabilities whose growth rate peaks 
just above the ion gyroscale. As was pointed out by \citet{SCKHS_brag}, 
these are, in fact, an inevitable consequence of 
any large-scale fluid motions that change the strength 
of the magnetic field. 
Although a number of interesting and plausible arguments 
can be made about the way the instabilities might 
determine the magnetic-field structure \citep{SC_dpp05,SCKRH_firehose,Rosin_etal_firehose,Rincon_etal_mirror}, 
it is not currently understood how the small-scale fluctuations resulting from these 
instabilities coexist with the Alfv\'enic cascade. 

The uncertainties that result from this 
imperfect understanding of the nature of the intracluster 
medium are exemplified by the problem of its thermal conductivity. 
The magnetic-field reversal scale in clusters is certainly 
not larger than the electron diffusion scale, $(m_i/m_e)^{1/2}\mfp$,  
which varies from a few kpc in the cores to a few hundred kpc in the bulk. 
Therefore, one would expect that the approximation of 
isothermal electron fluid (\secref{sec_els}) should certainly apply 
at all scales below the reversal scale, where 
$\dB\ll B_0$ presumably holds. 
Even this, however, is not absolutely clear. 
One could imagine the electrons being effectively adiabatic 
if (or in the regions where) the plasma instabilities give rise 
to large fluctuations of the magnetic field ($\dB/B_0\sim1$) 
at the ion gyroscale reducing the mean free path to $\mfp\sim\rho_i$
\citep{SCKRH_firehose,Rosin_etal_firehose,Rincon_etal_mirror}. 
Such fluctuations cannot be described by the gyrokinetics in its current form. 
The current state of the observational evidence does not 
allow one to exclude either of these possibilities. 
Both isothermal \citep{Fabian_etal_Perseus3,Sanders_Fabian} 
and non-isothermal \citep{Markevitch_Vikhlinin_review} 
coherent structures that appear to be shocks are observed. 
Disordered fluctuations of temperature can also be detected, 
which allows one to infer an upper limit for the scale at 
which the isothermal approximation can start being valid: 
thus, \citet{Markevitch_etal03} find temperature 
variations at all scales down to $\sim100$~kpc, which is the
statistical limit that defines the spatial resolution 
of their temperature map. In none of these or similar measurements 
is the magnetic field data available that would make 
possible a pointwise comparison of the magnetic and thermal structure. 

Because of this lack of information about the state of the 
magnetized plasma in clusters, theories of the 
intracluster medium are not sufficiently constrained by observations, 
so no one theory is in a position to prevail. 
This uncertain state of affairs  
might be improved by analyzing the 
observationally much better resolved case of the solar wind, which 
should be quite similar to the intracluster medium at very small 
scales (except for somewhat lower values of $\beta_i$ in the solar wind).  

\section{Conclusion}
\label{sec_conc}

In this paper, we have considered magnetized plasma turbulence in 
the astrophysically prevalent regime of weak collisionality. 
We have shown how the energy injected at the outer scale 
cascades in phase space, eventually to increase the entropy 
of the system and heat the particles. In the process, we have 
explained how one combines plasma physics tools---in particular, 
the gyrokinetic theory---with the ideas of a turbulent cascade of energy 
to arrive at a hierarchy of tractable models of turbulence in various 
physically distinct scale intervals. These models represent 
the branching pathways of a generalized energy cascade 
in phase space (the ``kinetic cascade''; see \figref{fig_cascade_channels}) 
and make explicit the ``fluid'' and ``kinetic'' aspects of plasma turbulence. 

A detailed outline of these developments 
was given in the Introduction. Intermediate technical summaries 
were provided in \secref{sec_els_sum}, \secref{sec_KRMHD_sum}, 
and \secref{sec_ERMHD_sum}. An astrophysical summary 
and discussion of the observational evidence was given 
in \secref{sec_astro}, with a particular emphasis on  
space plasmas (\secsdash{sec_SW_ir}{sec_pressure_aniso}). 
Our view of how the transformation of the large-scale 
turbulent energy into heat occurs was encapsulated in the concept of 
a kinetic cascade of generalized energy. 
It was previewed in \secref{sec_kinetic} and developed 
quantitatively in \secsdash{sec_en_GK}{sec_heating}, 
\secref{sec_en_els}, \secref{sec_en_KRMHD}, \secsdash{sec_inv_compr}{sec_en_compr},
\secsdash{sec_en_ERMHD}{sec_ent_els}, 
\apsand{ap_en_RMHD}{ap_Hall_en}. 

\lastpagefootnotes

Following a series of analytical contributions that set up a 
theoretical framework for astrophysical gyrokinetics 
\citep[][and this paper]{Howes_etal,Howes_etal2,SCD_kiev,SCDHHPQT_crete}, 
an extensive program of fluid, hybrid fluid-kinetic, 
and fully gyrokinetic\footnote{Using the publicly available {\tt GS2} code 
(developed originally for fusion applications; see http://gs2.sourceforge.net)
and the purpose-built {\tt AstroGK} code 
(see http://www.physics.uiowa.edu/\textasciitilde ghowes/astrogk/).\label{fn_GS2}} 
numerical simulations of magnetized plasma turbulence is now underway
\citep[for the first results of this program, see][]{Howes_etal3,Tatsuno_etal1,Tatsuno_etal2}. 
Careful comparisons of the fully gyrokinetic 
simulations with simulations based on the more readily computable 
models derived in this paper (RMHD---\secref{sec_RMHD}, 
isothermal electron fluid---\secref{sec_els}, 
KRMHD---\secref{sec_KRMHD}, 
ERMHD---\secref{sec_ERMHD},
HRMHD---\apref{ap_Hall}) as well as with 
the numerical studies based on various Landau fluid 
\citep{Snyder_Hammett_Dorland,Goswami_Passot_Sulem,Ramos,Sharma_etal06,Sharma_etal07,Passot_Sulem07} 
and gyrofluid \citep{Hammett_Dorland_Perkins,Dorland_Hammett,Snyder_Hammett,Scott} closures 
appear to be the way forward in developing a comprehensive 
numerical model of the kinetic turbulent cascade from the outer scale to the 
electron gyroscale. Of the many astrophysical 
plasmas to which these results apply, the solar wind and, perhaps, the magnetosheath, 
due to the high quality of turbulence measurements possible in them, appear 
to be the most suitable test beds for direct and detailed quantitative comparisons 
of the theory and simulation results with observational evidence. 
The objective of all this work remains a quantitative characterization 
of the scaling-range properties (spectra, anisotropy, nature  
of fluctuations and their interactions), the ion and electron heating, and 
the transport properties of the magnetized plasma turbulence.

\acknowledgements

We thank O.~Alexandrova, S.~Bale, J.~Borovsky, T.~Carter, S.~Chapman, C.~Chen, 
E.~Churazov, T.~En{\ss}lin, A.~Fabian, A.~Finoguenov, A.~Fletcher, 
M.~Haverkorn, B.~Hnat, T.~Horbury, K.~Issautier, C.~Lacombe, M.~Markevitch, 
K.~Osman, T.~Passot, F.~Sahraoui, A.~Shukurov, 
and A.~Vikhlinin for helpful discussions of experimental and observational data; 
I.~Abel, M.~Barnes, D.~Ernst, J.~Hastie, P.~Ricci, C.~Roach, and B.~Rogers 
for discussions of collisions in gyrokinetics; 
and G.~Plunk for discussions of the theory of gyrokinetic turbulence in 
two spatial dimensions. 
The authors' travel was supported by the US DOE Center for 
Multiscale Plasma Dynamics and by the Leverhulme Trust (UK) 
International Academic Network for Magnetized Plasma Turbulence. 
A.A.S.\ was supported in part by a PPARC/STFC Advanced Fellowship and by 
the STFC Grant ST/F002505/1. He also thanks 
the UCLA Plasma Group for its hospitality on several occasions. 
S.C.C.\ and W.D.\ thank the Kavli Institute for Theoretical Physics 
and the Aspen Center for Physics for their hospitality. 
G.W.H.\ was supported by the US DOE contract DE-AC02-76CH03073. 
G.G.H.\ and T.T.\ were supported by the US DOE Center for Multiscale Plasma Dynamics. 
E.Q.\ and G.G.H.\ were supported in part by the David and Lucille Packard Foundation. 

\begin{appendix}

\section{Braginskii's Two-Fluid Equations and Reduced MHD} 
\label{ap_Brag}

Here we explain how the standard one-fluid MHD equations used in \secref{sec_RMHD} 
and the collisional limit of the KRMHD system (\secref{sec_visc}, derived 
in \apref{ap_visc}) both emerge as limiting cases of the two-fluid theory. 
For the case of anisotropic fluctuations, $\kpar/\kperp\ll1$, all of this 
can, of course, be derived from gyrokinetics, but it is useful to provide 
a connection to the more well known fluid description of collisional plasmas. 

\subsection{Two-Fluid Equations}
\label{ap_two_fluid}

The rigorous derivation of the fluid equations for a collisional plasma was 
done in the classic paper of \citet{Braginskii}. His equations, valid 
for $\omega/\nui\ll1$, $\kpar\mfp\ll1$, $\kperp\rho_i\ll1$ (see \figref{fig_validity_gk}), 
evolve the densities $n_s$, mean velocities $\vu_s$ and temperatures $T_s$ 
of each plasma species ($s=i,e$): 
\bea
\label{Brag_ns}
\lt({\dd\over\dd t} + \vu_s\cdot\vdel\rt)n_s &=& - n_s\vdel\cdot\vu_s,\\
\label{Brag_us}
m_s n_s \lt({\dd\over\dd t} + \vu_s\cdot\vdel\rt)\vu_s &=&
- \vdel p_s - \vdel\cdot\vPi_s + \qs n_s\lt(\vE + {\vu_s\times\vB\over c}\rt) + \vF_s,\\
\label{Brag_Ts}
{3\over2}\,n_s \lt({\dd\over\dd t} + \vu_s\cdot\vdel\rt)T_s &=&
- p_s\vdel\cdot\vu_s - \vdel\cdot\vGamma_s - \vPi_s:\vdel\vu_s + Q_s,
\eea
where $p_s=n_s T_s$ and the expressions for the viscous stress tensor $\vPi_s$, 
the friction force $\vF_s$, the heat flux $\vGamma_s$ and the interspecies heat 
exchange $Q_s$ are given in \citet{Braginskii}. \Eqsdash{Brag_ns}{Brag_Ts} are 
complemented with the quasi-neutrality condition, $n_e=Zn_i$, and the 
Faraday and Amp\`ere laws, which are (in the non-relativistic limit)
\bea
{\dd\vB\over\dd t} = - c\vdel\times\vE,\quad
\vj = en_e(\vu_i-\vu_e) = {c\over4\pi}\vdel\times\vB.
\eea
Because of quasi-neutrality, we only need one of the continuity equations, 
say the ion one. We can also use the electron momentum equation [\eqref{Brag_us}, $s=e$] 
to express $\vE$, which we then substitute into the ion momentum equation 
and the Faraday law. The resulting system is 
\bea
\label{Brag_rho}
{d\rho\over dt} &=& - \rho\vdel\cdot\vu,\\
\label{Brag_u}
\rho {d\vu\over dt} &=&
- \vdel\lt(p+{B^2\over 8\pi}\rt) - \vdel\cdot\vPi 
+ {\vB\cdot\vdel\vB\over4\pi}
- {Zm_e\over m_i}\rho\lt({\dd\over\dd t} + \vu_e\cdot\vdel\rt)\vu_e,\\
\label{Brag_B}
{\dd\vB\over\dd t} &=& \vdel\times\lt[\vu\times\vB - {\vj\times\vB\over en_e} 
+ {c\vdel p_e\over en_e} + {c\vdel\cdot\vPi_e\over en_e} 
- {c\vF_e\over en_e} + {cm_e\over e}\lt({\dd\over\dd t} + \vu_e\cdot\vdel\rt)\vu_e\rt],
\eea
where $\rho=m_in_i$, $\vu=\vu_i$, $p=p_i+p_e$, $\vPi=\vPi_i+\vPi_e$, 
$\vu_e=\vu-\vj/en_e$, $n_e=Zn_i$, $d/dt=\dd/\dd t + \vu\cdot\vdel$. 
The ion and electron temperatures continue to satisfy \eqref{Brag_Ts}. 

\subsection{Strongly Magnetized Limit}
\label{ap_strongly_mag}

In this form, the two-fluid theory starts resembling the standard one-fluid MHD, 
which was our starting point in \secref{sec_RMHD}:
\eqsdash{Brag_rho}{Brag_B} already look similar to the continuity, momentum and 
induction equations. The additional terms that appear in these equations 
and the temperature equations \exref{Brag_Ts} are brought under control 
by considering how they depend on a number of dimensionless parameters: 
$\omega/\nui$, $\kpar\mfp$, $\kperp\rho_i$, $(m_e/m_i)^{1/2}$.  
While all these are small in Braginskii's calculation, 
no assumption is made as to how they compare to each other. 
We now specify that 
\bea
{\omega\over\nui}\sim {\kpar\mfp\over\sqrt{\beta_i}},\quad
\kperp\rho_i\ll\kpar\mfp\sim\sqrt{m_e\over m_i}\ll1
\label{Brag_small}
\eea
(see \figref{fig_validity_isoth}). 
Note that the first of these relations is equivalent to assuming 
that the fluctuation frequencies are Alfv\'enic---the same assumption 
as in gyrokinetics [\eqref{omega_vs_nu}]. 
The second relation in \eqref{Brag_small} will be referred to by us 
as the {\em strongly magnetized limit}. 
Under the assumptions~\exref{Brag_small}, 
the two-fluid equations reduce to the following 
closed set:\footnote{The structure of the momentum equation \exref{Brag_u2} 
is best understood by realizing 
that $\rho\nupar\lt(\vb\vb:\vdel\vu - \vdel\cdot\vu/3\rt)=\pperp-\ppar$, the difference 
between the perpendicular and parallel (ion) pressures. Since the total pressure is 
$p=(2/3)\pperp + (1/3)\ppar$, \eqref{Brag_u2} can be written 
\bea
\label{Brag_u3}
\rho {d\vu\over dt} =
- \vdel\lt(\pperp+{B^2\over 8\pi}\rt) 
+ \vdel\cdot\lt[\vb\vb\lt(\pperp-\ppar\rt)\rt]
+ {\vB\cdot\vdel\vB\over4\pi}.
\eea
This is the general form of the momentum equation that is also valid 
for collisionless plasmas, when $\kperp\rho_i\ll1$ but $\kpar\mfp$ is 
order unity or even large. \Eqref{Brag_u3} 
together with the continuity equation \exref{Brag_u2}, 
the induction equation \exref{Brag_B2} and a kinetic equation for the 
particle distribution function (from the solution of which $\pperp$ and $\ppar$ 
are determined) form the system known as Kinetic MHD \citep[KMHD, see][]{Kulsrud_Varenna,Kulsrud_HPP}. 
The collisional limit, $\kpar\mfp\ll1$, of KMHD is again \eqsdash{Brag_rho2}{Brag_Te}.
}
\bea
\label{Brag_rho2}
{d\rho\over dt} &=& - \rho\vdel\cdot\vu,\\
\label{Brag_u2}
\rho {d\vu\over dt} &=&
- \vdel\lt[p+{B^2\over 8\pi} 
+ {1\over3}\,\rho\nupar\lt(\vb\vb:\vdel\vu - {1\over3}\,\vdel\cdot\vu\rt)\rt] 
+ \vdel\cdot\lt[\vb\vb\rho\nupar\lt(\vb\vb:\vdel\vu - {1\over3}\,\vdel\cdot\vu\rt)\rt]
+ {\vB\cdot\vdel\vB\over4\pi},\\
\label{Brag_B2}
{d\vB\over d t} &=& \vB\cdot\vdel\vu - \vB\vdel\cdot\vu,\\
\label{Brag_Ti}
{d T_i\over dt} &=& - {2\over 3}T_i\vdel\cdot\vu 
+ {1\over\rho}\vdel\cdot\lt(\vb\rho\kappar\vb\cdot\vdel T_i\rt)
- \nuie\lt(T_i-T_e\rt)
+ {2\over3}\,m_i\nupar\lt(\vb\vb:\vdel\vu - {1\over3}\,\vdel\cdot\vu\rt)^2,\\
\label{Brag_Te}
{d T_e\over dt} &=& - {2\over 3}T_e\vdel\cdot\vu 
+ {1\over\rho}\vdel\cdot\lt(\vb\rho\kappare\vb\cdot\vdel T_e\rt)
- {1\over Z}\,\nuie\lt(T_e-T_i\rt),
\eea
where $\nupar=0.90 \vthi\mfp$ is the parallel ion viscosity, 
$\kappar=2.45 \vthi\mfp$ parallel ion thermal diffusivity, 
$\kappare =1.40 \vthe \mfpe \sim \lt(Z^2/\tau^{5/2}\rt)(m_i/m_e)^{1/2}\kappar$ 
parallel electron thermal diffusivity [here $\mfp=\vthi/\nui$ with 
$\nui$ defined in \eqref{nui_def}], 
and $\nuie$ ion--electron collision rate 
[defined in \eqref{nuie_def}]. 
Note that the last term in \eqref{Brag_Ti} represents the viscous heating of the ions. 

\subsection{One-Fluid Equations (MHD)}
\label{ap_MHD}

If we now restrict ourselves to the low-frequency regime where ion--electron 
collisions dominate over all other terms in the ion-temperature equation \exref{Brag_Ti},
\bea
\label{one_fluid}
{\omega\over\nuie} \sim {\kpar\mfp\over\sqrt{\beta_i}}\sqrt{m_i\over m_e} \ll1
\eea
[see \eqsand{Brag_small}{nuie_def}], we have, to lowest order in this new subsidiary 
expansion, $T_i=T_e=T$. We can now write $p=(n_i+n_e)T=(1+Z)\rho T/m_i$ and, adding 
\eqsand{Brag_Ti}{Brag_Te}, find the equation for pressure:
\bea
{d p\over dt} + {5\over3}\,p\vdel\cdot\vu = 
\vdel\cdot\lt(\vb n_e\kappare\vb\cdot\vdel T\rt)
+ {2\over3}\,m_i\nupar\lt(\vb\vb:\vdel\vu - {1\over3}\,\vdel\cdot\vu\rt)^2,
\label{Brag_p}
\eea
where we have neglected the ion thermal diffusivity compared to the electron one, 
but kept the ion heating term to maintain energy conservation. 
\Eqref{Brag_p} together with \eqsdash{Brag_rho2}{Brag_B2} constitutes the 
conventional one-fluid MHD system. With the dissipative terms [which are 
small because of \eqref{one_fluid}] neglected, this was the starting point for our 
fluid derivation of RMHD in \secref{sec_RMHD}. 

Note that the electrons in this regime are adiabatic because 
the electron thermal diffusion is small 
\bea
\label{ad_cond}
{\kappare\kpar^2\over\omega}\sim \kpar\mfp\sqrt{\beta_i}\sqrt{m_i\over m_e} \ll 1,
\eea
provided \eqref{one_fluid} holds and $\beta_i$ is order unity. If
we take $\beta_i\gg1$ instead, we can still satisfy \eqref{one_fluid}, 
so $T_i=T_e$ follows from the ion temperature equation \exref{Brag_Ti} 
and the one-fluid equations emerge as an expansion in 
high $\beta_i$. However, these equations now describe two 
physical regimes: the adiabatic long-wavelength regime 
that satisfies \eqref{ad_cond} and the shorter-wavelength 
regime in which  $(m_e/m_i)^{1/2}/\sqrt{\beta_i}
\ll \kpar\mfp \ll (m_e/m_i)^{1/2}\sqrt{\beta_i}$, 
so the fluid is isothermal, $T=T_0=\const$, 
$p=[(1+Z)T_0/m_i]\rho = c_s^2\rho$ [\eqref{MHD_p} holds with $\gamma=1$]. 

\subsection{Two-Fluid Equations with Isothermal Electrons}
\label{ap_isoth_els}

Let us now consider the regime in which the coupling between 
the ion and electron temperatures is small and the electron diffusion 
is large [the limit opposite to \eqsand{one_fluid}{ad_cond}]:
\bea
{\omega\over\nuie} \sim {\kpar\mfp\over\sqrt{\beta_i}}\sqrt{m_i\over m_e} \gg 1,\quad
{\kappare\kpar^2\over\omega}\sim \kpar\mfp\sqrt{\beta_i}\sqrt{m_i\over m_e} \gg 1,
\eea
Then the electrons are isothermal, $T_e=\Te=\const$ (with the usual assumption
of stochastic field lines, so $\Dpar T_e=0$ implies $\vdel T_e=0$, 
as in \secref{sec_dTe}), while the ion temperature satisfies 
\bea
\label{Brag_Ti2}
{d T_i\over dt} = - {2\over 3}T_i\vdel\cdot\vu 
+ {1\over\rho}\vdel\cdot\lt(\vb\rho\kappar\vb\cdot\vdel T_i\rt)
+ {2\over3}\,m_i\nupar\lt(\vb\vb:\vdel\vu - {1\over3}\,\vdel\cdot\vu\rt)^2.
\eea
\Eqref{Brag_Ti2} together with \eqsdash{Brag_rho2}{Brag_B2} 
and $p=\rho(T_i + Z\Te)/m_i$ are a closed system that describes 
an MHD-like fluid of adiabatic ions and isothermal electrons. 
Applying the ordering of \secref{sec_RMHDordering} to these equations 
and carrying out an expansion in $\kpar/\kperp\ll1$ entirely analogously to 
the way it was done in \secref{sec_RMHD}, we arrive at the 
RMHD equations \exsdash{RMHD_Psi}{RMHD_Phi} for the Alfv\'en waves 
and the following system for the compressive fluctuations (slow and entropy modes):
\bea
\label{Brag_cont}
&&{d\over dt}\lt({\drho\over\rho_0} - {\dBpar\over B_0}\rt) + \Dpar\upar = 0,\\ 
\label{Brag_mom}
&&{d\upar\over dt} - v_A^2\Dpar{\dBpar\over B_0}
= \nupar\,\Dpar\lt(\Dpar\upar + {1\over3}{d\over dt}{\drho\over\rho_0}\rt),\\
&&{d\over dt}{\dTi\over\Ti} - {2\over3}{d\over dt}{\drho\over\rho_0} 
= \kappar\Dpar\lt(\Dpar{\dTi\over\Ti}\rt), 
\label{Brag_en}
\eea
and the pressure balance 
\bea
\label{Brag_pr_bal}
\lt(1+ {Z\over\tau}\rt){\drho\over\rho_0} = - {\dTi\over\Ti} 
- {2\over\beta_i}\lt[{\dBpar\over B_0} + 
{1\over 3v_A^2}\nupar\lt(\Dpar\upar + {1\over3}{d\over dt}{\drho\over\rho_0}\rt)\rt]. 
\eea
Recall that these equations, being the consequence of Braginskii's two-fluid 
equations (\secref{ap_two_fluid}), are an expansion in $\kpar\mfp\ll1$ correct 
up to first order in this small parameter. Since the dissipative terms 
are small, we can replace $(d/dt)\drho/\rho_0$ in 
the viscous terms of \eqsand{Brag_mom}{Brag_pr_bal} by 
its value computed from \eqsref{Brag_cont}, \exref{Brag_en} and \exref{Brag_pr_bal} 
in neglect of dissipation: $(d/dt)\drho/\rho_0 = -\Dpar\upar/(1+c_s^2/v_A^2)$
[cf.\ \eqref{eq_drho}], where the speed of sound $c_s$ is defined by \eqref{cs_def}. 
Substituting this into \eqsand{Brag_mom}{Brag_pr_bal}, we recover 
the collisional limit of KRMHD derived in \apref{ap_visc}, see 
\eqsdash{cont_eq_diff}{en_eq_diff} and \exref{pr_bal_diff}. 

\section{Collisions in Gyrokinetics}
\label{ap_coll}

The general collision operator that appears in \eqref{Vlasov_eq} is 
\citep{Landau_co} 
\bea
\label{Landau_C}
\({\dd f_s\over\dd t}\)_{\rm c} = 2\pi\ln\Lambda\sum_{s'} {\qs^2\qsp^2\over m_s} 
{\dd\over\dd\vv}\cdot\int d^3\vv'{1\over w}\(\phantfrac\unity - {\vw\vw\over w^2}\)\cdot
\lt[{1\over m_s}\,f_{s'}(\vv')\,{\dd f_s(\vv)\over\dd\vv}
-{1\over m_{s'}}\,f_s(\vv)\,{\dd f_{s'}(\vv')\over\dd\vv'} 
\rt],
\eea
where $\vw=\vv-\vv'$ and $\ln\Lambda$ is the Coulomb logarithm. 
We now take into account the expansion of the distribution 
function~\exref{fs_exp}, use the fact that the collision operator vanishes 
when it acts on a Maxwellian, and retain only first-order terms in the gyrokinetic 
expansion. This gives us the general form of the collision term in \eqref{GK_eq}:
it is the ring-averaged linearized form of the Landau collision operator~\exref{Landau_C}, 
$(\dd\hs/\dd t)_{\rm c} = \lt<\dC_s[\hh]\rt>_{\vR_s}$, where  
\bea
\label{dC_def}
\dC_s[\hh] = 2\pi\ln\Lambda\sum_{s'} {\qs^2\qsp^2\over m_s} 
{\dd\over\dd\vv}\cdot\int d^3\vv'{1\over w}\(\phantfrac\unity - {\vw\vw\over w^2}\)\cdot
\lt[\fMsp(v')\({\vv'\over T_{0s'}} + {1\over m_s}{\dd\over\dd\vv}\)\hs(\vv)
- \fMs(v)\({\vv\over T_{0s}} + {1\over m_{s'}}{\dd\over\dd\vv'}\)\hh_{s'}(\vv')\rt].
\eea
Note that the velocity derivatives are taken at constant $\vr$, i.e., 
the gyrocenter distribution functions that appear in the integrand 
should be understood as 
$\hs(\vv)\equiv\hs(t,\vr+{\vvperp\times\vz/\Omega_s},\vperp,\vpar)$. 
The explicit form of the gyrokinetic collision operator can be derived 
in $k$ space as follows:
\bea
\dtcolls = \lt<\dC_s\lt[\sum_\vk e^{i\vk\cdot\vR}\hk\rt]\rt>_{\vR_s} 
= \sum_\vk\lt\<e^{i\vk\cdot\vr}\dC_s\lt[e^{-i\vk\cdot\vrho}\hk\rt]\rt\>_{\vR_s} 
= \sum_\vk e^{i\vk\cdot\vR_s}\lt\<e^{i\vk\cdot\vrho_s(\vv)}\dC_s\lt[e^{-i\vk\cdot\vrho}\hk\rt]\rt\>,
\label{Cgk_eq}
\eea
where $\vrho_s(\vv)=-\vvperp\times\vz/\Omega_s$ and $\vR_s=\vr-\vrho_s(\vv)$. 
Angle brackets with no subscript refer to averages over the gyroangle $\gktheta$ 
of quantities that do not depend on spatial coordinates. 
Note that inside the operator 
$\dC_s[\dots]$, $\hh$ occurs both with index $s$ and velocity $\vv$ 
and with index $s'$ and velocity $\vv'$ (over which summation/integration 
is done). In the latter case, $\vrho=\vrho_{s'}(\vv')=-\vvperp'\times\vz/\Omega_{s'}$ in 
the exponential factor inside the operator. 

Most of the properties of the collision operator 
that are used in the main body of this paper to order the collision terms 
can be established in general, already on the basis of \eqref{Cgk_eq} 
(\secsdash{ap_int_coll}{ap_ii}). 
If the explicit form of the collision operator is required, 
we could, in principle, perform 
the ring average on the linearized operator $\dC$ [\eqref{dC_def}] and derive an 
explicit form of $(\dd\hs/\dd t)_{\rm c}$. In practice, in gyrokinetics, as 
in the rest of plasma physics, the full collision operator 
is only used when it is absolutely unavoidable. 
In most problems of interest, further simplifications are possible: 
the same-species 
collisions are often modeled by simpler operators that share the full 
collision operator's conservation properties (\secref{ap_ss}), 
while the interspecies collision operators are expanded in the 
electron--ion mass ratio (\secref{ap_ei}).

\subsection{Velocity-Space Integral of the Gyrokinetic Collision Operator}
\label{ap_int_coll}

Many of our calculations involve integrating the gyrokinetic equation~\exref{GK_eq} 
over the velocity space while keeping $\vr$ constant. 
Here we estimate the size of the integral of the collision term 
when $\kperp\rho_s\ll1$. 
Using \eqref{Cgk_eq},
\bea
\nonumber
\int d^3\vv \lt<\dtcolls\rt>_\vr
&=& \sum_\vk\int d^3\vv\,e^{i\vk\cdot\vr - i\vk\cdot\vrho_s(\vv)}
\lt\<e^{i\vk\cdot\vrho_s(\vv)}\dC_s\lt[e^{-i\vk\cdot\vrho}\hk\rt]\rt\>\\ 
\nonumber
&=& \sum_\vk e^{i\vk\cdot\vr} 2\pi\int_0^\infty d\vperp\,\vperp\int_{-\infty}^{+\infty}d\vpar
\lt\<e^{-i\vk\cdot\vrho_s(\vv)}\rt\>
\lt\<e^{i\vk\cdot\vrho_s(\vv)}\dC_s\lt[e^{-i\vk\cdot\vrho}\hk\rt]\rt\>\\
\nonumber
&=& \sum_\vk e^{i\vk\cdot\vr} \int d^3\vv
\lt\<e^{-i\vk\cdot\vrho_s(\vv)}\rt\>
e^{i\vk\cdot\vrho_s(\vv)}\dC_s\lt[e^{-i\vk\cdot\vrho}\hk\rt]
= \sum_\vk e^{i\vk\cdot\vr} \int d^3\vv\,
J_0(\kr_s)\, e^{i\vk\cdot\vrho_s(\vv)}\dC_s\lt[e^{-i\vk\cdot\vrho}\hk\rt]\\
&=& \sum_\vk e^{i\vk\cdot\vr} \int d^3\vv\,
\lt[1-i\vk\cdot{\vvperp\times\vz\over\Omega_s}
-{1\over2}\,\(\vk\cdot{\vvperp\times\vz\over\Omega_s}\)^2 - 
{1\over4}\({\kperp\vperp\over\Omega_s}\)^2 
+ \dots\rt]\dC_s\lt[e^{-i\vk\cdot\vrho}\hk\rt].
\label{int_C_exp}
\eea
Since the (linearized) collision operator $\dC_s$ conserves particle number, 
the first term in the expansion vanishes. 
The operator $\dC_s=\dC_{ss}+\dC_{ss'}$ is a sum of the same-species collision operator 
[the $s'=s$ part of the sum in \eqref{dC_def}] 
and the interspecies collision operator (the $s'\neq s$ part). 
The former conserves total momentum of the particles of species $s$, 
so it gives no contribution to the second term in the expansion in \eqref{int_C_exp}. 
Therefore, 
\bea
\int d^3\vv \lt\<\<\dC_{ss}[\hs]\>_{\vR_s}\rt\>_\vr \sim \nu_{ss}\kperp^2\rho_s^2\dn_s. 
\eea 
The interspecies collisions do contribute to the second term in \eqref{int_C_exp}
due to momentum exchange 
with the species $s'$. This contribution is readily inferred from the standard 
formula for the linearized friction force \citep[see, e.g.,][]{Helander_Sigmar}:
\bea
m_s\int d^3\vv\,\vv\,\dC_{ss'}\lt[e^{-i\vk\cdot\vrho}\hk\rt] &=& 
- \int d^3\vv\,\vv\lt[m_s\nuS^{ss'}(v)e^{-i\vk\cdot\vrho_s(\vv)}\hks 
+ m_{s'}\nuS^{s's}(v)e^{-i\vk\cdot\vrho_{s'}(\vv)}\hksp\rt],\\
\nuS^{ss'}(v) &=& {\sqrt{2}\pi\nsp \qs^2\qsp^2\ln\Lambda\over m_s^{1/2}\Ts^{3/2}} 
\({\vths\over v}\)^3\(1+{m_s\over m_{s'}}\)
\lt[\erf\lt({v\over\vthsp}\rt) - {v\over\vthsp}\,\erf'\lt({v\over\vthsp}\rt)\rt],
\eea
where $\erf(x) = (2/\sqrt{\pi})\int_0^x dy\,\exp(-y^2)$ is the error function. 
From this, via a calculation of ring averages analogous to \eqref{vperp_avg}, 
we get 
\bea
\nonumber
\int d^3\vv\(-i\vk\cdot{\vvperp\times\vz\over\Omega_s}\)\dC_{ss'}\lt[e^{-i\vk\cdot\vrho}\hk\rt] 
&=& -\int d^3\vv 
\lt[\nuS^{ss'}(v)\Bigl\<i\vk\cdot\vrho_s(\vv)\,e^{-i\vk\cdot\vrho_s(\vv)}\Bigr\>\hks 
+ {m_{s'}\over m_s}{\Omega_{s'}\over\Omega_s}
\nuS^{s's}(v)\Bigl\<i\vk\cdot\vrho_{s'}(\vv)\,e^{-i\vk\cdot\vrho_{s'}(\vv)}\Bigr\>\hksp\rt]\\
&=& -\int d^3\vv
\lt[\nuS^{ss'}(v)\kr_s J_1(\kr_s)\hks 
+ {\qsp\over \qs}\,\nuS^{s's}(v)\kr_{s'}J_1(\kr_{s'})\hksp\rt]
\label{int_Cinter}
\sim \nu_{ss'}\kperp^2\rho_s^2\dn_s 
+ \nu_{s's}\kperp^2\rho_{s'}^2\dn_{s'}.\qquad
\eea
For the ion--electron collisions ($s=i$, $s'=e$), using \eqsand{rho_ratio}{nuie_def}, 
we find that both terms are $\sim(m_e/m_i)^{1/2}\nui\kperp^2\rho_i^2\dni$. 
Thus, besides an extra factor of $\kperp^2\rho_i^2$, the ion--electron 
collisions are also subdominant by one order in the mass-ratio expansion 
compared to the ion--ion collisions. 
The same estimate holds for the interspecies contributions to the 
third and fourth terms in \eqref{int_C_exp}. In a similar fashion, 
the integral of the electron--ion collision operator ($s=e$, $s'=i$), 
is $\sim\nue\kperp^2\rho_e^2\dne$, which is the same order as the 
integral of the electron--electron collisions. 

The conclusion of this section is that, both for ion and for 
electron collisions, the velocity-space integral 
(at constant $\vr$) of the gyrokinetic collision operator 
is higher order than the collision operator itself by two orders of $\kperp\rho_s$. 
This is the property that we relied on in neglecting collision 
terms in \eqsand{dne_eq}{dni_eq}. 

\subsection{Ordering of Collision Terms in \eqsand{g_eq}{dni_eq}} 
\label{ap_ii} 

In \secref{sec_KRMHD}, we claimed that the contribution to the 
ion--ion collision term due to the $(Ze\<\ephi\>_{\vR_i}/\Ti)\fMi$ 
part of the ion distribution function [\eqref{g_ansatz}] was one order 
of $\kperp\rho_i$ smaller than the contributions from the rest of $\hi$.  
This was used to order collision terms in \eqsand{g_eq}{dni_eq}. 
Indeed, from \eqref{Cgk_eq}, 
\bea
\nonumber
\lt<\dC_{ii}\lt[{Ze\<\ephi\>_{\vR_i}\over\Ti}\,\fMi\rt]\rt>_{\vR_i} 
&=& \sum_\vk e^{i\vk\cdot\vR_i}\lt\<e^{i\vk\cdot\vrho_i}
\dC_{ii}\lt[e^{-i\vk\cdot\vrho_i}J_0(\kr_i)\fMi\rt]\rt\>{Ze\ephi_\vk\over\Ti}\\
&=& \sum_\vk e^{i\vk\cdot\vR_i}\lt\<e^{i\vk\cdot\vrho_i}
\dC_{ii}\lt[\(1-i\vk\cdot\vrho_i - {1\over2}\(\vk\cdot\vrho_i\)^2 
- {\kr_i^2\over4}+\cdots\)\fMi\rt]\rt\>{Ze\ephi_\vk\over\Ti} 
\sim \nui\kperp^2\rho_i^2\,{Ze\ephi\over\Ti}\,\fMi.
\label{Cii_phi}
\eea
This estimate holds because, as it is easy to ascertain using \eqref{dC_def}, 
the operator $\dC_{ii}$ annihilates the first two terms in the expansion 
and only acts non-trivially on an expression that is second order 
in $\kperp\rho_i$. With the aid of \eqref{phi_order}, 
the desired ordering of the term~\exref{Cii_phi} in \eqref{g_eq} follows. 
When \eqref{Cii_phi} is integrated over velocity space, 
the result picks up two extra orders in $\kperp\rho_i$ 
[a general effect of integrating the gyroaveraged collision operator over 
the velocity space; see \eqref{int_C_exp}]:
\bea
\intvi \lt<\dC_{ii}\lt[{Ze\<\ephi\>_{\vR_i}\over\Ti}\,\fMi\rt]\rt>_{\vR_i} 
\sim \nui \kperp^4\rho_i^4\, {Ze\ephi\over\Ti},
\eea 
so the resulting term in \eqref{dni_eq} is third order, as stated in~\secref{sec_AW}. 

\subsection{Model Pitch-Angle-Scattering Operator for Same-Species Collisions}
\label{ap_ss}

A popular model operator for same-species collisions that conserves 
particle number, momentum, and energy is constructed by taking the test-particle 
pitch-angle-scattering operator and correcting it with an additional term 
that ensures momentum conservation (\citealt{Rosenbluth_Hazeltine_Hinton}; 
see also \citealt{Helander_Sigmar}): 
\bea
\Cpa[\hs] &=& 
\nuDs(v)\lt\{{1\over2}\lt[{\dd\over\dd\xi}\(1-\xi^2\){\dd\hs\over\dd\xi} 
+ {1\over1-\xi^2}{\dd^2\hs\over\dd\gktheta^2}\rt] 
+ {2\vv\cdot\vU[\hs]\over\vths^2}\,\fMs\rt\},\quad
\vU[\hs] = {3\over2}{\int d^3\vv\,\vv\,\nuDs(v)\,\hs
\over\int d^3\vv\,(v/\vths)^2\nuDs(v)\fMs(v)},\\
&&\nuDs(v) = \nuss\({\vths\over v}\)^3 
\lt[\(1-{1\over2}{\vths^2\over v^2}\)\erf\lt({v\over\vths}\rt)
+ {1\over2}{\vths\over v}\,\erf'\lt({v\over\vths}\rt)\rt],\quad
\nuss = {\sqrt{2}\pi\ns \qs^4\ln\Lambda\over m_s^{1/2}\Ts^{3/2}},
\label{nuD_def}
\eea
where the velocity derivatives are at constant $\vr$. 
The gyrokinetic version of this operator 
is \citep[cf.][]{Catto_Tsang,Dimits_Cohen}
\bea
\label{Cgk_formula}
\lt<\Cpa[\hs]\rt>_{\vR_s} &=& \sum_\vk e^{i\vk\cdot\vR_s} 
\nuDs(v)\lt\{{1\over2}{\dd\over\dd\xi}\(1-\xi^2\){\dd\hks\over\dd\xi}
- {v^2(1+\xi^2)\over4\vths^2}\,\kperp^2\rho_s^2\hks 
+ 2\,{\vperp J_1(\kr_s)\Uperp[\hks] 
+ \vpar J_0(\kr_s)\Upar[\hks]\over\vths^2}\,\fMs\rt\},\\
\nonumber
&&\Uperp[\hks] = 
{3\over2}{\int d^3\vv\,\vperp J_1(\kr_s)\,\nuDs(v)\hks(\vperp,\vpar) 
\over\int d^3\vv\,(v/\vths)^2\nuDs(v)\fMs(v)},\quad
\Upar[\hks] = 
{3\over2}{\int d^3\vv\,\vpar J_0(\kr_s)\,\nuDs(v)\hks(\vperp,\vpar) 
\over\int d^3\vv\,(v/\vths)^2\nuDs(v)\fMs(v)},
\eea
where $\kr_s=\kperp\vperp/\Omega_s$. 
The velocity derivatives are now at constant $\vR_s$. 
The spatial diffusion term appearing in the ring-averaged collision 
operator is physically due to the fact that a change in a particle's 
velocity resulting from a collision can lead to a 
change in the spatial position of its gyrocenter. 

In order to derive \eqref{Cgk_formula}, we use \eqref{Cgk_eq}. 
Since, $\vrho_s(\vv)=\(-\vx v\sqrt{1-\xi^2}\sin\gktheta 
+ \vy v\sqrt{1-\xi^2}\cos\gktheta\)/\Omega_s$, 
it is not hard to see that 
\bea
{\dd\over\dd\xi}\,e^{-i\vk\cdot\vrho_s(\vv)}\hks = 
e^{-i\vk\cdot\vrho_s(\vv)}\lt[{\dd\over\dd\xi} - 
{\xi\over1-\xi^2}{i\vkperp\cdot\bl(\vvperp\times\vz\br)\over\Omega_s}\rt]\hks,\quad
{\dd\over\dd\gktheta}\,e^{-i\vk\cdot\vrho_s(\vv)}\hks = 
e^{-i\vk\cdot\vrho(\vv)}\({\dd\over\dd\gktheta} + {i\vkperp\cdot\vvperp\over\Omega_s}\)\hks.
\eea
Therefore,
\beq
\lt\<e^{i\vk\cdot\vrho_s(\vv)}{\dd\over\dd\xi}\(1-\xi^2\){\dd\over\dd\xi}\,e^{-i\vk\cdot\vrho_s(\vv)}\hks\rt\>
= {\dd\over\dd\xi}\(1-\xi^2\){\dd\hks\over\dd\xi} - {v^2\xi^2\over2\Omega_s^2}\,\kperp^2\hks,
\quad
\lt\<e^{i\vk\cdot\vrho_s(\vv)}{\dd^2\over\dd\gktheta^2}\,e^{-i\vk\cdot\vrho_s(\vv)}\hks\rt\> 
= -{v^2\(1-\xi^2\)\over2\Omega_s^2}\,\kperp^2\hks.
\eeq
Combining these formulae, we obtain the first two terms in \eqref{Cgk_formula}.
Now let us work out the $\vU$ term:
\beq
\lt\<e^{i\vk\cdot\vrho_s(\vv)}\vv\cdot\int d^3\vv'\,\vv'\nuDs(v')e^{-i\vk\cdot\vrho_s(\vv')}
\hks\bl(\vperp',\vpar'\br)\rt\> 
= \Bigl\<\vv\,e^{i\vk\cdot\vrho_s(\vv)}\Bigr\>\cdot
2\pi\int_0^\infty d\vperp'\,\vperp'\int_{-\infty}^{+\infty}d\vpar'
\nuDs(v')\Bigl\<\vv'e^{-i\vk\cdot\vrho_s(\vv')}\Bigr\>
\hks\bl(\vperp',\vpar'\br).
\eeq
Since $\lt\<\vv\,e^{\pm i\vk\cdot\vrho_s(\vv)}\rt\> = 
\vz\vpar\lt\<e^{\pm i\vk\cdot\vrho_s(\vv)}\rt\> + \lt\<\vvperp e^{\pm i\vk\cdot\vrho_s(\vv)}\rt\>$, 
where $\lt\<e^{\pm i\vk\cdot\vrho_s(\vv)}\rt\> = J_0(\kr_s)$ and 
\beq
\lt\<\vvperp e^{\pm i\vk\cdot\vrho_s(\vv)}\rt\> =
\vz\times\lt\<\bl(\vvperp\times\vz\br)\exp\(\mp i\vkperp\cdot{\vvperp\times\vz\over\Omega_s}\)\rt\> 
= \pm i\Omega_s \vz\times{\dd\over\dd\vkperp}
\lt\<\exp\(\mp i\vkperp\cdot{\vvperp\times\vz\over\Omega_s}\)\rt\> 
= \pm i\, {\vz\times\vkperp\over\kperp}\,\vperp J_1(\kr_s),
\label{vperp_avg}
\eeq
we obtain the third term in \eqref{Cgk_formula}.

It is useful to give the lowest-order form of the operator~\exref{Cgk_formula} 
in the limit $\kperp\rho_s\ll1$: 
\bea
\lt<\Cpa[\hs]\rt>_{\vR_s} = \nuDs(v)\lt[{1\over2}{\dd\over\dd\xi}\(1-\xi^2\){\dd\hs\over\dd\xi}
+ {3\vpar\int d^3\vv'\vpar'\nuDs(v')\hs(\vperp',\vpar') 
\over\int d^3\vv'v^{\prime2}\nuDs(v')\fMs(v')}\,\fMs\rt] + O(\kperp^2\rho_s^2).
\label{Cgk_lowest}
\eea
This is the operator that can be used in the right-hand side of \eqref{sw_g}
(as, e.g., is done in the calculation of collisional transport 
terms in \apref{ap_transport}). 

In practical numerical computations of gyrokinetic turbulence, the pitch-angle scattering 
operator is not sufficient because the distribution function develops small scales 
not only in $\xi$ but also in $v$ (M.~Barnes, W.~Dorland and T.~Tatsuno 2006, unpublished). 
This is, indeed, expected because the phase-space entropy cascade produces 
small scales in $\vperp$, rather than just in $\xi$ (see \secref{sec_small_scales}). 
In order to provide a cut off in $v$, an energy-diffusion operator must be 
added to the pitch-angle-scattering operator derived above. 
A numerically tractable model gyrokinetic energy-diffusion operator was proposed  
by \citet{Abel_etal,Barnes_etal}.\footnote{The collision 
operator now used the {\tt GS2} and {\tt AstroGK} codes (see footnote \ref{fn_GS2}) 
is their energy-diffusion operator plus 
the pitch-angle-scattering operator~\exref{Cgk_formula}.} 

\subsection{Electron--Ion Collision Operator}
\label{ap_ei}

This operator can be expanded in $m_e/m_i$ and to the lowest order 
is \citep[see, e.g.,][]{Helander_Sigmar}
\bea
\dC_{ei}[\hh] = 
\nuDei(v)\lt\{{1\over2}\lt[{\dd\over\dd\xi}\(1-\xi^2\){\dd\he\over\dd\xi} 
+ {1\over1-\xi^2}{\dd^2\he\over\dd\gktheta^2}\rt] 
+ {2\vv\cdot\vu_i\over\vthe^2}\,\fMe\rt\},\quad
\nuDei(v) = \nue\({\vthe\over v}\)^3.
\label{Cei_exp}
\eea
The corrections to this form are $O(m_e/m_i)$. This is second order in the expansion 
of \secref{sec_els} and, therefore, we need not keep these corrections. 
The operator \exref{Cei_exp} is mathematically similar to the model operator 
for the same-species collisions [\eqref{Cgk_formula}]. The gyrokinetic 
version of this operator is derived in the way analogous to 
the calculation in \apref{ap_ss}. The result is 
\bea
\nonumber
\lt<\dC_{ei}[\hh]\rt>_{\vR_e} &=& \sum_\vk e^{i\vk\cdot\vR_e}
\nuDei(v)\lt[{1\over2}{\dd\over\dd\xi}\(1-\xi^2\){\dd\hke\over\dd\xi}
- {v^2(1+\xi^2)\over4\vthe^2}\,\kperp^2\rho_e^2\hke\rt.\\ 
&&-\lt. {Zm_e\over m_i}{\vperp^2\over\vthe^2}{J_1(\kr_e)\over\kr_e}\fMe 
\kperp^2\rho_i^2 
{1\over\ni}\int d^3\vv' {2\vperp^{\prime2}\over\vthi^2} 
{J_1(\kr_i')\over\kr_i'}\,\hki
+ {2\vpar J_0(\kr_e)\uparik\over\vthe^2}\fMe\rt].
\label{Cgk_ei}
\eea
At scales not too close to the electron gyroscale, namely, 
such that $\kperp\rho_e\sim(m_e/m_i)^{1/2}$, 
the second and third terms are manifestly second order in $(m_e/m_i)^{1/2}$, 
so have to be neglected along with other $O(m_e/m_i)$ contributions 
to the electron--ion collisions.\footnote{The third term in \eqref{Cgk_ei} 
is, in fact, never important: at the electron scales, $\kperp\rho_e\sim1$, 
it is negligible because of the Bessel function in the velocity 
integral \citep{Abel_etal}.} 
The remaining two terms are first order in the mass-ratio expansion:  
the first term vanishes for $\he=\hezero$ 
[\eqref{hezero_formula}], so its contribution is first order; 
in the fourth term, we can use \eqref{upare_eq} 
to express $\upari$ in terms of quantities that are also first order. 
Keeping only the first-order terms, the gyrokinetic electron--ion collision operator is 
\bea
\lt<\dC_{ei}[\hh]\rt>_{\vR_e} = 
\nuDei(v)\lt[{1\over2}{\dd\over\dd\xi}\(1-\xi^2\){\dd\heone\over\dd\xi}
+ {2\vpar\upari\over\vthe^2}\fMe\rt].
\label{Cgk_ei_first}
\eea
Note that the ion drag term is essential to represent the ion--electron friction 
correctly and, therefore, to capture the Ohmic resistivity (which, however, is 
rarely more important for unfreezing flux than the electron inertia and the 
finiteness of the electron gyroradius; see \secref{sec_unfreezing}). 

\section{A Heuristic Derivation of the Electron Equations}
\label{ap_nongyro}

Here we show how the equations~\exsdash{Apar_eq_sum}{dne_eq_sum} 
of \secref{sec_els} and the ERMHD equations~\exsdash{EMHD_Psi}{EMHD_Phi} 
of \secref{sec_ERMHD} can be derived heuristically from electron fluid dynamics
and a number of physical assumptions, without the use of 
gyrokinetics (\secref{ap_el_eqns}). This 
derivation is {\em not} rigorous. Its role is to provide an 
intuitive route to the isothermal electron fluid and ERMHD approximations. 

\subsection{Derivation of \eqsdash{Apar_eq_sum}{dne_eq_sum}}
\label{ap_el_eqns}

We start with the following three equations:
\bea
\label{el_fluid}
{\dd \vB\over\dd t} = -c \vdel\times\vE,\qquad
{\dd n_e\over\dd t} + \vdel\cdot\(n_e\vu_e\) = 0,\qquad
\vE + {\vu_e\times\vB\over c} = - {\vdel p_e\over e n_e}.
\eea
These are Faraday's law, the electron continuity equation, and 
the generalized Ohm's law, which is the electron momentum 
equation with all electron inertia terms neglected (i.e., effectively, 
the lowest order in the expansion in the electron mass~$m_e$). 
The electron pressure is assumed to be scalar by {\em fiat} 
(this can be justified in certain limits: for example in the collisional 
limit, as in \apref{ap_Brag}, or for the isothermal electron fluid 
approximation derived in \secref{sec_els}). 
The electron-pressure term in the right-hand side of Ohm's law is 
sometimes called the thermoelectric term. 
We now assume the same static uniform equilibrium, $\vE_0=0$, $\vB_0=B_0\vz$, 
that we have used throughout this paper and 
apply to \eqsref{el_fluid} the fundamental ordering discussed 
in \secref{sec_params}. 

First consider the projection of Ohm's law onto the {\em total} 
magnetic field $\vB$, use the definition of $\vE$ [\eqref{E_B_def}], 
and keep the leading-order terms in the $\epsilon$ expansion: 
\bea
\vE\cdot\vb = -{1\over en_e}\,\Dpar p_e
\quad\Rightarrow\quad
{1\over c}{\dd\Apar\over\dd t} + \Dpar\ephi = \Dpar {\dpe\over e\ne}.
\label{Ohm_par}
\eea
This turns into \eqref{Apar_eq_sum} if we also assume isothermal 
electrons, $\dpe=\Te\dne$ [see \eqref{dpe_eq}].

With the aid of Ohm's law, Faraday's law turns into 
\bea
\label{ind_eq_els}
{\dd\vB\over\dd t} = \vdel\times\(\vu_e\times\vB\) = 
-\vu_e\cdot\vdel\vB + \vB\cdot\vdel\vu_e - \vB\vdel\cdot\vu_e.
\eea
Keeping the leading-order terms, 
we find, for the components of \eqref{ind_eq_els} perpendicular and parallel 
to the mean field, 
\bea
\label{ind_eq_reduced}
\({\dd\over\dd t} + \vuperpe\cdot\vdperp\){\dvBperp\over B_0}
= \Dpar\vuperpe,\qquad
\({\dd\over\dd t} + \vuperpe\cdot\vdperp\)\({\dBpar\over B_0} - {\dne\over\ne}\) 
= \Dpar\upare.
\eea
In the last equation, 
we have used the electron continuity equation to write 
\bea
\vdel\cdot\vu_e = -\({\dd\over\dd t} + \vuperpe\cdot\vdperp\){\dne\over\ne}. 
\label{div_ue}
\eea
From Ohm's law, we have, to lowest order, 
\bea
\label{uperpe_eq}
\vuperpe = - \vz\times{c\over B_0}\(\vEperp + \vdperp{\dpe\over e\ne}\) 
= \vz\times\vdperp {c\over B_0}\(\ephi - {\dpe\over e\ne}\).
\eea
Using this expression in the second of the equations \exref{ind_eq_reduced} gives 
\bea
{d\over dt}\({\dBpar\over B_0} - {\dne\over\ne}\) - \Dpar\upare 
= {c\over B_0}\lt\{{\dpe\over e\ne},{\dBpar\over B_0}\rt\} 
- {c\over B_0}\lt\{{\dpe\over e\ne},{\dne\over\ne}\rt\}, 
\label{dBdn_eq}
\eea
where $d/dt$ is defined in the usual way [\eqref{def_ddt}]. 
Assuming isothermal electrons ($\dpe = \Te\dne$) annihilates 
the second term on the right-hand side and turns the above 
equation into \eqref{dne_eq_sum}. As for the first of the equations 
\exref{ind_eq_reduced}, the use 
of \eqref{uperpe_eq} and substitution of $\dvBperp = -\vz\times\vdperp\Apar$ 
turns it into the previously derived \eqref{Ohm_par}, 
whence follows \eqref{Apar_eq_sum}. 

Thus, we have shown that \eqsdash{Apar_eq_sum}{dne_eq_sum} can be 
derived as a direct consequence of Faraday's law, electron fluid dynamics 
(electron continuity equation and the electron force balance, 
a.~k.~a.\ the generalized Ohm's law), and the assumption of isothermal 
electrons---all taken to the leading order in the gyrokinetic 
ordering given in \secref{sec_params} (i.e., assuming strongly 
interacting anisotropic fluctuations with $\kpar\ll\kperp$). 

We have just proved that \eqsand{Apar_eq_sum}{dne_eq_sum} 
are simply the perpendicular and parallel part, respectively, 
of \eqref{ind_eq_els}. The latter equation means that 
the magnetic-field lines are frozen into the electron flow 
velocity $\vu_e$, i.e., the flux is conserved, 
the result formally proven in \secref{sec_flux} 
[see \eqref{Far_law}]. 

\subsection{Electron MHD and the Derivation of \eqsdash{EMHD_Psi}{EMHD_Phi}} 
\label{ap_EMHD}

One route to \eqsdash{EMHD_Psi}{EMHD_Phi},  
already explained in \secref{sec_ERMHD_eqns}, is to start with  
\eqsand{Ohm_par}{dBdn_eq} and assume Boltzmann electrons and ions and  
the total pressure balance. Another approach, more standard in 
the literature on the Hall and Electron MHD, is to start 
with \eqref{ind_eq_els}, which 
states that the magnetic field is frozen into the electron flow. 
The electron velocity can be written in terms of the ion velocity and 
the current density, and the latter then related to the magnetic field 
via Amp\`ere's law:
\beq
\vu_e = \vu_i - {\vj\over en_e} = 
\vu_i - {c\over4\pi e n_e}\,\vdel\times\vB.
\label{Hall}
\eeq
To the leading order in $\epsilon$, the perpendicular and parallel parts of \eqref{ind_eq_els} 
are \eqsref{ind_eq_reduced}, respectively, where the perpendicular 
and parallel electron velocities are [from \eqref{Hall}] 
\beq
\vuperpe = \vuperpi + {c\over4\pi e\ne}\,\vz\times\vdperp\dBpar,
\quad
\upare = \upari + {c\over4\pi e\ne}\,\vdperp^2\Apar.
\label{Hall_perp_par}
\eeq
The relative size of the two terms in each of these expressions is controlled 
by the size of $\kperp d_i$, where $d_i=\rho_i/\sqrt{\beta_i}$ is the ion 
inertial scale. When $\kperp d_i\gg1$, we may set $\vu_i=0$. Note, however, that 
the ion motion is not totally neglected: indeed, in the second of the equations 
\exref{ind_eq_reduced}, the $\dne/ne$ terms comes, via \eqref{div_ue}, from the divergence 
of the ion velocity [from \eqref{Hall}, $\vdel\cdot\vu_i=\vdel\cdot\vu_e$]. 
To complete the derivation, we relate $\dne$ to $\dBpar$ via 
the assumption of total pressure balance, as explained in \secref{sec_ERMHD_eqns}, 
giving us \eqref{dBpar_via_dne}. Substituting this equation and \eqsref{Hall_perp_par}
into \eqsref{ind_eq_reduced}, we obtain 
\bea
\label{ERMHD_eqns_ap}
{\dd\Psi\over\dd t} = v_A^2 d_i\,\Dpar{\dBpar\over B_0},
\qquad
{\dd\over\dd t}{\dBpar\over B_0} =
- {d_i\over1+2/\beta_i(1+Z/\tau)}\,\Dpar\vdperp^2\Psi,
\eea
where $\Psi=-\Apar/\sqrt{4\pi m_i\ni}$. \Eqsref{ERMHD_eqns_ap} evolve the perturbed magnetic field. 
These equations become the ERMHD equations \exsdash{EMHD_Psi}{EMHD_Phi} 
if $\dBpar/B_0$ is expressed in terms of the scalar potential via \eqref{EMHD_dBpar}. 

Note that there are two special limits in which the assumption of immobile ions 
suffices to derive \eqsref{ERMHD_eqns_ap} from \eqref{ind_eq_els} 
without the need for the pressure balance: $\beta_i\gg1$ (incompressible ions) 
or $\tau=\Ti/\Te\ll1$ (cold ions) but $\beta_e=\beta_iZ/\tau\gg1$.
In both cases, \eqref{dBpar_via_dne} shows that 
$\dne/\ne\ll\dBpar/B_0$, so the density perturbation can be ignored and 
the coefficient of the right-hand side of the second of the equations~\exref{ERMHD_eqns_ap} 
is equal to~1. The limit of cold ions is discussed further in \apref{ap_Hall}. 

\section{Fluid Limit of the Kinetic RMHD} 
\label{ap_visc}

Taking the fluid (collisional) limit of the KRMHD system 
(summarized in \secref{sec_KRMHD_sum})
means carrying out another subsidiary expansion---this time in $\kpar\mfp\ll1$. 
The expansion only affects the equations for the 
density and magnetic-field-strength fluctuations (\secref{sec_sw}) 
because the Alfv\'en waves are indifferent to collisional effects. 

The calculation presented below follows a standard 
perturbation algorithm used in the kinetic theory of gases and in 
plasma physics to derive fluid equations with 
collisional transport coefficients \citep{Chapman_Cowling}. 
For magnetized plasma, this calculation was carried out in full generality by 
\citet{Braginskii}, whose starting point was the full plasma kinetic theory 
[\eqsdash{Vlasov_eq}{Max_Ampere}]. While what we do below is, strictly speaking, 
merely a particular case of his calculation 
(see \apref{ap_Brag}), it has the advantage of relative 
simplicity and also serves to show how the fluid limit is recovered from the 
gyrokinetic formalism---a demonstration that we believe to be of value. 

It will be convenient to use the KRMHD system written in terms of 
the function $\tdfi = g + (\vperp^2/\vthi^2)(\dBpar/B_0)\fMi$, 
which is the perturbation of the local Maxwellian in the frame of the 
Alfv\'en waves [\eqsdash{KRMHD_dfi}{KRMHD_dfi_Bpar}]. 
We want to expand \eqref{KRMHD_dfi} in powers of $\kpar\mfp$, so we let 
$\tdfi = \dfzero + \dfone + \dots$, 
$\dBpar = \dBzero + \dBone + \dots$, etc. 

\subsection{Zeroth Order: Ideal Fluid Equations} 

Since [see \eqref{omega_vs_nu}]
\bea
{\omega\over\nui}\sim {\kpar v_A\over\nui}\sim {\kpar\mfp\over\sqrt{\beta_i}},\quad
{\kpar\vpar\over\nui}\sim {\kpar\vthi\over\nui}\sim \kpar\mfp, 
\eea 
to zeroth order \eqref{KRMHD_dfi} becomes
$\lt<\dC_{ii}\lt[\dfzero\rt]\rt>_{\vR_i} = 0$. 
The zero mode of the collision operator is a Maxwellian. 
Therefore, we may write the full ion distribution function 
up to zeroth order in $\kpar\mfp$ as follows [see \eqref{fi_AW}]
\bea
f_i= {n_i\over\(2\pi T_i/m_i\)^{3/2}}\,\exp\lt\{
- {m_i[(\vvperp-\vu_E)^2+(\vpar-\upar)^2]\over2 T_i}\rt\},
\label{Max_unexpanded}
\eea
where $n_i=\ni + \dni$ and $T_i=\Ti+\dTi$ include both the unperturbed quantities and 
their perturbations. 
The $\vE\times\vB$ drift velocity $\vu_E$ comes from the Alfv\'en waves 
(see \secref{sec_AW_coll}) and does not concern us here. 
Since the perturbations 
$\dni$, $\upar$ and $\dTi$ are small in the original gyrokinetic expansion, 
\eqref{Max_unexpanded} is equivalent to
\bea
\dfzero = \lt[{\dnzero\over\ne} 
+ \({v^2\over\vthi^2}-{3\over2}\){\dTzero\over\Ti} 
+ {2\vpar\over\vthi^2}\,\uzero\rt]\fMi,
\label{dfzero_expr}
\eea
where we have used quasi-neutrality to replace $\dni/\ni=\dne/\ne$. 
This automatically satisfies \eqref{KRMHD_dfi_n}, 
while \eqref{KRMHD_dfi_Bpar} gives us an expression for the ion-temperature 
perturbation:
\bea
{\dTzero\over\Ti} = -\(1+{Z\over\tau}\){\dnzero\over\ne} - {2\over\beta_i}{\dBzero\over B_0}. 
\label{dTi_eq}
\eea 
Note that this is consistent with the interpretation of 
the perpendicular Amp\`ere's law [\eqref{Amp_perp}, 
which is the progenitor of \eqref{KRMHD_dfi_Bpar}] 
as the pressure balance [see \eqref{GK_pr_bal}]: indeed, 
recalling that the electron pressure perturbation is $\dpe=\Te\dne$ [\eqref{dpe_eq}], 
we have 
\bea
\delta\,{B^2\over8\pi} = {B_0^2\over4\pi}{\dBpar\over B_0} 
= -\dpe - \dpi = 
-\dne\Te -\dni\Ti - \ni\dTi,
\eea
whence follows \eqref{dTi_eq} by way of quasi-neutrality ($Zn_i=n_e$) 
and the definitions of $Z$, $\tau$, $\beta_i$ [\eqsdash{Z_def}{betai_def}]. 

Since the collision operator conserves particle number, 
momentum and energy, we can obtain evolution equations for 
$\dnzero/\ne$, $\uzero$ and $\dBzero/B_0$ by multiplying \eqref{KRMHD_dfi} 
by $1$, $\vpar$, $v^2/\vthi^2$, respectively, and integrating over 
the velocity space. The three moments that emerge this way are 
\bea
\intvi \dfzero = {\dnzero\over\ne},\qquad 
\intvi \vpar\dfzero = \uzero,\qquad
\intvi {v^2\over\vthi^2}\,\dfzero = 
{3\over2}\lt({\dnzero\over\ne} + {\dTzero\over\Ti}\rt).
\label{moments}
\eea
The three evolution equations for these moments are 
\bea
\label{cont_eq}
&&{d\over dt}\({\dnzero\over\ne} - {\dBzero\over B_0}\) + 
\Dpar\uzero = 0,\\
\label{mom_eq}
&&{d\uzero\over dt} - v_A^2\,\Dpar{\dBzero\over B_0} = 0,\\ 
\label{en_eq}
&&{d\over dt}\lt[{3\over2}\lt({\dnzero\over\ne} + {\dTzero\over\Ti}\rt)
- {5\over2}{\dBzero\over B_0}\rt] 
+ {5\over2}\,\Dpar\uzero = 0.
\eea
These allow us to recover the fluid equations we derived in \secref{sec_sw_fluid}: 
\eqref{mom_eq} is the parallel component of the MHD momentum 
equation~\exref{eq_upar}; 
combining \eqsref{cont_eq}, \exref{en_eq} and \exref{dTi_eq}, 
we obtain the continuity equation 
and the parallel component of the induction equation---these 
are the same as \eqsand{eq_drho}{eq_Bpar}:
\bea
\label{den_eq}
{d\over dt}{\dnzero\over\ne} = - {1\over1+ c_s^2/v_A^2}\,\Dpar\uzero,\quad
{d\over dt}{\dBzero\over B_0} = {1\over1+ v_A^2/c_s^2}\,\Dpar\uzero,
\eea
where the sound speed $c_s$ is defined by \eqref{cs_def}. 
From \eqsand{cont_eq}{en_eq}, we also find the analog of 
the entropy equation~\exref{eq_ds}: 
\bea
{d\over dt}{\dTzero\over\Ti} = {2\over3}{d\over dt}{\dnzero\over\ne}
\quad\Leftrightarrow\quad
{d\over dt}{\dszero\over s_0} = 0,\quad
{\dszero\over s_0} = {\dTzero\over\Ti} - {2\over3}{\dnzero\over\ne} 
= -\({5\over3}+{Z\over\tau}\)\({\dnzero\over\ne}+{v_A^2\over c_s^2}{\dBzero\over B_0}\).
\label{comp_heat_eq}
\eea
This implies that the temperature changes due to compressional heating only. 

\subsection{Generalized Energy: Five RMHD Cascades Recovered}
\label{ap_en_RMHD} 

We now calculate the generalized energy by substituting $\tdfi$ from 
\eqref{dfzero_expr} into \eqref{W_KRMHD} and using \eqsand{dTi_eq}{comp_heat_eq}:
\bea
\nonumber
W &=& \int d^3\vr\lt[{m_i\ni u_E^2\over2} + {\dBperp^2\over8\pi}
+ {m_i\ni\upar^2\over2} + {\dBpar^2\over8\pi} \(1+{v_A^2\over c_s^2}\)
+ {3\over4}\,\ni\Ti\,{1+Z/\tau\over5/3+Z/\tau}{\ds^2\over s_0^2}\rt]\\
&=& \Wperp^+ + \Wperp^- + \Wpar^+ + \Wpar^- + 
{3\over2}\,\ni\Ti\,{1+Z/\tau\over5/3+Z/\tau}\,\Ws.
\label{W_RMHD}
\eea
The first two terms are the Alfv\'en-wave energy [\eqref{W_AW}]. 
The following two terms are the slow-wave energy, which 
splits into the independently cascaded energies of ``$+$'' and ``$-$'' waves 
(see \secref{sec_elsasser_SW}): 
\bea
W_{\rm SW} = \Wpar^+ + \Wpar^-
= \intr{m_i\ni\over2}\lt(|\zpar^+|^2 + |\zpar^-|^2\rt).
\eea
The last term is the total variance of the entropy mode. 
Thus, we have recovered the five cascades of the RMHD system 
(\secref{sec_RMHD_cascades}; \figref{fig_cascade_channels} 
maps out the fate of these cascades at kinetic scales). 

\subsection{First Order: Collisional Transport} 
\label{ap_transport}

Now let us compute the collisional transport terms for the 
equations derived above. In order to do this, we have to 
determine the first-order perturbed 
distribution function $\dfone$, which satisfies [see \eqref{KRMHD_dfi}] 
\bea
\label{dfone_eq}
\lt<\dC_{ii}\lt[\dfone\rt]\rt>_{\vR_i} 
= {d\over dt}\lt(\dfzero - {\vperp^2\over\vthi^2}{\dBzero\over B_0}\rt)  
+ \vpar\,\Dpar\lt(\dfzero+{Z\over\tau}{\dnzero\over\ne}\fMi\rt).
\eea
We now use \eqref{dfzero_expr} to substitute for $\dfzero$ 
and \eqsdash{den_eq}{comp_heat_eq} and \exref{mom_eq} to compute the time derivatives. 
\Eqref{dfone_eq} becomes 
\bea
\lt<\dC_{ii}\lt[\dfone\rt]\rt>_{\vR_i} 
= \lt[- \(1-3\xi^2\){v^2\over\vthi^2} 
{2/3 + c_s^2/v_A^2\over 1 + c_s^2/v_A^2}\, 
\Dpar\uzero + \xi v\({v^2\over\vthi^2}-{5\over2}\)
\Dpar{\dTzero\over\Ti}\rt]\fMi(v),
\label{dfone_eq2}
\eea
where $\xi=\vpar/v$. 
Note that the right-hand side gives zero when multiplied by 
$1$, $\vpar$ or $v^2$ and integrated over the velocity space, 
as it must do because the collision operator in the left-hand side 
conserves particle number, momentum and energy. 

Solving \eqref{dfone_eq2} requires inverting the collision 
operator. While this can be done for the general Landau 
collision operator \citep[see][]{Braginskii},  
for our purposes, it is sufficient to use the model operator 
given in \apref{ap_ss}, \eqref{Cgk_lowest}. This simplifies calculations 
at the expense of an order-one inaccuracy in the numerical 
values of the transport coefficients. As the exact value of these 
coefficients will never be crucial for us, this is an acceptable loss 
of precision. Inverting the collision operator in \eqref{dfone_eq2} then 
gives  
\bea
\dfone = {1\over\nuDi(v)}\lt[{1-3\xi^2\over 3}{v^2\over\vthi^2} 
{2/3 + c_s^2/v_A^2\over 1 + c_s^2/v_A^2}\, 
\Dpar\uzero - \xi v\({v^2\over\vthi^2}-{5\over2}\)
\Dpar{\dTzero\over\Ti}\rt]\fMi(v),
\label{dfone_sln}
\eea
where $\nuDi(v)$ is a collision frequency defined in \eqref{nuD_def} 
and we have chosen the constants of integration in such a way that 
the three conservation laws are respected: $\int d^3\vv\,\dfone=0$, 
$\int d^3\vv\,\vpar\dfone=0$, $\int d^3\vv\,v^2\dfone=0$. These relations 
mean that $\dnone=0$, $\uone=0$, $\dTone=0$ 
and that, in view of \eqref{KRMHD_dfi_Bpar}, we have 
\bea
{\dBone\over B_0} = 
- {1\over 3v_A^2}{2/3 + c_s^2/v_A^2\over 1 + c_s^2/v_A^2}\,
\nupar\Dpar\upar,
\label{dBone_sln}
\eea
where $\nupar$ is defined below [\eqref{nupar_def}]. 
\Eqsdash{dfone_sln}{dBone_sln} are now used 
to calculate the first-order corrections to the moment equations 
\exsdash{cont_eq}{en_eq}. They become 
\bea
\label{cont_eq_diff}
&&{d\over dt}\lt({\dne\over\ne} - {\dBpar\over B_0}\rt) + \Dpar\upar = 0,\\ 
\label{mom_eq_diff}
&&{d\upar\over dt} - v_A^2\Dpar{\dBpar\over B_0}
= {2/3 + c_s^2/v_A^2\over 1 + c_s^2/v_A^2}\,\nupar\,\Dpar\lt(\Dpar\upar\rt),\\
&&{d\over dt}{\dTi\over\Ti} - {2\over3}{d\over dt}{\dne\over\ne} 
= \kappar\Dpar\lt(\Dpar{\dTi\over\Ti}\rt), 
\label{en_eq_diff}
\eea
where we have introduced the coefficients of parallel viscosity
and parallel thermal diffusivity:
\bea
\label{nupar_def}
\nupar = {2\over15}\intvi {v^4\over\nuDi(v)\vthi^2}\,\fMi(v),\qquad
\kappar = {2\over9}\intvi {v^4\over\nuDi(v)\vthi^2}
\({v^2\over\vthi^2}-{5\over2}\)\fMi(v). 
\label{kappar_def}
\eea
All perturbed quantities are now accurate up to first order in $\kpar\mfp$. 
Note that in \eqref{mom_eq_diff}, we used \eqref{dBone_sln} to express 
$\dBzero = \dBpar - \dBone$. We do the same in \eqref{dTi_eq} and obtain 
\bea
\lt(1+ {Z\over\tau}\rt){\dne\over\ne} = - {\dTi\over\Ti} 
- {2\over\beta_i}\lt({\dBpar\over B_0} + 
{1\over 3v_A^2}{2/3 + c_s^2/v_A^2\over 1 + c_s^2/v_A^2}\,
\nupar\Dpar\upar\rt). 
\label{pr_bal_diff}
\eea
This equation completes the system \exsdash{cont_eq_diff}{en_eq_diff}, which 
allows us to determine $\dne$, $\upar$, $\dTi$ and $\dBpar$. In \secref{sec_visc}, 
we use the equations derived above, but absorb the prefactor 
$(2/3 + c_s^2/v_A^2)/(1+c_s^2/v_A^2)$ into the definition of $\nupar$. 
The same system of equations can also be derived from Braginskii's 
two-fluid theory (\apref{ap_isoth_els}), from which we can borrow 
the quantitatively correct values of the viscosity and ion thermal 
diffusivity: $\nupar=0.90\vthi^2/\nui$, $\kappar=2.45\vthi^2/\nui$, 
where $\nui$ is defined in \eqref{nui_def}. 

\section{Hall Reduced MHD}
\label{ap_Hall}

The popular Hall MHD approximation consists in assuming that the magnetic field 
is frozen into the electron flow velocity [\eqref{ind_eq_els}]. 
The latter is calculated from the ion flow velocity and the 
current determined by Amp\`ere's law [\eqref{Hall}]: 
\bea
\label{Hall_MHD}
{\dd\vB\over\dd t} = \vdel\times\lt[\lt(\vu_i-{c\over4\pi e\ne}\vdel\times\vB\rt)\times\vB\rt],
\eea
where the ion flow velocity $\vu_i$
satisfies the conventional MHD momentum equation \exref{MHD_u}. 
The Hall MHD is an appealing theoretical model that appears to capture 
both the MHD behavior at long wavelengths (when $\vu_e\simeq\vu_i$) and 
some of the kinetic effects that become important at small scales 
due to decoupling between the electron and ion flows 
(the appearance of dispersive waves) without bringing in the full complexity 
of the kinetic theory. However, unlike the kinetic theory, it completely 
ignores the collisionless damping effects and suggests that the key 
small-scale physical change is associated with the ion inertial 
scale $d_i=\rho_i/\sqrt{\beta_i}$ 
(or, when $\beta_e\ll1$, the ion sound scale $\rho_s=\rho_i\sqrt{Z/2\tau}$; 
see \secref{ap_Hall_lin}), 
rather than the ion gyroscale $\rho_i$. Is this an acceptable 
model for plasma turbulence? 
\Figref{fig_omegas} illustrates 
the fact that at $\tau\sim1$, the ion inertial scale does {\em not} play 
a special role linearly, the MHD Alfv\'en wave becomes dispersive 
at the ion gyroscale, not at $d_i$, and that the collisionless damping 
cannot in general be neglected. 
A detailed comparison of the Hall MHD linear dispersion 
relation with full hot plasma dispersion relation leads to the conclusion 
that Hall MHD is only a valid approximation 
in the limit of cold ions, namely, $\tau=\Ti/\Te\ll1$ \citep{Ito_etal,Hirose_etal}. 
In this Appendix, we show that 
a reduced (low-frequency, anisotropic) version of Hall MHD can, indeed, 
be derived from gyrokinetics in the limit $\tau\ll1$.\footnote{Note that, strictly speaking, 
our ordering of the collision frequency does not allow us to take this limit 
(see footnote~\ref{fn_temperatures}), but this is a minor betrayal of rigor, 
which does not, in fact, invalidate the results.} 
This demonstrates that the Hall 
MHD model fits into the theoretical framework proposed in this paper
as a special limit. However, the parameter regime that gives rise 
to this special limit is not common in space and astrophysical plasmas of interest. 

\subsection{Gyrokinetic Derivation of Hall Reduced MHD}
\label{ap_HRMHD}

Let us start with the equations of isothermal electron fluid, 
\eqsdash{Apar_eq_sum}{GK_ions_sum}, i.e., work within the assumptions that 
allowed us to carry out the mass-ratio expansion (\secref{sec_els_validity}). 
In \eqref{dBpar_eq_sum} (perpendicular Amp\`ere's law, or gyrokinetic 
pressure balance), taking the limit $\tau\ll1$ gives
\bea
{\dBpar\over B_0} = {\beta_i\over2}{Z\over\tau}\Biggl\{{Ze\ephi\over\Ti}
-\sum_\vk e^{i\vk\cdot\vr}\intvi J_0(\kr_i)\hki\Biggr\}
= -{\beta_e\over2}{\dne\over\ne},
\label{dne_Hall}
\eea
where we have used \eqref{quasineut_sum} to express the $\hi$ integral 
and the expression for the electron beta $\beta_e=\beta_i Z/\tau$. 
Note that the above equation is simply 
the statement of a balance between the magnetic and electron thermal pressure 
(the ions are relatively cold, so they have fallen out of the pressure balance). 
Using \eqref{dne_Hall} to express $\dne$ in terms of $\dBpar$ in 
\eqsand{Apar_eq_sum}{dne_eq_sum} and also substituting for $\upare$ from 
\eqref{upare_eq_sum} [or, equivalently, \eqref{upare_eq}], we get
\bea
\label{Hall_B}
{\dd\Psi\over\dd t} = v_A\Dpar\lt(\Phi + v_A d_i\,{\dBpar\over B_0}\rt),
\qquad
{d\over dt}{\dBpar\over B_0} = {1\over 1+{2/\beta_e}}\,
\Dpar\lt(\upari - d_i\dperp^2\Psi\rt),
\eea
where we have used our usual definitions of the stream and flux 
functions [\eqref{Phi_Psi_def2}] and of the full derivatives [\eqref{dd_def_sum}]. 
These equations determine the evolution of the magnetic field, but 
we still need the ion gyrokinetic equation \exref{GK_ions_sum} to 
calculate the ion motion ($\Phi=c\ephi/B_0$ and $\upari$) 
via \eqsand{quasineut_sum}{upari_def}. There are two limits 
in which the ion kinetics can be reduced to simple fluid models. 

\subsubsection{High-Ion-Beta Limit, $\beta_i\gg1$}
\label{ap_Hall_high_beta}

In this limit, $\kperp\rho_i=\kperp d_i\sqrt{\beta_i}\gg1$ as long 
as $\kperp d_i$ is not small. Then the ion motion can be neglected 
because it is averaged out by the Bessel functions in 
\eqsand{quasineut_sum}{upari_def}---in the same way as in \secref{sec_ERMHD_eqns}. 
So we get $\Phi=(\tau/Z)v_Ad_i\dBpar/B_0$ [using \eqref{dne_Hall}; this is the 
$\tau\ll1$ limit of \eqref{EMHD_dBpar}] and $\upari=0$. 
Noting that $\beta_e=\beta_i Z/\tau\gg1$ in this limit, we find that 
\eqsref{Hall_B} reduce to 
\bea
{\dd\Psi\over\dd t} = v_A^2d_i\,\Dpar{\dBpar\over B_0},
\qquad
{\dd\over\dd t}{\dBpar\over B_0} = -d_i\,\Dpar\dperp^2\Psi,
\label{Hall_high_beta}
\eea
which is the $\tau\ll1$ limit of our ERMHD equations \exsdash{EMHD_Psi}{EMHD_Phi} 
[or, equivalently, \eqsref{ERMHD_eqns_ap}]. 

\subsubsection{Low-Ion-Beta Limit, $\beta_i\sim\tau\ll1$ (the Hall Limit)}
\label{ap_Hall_low_beta} 

This limit is similar to the RMHD limit worked out in \secref{sec_KRMHD}: 
we take, for now, $\kperp d_i\sim1$ and $\beta_e\sim1$ (in which subsidiary expansions 
can be carried out later), and expand the ion gyrokinetics 
in $\kperp\rho_i=\kperp d_i\sqrt{\beta_i}\ll1$. 
Note that ordering $\beta_e\sim1$ means that we have ordered 
$\beta_i\sim\tau\ll1$. 
We now proceed analogously to the way we did in \secref{sec_KRMHD}: 
express the ion distribution in terms of the $\gi$ function defined 
by \eqref{g_ansatz} and, using the relation \exref{dne_Hall}
between $\dBpar/B_0$ and $\dne/\ne$, write \eqsdash{g_eq}{uparek_from_g} as follows:
\bea
\nonumber
\order{\!\!\!-1}{{\dd\gi\over\dd t}} &+& \order{0}{\vpar{\dd\gi\over\dd z}} 
+ {c\over B_0}\Biggl\{\biggl\<\order{\!\!\!-1}{\ephi} 
- \order{0}{{\vpar\Apar\over c}} 
- \order{1}{{\vvperp\cdot\vAperp\over c}}\Biggr\>_{\vR_i},\gi\Biggr\} 
- \order{0}{\lt<\dC_{ii}[\gi]\rt>_{\vR_i}}\\ 
\label{Hall_g}
&=& {Ze\over\Ti}\Biggl[\vpar\lt\<\order{1}{-{1\over B_0}\lt\{\Apar,\ephi-\<\ephi\>_{\vR_i}\rt\}}
+ \Dpar\Biggl(\order{\!\!\!-1}{{\Te\over e}{2\over\beta_e}{\dBpar\over B_0}}
+\order{1}{\lt\<{\vvperp\cdot\vAperp\over c}\rt\>_{\vR_i}}\Biggr)\rt\>{\vphantom{\Biggr>}}_{\vR_i}\,\fMi 
+ \lt<\dC_{ii}\biggl[\biggl\<\order{1}{\ephi}
-\order{1}{{\vvperp\cdot\vAperp\over c}}\biggr\>{\vphantom{\biggr>}}_{\vR_i}\,\fMi\biggr]\rt>{\vphantom{\Biggr>}}_{\vR_i}\Biggr],\quad
\eea
\bea
\label{Hall_fields}
-\order{0}{\lt[\Gamma_1(\krsq_i)+{2\over\beta_e}\rt]{\dBpark\over B_0}} 
+ \order{0}{\bl[1-\Gamma_0(\krsq_i)\br]{Ze\ephi_\vk\over\Ti}} 
= \order{\!\!\!-1}{\intvi J_0(\kr_i)\gki},\qquad
\order{\!\!\!-1}{\uparik} = \order{\!\!\!-1}{\intvi \vpar J_0(\kr_i)\gki}.
\eea
All terms in these equations can be ordered with respect to the small 
parameter $\sqrt{\beta_i}$ (an expansion subsidiary to the 
gyrokinetic expansion in $\epsilon$ and the Hall expansion in $\tau\ll1$). 
The lowest order to which they enter is indicated underneath each term.  
The ordering we use is the same as in \secref{sec_sub_order}, but now we count 
the powers of $\sqrt{\beta_i}$ and order formally $\kperp d_i\sim1$ and $\beta_e\sim1$. 
It is easy to check that this ordering can be summarized as follows 
\bea
{Ze\ephi\over\Ti}\sim{1\over\beta_i}{\dBpar\over B_0},\quad
{\dBperp\over B_0}\sim{\dBpar\over B_0},\quad
{g\over\fMi}\sim{\upar\over\vthi}\sim{1\over\sqrt{\beta_i}}{\dBpar\over B_0}
\eea
and that the ion and electron terms in \eqsref{Hall_B} are comparable 
under this ordering, so their competition is retained 
(in fact, this could be used as the underlying assumption behind 
the ordering). The fluctuation frequency continues to be 
ordered as the Alfv\'en frequency, $\omega\sim\kpar v_A$. The collision terms 
are ordered via $\omega/\nui\sim \kpar\mfp/\sqrt{\beta_i}$ and $\kpar\mfp\sim1$, 
although the latter assumption is not essential for what follows, because 
collisions turn out to be negligible and it is fine to take $\kpar\mfp\gg1$ 
from the outset and neglect them completely. 

In \eqsref{Hall_fields}, we use \eqsand{G0_def}{G1_def} to 
write $1-\Gamma_0(\krsq_i)\simeq\krsq_i=\kperp^2\rho_i^2/2$ 
and $\Gamma_1(\krsq_i)\simeq1$. These equations 
imply that if we expand $\gi = \gi^{(-1)}+\gi^{(0)}+\dots$, 
we must have $\int d^3\vv\gi^{(-1)}=0$, so the contribution to the right-hand side 
of the first of the equations \exref{Hall_fields} (the quasi-neutrality equation) 
comes from $\gi^{(0)}$, while the parallel ion flow 
is determined by $\gi^{(-1)}$. Retaining only the lowest (minus first) 
order terms in \eqref{Hall_g}, we find the equation for $\gi^{(-1)}$, the $\vpar$ moment 
of which gives an equation for $\upari$: 
\bea
{\dd\gi^{(-1)}\over\dd t} + {c\over B_0}\,\{\ephi,\gi^{(-1)}\} 
= {2\over\beta_i}\,\vpar\Dpar{\dBpar\over B_0}\,\fMi
\quad\Rightarrow\quad
{d\upari\over dt} = v_A^2\Dpar{\dBpar\over B_0}.
\label{Hall_upar_ap}
\eea
Now integrating \eqref{Hall_g} over the velocity space (at constant $\vr$), 
using the first of the equations \exref{Hall_fields} to express the integral of $\gi^{(0)}$, 
and retaining only the lowest (zeroth) order terms, we find 
\bea
{d\over dt}\lt[-{1\over2}\rho_i^2\dperp^2{Ze\ephi\over\Ti}
-\lt(1+{2\over\beta_e}\rt){\dBpar\over B_0}\rt] + \Dpar\upari = 0 
\quad\Rightarrow\quad
{d\over dt}\dperp^2\Phi = v_A\Dpar\dperp^2\Psi,
\label{Hall_Phi_ap}
\eea 
where we have used the second 
of the equations~\exref{Hall_B} to express the time derivative of $\dBpar/B_0$.

Together with \eqsref{Hall_B}, \eqsand{Hall_upar_ap}{Hall_Phi_ap} form a closed 
system, which it is natural to call {\em Hall Reduced MHD (HRMHD)} because these equations 
can be straightforwardly derived by applying the RMHD ordering (\secref{sec_RMHDordering}) 
to the MHD equations \exsdash{MHD_u}{MHD_B} with the induction equation \exref{MHD_B} replaced 
by \eqref{Hall_MHD}. Indeed, \eqsand{Hall_upar_ap}{Hall_Phi_ap} exactly coincide with 
\eqsand{eq_upar}{RMHD_Phi}, which are the parallel and perpendicular 
components of the MHD momentum equation \exref{MHD_u} under the 
RMHD ordering; \eqsref{Hall_B} should be compared 
\eqsand{RMHD_Psi}{eq_Bpar} while noticing that, in the limit $\tau\ll1$, 
the sound speed is $c_s=v_A\sqrt{\beta_e/2}$ [see \eqref{cs_def}]. 
The incompressible case \citep{Mahajan_Yoshida} 
is recovered in the subsidiary limit $\beta_e\gg1$ (i.e., $1\gg\beta_i\gg\tau$). 

\subsection{Generalized Energy for Hall RMHD and the Passive Entropy Mode}
\label{ap_Hall_en}

To work out the generalized energy (\secref{sec_en_GK}) for the HRMHD regime, 
we start with the generalized energy for the isothermal electron fluid 
[\eqref{W_els}] and use \eqref{dne_Hall} to express the density perturbation:
\bea
W = \int d^3\vr\lt[\int d^3\vv\,{\Ti\dfi^2\over2\fMi}  
+ {\dBperp^2\over8\pi} 
+ \lt(1+{2\over\beta_e}\rt){\dBpar^2\over8\pi}\rt],
\label{W_Hall} 
\eea
where $\dvBperp=\vz\times\vdperp\Psi$. 
The perturbed ion distribution function can be written 
in the same form as it was done in \secref{sec_AW_coll} [\eqref{fi_KRMHD}]: 
to lowest order in the $\sqrt{\beta_i}$ expansion (\secref{ap_Hall_low_beta}), 
\bea
\dfi^{(-1)} = {2\vvperp\cdot\vuperp\over\vthi^2}\,\fMi + \gi^{(-1)}
= {2\vvperp\cdot\vuperp\over\vthi^2}\,\fMi + {2\vpar\upari\over\vthi^2}\,\fMi 
+ \tilde\gi,
\label{dfi_HRMHD}
\eea
where $\vuperp=\vz\times\vdperp\Phi$. The last equality above is achieved 
by noticing that, since $\gi^{(-1)}$ satisfies \eqref{Hall_upar_ap}, we may 
split it into a perturbed Maxwellian with parallel velocity $\upari$ 
and the remainder: $\gi^{(-1)} = 2\vpar\upari\fMi/\vthi^2 + \tilde\gi$. 
Then $\tilde\gi$ is the homogeneous solution of the leading-order kinetic 
equation [see \eqref{Hall_upar_ap}]:
\bea
{\dd\tilde\gi\over\dd t} + \{\Phi,\tilde\gi\} = 0,
\quad
\int d^3\vv\,\tilde\gi=0.
\label{tgi_eq}
\eea
Substituting \eqref{dfi_HRMHD} into \eqref{W_Hall} and keeping only the leading-order 
terms in the $\sqrt{\beta_i}$ expansion, we get
\bea
W = \int d^3\vr\lt[{m_i\ni \uperp^2\over2} + {\dBperp^2\over8\pi}
+ {m_i\ni\upar^2\over2} + {\dBpar^2\over8\pi} \(1+{2\over\beta_e}\)
+ \int d^3\vv\,{\Ti\tilde\gi^2\over2\fMi}\rt].
\label{W_HRMHD} 
\eea
The first four terms are the energy of the Alfv\'enic and slow-wave-polarized 
fluctuations [cf.\ \eqref{W_RMHD}]. Unlike in RMHD, these are not decoupled in HRMHD, 
unless a further subsidiary long-wavelength limit is taken (see \secref{ap_Hall_sum}).
It is easy to verify that the sum of these four terms is indeed conserved by 
\eqsref{Hall_B}, \exsand{Hall_upar_ap}{Hall_Phi_ap}.  
The last term in \eqref{W_HRMHD} is an individually conserved kinetic quantity.  
Its conservation reflects the fact that $\tilde\gi$ is decoupled from the 
wave dynamics and passively advected by the Alfv\'enic velocities 
via \eqref{tgi_eq}.\footnote{A similar splitting of the generalized energy cascade 
into a fluid-like cascade plus a passive cascade of a zero-density part of the 
distribution function occurs in the Hasegawa--Mima regime, which is the electrostatic 
version of the Hall limit \citep{Plunk_etal}.} 

The passive kinetic mode $\tilde\gi$ can be thought of as a kinetic version of 
the MHD entropy mode and, indeed, reduces to it if the collision 
operator in \eqref{Hall_g} is upgraded to the leading order by ordering 
$\omega/\nui\sim1$ (i.e., by considering long parallel wavelengths, 
$\kpar\mfp\sim\sqrt{\beta_i}$). In such a collisional limit, $\tilde\gi$ has 
to be a perturbed Maxwellian with no density or velocity perturbation 
[because $\int d^3\vv\tilde\gi = 0$, while the velocity perturbation is 
explicitly separated from $\tilde\gi$ in \eqref{dfi_HRMHD}]. Therefore, 
\bea
\tilde\gi = \lt({v^2\over\vthi^2} - {3\over2}\rt){\dTi\over\Ti}\,\fMi
\quad\Rightarrow\quad 
{d\over dt}{\dTi\over\Ti} = 0,\quad
\int d^3\vr\int d^3\vv\,{\Ti\tilde\gi^2\over2\fMi}
= \int d^3\vr\, {3\over 4}\,\ni\Ti\,{\dTi^2\over\Ti^2}.
\label{Hall_ds}
\eea
This is to be compared with the $\beta_i\sim\tau\ll1$ limit of \eqsand{comp_heat_eq}{W_RMHD}. 
As we have established, in the $\sqrt{\beta_i}$ expansion, $\dTi=\dTi^{(-1)}$, $\dni=\dni^{(0)}$, 
$\dBpar = \dBpar^{(0)}$, so to lowest order $\ds/s_0 = \dTi/\Ti$ and \eqref{Hall_ds} 
describes the entropy mode in the Hall limit. 

\subsection{Hall RMHD Dispersion Relation}
\label{ap_Hall_lin}

Linearizing the Hall RMHD equations \exref{Hall_B}, 
\exref{Hall_upar_ap} and \exref{Hall_Phi_ap} (derived in 
\secref{ap_Hall_low_beta} assuming the ordering $\beta_i\sim\tau\ll1$), 
we obtain the following dispersion relation:\footnote{The full 
gyrokinetic dispersion relation in a similar limit was 
worked out in \citet{Howes_etal}, Appendix D.2.1.}
\bea
\lt(\omega^2-\kpar^2v_A^2\rt)\lt(\omega^2-{\kpar^2v_A^2\over1+2/\beta_e}\rt) 
= \omega^2\kpar^2 v_A^2{\kperp^2 d_i^2\over1+2/\beta_e}.
\label{HRMHD_disp_rln}
\eea
When the coupling term on the right-hand side is negligible, 
$\kperp d_i/\sqrt{1+2/\beta_e}\ll1$, we recover the MHD Alfv\'en wave, 
$\omega^2=\kpar^2v_A^2$, and the MHD slow wave, 
$\omega^2=\kpar^2v_A^2/(1+v_A^2/c_s^2)$ [\eqref{sw_disp_rln}], 
where $c_s=v_A\sqrt{\beta_e/2}$ in the limit $\tau\ll1$ [\eqref{cs_def}].
In the opposite limit, we get the kinetic Alfv\'en wave, 
$\omega^2=\kpar^2v_A^2\kperp^2d_i^2/(1+2/\beta_e)$ 
[same as \eqref{omega_KAW} with $\tau\ll1$]. 

The solution of the dispersion relation \exref{HRMHD_disp_rln} is
\bea
\omega^2 = {\kpar^2 v_A^2\over 1+{2/\beta_e}}
\lt[1+{1\over\beta_e} + {\kperp^2 d_i^2\over2}
\pm \sqrt{{1\over\beta_e^2} + \lt(1+{1\over\beta_e}\rt)\kperp^2 d_i^2 + {\kperp^4d_i^4\over 4}}\rt].
\label{HRMHD_omega}
\eea
The corresponding eigenfunctions then satisfy\footnote{Note that wave packets 
with $|\vkperp|=\kperp$ and satisfying \eqref{HRMHD_sln} with $\kpar v_A/\omega$ 
as a function of $\kperp$ given by \eqref{HRMHD_omega} 
are exact nonlinear solutions of the HRMHD equations 
\exref{Hall_B} and \exsdash{Hall_upar_ap}{Hall_Phi_ap}. 
This can be shown via a calculation analogous to that in \secref{sec_KAW} 
\citep[for the incompressible Hall MHD, this was done by][]{Mahajan_Krishan}.} 
\bea
\Psi = -{\kpar v_A\over\omega}\lt(\Phi + v_A d_i\,{\dBpar\over B_0}\rt),\quad
\upari = -{\kpar v_A^2\over\omega}{\dBpar\over B_0},\quad
\Phi = -{\kpar v_A\over\omega}\,\Psi.
\label{HRMHD_sln}
\eea
\Eqref{HRMHD_omega} takes a particularly simple form in the subsidiary 
limits of high and low electron beta $\beta_e=\beta_iZ/\tau$:
\bea
\beta_e\gg1: \ \omega^2 = \kpar^2v_A^2\lt[1+{\kperp^2 d_i^2\over2} 
\pm \sqrt{\lt(1+{\kperp^2 d_i^2\over2}\rt)^2-1}\rt],\qquad
\beta_e\ll1: \ \omega^2 = \kpar^2v_A^2\lt(1+\kperp^2\rho_s^2\rt) 
\ {\rm and}\ 
\omega^2 = {\kpar^2 c_s^2\over1+\kperp^2\rho_s^2},
\eea
where $\rho_s=d_i\sqrt{\beta_e/2}=\rho_i\sqrt{Z/2\tau}=c_s/\Omega_i$ is called 
the ion sound scale. The Alfv\'en wave and the slow wave 
(known as the ion acoustic wave in the limit of $\tau\ll1$, $\beta_e\ll1$) 
become dispersive at the ion inertial scale ($\kperp d_i\sim1$) when $\beta_e\gg1$ 
and at the ion sound scale ($\kperp\rho_s\sim1$) when $\beta_e\ll1$. 

\subsection{Summary of Hall RMHD and the Role of the Ion Inertial and Ion Sound Scales}
\label{ap_Hall_sum}

We have shown that in the limit of cold ions and low ion 
beta ($\beta_i\sim\tau\ll1$, ``the Hall limit''), gyrokinetic turbulence can be described 
by five scalar functions: the stream and flux functions $\Phi$ and $\Psi$ for the 
Alfv\'enic fluctuations, the parallel velocity and magnetic-field perturbations 
$\upari$ and $\dBpar$ for the slow-wave-polarized fluctuations, and $\tilde\gi$,
the zero-density, zero-velocity part of the ion distribution function, which is 
the kinetic version of the MHD entropy mode. 
The first four of these functions satisfy a closed set of four fluid-like equations, 
derived in \secref{ap_HRMHD} and collected here:
\bea
\label{B_HRMHD_sum}
&&{\dd\Psi\over\dd t} = v_A\Dpar\lt(\Phi + v_A d_i\,{\dBpar\over B_0}\rt),
\qquad
{d\over dt}{\dBpar\over B_0} = {1\over 1+{2/\beta_e}}\,
\Dpar\lt(\upari - d_i\dperp^2\Psi\rt),\\
\label{u_HRMHD_sum}
&&{d\over dt}\dperp^2\Phi = v_A\Dpar\dperp^2\Psi,
\qquad\qquad\quad
{d\upari\over dt} = v_A^2\Dpar{\dBpar\over B_0}.
\eea
We call these equations the {\em Hall Reduced Magnetohydrodynamics (HRMHD)}.
To fully account for the generalized energy cascade, one must append to the four 
HRMHD equations the fifth, kinetic equation \exref{tgi_eq} for $\tilde\gi$, which is 
energetically decoupled from HRMHD and slaved to the Alfv\'enic velocity 
fluctuations (\secref{ap_Hall_en}). 

The equations given above are valid above the ion gyroscale, $\kperp\rho_i\ll1$. 
They contain a special scale, $d_i/\sqrt{1+2/\beta_e}$, which is the ion 
inertial scale $d_i$ for $\beta_e\gg1$ and 
the ion sound scale $\rho_s=c_s/\Omega_i$ for $\beta_e\ll1$. 
As becomes clear from the linear theory (\secref{ap_Hall_lin}), 
the Alfv\'en and slow waves become dispersive at this scale.  
Nonlinearly, this scale marks the transition from the regime in which 
the Alfv\'enic and slow-wave-polarized fluctuations 
are decoupled to the regime in which they are mixed. 
Namely, when $\kperp d_i/\sqrt{1+2/\beta_e}\ll1$, 
HRMHD turns into RMHD: \eqsref{B_HRMHD_sum} 
become \eqsand{RMHD_Psi}{eq_Bpar}, while \eqsref{u_HRMHD_sum} 
remain unchanged and identical to \eqsand{RMHD_Phi}{eq_upar}; 
in the opposite limit, $\kperp d_i/\sqrt{1+2/\beta_e}\gg1$, 
the ion motion decouples from the magnetic-field evolution and 
\eqsref{B_HRMHD_sum} turn into the ERMHD equations~\exsdash{EMHD_Psi}{EMHD_Phi}. 


Since we are considering the case 
$\beta_i\ll1$, both $d_i$ and $\rho_s$ are much larger than the 
ion gyroscale $\rho_i$. In the opposite limit 
of $\beta_i\gg1$ (\secref{ap_Hall_high_beta}), while $d_i$ is the only 
scale that appears explicitly in \eqsref{Hall_high_beta}, we have $d_i\ll\rho_i$
and the equations themselves represent the dynamics at scales much smaller 
than the ion gyroscale, so the transition between the RMHD and ERMHD 
regimes occurs at $\kperp\rho_i\sim1$. The same is true for $\beta_i\sim1$, 
when $d_i\sim\rho_i$. 
The ion sound scale $\rho_s\gg\rho_i$ does not play a special role
when $\beta_i$ is not small: 
it is not hard to see that for $\kperp\rho_s\sim1$, the ion motion terms 
in \eqsref{B_HRMHD_sum} dominate and we simply recover 
the inertial-range KRMHD model (\secref{sec_KRMHD}) 
by expanding in $\kperp\rho_i=\kperp\rho_s\sqrt{2\tau/Z}\ll1$. 

Various theories of the dissipation-range turbulence based on Hall and 
Electron MHD are further discussed in \secref{sec_dr_alt}. 

\section{Two-Dimensional Invariants in Gyrokinetics}
\label{ap_inv}

Since gyrokinetics is in a sense a ``quasi-two-dimensional'' approximation, 
it is natural to inquire if this gives rise to additional conservation 
properties (besides the conservation of the generalized energy discussed in 
\secref{sec_en_GK}) and how they are broken by the presence of parallel propagation 
terms. It is important to emphasize that, except in 
a few special cases, these invariants are only invariants in 2D, so gyrokinetic 
turbulence in 2D and 3D has fundamentally different properties, despite 
its seemingly ``quasi-2D'' nature. It is, therefore, generally not correct 
to think of the gyrokinetic turbulence (or its special case the MHD 
turbulence) as essentially a 2D turbulence with an admixture 
of parallel-propagating waves \citep{Fyfe_Joyce_Montgomery,Montgomery_Turner}. 

In this Appendix, we work out the 2D invariants. 
Without attempting to present a complete analysis of the 2D conservation 
properties of gyrokinetics, we limit our discussion to showing how 
some more familiar fluid invariants (most notably, magnetic helicity) 
emerge from the general 2D invariants in the appropriate asymptotic limits. 

\subsection{General 2D Invariants}
\label{ap_inv_gen}

In deriving the generalized energy invariant, we used 
the fact that $\intRs\hs\{\avchi,\hs\}=0$, 
so \eqref{GK_eq} after multiplication by $\Ts\hs/\fMs$ and integration 
over space contains no contribution from the Poisson-bracket nonlinearity. 
Since we also have $\intRs\avchi\{\avchi,\hs\}=0$, 
multiplying \eqref{GK_eq} by $\qs\avchi$ and integrating over space has 
a similar outcome. Subtracting the latter integrated equation from the 
former and rearranging terms gives
\bea
{\dd\Is\over\dd t}\equiv
{\dd\over\dd t}{\Ts\over2\fMs}\intRs\lt(\hs - {\qs\avchi\over\Ts}\,\fMs\rt)^2 
= \qs\vpar\intRs\avchi{\dd\hs\over\dd z}
+ {\Ts\over\fMs}\intRs\lt(\hs - {\qs\avchi\over\Ts}\,\fMs\rt)\dtcolls.
\label{Is_def}
\eea
We see that in a purely 2D situation, when $\dd/\dd z=0$, we have an infinite 
family of invariants $\Is=\Is(\vperp,\vpar)$ whose conservation 
(for each species and for every value of $\vperp$ and $\vpar$!) is broken 
only by collisions. In 3D, the parallel particle streaming (propagation) term 
in the gyrokinetic equation generally breaks these invariants, although 
special cases may arise in which the first term on the right-hand side 
of \eqref{Is_def} vanishes and a genuine 3D invariant appears. 

\subsection{``$\Apar^2$-Stuff''}
\label{ap_Aparsq}

Let apply the mass-ratio expansion (\secref{IEF_ordering}) 
to \eqref{Is_def} for electrons. Using the solution \exref{hezero_formula} for 
the electron distribution function, we find 
\bea
\nonumber
{\dd\Ie\over\dd t} &=& {\dd\over\dd t}{\Te\fMe\over2}
\intr\lt({\dne\over\ne} - {e\over\Te}{\vpar\Apar\over c} 
- {\vperp^2\over\vthe^2}\,{\dBpar\over B_0}\rt)^2
= {\dd\over\dd t}\lt[
{e^2\vpar^2\over c^2}{\fMe\over\Te}\intr{\Apar^2\over2}
- {e\vpar\over c}\,\fMe\intr\Apar\lt({\dne\over\ne}-{\vperp^2\over\vthe^2}{\dBpar\over B_0}\rt) 
+ \cdots\rt]\\ 
&=&-e\vpar\intr\lt[\lt(\ephi - {\vpar\Apar\over c} 
- {\Te\over e}{\vperp^2\over\vthe^2}\,{\dBpar\over B_0}\rt)
{\dd\over\dd z}\lt({\dne\over\ne} - {e\ephi\over\Te}\rt)\fMe 
- {\vpar\Apar\over c}{\dd\heone\over\dd z}\rt]
-{e\vpar\over c}\intr\Apar\dtcolle,
\label{Ie_IEF}
\eea
where we have kept terms to two leading orders in the expansion. 
To lowest order, the above equation reduces to
\bea
\label{Apar_stuff}
{d\over dt}\intr{\Apar^2\over2} = 
c\intr\Apar{\dd\over\dd z}\lt({\Te\over e}{\dne\over\ne} - \ephi\rt).
\eea 
This equation can also be obtained directly from \eqref{Apar_eq_sum} 
(multiply by $\Apar$ and integrate). In 2D, it expresses a well known 
conservation law of the ``$\Apar^2$-stuff.'' As this 2D invariant 
exists already on the level of the mass-ratio expansion of the electron 
kinetics, with no assumptions about the ions, it is inherited both 
by the RMHD equations in the limit of $\kperp\rho_i\ll1$ (\secref{sec_AW}) and 
by the ERMHD equations in the limit of $\kperp\rho_i\gg1$ (\secref{sec_ERMHD_eqns}). 
In the former limit, $\dne/\ne$ on the right-hand side of \eqref{Apar_stuff} 
is negligible (under the ordering explained in \secref{sec_sub_order}); 
in the latter limit, it is expressed in terms of $\ephi$ via \eqref{EMHD_dne}. 
The conservation of ``$\Apar^2$-stuff'' is a uniquely 2D feature, 
broken by the parallel propagation term in 3D. 

\subsection{Magnetic Helicity in the Electron Fluid}
\label{ap_hel_els}

If we now divide \eqref{Ie_IEF} through by $e\vpar/c$ and integrate over 
velocities, we get, after some integrations by parts, 
another relation that becomes a conservation law in 2D and 
that can also easily be derived directly from the 
equations of the isothermal electron fluid \exsdash{Apar_eq_sum}{dne_eq_sum}: 
\bea
{d\over dt}\intr\Apar\lt({\dne\over\ne}-{\dBpar\over B_0}\rt)
= -\,c\intr\lt[{\dne\over\ne}{\dd\ephi\over\dd z} + 
{\dBpar\over B_0}{\dd\over\dd z}\lt({\Te\over e}{\dne\over\ne}-\ephi\rt) 
+ \Apar\,{\dd\upare\over\dd z}\rt].
\label{IEF_helicity}
\eea
In the ERMHD limit $\kperp\rho_i\gg1$ (\secref{sec_ERMHD_eqns}), 
we use \eqsdash{EMHD_dne}{EMHD_dBpar} to simplify the above equation 
and find that the integral on the right-hand side vanishes and we get 
a genuine 3D conservation law:
\bea
{d\over dt}\intr\Apar\dBpar = 0.
\eea 
This can also be derived directly from the ERMHD 
equations \exsdash{EMHD_Psi}{EMHD_Phi} [using \eqref{EMHD_dBpar}]. 
The conserved quantity is readily seen to be the helicity 
of the perturbed magnetic field: 
\bea
\intr\vA\cdot\dvB = \intr\lt[\vAperp\cdot\lt(\vdperp\times\Apar\vz\rt) + \Apar\dBpar\rt]
= \intr\lt[\Apar\vz\cdot\lt(\vdperp\times\vAperp\rt) + \Apar\dBpar\rt] 
= 2\intr\Apar\dBpar.
\eea

\subsection{Magnetic Helicity in the RMHD Limit}
\label{ap_hel_RMHD}

Unlike in the case of ERMHD, the helicity of the perturbed magnetic field 
in RMHD is conserved only in 2D. This is because the induction equation 
for the perturbed field has an inhomogeneous term associated with the mean field
[\eqref{MHD_B} with $\vB=B_0\vz+\dvB$] \citep[this issue has been extensively discussed in the 
literature; see][]{Matthaeus_Goldstein,Stribling_Matthaeus_Ghosh,Berger,Montgomery_Bates,Brandenburg_Matthaeus}. 
Directly from the induction equation or from its RMHD descendants \eqsand{RMHD_Psi}{eq_Bpar}, 
we obtain [note the definitions \exref{Phi_Psi_def2}]
\bea
{d\over dt}\intr\Apar\dBpar = 
\intr\lt(c\ephi\,{\dd\dBpar\over\dd z} + {B_0\Apar\over1+v_A^2/c_s^2}{\dd\upar\over\dd z}\rt),
\label{RMHD_helicity}
\eea
so helicity is conserved only if $\dd/\dd z=0$. 

For completeness, let us now show that this 2D conservation law is 
a particular case of \eqref{Is_def} for ions. 
Let us consider the inertial range ($\kperp\rho_i\ll1$). 
We substitute \eqref{g_ansatz} into \eqref{Is_def} for ions 
and expand to two leading orders in $\kperp\rho_i$ using 
the ordering explained in \secref{sec_sub_order}:  
\bea
\nonumber
{\dd\Ii\over\dd t} &=& {\dd\over\dd t}{\Ti\over2\fMi}\intRi
\lt(\gi + {Ze\over\Ti}{\vpar\avApari\over c}\,\fMi\rt)^2 
= {\dd\over\dd t}\lt({Z^2e^2\vpar^2\over c^2}{\fMi\over\Ti}\intr{\Apar^2\over2} 
+ {Ze\vpar\over c}\intr\Apar\gi + \cdots\rt)\\
&=& 
-{Z^2e^2\vpar^2\over c}{\fMi\over\Ti}\intr\Apar
{\dd\over\dd z}\lt(\ephi+{\Ti\over Ze}{\vperp^2\over\vthi^2}{\dBpar\over B_0}\rt)
+ Ze\vpar\intr\lt(\ephi - {\vpar\Apar\over c}\rt){\dd\gi\over\dd z} 
+ {Ze\vpar\over c}\intr\Apar\dtcolli.
\label{Ii_g}
\eea
The lowest-order terms in the above equations (all proportional to 
$\vpar^2\fMi$) simply reproduce the 2D conservation of ``$\Apar^2$-stuff,'' 
given by \eqref{Apar_stuff}. 
We now subtract \eqref{Apar_stuff} multiplied by $(Ze\vpar/c)^2\fMi/\Ti$ 
from \eqref{Ii_g}. This leaves us with 
\bea
{\dd\over\dd t}\intr\Apar\gi 
= c\intr\lt(\ephi-{\vpar\Apar\over c}\rt){\dd\gi\over\dd z}
+\vpar\fMi\intr\lt({Z\over\tau}{\dne\over\ne}
+{\vperp^2\over\vthi^2}{\dBpar\over B_0}\rt){\dd\Apar\over\dd z}
+\intr\Apar\dtcolli.
\label{KRMHD_helicity}
\eea
This equation is a general 2D conservation law of the KRMHD 
equations (see \secref{sec_KRMHD_sum}) and can also be derived 
directly from them. If we integrate it over velocities 
and use \eqsand{sw_n}{sw_upar}, we simply recover \eqref{IEF_helicity}. 
However, since \eqref{KRMHD_helicity} holds for every value of 
$\vpar$ and $\vperp$, it carries much more information than \eqref{IEF_helicity}. 

To make connection to MHD, let us consider the fluid (collisional) limit of KRMHD 
worked out in \apref{ap_visc}. The distribution function to lowest order in the 
$\kpar\mfp\ll1$ expansion is $g=-(\vperp^2/\vthi^2)\dBpar/B_0 + \dfzero$, 
where $\dfzero$ is the perturbed Maxwellian given by \eqref{dfzero_expr}. 
We can substitute this expression into \eqref{KRMHD_helicity}. Since in this expansion 
the collision integral is applied to $\dfone$ and is the same order as the 
rest of the terms (see \secref{ap_transport}), conservation laws are best derived 
by taking $1$, $\vpar$, and $v^2/\vthi^2$ moments of \eqref{KRMHD_helicity} 
so as to make the collision term vanish. In particular, 
multiplying \eqref{KRMHD_helicity} by $1+(2\tau/3Z)v^2/\vthi^2$, integrating 
over velocities and using \eqsand{dTi_eq}{moments}, we obtain the evolution equation 
for $\intr\Apar\dBpar$, which coincides with \eqref{RMHD_helicity}. 
Note that, either proceeding in an analogous way, one can derive similar equations  
for $\intr\Apar\dne$ and $\intr\Apar\upar$---these are also 2D invariants of the 
RMHD system, broken in 3D by the presence of the propagation terms. 
The same result can be derived directly from the evolution 
equations \exsand{mom_eq}{den_eq}. 

\subsection{Electrostatic Invariant}
\label{ap_inv_ES}

Interestingly, the existence of the general 2D invariants introduced in 
\secref{ap_inv_gen} alongside the generalized energy invariant given by \eqref{W_cons} 
means that one can construct a 2D invariant of gyrokinetics that 
does not involve any velocity-space quantities. 
In order to do that, one must integrate \eqref{Is_def} 
over velocities, sum over species, and subtract \eqref{W_cons} 
from the resulting equation (thus removing the $\hs^2$ integrals). 
The result is not particularly edifying 
in the general case, but it takes a simple form if one considers 
electrostatic perturbations ($\dvB=0$). In this case, $\chi=\ephi$, and 
the manipulations described above lead to the following equation
\beq
{dY\over dt}\equiv
{d\over dt}\lt(\sum_s\intv\,\Is - W\rt) 
= -{d\over dt}\sum_s\sum_\vk{\qs^2\ns\over2\Ts}\bl[1-\Gamma_0(\krsq_s)\br]|\ephi_\vk|^2
= \intr\Epar\jpar -\sum_s\qs\intv\intRs\lt\<\ephi\rt>_{\vR_s}\dtcolls,
\label{Y_def}
\eeq
where $\Epar=-{\dd\ephi/\dd z}$, 
$\krsq_s=\kperp^2\rho_s^2/2$ and $\Gamma_0$ is defined by \eqref{G0_def}. 
In 2D, $\Epar=0$ and the above equation expresses a conservation law 
broken only by collisions. The complete derivation and analysis 
of 2D conservation properties of gyrokinetics in the electrostatic limit, including 
the invariant \exref{Y_def}, the electrostatic version of \eqref{Is_def}, and their 
consequences for scalings and cascades, was given by \citet{Plunk_etal}. 
Here we briefly consider a few relevant limits. 

For $\kperp\rho_i\ll1$, we have $\Gamma_0(\krsq)= 1-\krsq_s + \dots$, 
so the invariant given by \eqref{Y_def} is simply the kinetic energy of the $\vE\times\vB$ flows:
$Y=\sum_s (m_s\ns/2)\intr|\vdperp\Phi|^2$, where $\Phi=c\ephi/B_0$. 
In the limit $\kperp\rho_i\gg1$, $\kperp\rho_e\ll1$, we have $Y=-\ni\intr Z^2e^2\ephi^2/2\Ti$. 
In the limit $\kperp\rho_e\gg1$, we have $Y=-(1+Z/\tau)\ne\intr e^2\ephi^2/2\Te$. 
Whereas we are not interested in electrostatic fluctuations in the inertial 
range, electrostatic turbulence in the dissipation range was discussed 
in \secsand{sec_ent_no_KAW}{sec_ent_els}. The electrostatic 2D invariant 
in the limits $\kperp\rho_i\gg1$, $\kperp\rho_e\ll1$ and $\kperp\rho_e\gg1$ 
can also be derived directly from the equations given there 
[in the former limit, use \eqref{Apar_no_KAW} to express 
$\upari$ in terms of $\jpar$ in order to get \eqref{Y_def}]. 

Note that, taken separately and integrated over velocities, \eqref{Is_def} for ions 
(when $\kperp\rho_i\gg1$, $\kperp\rho_e\ll1$) and for electrons (when $\kperp\rho_e\gg1$), 
reduces to lowest order to the statement of 3D conservation 
of $\intv\intRi\Ti\hi^2/2\fMi$ [$\Whi$ in \eqref{W_ERMHD}]
and $\intv\intRe\Te\he^2/2\fMe$ [\eqref{W_last}], respectively. 

\subsection{Implications for Turbulent Cascades and Scalings} 
\label{ap_inv_scalings}

Since invariants other than the generalized energy or its constituent parts 
are present in 2D and, in some limits, also in 3D, one might ask how their presence 
affects the turbulent cascades and scalings. As an example, let us consider 
the magnetic helicity in KAW turbulence, which is a 3D invariant of the 
ERMHD equations (\secref{ap_hel_els}).  

A Kolmogorov-style analysis of a local KAW cascade based on a constant flux of 
helicity gives (proceeding as in \secref{sec_KAW_turb}):
\bea
{\Psil\Phil\over\tKAW}
\sim\sqrt{1+\beta_i}\,{\lambda\over\rho_i}{\Phil^2\over\tKAW}
\sim\sqrt{1+\beta_i}\,{\Phil^3\over\rho_i\lambda}\sim\epshel=\const
\quad\Rightarrow\quad
\Phil\sim {\epshel\over(1+\beta_i)^{1/6}}\,\rho_i^{1/3}\lambda^{1/3},
\label{KAW_hel_scaling}
\eea 
where $\epshel$ is the helicity flux 
(omitting constant dimensional factors, the helicity is now defined as $\intr\Psi\Phi$
and assumed to be non-zero). 
This corresponds to a $\kperp^{-5/3}$ spectrum of magnetic energy. 

In order to decide whether we expect the scalings to be determined by 
the constant-helicity flux or by the constant-energy flux (as assumed in \secref{sec_KAW_turb}), 
we adapt a standard argument originally 
due to \citet{Fjortoft}. If the helicity flux of the KAW turbulence originating at the 
ion gyroscale (via partial conversion from the inertial-range turbulence; see \secref{sec_ERMHD})
is $\epshel$, its energy flux is $\epsB\sim\epshel$ 
[set $\lambda=\rho_i$ in \eqref{KAW_hel_scaling} and compare with \eqref{const_flux_KAW}]. 
If the cascade between the ion and electron gyroscales is 
controlled by maintaining a constant flux of helicity, 
then the helicity flux arriving to the electron gyroscale 
is still $\epshel$, while the associated energy flux is 
$\epshel\rho_i/\rho_e\gg\epsB$, i.e., 
more energy arrives to $\rho_e$ than there was at $\rho_i$! 
This is clearly impossible in a stationary state. 
The way to resolve this contradiction is to conclude that 
the helicity cascade is, in fact, inverse (i.e., directed towards larger scales), 
while the energy cascade is direct (to smaller scales). A similar argument based 
on the constancy of the energy flux $\epsB$ then leads 
to the conclusion that the helicity flux arriving to the 
electron gyroscale is $\epsB\rho_e/\rho_i\ll\epshel\sim\epsB$, i.e., 
the helicity indeed does not cascade to smaller scales. 
It does not, in fact, cascade to large scales either because 
the ERMHD equations are not valid above the ion gyroscale and 
the helicity of the perturbed magnetic field in the inertial range 
is not a 3D invariant (\secref{ap_hel_RMHD}). 
The situation would be different if an energy source existed 
either at the electron gyroscale or somewhere in between 
$\rho_e$ and $\rho_i$. In such a case, one would expect an 
inverse helicity cascade and the consequent shallower 
scaling [\eqref{KAW_hel_scaling}] between the energy-injection 
scale and the ion gyroscale. 

Other invariants introduced above can in a similar fashion 
be argued to give rise to inverse cascades in the hypothetical 
2D situations where they are valid and provided there is energy injection 
at small scales (for the electrostatic case, see \citealt{Plunk_etal}
and numerical simulations by \citealt{Tatsuno_etal2}). 
The view of turbulence advanced in this paper does 
not generally allow for this to happen. First, the fundamentally 3D nature 
of the turbulence is imposed via the critical balance conjecture 
and supported by the argument that ``two dimensionality'' can only 
be maintained across parallel distances that do not exceed the distance 
a parallel-propagating wave (or parallel-streaming particles) 
travels over one nonlinear decorrelation time 
(see \secref{sec_GS}, \secref{sec_KAW_turb} and \secref{sec_par_no_KAW}). 
Secondly, the lack of small-scale energy injection was assumed at the outset. 
This can, however, be violated in real astrophysical plasmas 
by various small-scale plasma instabilities (e.g., triggered 
by pressure anisotropies; see discussion in \secref{sec_pressure_aniso}). 
Treatment of such effects falls outside the scope of this paper 
and remains a matter for future work. 

\end{appendix}

\end{document}